\documentclass[a4paper,12pt,leqno,openbib,oldfontcommands]{memoir}
\usepackage{datetime}
\usepackage{pdfpages}
\usepackage{url}
\usepackage{comment}

\usepackage{ifpdf}
\ifpdf
\fi
\usepackage{datetime}
\usepackage{ifpdf}
\ifpdf
\fi
\ifdraftdoc 
	\usepackage{draftwatermark}				
	\SetWatermarkScale{0.3}
	\SetWatermarkText{\bf Draft:\today}
\fi
\settrimmedsize{297mm}{210mm}{*}
\settypeblocksize{634pt}{448.13pt}{*} 
\setulmargins{4cm}{*}{*} 
\setlrmargins{*}{*}{1.5} 

\setmarginnotes{17pt}{51pt}{\onelineskip} 
\setheadfoot{\onelineskip}{2\onelineskip} 
\setheaderspaces{*}{2\onelineskip}{*} 
\checkandfixthelayout
\usepackage{fouriernc}
\usepackage[T1]{fontenc}
\OnehalfSpacing 
\setsecnumdepth{subsection} 
\maxsecnumdepth{subsubsection}
\usepackage{calc,soul}
\makeatletter 
\newlength\dlf@normtxtw 
\newsavebox{\feline@chapter} 
\newcommand\feline@chapter@marker[1][4cm]{%
	\sbox\feline@chapter{%
		\resizebox{!}{#1}{\fboxsep=1pt%
			\colorbox{gray}{\color{white}\thechapter}%
		}}%
		\rotatebox{90}{%
			\resizebox{%
				\heightof{\usebox{\feline@chapter}}+\depthof{\usebox{\feline@chapter}}}%
			{!}{\scshape\so\@chapapp}}\quad%
		\raisebox{\depthof{\usebox{\feline@chapter}}}{\usebox{\feline@chapter}}%
} 
\newcommand\feline@chm[1][4cm]{%
	\sbox\feline@chapter{\feline@chapter@marker[#1]}%
	\makebox[0pt][c]{
		\makebox[1cm][r]{\usebox\feline@chapter}%
	}}
\makechapterstyle{daleifmodif}{

	\renewcommand\printchapternum{\null\hfill\feline@chm[2.5cm]\par}

} 
\makeatother 
\chapterstyle{daleifmodif}
\makepagestyle{myvf} 
\makeoddfoot{myvf}{}{\thepage}{} 
\makeevenfoot{myvf}{}{\thepage}{} 
\makeheadrule{myvf}{\textwidth}{\normalrulethickness} 
\makeevenhead{myvf}{\small\textsc{\leftmark}}{}{} 
\makeoddhead{myvf}{}{}{\small\textsc{\rightmark}}
\newcommand{\clearemptydoublepage}{\newpage{\thispagestyle{empty}\cleardoublepage}}
\makeindex
\usepackage{import}

\usepackage{lipsum}					
\usepackage{amsfonts} 					
\usepackage[centertags]{amsmath}			
\usepackage{stmaryrd}					
\usepackage{amssymb}					
\usepackage{amsthm}					
\usepackage{newlfont}					
\usepackage{graphicx}					

\usepackage{subcaption}
\usepackage{longtable,rotating}			
\usepackage{colortbl}					
\usepackage{wasysym}					
\usepackage{mathrsfs}					
\usepackage{float}						
\usepackage{verbatim}					
\usepackage{upgreek}					
\usepackage{latexsym}					
\usepackage[square,numbers,
		     sort&compress]{natbib}		
\usepackage{url}						
\usepackage[spanish,english]{babel}		
\usepackage{color}                    				
\usepackage[pdfencoding=auto,
            colorlinks=true,
		     allcolors=black]{hyperref}              
\usepackage{memhfixc}					
\usepackage{enumerate}					
\usepackage{footnote}					
\usepackage{microtype}					
\usepackage{rotfloat}					
\usepackage{alltt}						
\usepackage{tikz}						

\usepackage{graphicx}\usepackage{color}
\usepackage[utf8]{inputenc}
\usepackage{graphicx}
\usepackage{booktabs}
\usepackage{enumerate}
\usepackage{balance}
\usepackage{flushend}
\usepackage{soul}
\usepackage{multirow}
\usepackage{algorithm}
\usepackage[noend]{algpseudocode}
\usepackage{color}
\usepackage{listings}
\definecolor{codeblack}{rgb}{0.0,0.0,0.0}
\definecolor{codegray}{rgb}{0.5,0.5,0.5}
\definecolor{backcolour}{rgb}{0.95,0.95,0.92}
\definecolor{codegreen}{rgb}{0,0.6,0}
\definecolor{codepurple}{rgb}{0.58,0,0.82}
\definecolor{codeblue}{rgb}{0,0,0.82}
\lstdefinestyle{mystyle}{
    backgroundcolor=\color{backcolour},   
    commentstyle=\color{codegreen},
    keywordstyle=\color{blue},
    numberstyle=\tiny\color{codegray},
    stringstyle=\color{codepurple},
    basicstyle=\footnotesize,
    breakatwhitespace=false,         
    breaklines=true,                 
    captionpos=b,                    
    keepspaces=true,                 
    numbers=left,                    
    numbersep=5pt,                  
    showspaces=false,                
    showstringspaces=false,
    showtabs=false,                  
    tabsize=2
}
\usepackage{array}
\newcolumntype{P}[1]{>{\centering\arraybackslash}p{#1}}
\usepackage{amssymb}

 \usepackage{amsmath}

\usepackage{dblfloatfix}

\usepackage{todonotes}
\usepackage{easyReview}
\setreviewson

\usetikzlibrary{arrows,shapes,decorations,
		       automata,backgrounds,
		       petri,topaths}				
\newcommand{\pgftextcircled}[1]{                                                                    
    \setbox0=\hbox{#1}%
    \dimen0\wd0%
    \divide\dimen0 by 2%
    \begin{tikzpicture}[baseline=(a.base)]%
        \useasboundingbox (-\the\dimen0,0pt) rectangle (\the\dimen0,1pt);
        \node[circle,draw,outer sep=0pt,inner sep=0.1ex] (a) {#1};
    \end{tikzpicture}
}
\newcommand{\blackged}{\hfill$\blacksquare$}
\newcommand{\whiteged}{\hfill$\square$}
\newcounter{proofcount}

\let\oldsqrt\sqrt
\def\sqrt{\mathpalette\DHLhksqrt}
\def\DHLhksqrt#1#2{%
\setbox0=\hbox{$#1\oldsqrt{#2\,}$}\dimen0=\ht0
\advance\dimen0-0.2\ht0
\setbox2=\hbox{\vrule height\ht0 depth -\dimen0}%
{\box0\lower0.4pt\box2}}
\newcommand{\mycaption}[2][\@empty]{
	\captionnamefont{\scshape} 
	\changecaptionwidth
	\captionwidth{0.9\linewidth}
	\captiondelim{.\:} 
	\indentcaption{0.75cm}
	\captionstyle[\centering]{}
	\setlength{\belowcaptionskip}{10pt}
	\ifx \@empty#1 \caption{#2}\else \caption[#1]{#2}
}
\DeclareMathOperator*{\argmax}{argmax}
\usepackage{lettrine}
\newcommand{\initial}[1]{%
	\lettrine[lines=3,lhang=0.33,nindent=0em]{
		\color{gray}
     		{\textsc{#1}}}{}}
\theoremstyle{plain}

\theoremstyle{plain}

\theoremstyle{plain}
\theoremstyle{definition}

\theoremstyle{plain}

\theoremstyle{plain}

\theoremstyle{plain}

\theoremstyle{plain}
   
\begin{document}
%
%
%
%
\frontmatter
\pagenumbering{roman}
%
	\begin{center}
		{
		\vspace{2 cm}
			{\Large \bf Prediction of Performance and Power Consumption of GPGPU Applications} \\[0.5cm] 
			\textsc{\Large \bfseries THESIS}\\ \vspace{0.5cm} 
			{\large  submitted in partial fulfillment \\ of the requirements for the degree of
			}\\~\\ 
			\textbf{\large \bfseries DOCTOR OF PHILOSOPHY}\\ \vspace{0.5cm}
			{\Large by} \\ \vspace{0.5cm}
			\textbf{\large \bfseries ALAVANI GARGI KABIRDAS}\\ \vspace{1cm}
			{\large  under the supervision of	
			}\\ \vspace{0.25cm}
		\textbf{\large \bfseries Prof. SANTONU SARKAR }\\ \vspace{0.5cm}
		{\large and co-supervision of }\\ \vspace{0.25cm}
			\textbf{\large \bfseries Prof. NEENA GOVEAS }\\ \vspace{0.5cm}
		\vspace{2 cm}
			\begin{figure}[htbp]
				\centering		
				\def\svgwidth{100\unitlength}
			\includegraphics[scale=0.1]{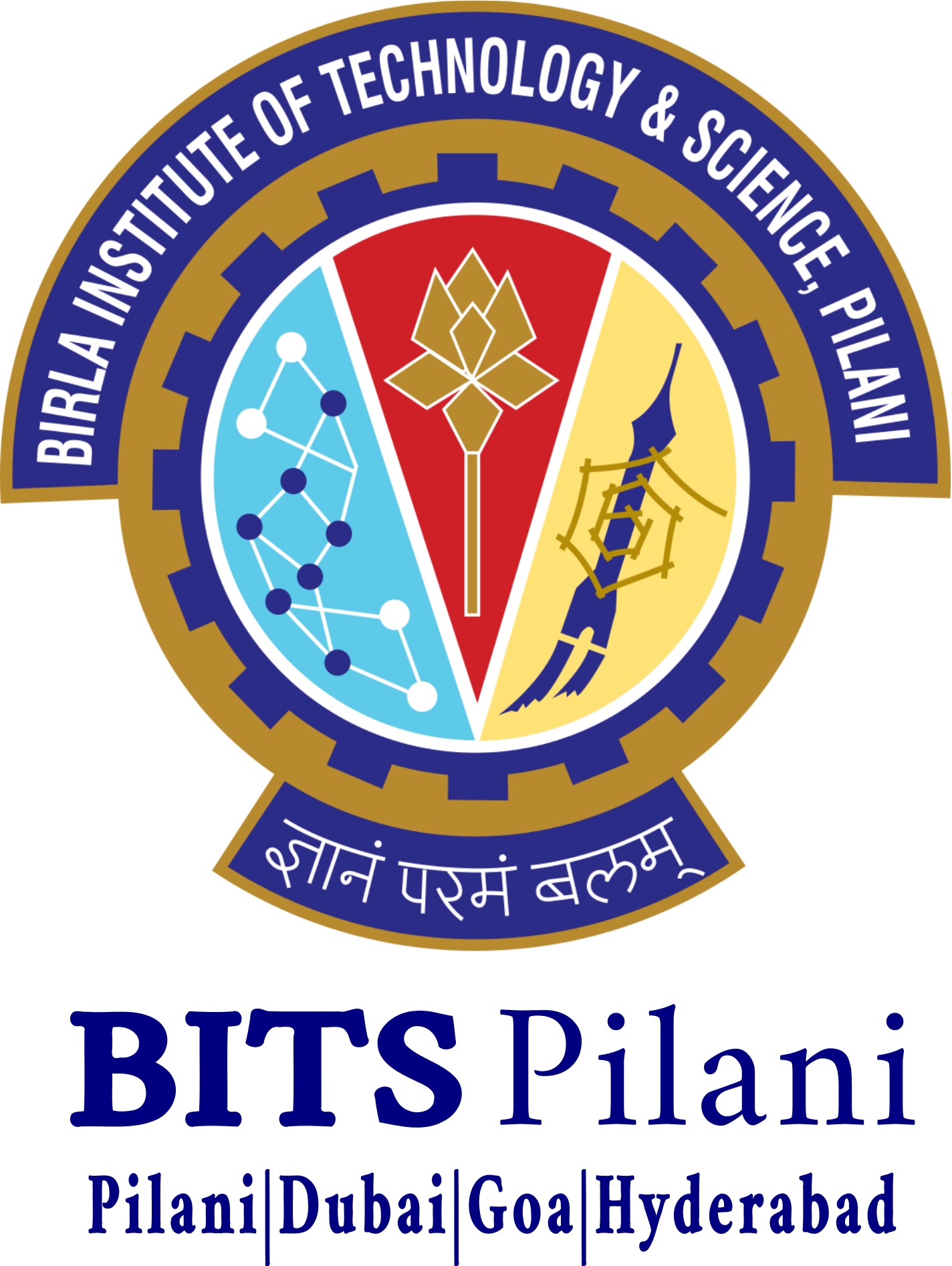}
			\end{figure}
			\textsc{\large \bfseries BIRLA INSTITUTE OF TECHNOLOGY AND SCIENCE, PILANI}\\
			{\large \bfseries 2023 }\\ 
		}
	\end{center}
  \thispagestyle{empty} 
\clearpage

\let\cleardoublepage\clearpage
\let\cleardoublepage\clearpage
\clearemptydoublepage
 \chapter*{Abstract}
\begin{SingleSpace}
\initial{G}raphics Processing Units (GPUs) have become an integral part of High-Performance Computing to achieve an Exascale performance. 
The main goal of application developers of GPU is to tune their code extensively to obtain optimal performance,  making efficient use of different resources available on the GPU. While extracting optimal performance of applications on an HPC infrastructure, developers should also ensure the applications have the least energy usage considering the massive power consumption of data centers and HPC servers. In order to do this, a developer must be equipped with tools to understand the energy consumption of applications before launching them. We firmly believe that understanding and estimating GPU performance and power consumption will aid developers in designing performance-driven as well as energy-aware applications for a given architecture.

This thesis presents two models developed using static and dynamic analysis of CUDA code, which helps developers in analysing the CUDA kernel's energy dissipation. The first one is a model that predicts the CUDA kernel's execution time. Here a PTX code is statically analysed to extract instruction features, control flow, and data dependence. We propose two scheduling algorithm approaches that satisfy the performance and hardware constraints. The first approach predicts the performance constraints by considering Instruction Level Parallelism (ILP) and Thread Level Parallelism (TLP), which are two types of parallelism that GPUs exploit in order to achieve higher performance. The second approach utilizes resource reservation constraints to schedule these instructions in threads across Streaming Multiprocessors (SMs). We use dynamic analysis to build a set of memory access penalty models and use these models in conjunction with the scheduling information to estimate the execution time of the code.

The second model is a static analysis-based power prediction built by utilizing machine learning techniques. Features used for building the model are derived by static analysis of PTX code. These features are chosen to understand the relationship between GPU power consumption and program features that can aid developers in building energy-efficient, sustainable applications. The dataset used for validating both models include kernels from different benchmarks suits, sizes, nature (e.g., compute-bound, memory-bound), and complexity (e.g., control divergence, memory access patterns). Developers can refactor their code to build energy-aware GPU applications by employing these two models. We also present a tool that has practically validated the effectiveness and ease of using the two models as design assistance tools for GPU.

\end{SingleSpace}
\clearpage

\clearemptydoublepage\clearpage
\let\cleardoublepage\clearpage
\let\cleardoublepage\clearpage

\renewcommand{\contentsname}{
Table of Contents}
\maxtocdepth{subsection}
\tableofcontents*
\addtocontents{toc}{\par\nobreak \mbox{}\hfill{\bf Page}\par\nobreak}
\clearemptydoublepage
\listoftables
\addtocontents{lot}{\par\nobreak\textbf{{\scshape Table} \hfill Page}\par\nobreak}
\clearemptydoublepage
\listoffigures
\addtocontents{lof}{\par\nobreak\textbf{{\scshape Figure} \hfill Page}\par\nobreak}
\clearemptydoublepage

\chapter*{List of Abbreviations}
\begin{SingleSpace}
\begin{table} [ht]
	\centering
	\begin{tabular}{|p{1.0 in}||p{4.0 in}|}
		\toprule
		\midrule
		$HPC$ & High Performance Computing\\
		$GPU$ &  Graphics Processing Unit \\
		$GPGPU$ & General Purpose Graphics Processing Unit. By GPU and GPGPU, we mean NVIDIA GPU in this thesis\\
		$CUDA$ & Compute Unified Device Architecture \\
		$SM$ & Streaming Multiprocessor \\
		$PTX$ & Parallel Thread Execution \\
		$SP$ & single precision cores \\
		$ILP$ & Instruction Level Parallelism \\
		$TLP$ & Thread Level Parallelism \\ 
 		$WS$ & warp schedulers \\
		$SFU$ & special function units \\
		$DFU$ & double precision units \\
		$LSU$ & load-store units \\
		$SM$  & Streaming Multiprocessor \\
		$RF$  &  Random Forest \\
	    $SVR$  & Support Vector Regression \\
	    $ANN$ &  Artificial Neural Network \\
	    $XGBoost$ & Extreme Gradient Boosting \\
	    $CatBoost$ & Categorical Boosting \\
	    $Extra Trees$ & Extremely Randomized Trees \\
		\bottomrule
	\end{tabular}
	\label{tab:thesis-abbreviations}
	\vspace{-5pt}
\end{table}
\end{SingleSpace}
\clearpage
\clearemptydoublepage\clearpage
%
\mainmatter
\DoubleSpacing
%

\let \textcircled=\pgftextcircled
\chapter{Introduction}
\label{ch:introduction}
\initial{S}cientific computing in the early 90's needed data parallel computation, due to which the early High Performance Computing (HPC) machine, CM-5, comprising of an array of co-processors, was built. With CM-5, started the idea of offloading vector and matrix computations onto a co-processor array. The HPC programming community during that time was small and highly specialized in hand-crafting programs to exploit the hardware infrastructure with some support from the compiler such as automatic vectorization of some part of the code. The HPC hardware and software infrastructure were never accessible to the general programming community.

The scenario has changed over the last two decades with the introduction of cell processors, GPGPU, or Xeon Phi accelerators. The combination of a large number of scalar processors and the accelerators have been able to achieve Exa-FLOP-level performance under the umbrella of Heterogeneous Computing \footnote{https://www.top500.org/lists/}. Furthermore, these processors and accelerators are now available as commodity hardware. Simultaneously, the scalar CPUs have become multi and many-core.  As a result, the High-Performance Computing (HPC) capability today is not just restricted within the scientific community but became affordable for the programming community at large. Consequently, there have been a shift in computing paradigm from serial to parallel \cite{asanovic2006}. Furthermore, these processors and accelerators are now available as a commodity hardware in the form of Graphics Processing Units (GPUs) . 

GPUs are an integral part of most modern supercomputers, HPC clusters, and servers of different scales as well as a commodity hardware. Three out of five top supercomputers in the world utilize GPUs to achieve exa-flop performance.  Although GPUs were traditionally built for rendering graphical images because each pixel of an image can be rendered by each core since it is a multi-core machine. Over last two decades, this capability of GPU is harnessed to perform compute-intensive tasks which need high data processing parallelism for e.g. Deep learning \cite{Gianniti2018,Lattuada2022}.  Many scientific and industrial applications are written to harness the compute power of GPUs where the embarrassingly parallel parts of an application which run on large data sets are executed by the GPU \cite{Huang2009}. Since the same set of operations are to be performed on each data element independently, each GPU thread is capable of carrying out this task. For e.g. leading automobile company Tesla is using a supercomputer with NVIDIA GPU cores which are used to train deep neural networks for Tesla's Autopilot self-driving system \cite{Tesla}.

Although GPUs are popular for their programmability and high throughput, their energy consumption is a crucial concern \cite{Bridges2016,Mittal2014,Puig2018}. Currently, the world's fastest Supercomputer Fugaku consumes 717576 kWh energy in a day, almost triple its predecessor, Supercomputer Summit (242304 kWh/day) \footnote{https://www.top500.org/lists/}. If every new performance-hungry machine triples its consumption, it will become the biggest hurdle on the path toward sustainable and green computing. Besides, the increase in energy consumption gives rise to temperature management issues, which leads to employing cooling systems, which also contribute to energy consumption and an increase in maintenance cost. Therefore, understanding and predicting GPU energy consumption is the need of the hour. As GPUs evolve with every new architecture, this task becomes more complex and challenging. In this thesis, we make a contribution towards estimating the amount of energy that an NVIDIA GPU application can consume while developing the application. Such a support can be a useful tool to build energy-aware GPU applications, with a broader perspective of sustainable computing. We first discuss the motivation of this work and then delve into its background.  

\section{Motivation}
Energy consumption is a major challenge for HPC systems in their quest toward the holy grail of Exa-FLOP performance \cite{Bridges2016, Mittal2014}. 
In the past, improvements in the energy efficiency of CPUs came about as a byproduct of Moore's law, which accurately predicted that the number of transistors on a CPU would double every two years. However, the pace of improvement in the energy efficiency of CPUs has been slowing down. Because the scaling of CPUs has hit this power wall, the focus is shifting to multi-core and parallel architectures to improve energy efficiency. An important factor in choosing a GPU is that it offers better performance per watt ratio \cite{Mittal2014, Puig2018}, compared to the CPU for high-throughput and high-latency applications. However, if the GPU is not utilised efficiently, the energy consumption of GPUs, which is almost triple the amount of energy consumed by a CPU, is of grave concern \cite{Collange2009}.  

In order to circumvent energy debt \cite{couto2014,Couto2020}, which is the wastage of energy due to poorly written code, it is crucial to understand the energy consumption of applications at the code level. Green Software Foundation released Software Carbon Intensity (SCI), an initial specification for scoring a software system for carbon emissions \cite{SCI}. The purpose of  SCI is to help users and developers make informed choices about carbon emissions from the tools, approaches, architectures, and services used in software development. GSF states that there are three ways to reduce the carbon emissions of software:(a) Use less hardware, (b) Use less energy, and (c) Carbon awareness. Considering the importance of energy consumption from a carbon emission perspective, We must equip GPU developers with tools that can assist them in building energy-saving applications to achieve the goals of sustainable computing \cite{Madougou2016}. An accurate prediction approach for predicting the energy consumption of an application before running it is a significant research problem. Looking at the pace at which parallel programming is being adapted in various application domains \cite{Asanovic2009}, if this problem is not explored now, developers will have to run their programs to compute energy consumed, which is not always feasible, environment-friendly, and economical. Therefore, in this thesis, we attempt to propose a solution to this critical problem of energy prediction. 

Energy prediction of an application can be performed using static  \cite{nielson2004principles,kleimenov2021method} or dynamic analysis \cite{sahni1996performance,zhou2021gpa}. A purely static analysis-based prediction is done by analysing the application code without executing the application whereas dynamic analysis based approach involves the evaluation by running the application on the target hardware. A dynamic analysis-based model can be more accurate since the data collected represents the actual execution of the application most accurately. However, it defeats the purpose of energy prediction since multiple runs of application will lead to energy wastage. Also, a dynamic analysis model which uses special hardware cannot be architecture-agnostic since the hardware may not be compatible with newer architectures \cite{Bridges2016}. Chen and Shi \cite{Chen2012} advocated for software based prediction tools; although these tools provide lower degree of prediction accuracy, they allow better flexibility, specifically when the required specialized hardware is unavailable. 

Considering the architectural complexity of the hardware and the fact that crucial insights are not revealed by the vendors, a purely static analysis-based model is a challenging problem to solve. Architectural complexities involve factors such as GPU's complex internal memory hierarchy and thousands of processors, which lead to numerous impediments for energy modelling. Also, for GPU applications, one can perform static analysis of PTX (Parallel Thread Execution) code, an intermediate representation of CUDA (Compute Unified Device Architecture) kernel. All the instructions in PTX code may or may not be executed on hardware due to hardware level optimizations. 

The advantage of the static analysis-based model is that it can be used to design an energy-aware GPU application. It can assist application developers in using appropriate program features while developing an application so that the GPU consumes less energy when the application is deployed on the GPU. Without any application development support, such fine-tuning is not accessible during implementation. There is a dire need for a reasonably accurate static analysis-based energy prediction model, which can predict energy usage without running the application, and which can be modelled on different GPU hardware characteristics. 

Energy Modelling solutions for GPUs have been proposed using statistical techniques on hardware parameters \cite{Shuaiwen2013,Benedict2016}, simulators \cite{Bakhoda2009,Leng2013}, and instruction-level source code analysis \cite{Wang2011}. We observe that a significantly high amount of effort has been put into GPU application development in comparison to the development of supporting tools for performance analysis. Existing profilers \cite{NvidiaProfile,Purnomo2010}, and simulators \cite{Bakhoda2009,Malhotra2014} show only the execution profile data of an application. Developers are expected to be knowledgeable to analyze this information, understand the application flow, and reason about the performance bottlenecks. These are undoubtedly highly time-consuming and expertise-dependent tasks. 

These observations have motivated us to build an energy prediction model to estimate the energy usage of an NVIDIA CUDA kernel application without the need to execute it. We are proposing a GPU energy model based on two sub-models that, when combined, provide energy estimation. The first model is designed to predict the execution time (performance) of the application. The second model estimates the power consumption by analyzing the code. These models can be integrated in a design assistance tool to help GPU programmers develop energy-aware and energy-efficient applications. The proposed approach should be architecture-agnostic and needs to be extended to newer architectures by tuning some parameters.

Any GPU application developer can utilise the proposed models easily and find their usefulness in multiple domains. For e.g., one can use the performance prediction model in the Cloud-based GPU renting services~\cite{Malik2011}. In such a scenario, a developer, with a reasonable approximation, would like to foresee the required amount of time for which he needs to rent this hardware to plan his budget. 

Although we plan to design a static analysis model for energy consumption, energy dissipation is more intricately related to architecture components. Hence one cannot develop a solution for modeling energy consumption without understanding architectural constructs. Therefore in the next section, we present the background details of GPU architecture which are essential to understand our proposed solution for the performance and power prediction of the CUDA kernel. As mentioned earlier, we are presenting a static analysis based solution; hence one needs to understand the programming model used for GPUs. Therefore we also discuss the required background information on the programming model for GPUs in the next section. 

\section{Background - GPU Architecture \& Programming Model }
GPU works as co-processor to a CPU, where it accelerates the task for general-purpose computing applications. CPU offloads some of the compute-intensive and time consuming portions of the code on the GPU where it uses its massively parallel processing power to boost performance. It is extremely crucial to understand the architecture of a GPU for building an empirical model which captures its execution details. 

\begin{figure}
    \centering
    \includegraphics[width=13cm]{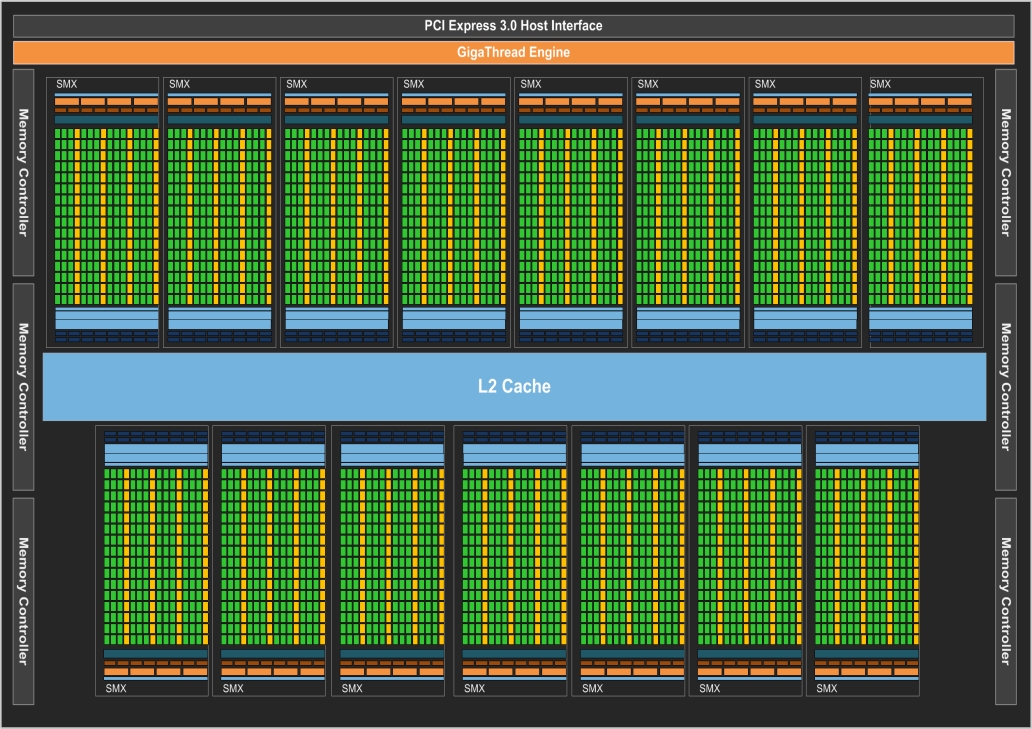}
    \caption{Kepler GK110 architecture}
    \label{fig:keplerarch}
\end{figure}

\begin{figure}[!htb]
    \centering
    \begin{minipage}{.5\textwidth}
        \centering
\includegraphics[width=\linewidth]{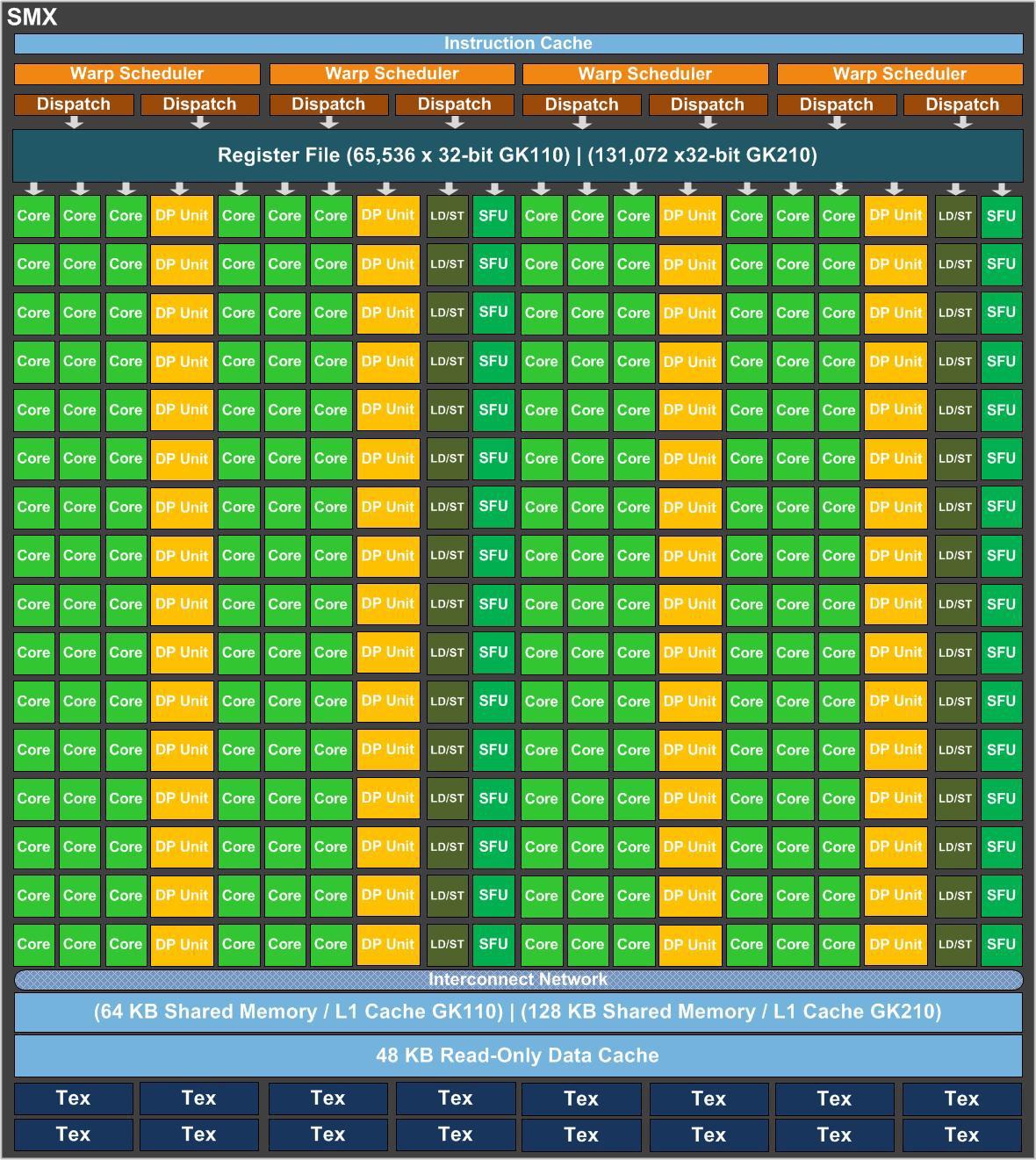}
\caption{Tesla K20 SM Architecture}
\label{fig:sm}
\end{minipage}%
    \begin{minipage}{0.5\textwidth} \centering
\includegraphics[width=\linewidth]{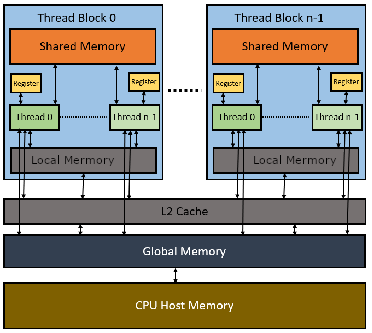}
\caption{GPU Memory Design}
\label{fig_memory_design}
\end{minipage}
\end{figure}

The model proposed in this thesis is architecture agnostic and works across multiple GPU architectures. The GPU which was extensively used for experimentation and validation; an NVIDIA Tesla K20 is described here for understanding the architecture details of a GPU. The NVIDIA Tesla K20 GPU (Figure \ref{fig:keplerarch} ) is based on Kepler GK110 architecture with compute capability of 3.5 and is composed of 13 computational components which are called streaming multiprocessors (SMs). Each SM consists of fully pipelined
 192 single-precision CUDA cores, 64 Double Precision units (DPUs), 32  special function units (SFUs) \footnote{https://www.nvidia.com/content/dam/en-zz/Solutions/Data-Center/tesla-product-literature/NVIDIA-Kepler-GK110-GK210-Architecture-Whitepaper.pdf}. The SM schedules threads in groups of 32 parallel threads called warps. Each SM features four warp schedulers and eight instruction dispatch units (Figure \ref{fig:sm}) , allowing four warps to be issued and executed concurrently. Kepler’s quad warp scheduler selects four warps, and two independent instructions per warp can be dispatched each cycle.   
 
GPU architecture offers multiple memory options such as global, shared, constant, and texture which developer can utilize as per applications demand. The constant and texture are read only memory spaces available on GPU.  Since we focus on General Purpose GPU Computing (GPGPU), we consider global and shared memory in our model because constant and texture memory is more concerned with graphics applications.  We have illustrated the memory design for a CUDA thread in Figure \ref{fig_memory_design}. The CUDA cores have access to a private set of registers allocated from a register file. The shared memory is assigned to a block, and global memory space is within a grid.  Hence, threads within a block can share data using shared memory, but to access data across blocks, the data must be accessed from global memory. L1 cache co-exists with shared memory and Global memory accesses are always cached in L2.

\subsection{CUDA Programming Model}
NVIDIA GPUs can be programmed using CUDA (Compute Unified Device Architecture) for GPGPU applications. CUDA is an extension to standard ANSI C where the part of the code that's supposed to be executed by thousands of threads in the GPU is expressed as a {\em kernel} function. Although CUDA is an extension to ANSI C, it offers very  low level constructs which are related to architecture components of GPU. For e.g. thread level operations, choice of different types of memory while programming. Through these features CUDA offers a direct extension to GPU architecture. In the past studies, strong correlation is observed between CUDA constucts and energy consumption for constructs such as global memory instructions \cite{Hong2010} , number of blocks \cite{ikram2022investigating,boughzala2020predicting}  etc. Therefore, in this section, we discuss only those features of CUDA programming model which will be utilised which are linked to hardware and are utilised in building the proposed model.

The {\em kernel} function which executes on a GPU is declared using the \_\_global\_\_ keyword. All the instructions to be executed on GPU are incorporated in this function. The kernel is launched using a hierarchy of thread groups: grid, block, and thread. A grid is composed of a set of thread blocks that executes a kernel function, and each block is composed of hundreds of threads. 

CUDA provides a device abstraction by defining a logical architecture of the device, which remains unchanged even when the GPU is changed~\cite{CUDA2017}. CUDA also provides flexibility in assigning resources. For instance, the programmer can use a thread block, which shares their data through shared memory and synchronize their actions by utilizing built-in primitives. The CUDA programming model assumes a Single Instruction Multiple Threads (SIMT) computation where instructions are executed in a lockstep fashion. The threads are synchronized with a function \verb+sync_threads()+, which is a barrier for threads. Furthermore, the programmer has the design option to share the registers among a group of threads.

CUDA provides an abstraction in memory usage by providing keywords such as {\tt global, local} and {\tt shared}. In addition, there are two programming constructs- namely "{\tt constant}" and "{\tt texture}" for two particular memory types. The "constant" memory is visible to all threads and stores data that will not be changed over the course of kernel execution.

The next important aspect is to investigate how a CUDA kernel is executed in a GPU. This is relevant for our work since we develop a scheduling algorithm in this thesis which tries to imitate instruction scheduling of a GPU. 

\subsection{Mapping of CUDA Programming model to the GPU Architecture}
The host program specifies the parameters for each kernel invocation at runtime. The required parameters represent the grid and block dimensions. Optionally the user can also specify how much shared memory
is available per-block. Threads within the same block are assigned unique identifiers based on their coordinates. In a similar manner, the blocks which form the grid are assigned identifiers based on their coordinates. A thread is uniquely identified in a grid by combining its thread ID with the ID of the block to which it belongs. Threads within a block can share data through shared memory and can perform barrier synchronization using special CUDA API calls. CUDA threads can access data from different memory spaces as illustrated in Figure \ref{fig_memory_design}. 

The CUDA execution model achieves {\em throughput} by assigning thread blocks of a kernel to SMs. When a kernel is launched, the blocks of the grid are distributed to the SMs in parallel based on the hardware limit.  The registers are allocated from the register file which is partitioned among the warps, and shared memory is allocated for each of the thread blocks. Since these resources are limited, there is a hardware limit on number of blocks and warps that can be executed together on a SM. 

A very important parameter in GPU execution is {\em Warp} which is a group of 32 thread for scheduling threads on GPU architecture. SM schedules instruction at warp granularity. A Warp Scheduler Unit on GPU schedules warps for execution, and Dispatch Unit issues the same instruction for all the warp’s threads. Threads may have diverged path because of control-flow. In such a scenario, all the divergent paths are executed serially and the path is disabled for inactive threads. 

With this background information on GPU architecture and CUDA programming model we now move to discussing how the energy modelling of GPU is proposed in this thesis. %

\subsection{Energy Modelling of GPU}

Energy modelling of a machine involves measuring its performance in execution time and the power consumed during its runtime.  Hence  one can estimate energy consumption of a kernel based on two sub-models which independently compute the power level and execution time. We can combine the results obtained from the performance model and the power-occupancy model to estimate the energy consumption of GPU.

$$E = P \times t_{kernel}$$

where $E$ is the energy estimate of the kernel in Joules, $P$ is the estimated average power level during the execution of the kernel in Watts, and $t_{kernel}$ is the execution time of the kernel in seconds.

In order to calculate the energy consumption, it must be noted that there may be several sources of error in the computation model, which involve a very deep knowledge of the NVIDIA infrastructure. NVIDIA's GPU architecture is proprietary information and cannot be easily found, except through several experiments conducted on the GPUs and rigorous microbenchmarking \cite{Wong2010}. Because hardware implementation details are not publicly available, calculating the execution time and power consumption of a kernel is an ambitious and challenging task.

As seen in a K20 GPU power measurement paper by Burtscher et al. \cite{k20builtinsensor}, the actual power level of the GPU during kernel execution was found to be mostly constant. This is because the components of the GPU's SMs were in active state. If the components are idle, the power level drops to a much lower value. The power level of the GPU needs to be estimated accurately. For this, the possible approaches must be analyzed to find the one that works best. Possible approaches include a regression-based approach to predict the power, and statistical estimates based on average power level over several past usages of the GPU.

With respect to execution time prediction model, Amaris et al. \cite{Amaris2016} compared results of analytical and machine learning models and concluded that analytical models are more precise than the other approaches. In this work, we built analytical model for execution time prediction by creating a mathematical abstraction of the program’s execution. The computation is performed in the form of a function which takes into account the target architecture, the input program and the launch parameters. With this background, we now discuss the problem statement of our work in further detail.

\section{Problem Statement}
Based on our plan to model static analysis-based energy prediction using performance and power estimation, we present following objectives that we fulfill in this thesis. 
\subsection{Objective 1} We aim to design an analytical model to predict execution time of CUDA kernel. For this, we plan the following activities
\begin{enumerate}
    \item \textit{ Identifying microbenchmarks for computing intrinsic details of GPU hardware}
    
Instruction execution is governed by some critical parameters such as latency, througput etc. Very little information about these parameters (such as theoretical peak performance) are provided by the vendor \cite{CUDA2017}. These details may not precisely match the real-time metrics due to hardware complexities and hence we have to recompute it by microbenchmarking the hardware. 
\item \textit{Creating an accurate GPU execution model}

Due to the increasing popularity of GPUs in main stream computing, it is becoming exceedingly important to analyze and understand GPU application performance. We observe that a significantly high amount of effort has been put into GPU application development compared to supporting tools for performance analysis. Existing profilers \cite{NvidiaProfile,Purnomo2010} and simulators\cite{Bakhoda2009,Malhotra2014} show only the execution profile data of an application. Developers are expected to be knowledgeable to analyze this information, understand the application flow, and reason about the performance bottlenecks. These are undoubtedly highly time-consuming and expertise-dependent tasks. 

The execution time of a CUDA program, besides data dependency, is dependent on multiple hardware features such as memory and cache usage, branch prediction, pipeline stalls, instruction latency, and others. Predicting the execution time for a particular GPU architecture needs a profound understanding of the architectural details, which NVIDIA does not disclose in the public domain. In the absence of this knowledge, modeling various architectural characteristics of a GPU such as scheduling of instructions, estimating latency of every instruction, estimating memory throughput, and modeling launch overhead without any architectural details from the vendor is quite challenging.

We need a performance prediction model tool that can
\begin{itemize}
    \item Estimate the execution time of a CUDA kernel without having to run the kernel on an NVIDIA GPU
    \item Solution can involve extensive static analysis of code and if dynamic analysis is required it should be one time collection of data.
    \item The approach can be attuned to gain insights into program design issues to help the developer.
\end{itemize}
\end{enumerate}
\subsection{Objective 2} 
We aim to build a prediction model for modelling GPU power consumption of CUDA kernel. For this, we plan the following activities
\begin{enumerate}
\item \textit{Choosing features which can be generated using static analysis}
Power consumption of a computing device is intricately related to its hardware features. Consequently, there has been a significant body of work to investigate the relationship between hardware features and power consumption~\cite{Fan2019,Guerreiro2018,Lim2015,Lucas2013}. 

However, our goal is to develop a power prediction model based on code features and minimal runtime information so that the model can be used for power consumption prediction without running a CUDA program. To the best of our knowledge, not much work has been done that uses such an approach. Choosing the features which  contribute to power consumption and can be collected without running an application is a difficult task. 
\item \textit{Building a dataset of static analysis based features}
To the best of our knowledge, we did not come across any existing dataset which collects program analysis data and power consumption. Hence we have to build an in-house dataset for the same. To create the dataset we need to run CUDA benchmarks with different launch configurations (grid size, block size) and measure power consumption of each benchmark. We also need a method using which we can collect power consumption of benchmarks without using any external hardware. Since the proposed model should be easily usable by a application developer without a need of any specialised equipment. Therefore data collection is a tedious and laborious task involving multiple methods and resources to cover the spectrum of features extracted using static analysis. 
\item \textit{Choosing a suitable model for predicting power consumption}
We did not come across any algorithmic approaches to understand the relationship between power consumption and program characteristics for GPU applications. Hence we will have to use machine learning based model for building an accurate power prediction model. We need to systematically analyse and select the most appropriate model for power prediction from the pool of available machine learning algorithms. 
\end{enumerate}
\subsection{Objective 3} 
We expect that the insights gained while building these models will significantly help to create an architecture-agnostic model for energy consumption of CUDA kernel which can be tuned to multiple architecture.  For this, we plan the following activities
\begin{enumerate}

\item \textit{ Investigate different architectures which can be tuned by the model to adapt to newer architectures}

 Over the last two decades NVIDIA has introduced eleven architectures with increasing GPU's performance efficiency sometimes at the cost of its complexity and power dissipation \footnote{https://www.nvidia.com/en-in/about-nvidia/corporate-timeline/}\cite{Kerr2012,Bridges2016}. Therefore we need to consider the adaptability of our model to  multiple architectures since the GPU hardware evolves much faster than the lifespan of an application running on the hardware . The proposed solution should be designed in such a way that it can be tuned to newer architectures by taking its architecture details as input.  Hence we need to investigate different architecture and find out the features which can be tuned such that the model can be adapted to newer architectures. 
 
 \item \textit{ Validating the model across multiple architectures}
 
 Once we find the parameters for tuning the model to newer architecture, we need to validate if both performance and power prediction models work for different architectures. In order to do so, we have to identify the newer architectures and popular GPUs belonging to these architectures. Adapting the performance model to newer architecture will be easier than power prediction model since we wish to use an analytical model. Data collection for performance prediction will involve only getting actual execution time of benchmarks and microbenchmarking.  Whereas data collection for power prediction model will involve executing thousands of benchmarks with different launch configuration and hence will be more laborious task. 
 \end{enumerate}
\section{Contribution}
In order to address the research questions identified, we proposed solutions for building energy-aware GPU application. Our contributions in this work are as follows: 
\begin{itemize}
    \item We proposed a novel set of microbenchmarks to collect different architectural features of a GPU.
    \item We designed a lightweight approach that combines the static analysis of a CUDA application and architectural features of a GPU derived from the dynamic analysis performed only once for given hardware.
    \item We have created a dataset from a set of CUDA benchmark applications. We have defined a feature engineering methodology to identify a set of features that influences power consumption. This dataset and feature extraction code is made available \footnote{ https://github.com/santonus/energymodel}.
    \item We developed a machine-learning based power prediction model for NVIDIA GPGPU applications using program analysis and hardware attributes.
    \item We analyze the results of power prediction model to understand the impact of program features on the power consumption of NVIDIA GPU.
    \item We have considered the adaptability of our execution time and power prediction model to different architectures. We have found that our approach works across four architectures: Kepler, Maxwell, Pascal, and Volta.
    \item We also presented a design assistance tool which can assist heterogeneous application developers to build energy-aware applications. 
\end{itemize}

\section{Publications from this Research}
\begin{figure}
    \centering
    \includegraphics[width=\linewidth]{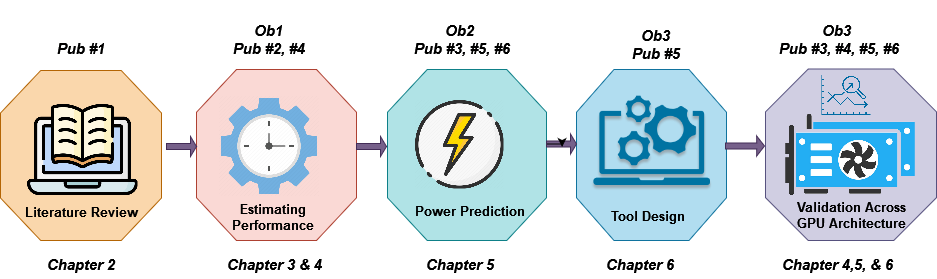}
    \caption{Topic flow of Thesis with publications mapped to Objectives}
    \label{fig:topicsandpublication}
\end{figure}

In Figure \ref{fig:topicsandpublication}, we have mapped the publications to the topics covered in this thesis. The following is the list of publications that resulted from our research. 

\begin{enumerate}
   \item Santonu Sarkar and Gargi Alavani. 2018. How Easy it is to Write Software for Heterogeneous Systems? SIGSOFT Softw. Eng. Notes 42, 4 (October 2017), 1–7. 
   \item Gargi Alavani, Kajal Varma and Santonu Sarkar, "Predicting Execution Time of CUDA Kernel Using Static Analysis," 2018 IEEE Intl Conf on Parallel \& Distributed Processing with Applications, Ubiquitous Computing \& Communications, Big Data \& Cloud Computing, Social Computing \& Networking, Sustainable Computing \& Communications, 2018, pp. 948-955.
   \item Gargi Alavani, Jineet Desai and Santonu Sarkar, "An Approach to Estimate Power Consumption of a CUDA Kernel," 2020 IEEE Intl Conf on Parallel \& Distributed Processing with Applications, Big Data \& Cloud Computing, Sustainable Computing \& Communications, Social Computing \& Networking , 2020, pp. 984-991.
   \item  Gargi Alavani, Santonu Sarkar, "Performance modeling of graphics processing unit application using static and dynamic analysis", Concurrency and Computation: Practice and Experience, 2021, Volume - 34.  
   \item Gargi Alavani, Jineet Desai and Santonu Sarkar, "GPPT: A Power Prediction Tool for CUDA Applications," 2021 36th IEEE/ACM International Conference on Automated Software Engineering Workshops (ASEW), 2021, pp. 247-250.
   \item Gargi Alavani, Jineet Desai, Snehanshu Saha and Santonu Sarkar, " Program Analysis and Machine Learning based Approach to Predict Power Consumption of CUDA Kernel" Communicated
\end{enumerate}

\section{Thesis Outline}
The rest of this dissertation is organized as follows. 
\textbf{Chapter 2} surveys the state of the art and contextualizes our contribution to prior work in this field.
\textbf{Chapter 3} presents the microbenchmarking approach and results.
\textbf{Chapter 4} describes the execution time prediction model.
\textbf{Chapter 5} proposes the power prediction model based on machine learning techniques.
\textbf{Chapter 6} presents the design assistance tool developed in this study.
\textbf{Chapter 7} concludes the dissertation and presents directions for future research.

\let \textcircled=\pgftextcircled
\chapter{Literature Survey}
\label{chap:ls}
\initial{W}ith the growing demand for performance-oriented problems, programmers routinely execute the embarrassing parallel part of the application in a GPU in order to achieve significant speedup. These
applications are becoming complex and long-running. Traditional programmers are not well-versed with programming for heterogeneous systems, and it is quite some time away for these programmers to become parallel programming experts. With evolving architectural complexities of multi-core systems, accelerators, and hybrid systems, the need for tools that can help application developers to harness their computing power efficiently is the need of the hour.

\section{Design Assistance}
Parallel programming is arduous compared to serial programming, and hence there is a need for different types of design assistance tools. As observed in Berkeley's technical report, \cite{Asanovic2009}, unless writing scalable parallel programs becomes as easy as writing programs for sequential computers, developers will find it difficult to switch to the new paradigm. The technical report by Berkeley and Illinois parallel processing group \cite{kumar2008} has highlighted that unless such support tools and methods are not in place, heterogeneous computing will not be accepted in the mainstream. The work by Wienke et al.~\cite{wienke2015} reports an empirical study to determine various factors that can influence programmer productivity. Though preliminary, the result shows that the prior knowledge of the parallel computing platform plays the most significant role in programmer productivity. However, it is noteworthy that influences of the environment and tools on productivity are ranked second and third, respectively. 

So far, the main focus areas in the parallel computing software landscape have been the following.\\\textit{a. Compiler infrastructure: }Which is mainly pioneered by the hardware manufacturers like Intel, AMD, and NVIDIA to build sophisticated compilers mostly.\\
\textit{b. Programming language support:} Which is also pioneered by the likes of Intel, AMD, NVIDIA, and so on to build programming language extensions (like CUDA, OpenCL) and libraries to write parallel programs.  \\
\textit{c. Algorithm development:} Many advanced parallel algorithms have been developed by trained developers to harness the computing power of mutlticore architectures e.g. CUDA toolkit \cite{NvidiaCUDAToolkit}.  \\
While these areas are essential, we posit that there are other critical research areas for parallel computing software, which are discussed as follows.

Software development for a heterogeneous computing environment comes with some inherent challenges:

\begin{enumerate}[(i)]
\itemsep0em
    \item How can software development be made simpler by hiding the underlying hardware complexities? \cite{Hwu2011,Rubin2014,You2015}
    \item What are the existing solutions available for common design problems? \cite{Steuwer2013,Marques2013}
    \item What kind of algorithmic skeletons can be used to implement efficient parallelism? \cite{Kurzak2013,anderson2012,Li2013}
    \item What reusable components can be used to develop software faster? \cite{Sarkar2012,collobert2002}
    \item  How software can be autotuned to gain maximum performance on hardware? 
    \item What are the techniques available for verifying the semantics and pointing out software bugs? \cite{Betts2015,Li2010,Sergey2015}
    \item Which tools can help developers build energy-aware sustainable computing applications? \cite{SCI}
\end{enumerate}

\begin{figure}
  \includegraphics[width=\linewidth]{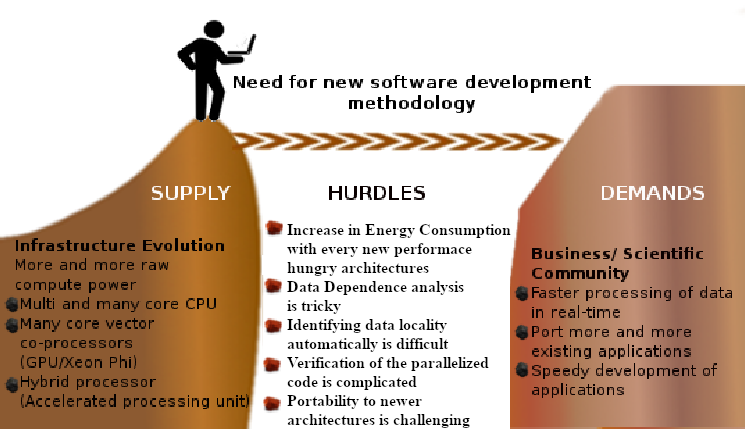}
  \caption{Crossing the Chasm}
  \label{fig:chasm}
\end{figure}

Considering these critical issues, developers find it challenging to switch from a sequential programming style to deal with the complexities of a parallel program (depicted as a chasm in Figure \ref{fig:chasm}). Proposing design assistance solutions for each of these issues can make the job of parallel developers efficient and easier. In this thesis, we focus on one of these research questions, i.e., " Which tools can help developers build energy-aware sustainable computing applications?". With the advancements in parallel programming architectures comes the dire need for building design assistance tools to help developers build energy-efficient tools that make these architectures a sustainable computing alternative. In this thesis, we are attempting to propose design assistance tools that will aid developers in building energy-aware sustainable computing GPU applications. Hence we survey the existing solutions for the prediction of energy consumption of GPUs in the following sections. 

\section{Energy Prediction of GPU applications}
While supercomputers and parallel computers generally augment our computation power manifold, they consume a lot of energy and dissipate a lot of heat. As a result, the cooling systems form a significant part of operating costs and the cost of electricity (enough to power a small city) \cite{Bridges2016}. The main goal of developers of GPU is to tune their code extensively to obtain optimal performance,  making efficient use of different resources available on the  GPU. However, an application running on an HPC infrastructure should perform optimally, and at the same time, it needs to consume the least energy required to run the application. 

\begin{figure}
  \includegraphics[width=\linewidth]{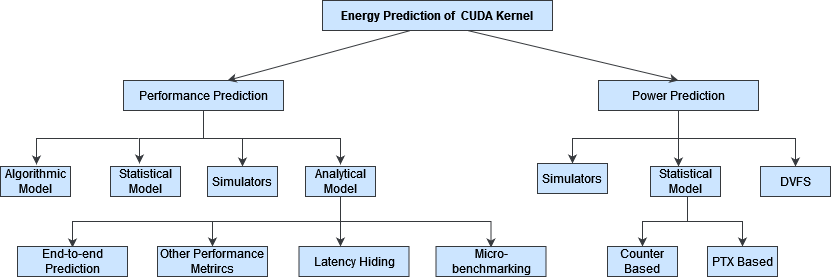}
  \caption{Outline of the Survey}
  \label{fig:literature_overview}
\end{figure}

A developer must be equipped with tools to understand the energy consumption of GPUs before launching the application. A significant body of work has attempted to address various aspects of the energy prediction model of GPU programs \cite{Mittal2014}. As discussed earlier, we are dividing the problem of energy consumption into performance and power estimation. Hence in this thesis, we survey prior work related to modelling performance and power consumption of GPU applications. Firstly, we discuss in detail existing literature on performance prediction. Then we review the existing literature on the power prediction of GPU. The overview of this literature survey is shown in Figure \ref{fig:literature_overview}.

\section{Performance Modelling of GPU}
Performance modelling is a broad and well-studied topic entailing various metrics such as latency, throughput, concurrency, memory access efficiency, and so on, at various levels of granularity, starting at the level of individual instruction upto the entire application \cite{Madougou2016}. Our work focuses on developing a complete GPU kernel's execution time prediction model rather than predicting latency and throughput at the instruction level. Existing studies on performance prediction can be classified as algorithmic\cite{Kothapalli2009,Kirtzic2012}, analytical models\cite{resios2011,Amaris2016,Hong2010,luo2011,Zhang2011}, statistical approaches\cite{Amaris2016,Baldini2014,ZhangY2011,Ardalani2015}, and simulation \cite{chapuis2016gpu}. There are some application-specific performance models developed which are tuned to computation needs, memory usage, access patterns of GPU application \cite{lehnert2016,konstantinidis2015,remmelg2020}. We present a survey of these performance prediction techniques covering a breadth of approaches and existing solutions.

\subsection{Algorithmic Approaches}
Early work in modelling parallel algorithm behaviour began with the PRAM model\cite{Fortune1978}, followed by the Network model\cite{Ryu1990}, the BSP model\cite{Valiant1990}, the  Log-P model \cite{Culler1993}, and the QRQW model\cite{GIBBONS1998,Gibbons1999}. These models help the programmer to understand parallelism in a multiprocessor system. 

Amaris et al. \cite{amaris2015} designed a simplistic and intuitive  Bulk Synchronous Parallel (BSP)-based design for execution time prediction. BSP is a model for parallel computation that allows algorithmic analysis of programs that consider the number of computations and memory accesses of the GPU, with supplementary information on cache usage collected from profiling. Model results are within 0.8 to 1.2 times the actual performance for the two benchmarks. In another work by Amaris et al. \cite{Amaris2016}, they compared the results of analytical and machine learning models and concluded that analytical models are more precise than the other approaches. However, they conducted the study on only two benchmarks and showed nearly 20\% prediction error. 

Since algorithmic approaches are not architecture-specific, one can neither perform an architecture-specific performance analysis nor predict the execution time of a given program for a complex architecture such as a GPU using these models. We now present the statistical/machine learning approaches for performance prediction.  

\subsection{Statistical Approaches}
The machine learning model by Amaris et al. \cite{Amaris2016} utilised Linear
Regression, Support Vector Machine and Random Forest to predict execution time of application.  Eiger\cite{Kerr2012} is an automated statistical methodology for performance model. They utilised 47 features from both the application and the hardware which are collected by executing the application on target machine. The approach is validated using 12 benchmarks.  Madougou et al. \cite{Madougou2016} utilised Eiger for GPU and found that the accuracy is not uniform for different types of benchmarks. 

The work by Shuaiwen et al. \cite{Shuaiwen2013} built an ANN model for execution time prediction using performance counters using the CUPTI library rather than using PTX code information. The authors claimed to restrict the average absolute prediction error within 6.7\% for 12 benchmarks. A major difference of their approach with ours is that their ANN model requires performance counter values which can only be obtained by running CUDA kernels. In our opinion, this defeats the purpose of predicting the application's execution time without the need for running the same. Also, their model does not work for the asymmetric workload, whereas our model takes care of those, as mentioned earlier.

Stargazer \cite{Wenhao2012} is an automated GPU performance exploration framework which randomly samples the parameter values of the
full GPU design space. It is based on stepwise regression modeling which helps to understand the most influential architectural parameters to the performance
of the application. Stargazer uses simulated data (GPGPU-Sim \cite{Bakhoda2009}) and only considers hardware characteristics. It is highly time-consuming approach and lacks the performance analysis from application perspective.

Zhang et al. \cite{ZhangY2011} proposed a Random Forest (RF) based model for ATI GPU. They utilize feature importance of RF to understand most influential variables to the execution performance (throughputs) of the target GPU.  The approach is validated on 22 kernels and runtime data for building the model is collected using 23 performance counter.  Counter-based approaches are not extendable to GPUs which do not support counters and also the current counters do not cover all performance factors. 

ScaleClass \cite{Wu2015} estimates the performance and power of GPUs across a range of hardware configurations for OpenCL kernels. A neural network classifier is utilised to predict cluster’s scaling behavior which is described by a new kernel based on its performance counters. However compared to earlier statistical approaches based on LR and RF, ScaleClass accuracy is lower which could be  because regression-based approaches usually perform better than neural networks in terms of accuracy \cite{Lee2007}. Existing statistical approaches are utilising performance counters, runtime profiling data, or simulation data for the performance prediction. We will now discuss how simulators contribute to performance prediction.    

\subsection{Simulators for Performance Evaluation}
Simulators are quite helpful for investigating GPU-based workloads' characteristics. Since simulators are primarily built to emulate machines hardware behaviour they do not directly perform execution time prediction. However, they play an important in performance prediction by being utilised to build a prediction model using the traces of GPU runtime hardware behaviour. For e.g. GPGPU-Sim \cite{Bakhoda2009} is used for building performance prediction tool \cite{ Wenhao2012}. Hence we briefly discuss their contribution in performance prediction. 


GPGPU-Sim \cite{Bakhoda2009}, GPU-Tejas \cite{Malhotra2014}, PPT-GPU \cite{Arafa2019} are popular simulators available for NVIDIA GPU. Though PPT-GPU \cite{Arafa2019} simulation framework claims to predict applications' performance in a fast and accurate manner; however they have not presented any result to support this claim. Simulators have reported their execution times to perform simulations. For instance, the mean execution time for the GPGPU-Sim tool is 14100.29 seconds, whereas GPU-Tejas reported an average execution time to be 32.80s~\cite{Malhotra2014}.   

\subsection{Analytical Model}
Analytical models represent a mathematical abstraction of the program’s execution. The result is usually represented in the form of a function which takes into account the target architecture, the input program and the input data.  Analytical models are built for end-to-end prediction of execution time as well as for representing performance of GPU using metrics such as throughput. We first discuss the existing approaches for end-to-end prediction of execution time for CUDA kernel followed by other performance evaluation metrics utilised for performance prediction. 

\subsubsection{End-to-end Execution Time Prediction}
An important work closely aligned to ours was by  Kothapalli et al. \cite{Kothapalli2009} where the authors proposed a GPU-specific prediction model, inspired by Queue-Read Queue-Write PRAM Model ~\cite{GIBBONS1998,Gibbons1999}, to give a rough estimate of the execution time of a CUDA kernel. Here the authors proposed a scheduling model that is based on two approaches. It considers either all instruction latency while scheduling or only the maximum of either memory or computation instruction latency. However, the results suggest that such an approach sometimes grossly overestimates and sometimes underestimates massively. The authors have considered three benchmarks for their case study: Matrix Multiplication (31.25\%  error), List Ranking (12.50\% error), and Histogram. 

Hong and Kim \cite{Hong2009} have proposed an early execution time prediction model for a GPU. The approach computes the maximum possible number of warps per SM and counts computing instructions (e.g., add, mul) which can be executed when waiting for one memory operation. A significant drawback of this model is that it does not consider computing instruction latency which is crucial for any performance prediction. They assume a flat four cycles issue latency for compute instruction irrespective of occupancy. Hence if a benchmark is compute-intensive with very few memory instructions, the execution time prediction will be underestimated.  

Sim et al.\cite{Sim2012} improved Hong and Kim's approach~\cite{Hong2009} by including arithmetic latency. The authors first computed the time taken by arithmetic instructions and memory instructions separately. Total time is then calculated by either choosing the max of the two or using their proposed equation. Along with the ILP and TLP, they also considered Memory Level Parallelism (MLP). However, this model continues to have a large overestimation error for kernels with high memory latency. 

Huang et al. \cite{Huang2014} presented an interval analysis based technique called GPUMech, for modelling GPU performance. GPUMech has modeled multithreading and resource allocation due to memory usage. It presents results for two scheduling approaches: Round Robin policy and greedy-then-oldest policy. They used GPUOcelot \cite{Diamos2010} that executes GPGPU  kernel and collect per-warp instruction traces. Observed error for GPUMech is  13.2\% the round-robin scheduling policy and 14.0\% error for the greedy-then-oldest policy.

A GPU kernel execution time prediction model based on compiler techniques that consider computing as well as memory access latency together was proposed for the first time by Baghsorkhi et al. \cite{Baghsorkhi2010}. They utilized an annotated program dependence graph called ``Work flow graph'', which is an extension of CFG. Later, Volkov et al. \cite{volkov2016} experimentally proved that this model works best at 100\% occupancy and fails in any other case. In reality, 100\% occupancy cannot be achieved in many non-trivial applications. We also observed that the instruction throughput shows a linear or sublinear growth when we plot it against number of warps per SM, which was not visible in their model \cite{Baghsorkhi2010}.

Resios et al. \cite{resios2011} formulated a parameterized model for delay calculation of computing and memory instructions. However, their usage of parameters such as Instruction Level Parallelism (ILP) as well as Thread Level Parallelism (TLP) makes this delay value too low to be true. Their thread scheduling model is overly simplified since they multiply the execution time by the number of threads. The authors have reported the precision of 1.64 for the Transpose benchmark and 1.89 for the 2D Convolution filter benchmark.

Williams et al.~\cite{Williams2009} proposed the roofline model concept for multicore systems. This model relates processor performance to off-chip memory traffic, where the execution time is computed using peak floating-point performance, peak memory bandwidth, and operational intensity. This model has influenced several research works related to GPU performance. Konstantinidis et al. \cite{Konstantinidis2017AQR} improvised the roofline model\cite{Williams2009} and used only runtime data collected by executing the CUDA kernel code multiple times to build their execution time prediction model.  To improve the accuracy of the model, the authors proposed include the utilization factor ($E_{util}$), as an option. However, $E_{util}$ is also computed using actual execution time for each kernel. 

An execution time prediction analytical model for GPU with instruction level and thread-level parallelism awareness was presented by Luo et al. \cite{luo2011}. The proposed model contains two components: memory sub-model and computation sub-model. The memory sub-model is estimating the cost of memory instructions by considering the number of active threads and GPU memory bandwidth. Correspondingly, the computation sub-model is estimating the cost of computation instructions by considering the number of active threads and the applications arithmetic intensity. They utilised Ocelot\cite{Diamos2010} to analysis PTX codes and obtain time cost of each PTX instructions. Their result show that the model can reach 90 percentage accuracy in average for the six benchmarks considered in this study. 

\subsubsection{Other Performance Evaluation Metrics}
We now evaluate a few important publications that focused on predicting performance metrics other than end-to-end execution time. Li et al. \cite{Li2015} developed a cross-roofline model called {\it Transit}, that predicts computation throughput and memory throughput, which is validated on Fermi and Kepler architecture with 90\% and 83\% prediction accuracy, respectively. However, unlike our approach, their model fails to predict complex kernels such as Gaussian, lavaMD, and ParticleFilter. 

Volkov et al. \cite{volkov2016} proposed an analytical model to predict arithmetic and memory throughput. Furthermore, the model is not tested for real-life applications and is tested for only one benchmark. The pointer chasing benchmark used by them is a mix of arithmetic and memory instructions, where all arithmetic instructions are identical, and all memory instructions are identical. As a result, instructions become evenly interleaved and back-to-back dependent. We have noticed another discrepancy in Volkov's article \cite{volkov2016} where the memory throughput model presented is not exponential, though the author acknowledges that the memory throughput should be exponential. Through extensive experimental study, we have observed that the actual memory throughput shows exponential growth. 

Huang et al. \cite{Huang2014} presented a GPUMech tool based on interval analysis technique to simulate multithreading and resource allocation due to memory usage in a GPU. They used GPUOcelot framework to collect per-warp instruction traces using run-time analysis. GPUMech provides Cycles Per Instruction(CPI) stacks for the memory hardware components analysed, which can assist developers in visualizing performance bottlenecks of a kernel.GPUMech simulation model\cite{Huang2014} presented the observed error to be  13.2\% for round-robin scheduling policy and 14.0\%  for the greedy-then-oldest policy. GPUMech utilizes runtime traces of warp execution along with instruction latency.  

Lemeire et al.\cite{Lemeire2016} proposed a set of microbenchmarks in OpenCL using occupancy roofline method which they claim to be sufficient in order to predict the performance of GPUs. In order to compute the cycles per instruction for each compute unit, they compute the characteristics of an instruction as: its issue and completion latency. They modelled each compute unit as a pipeline for computations and a pipeline for the memory access. They also measured the influence of independent instructions within a kernel and impact of thread divergence.

Branch divergence plays a important role in performance of an application. If the data, processed by a GPU application causes a periodic branch divergence which has a patterns, it is sometimes possible to refactor the code remove this divergence, if this pattern can be discovered through a dynamic analysis~\cite{sarkar2015profile}. 
However, it is extremely hard to predict any characteristics of a potential branch divergence phenomenon with purely static analysis based approach. 

We have tabulated the results of existing approaches in Table \ref{tab:existing approaches}. In this thesis, we propose an analytical model for performance prediction. In order to design an analytical model for performance, it is crucial to understand how latency hiding is modelled in the model. We now discuss some of the approaches used for the same.

\subsection{Latency Hiding}
Latency hiding for any GPU performance model is an important aspect that should model carefully. Latency hiding in a GPU is carried out by running independent instructions within multiple warps. We compare the latency hiding of our proposed algorithm with other approaches. Hong et al. \cite{Hong2009} assumed 4 cycles per instruction for arithmetic throughput irrespective of GPU occupancy.    
In many popular approaches \cite{Volkov2008,volkov2016,Lemeire2016} it is empirically shown that at 100\% occupancy, the GPU almost completely hides the latency of arithmetic instructions. In such cases, Hong et al. \cite{Hong2009} model will grossly overestimate the performance, and when the occupancy is very low, it will lead to underestimation (an arithmetic instruction takes more than 4 cycles). Volkov~\cite{volkov2016} model performs latency hiding using a concept called arithmetic intensity along with both arithmetic and global memory instruction latencies. However, their model is derived from one benchmark with identical arithmetic and memory instructions, with no independent instructions. In contrast, our approach considers independent instructions, resource allocation. Furthermore, we have evaluated our approach on 44 benchmarks from different benchmarks suites, sizes, nature (e.g., compute-bound, memory-bound), and complexity (e.g., control divergence, memory access patterns).

The idea of using data dependence and instruction latency in a graph-based model presented by Baghsorkhi et al. \cite{Baghsorkhi2010} is on similar lines as our model. However, while implementing latency hiding, they compute delay by subtracting non-blocked arithmetic instructions from the latency of memory instruction using analytical equations. 

As discussed earlier, microbenchmarking is crucial in designing a fairly accurate performance prediction model since the crucial hardware details are not publicly available. Hence we discuss some of the existing popular work on microbenchmarking GPUs.

\subsection{microbenchmarking}
The use of microbenchmarks is critical to understand the nature of GPU instruction execution\cite{Li2012,Wong2010,Mei2017,Jia2018} since the vendors are reluctant to share these crucial details.  We discuss here notable works that use microbenchmark extensively to analyze GPUs performance \cite{Lemeire2016}, understanding hardware characteristics \cite{Wong2010} and apply it for assessing energy utilization \cite{Lucas2019}.

Lemeire et al. \cite{Lemeire2016} presented OpenCL microbenchmarks to gather performance characteristics of GPU. The authors performed a study of the influence of independent instructions within a kernel and thread divergence. 

Wong et al. \cite{Wong2010} measured the latency and throughput of different types of arithmetic and logic operations performed in a kernel for Fermi architecture GPU. They studied functionality of branch divergence and of the barrier synchronization, and also measured the structure and performance of the memory caching hierarchies. 

Andersch et al. \cite{Andersch2015} analyzed the static as well as dynamic latency behavior of GPU microarchitectural components using GPGPU-Sim GPU timing simulator. They conclude that GPUs are not as effective in latency hiding as believed to be, and suggest that latency should be a major GPU design consideration besides throughput.

%


\section{Power Modelling of GPU}
 Over the last few years, the importance of understanding the power consumption of HPC machines has increased significantly. GPU power measurement and estimation can be briefly divided into direct methods, including using external and internal hardware for monitoring power and indirect techniques such as modeling and simulation \cite{Bridges2016}. Utilizing external hardware sensors is costly for profiling large-scale distributed systems, and internal hardware sensors are still being verified for their accuracy and reliability. Hence researchers are also working extensively on indirect power modeling and measurements. Indirect power modelling includes models based on hardware performance counters, simulators, and PTX-based power models. Our work is an attempt to contribute to indirect power modelling of GPUs. We built a PTX-based power model using machine learning techniques. We discuss some of the research efforts that propose and evaluate indirect power estimation. 
 
 Chen and Shi \cite{Chen2012} discuss the need for power profiling to optimize existing algorithms and lay down various hardware and software-based mechanisms existing in power modeling of computing systems. They believe that although software-based tools offer a lower degree of accuracy, they allow for flexible usage due to the lack of specialized hardware. The power consumption of a computing device is intricately related to its hardware features.  Consequently, there has been a significant body of work to investigate the relationship between hardware features of a machine and its power consumption~\cite{Fan2019,Guerreiro2018,Lim2015,Lucas2013}. Mittal and Vetter \cite{Mittal2014} surveyed the methods of analyzing GPU energy efficiency. They conclude that there is a need for using multiple approaches at the chip design level, architectural level, programming level, etc., to get the maximum increase in GPU energy efficiency. A survey by Bridges et al. \cite{Bridges2016} also promotes application-level analysis for power prediction especially PTX-based power models, which they believe can yield valuable and informative predictions. 
 However, not much work has been done to understand the relationship between code features and power consumption. To the best of our knowledge, we still  do not have any algorithmic approach to understand the relationship between power consumption and program characteristics.
 
 \subsection{Statistical Approaches}
Statistical approaches employ machine learning algorithm to perform GPU power prediction. Data utilised for these approaches is collected from hardware events which are captured using sensors and counters or application based features. One of the initial works on the statistical model for power prediction was carried out by Ma et al.\cite{ma2009}. They utilized Support Vector Machine (SVR) and Support Linear Machine (SLR) to estimates the power used by the GPU over a small time window. Training data was generated by creating benchmarks that stress different GPU sub-units. SVR outperformed SLR emphasizing a complex model works efficiently over a linear model for power prediction. 

\subsubsection{Counter Based Models}
Counter Based Modeling techniques predict power consumption of application by correlating each with hardware events. These events are accessible through performance counters. Performance counters are generated based on hardware actions, (e.g., number of issued instructions, number of memory accesses, cache operations). These operational events give users access to low-level hardware activities. Models designed using data from counters which are able to explain hardware events, exhibit strong
correlations to power consumption. Hence these models can yield accurate power models.

Nagasaka et al. \cite{Nagasaka2010} propose a statistical method to estimate a CUDA kernel's power consumption using performance counters. The approach uses GPU performance counters as the independent variable and power consumption as the dependent variable and trains a linear regression model. Both Ma et al. \cite{ma2009} and Nagasaka et al. \cite{Nagasaka2010} emphasized that the acquisition of global memory access counts is crucial for accurate GPU power modeling. Song et al. \cite{Shuaiwen2013} proposed the first counter-based model built using a neural network. 

Jia et al. \cite{Jia2015} presented Starchart, a design space partitioning tool that uses regression trees to perform
GPU Performance and Power Tuning.  The Performance API (PAPI) library \cite{Weaver2012} is extended to measure energy and power, in addition to performance counters. However, it produces some anomalies while measuring power. Song et al. \cite{Shuaiwen2013} believe that no accurate model combines simplicity, accuracy, and support for emergent GPU architectures in isolating energy and performance bottlenecks to identify their root causes. They proposed a neural network model to estimate the average power using performance metrics. They utilize hardware performance counters with BP-ANN based prediction model that is accurate to within 2.1\% for power prediction.

\subsubsection{PTX-based Model}
Parallel Thread Execution (PTX) is a pseudo-assembly code which is an intermediate step in CUDA compilation. This code includes the list of instructions to be executed by the GPU. If one wants to pedict power consumption a priori, PTX instructions can be leveraged to analyze the power of a CUDA program.  This approach provides powerful optimization capabilities to programmers before executing their code. 

A work by Hong et al. \cite{Hong2010} propose an integrated power and performance prediction model for a GPU architecture. They modelled the relationship between power and PTX instructions through mathematical equations. They designed a set of synthetic microbenchmarks which stress different architectural components in the GPU. They then map these components to PTX instructions to understand instructions impact on power consumption. They use this approach to predict the optimal number of active processors for a given application. However, their model fails to predict asymmetric and control-flow intensive applications. 

Singh et al. \cite{Singh2009} applied non-linear functions on data collected using micro benchmarking to predict the power consumption of GPU. Zhao et al. \cite{Zhao2013} present a model which gives power prediction for the GPU by counting each PTX instruction as input.
 
\subsection{Dynamic voltage and frequency scaling (DVFS) }
Dynamic voltage and frequency scaling (DVFS) is a popular approach for improving power efficiency \cite{Bridges2016,Fan2019,Guerreiro2018}. Abe et al. \cite{Abe2012} explore the domain of application of DVFS techniques to GPU accelerated systems which produce predictable results in CPU-only systems trivially, but owing to the complex nature of GPU architecture, fail to produce similar results. 

We cannot use DVFS for static-anaysis based power prediction. This approach is not suitable to build a model to predict the power consumed by a CUDA application because: Firstly, there is no established relationship (direct or latent) between the GPU frequency and the CUDA program structure. Moreover, DVFS is auto-enabled in GPUs and should not be disabled since it can lead to system failure. Moreover, the vendor does not provide any external control to manipulate DVFS in order to build a dynamic model which can capture system behaviour by executing only once.

\subsection{Simulators for power prediction}
Although counter-based models are popular and have proven to be effective, the execution of an application defeats its purpose. There are many simulation-based power prediction tools available which replace the need to execute an application yet help to find correlation between hardware features and power \cite{Sheaffer2004,Lim2015,Leng2013} . 

Chen et al. \cite{Chen2011} proposed a GPU power consumption model using the tree-based random forest method to understand and estimate power consumption based on performance variables collected by collecting runtime data from GPGPU-Sim. Prediction accuracy using leave-one-out cross-validation (LOOCV) for the random forest is 7.77\% with MSE 302.6, regression tree is 11.68\% with MSE 637.8, and linear regression is 11.70\% with MSE 2548.0. However, it fails to predict for benchmarks such as BlackScholes, which are compute-intensive. Results observed in this model emphasize that the relationship between GPU power and its subprocesses is non-linear and complex. Hence advanced machine learning algorithms are required to capture this relationship.

Lim et al. \cite{Lim2015} constructs a power model for GPUs by combining empirical data with data obtained from McPAT \cite{Song2014}, a CPU power tool. The model is trained using a particular set of benchmarks, after which it is used for predicting power. The reported average error is 12.8\% for Merge benchmarks. Although simulations provide useful power predictions, the modeling of the simulator requires a deep understanding of architecture and processes, which is quite complicated. 

A simulator called GPUSimPow \cite{Lucas2013} helps estimate the power consumption of a kernel on a given GPGPU architecture without physically running the kernel on the GPU, but by using hardware models for the internal components of GPUs. It breaks the overall power estimate to individual component power usage. Because the hardware is modelled directly, authors claim that simulation is possibly the most accurate method of estimating the energy consumption of a kernel. However, it is also a more time-consuming method as the entire execution of the kernel from start to finish needs to be modelled precisely. For a set of benchmarks, GPUSimPow shows an average relative error of 11.7\% for GT240 and 10.8\% for GTX580. Wang\cite{Wang2010} extended the popular GPU simulator GPGPU-Sim \cite{Bakhoda2009} to include power modelling. However, the accuracy of his results was not published.

\section{Limitations of Existing Work}
\subsection{Performance Prediction}
\begin{itemize}
    \item \textbf{\textit{Problem Solving Approach: }} Most of the existing approaches for performance modelling are derived from runtime data \cite{Hong2009,Baghsorkhi2010,Shuaiwen2013,Huang2014}. To utilize these approaches for understanding the performance of a long-running application is not feasible, costly, and energy inefficient. Hence, building a model that can predict applications' performance without execution is a significant contribution in this area. In the work by  Konstantinidis et al. \cite{Konstantinidis2017AQR},  to improve the accuracy of the model, the authors proposed inclusion of the utilization factor ($E_{util}$), as an option. However, $E_{util}$ is computed using actual execution time for each kernel which defeats the purpose of prediction for conserving energy.  
    
    Two important studies from the static analysis approach include Resios et al. \cite{resios2011}'s parameterized model, which provides a good foundation for an analytical model. However, their GPU simulation model is weak, with no scheduling of threads involved. Similarly, the work by Kothapalli et al. \cite{Kothapalli2009} should have been more intuitive since their approach sometimes grossly overestimates and sometimes underestimates. 
    
    \item \textbf{\textit{Need for microbenchmarking: }}
    Microbenchmarking is the key to peeping into the GPU BlackBox, and it should be the first step toward building a good prediction model. Although there are existing microbenchmark suits such as Wong et al. \cite{Wong2010}, their results are based on an older Tesla architecture NVIDIA GT200 GPU. Building a microbenchmark design that can be effectively utilized across new architectures is crucial. We also need new set of microbenchmarks which can capture critical characteristics such as kernel launch overhead, memory throughput, the effect of throughput for different ILP values etc. Baghsorkhi et al. \cite{Baghsorkhi2010} discussed kernel launch overhead and how they are significant for benchmarks with very low execution time. However, they did not present any methodology for computing and incorporating it in their work.
    
    \item \textbf{\textit {Validation: }} Although some of the existing approaches claim to observe decent accuracy, it should be noted that most of them are validated on a very small dataset of two to ten kernels \cite{amaris2015,luo2011,resios2011,Hong2009,Kothapalli2009}. Building a performance prediction model validated across a decent number of benchmarks is also crucial for it to be considered a reliable solution. 
    
    \item \textbf{\textit {Architecture-specific Model:}} It is observed that most of the existing approaches which is validated for one architecture \cite{resios2011,Hong2009,Kothapalli2009,amaris2015}. With newer architecture emerging within a year or two, building architecture-specific models seems an expensive approach in terms of resources and time. We need architecture-agnostic models which work across multiple architectures after tunning a few parameters.

\end{itemize}

\subsection{Power Prediction}
\begin{itemize}
\item  \textbf{\textit{Problem Solving Approach: }} 
The power consumption of a computing device is intricately related to its hardware features. Consequently, there has been a significant body of work to investigate the relationship between hardware features and power consumption \cite{Nagasaka2010,Pool2010,Shuaiwen2013,Weaver2012,ma2009,Chen2011}. This runtime hardware data is collected either using power meters, hardware counters, or by profiling the execution of the CUDA kernel. Although using a power meter or hardware counter-based approach is considered more reliable and accurate, multiple runs of an application are sometimes required to collect this data. As seen in the work of Nagasaka et al. \cite{Nagasaka2010}, some hardware counters may not be available, which affects the prediction model. Also, some architectures allow counters to a whole SM. In such a case, if there is an imbalanced number of cores utilized per SM, the prediction may get affected \cite{Bridges2016}.
   
We observe that very few related works are based on static input features \cite{Hong2010,Zhao2013}. It is crucial to propose a static analysis-based power prediction tool that saves energy for executing an application and results in accurate prediction. Such an approach is helpful for a priori analysis. A survey by Bridges et al. \cite{Bridges2016} suggests that a PTX-based power model can produce valuable and informative predictions when one wants to optimize an application. We need to equip developers with a tool to help programmers identify energy hotspots in their CUDA kernel and refactor the code for energy efficiency.

\item \textbf{\textit {Architecture-specific Model: }} Power computing model based on hardware counters cannot be architecture-agnostic since the number of hardware counters and type of counters utilized may not be uniform across architectures  \cite{Bridges2016}. Building a model based on only hardware details for power prediction may not work as efficiently on newer hardware. In such a scenario, there is a need for an approach which can be tuned for newer architectures without much time and efforts.
 
\item \textbf{\textit{Ease of Use: }}   Although direct methods are fundamental, they still exhibit a tradeoff between accuracy and ease of use. Also, in the age of cloud computing, hardware is not available on sight for direct power measurement. Internal sensors are easy to use, but vendor documentation for these sensors is unavailable, e.g., the NVML library for GPUs abstracts crucial sensor information. In this scenario, an easy-to-use program analysis-based approach that can be executed on commodity hardware can be essential.   
\end{itemize}

\section{Discussion}
In this chapter, we surveyed the field to precisely understand the state of the art of techniques for predicting the performance and power consumption of GPU applications. In this context, our literature survey covered the spectrum covering different types of existing approaches for tackling both problems. We have included several approaches based on runtime profiling, hardware data, simulators, and static analysis.

Statistical approaches for performance prediction are utilising performance counters, runtime profiling data, or simulation data. Using runtime data routs the purpose of prediction if one wants to use these results for a priori analysis. Existing static analysis based approaches for performance prediction are algorithmic and analytical. We observed that existing algorithmic methods reported in the literature are not architecture-specific. Hence one can use them for modelling an abstract parallel machine but not for an architecture-specific performance analysis or execution time prediction of a specific device such as a GPU. Most of the existing static analysis based analytical models predict performance using some metrics (e.g. throughput) and not end-to-end execution time. Also, many of these models are tested on only one architecture; hence re-usability efficiency of these models is uncertain.   

Since power consumption is more intrinsic to hardware behaviour, existing work is more focused on observing hardware behaviour while executing an application using performance counters and simulators. Current literature does not have sufficient studies around static analysis based power prediction approach. Existing studies that utilise hardware-specific runtime data cannot be extended to newer architectures and restrict usability without specialised hardware.      

We seek to overcome observed limitations in the literature by proposing a reasonably accurate performance and power prediction model that is more intuitive, static analysis-based, and architecture-agnostic. We also believe that a study that is exhaustively tested with more number of benchmarks will be more reliable and acceptable. 

\let \textcircled=\pgftextcircled
\chapter{Microbenchmarking}\label{ch:microbench}

\initial{A}s concluded in Chapter 2, microbenchmarking is a very crucial step in the quest of modeling GPU behaviour. This chapter discusses the microbenchmarks designed in this study for the various native instruction set.

The purpose of designing microbenchmarks is to compute the properties of some fundamental operations that influence the higher level system behavior. A microbenchmark attempts to measure the performance of a small unit of code with which one can determine how the code is going to behave as part of an application. This microbenchmarking data can be used for optimization, simulation and analysis of GPU software.

Microbenchmarking data has been proved to be a crucial input in performance models \cite{resios2011, Lemeire2016, Volkov2008}. It is a way to create knowledge and deeper understanding about a hardware architecture. It helps researchers to provide undocumented processor performance properties not disclosed by vendors. We can utilise microbenchmark to compute upper performance limits for sustained performance. Microbenchmarks can also help in finding performance bugs in architectures.

Every architecture differs from its predecessor and hence instruction execution characteristics of one architecture may vary from the other. Instruction execution is governed by some critical parameters such as latency, througput etc. Very little information about these parameters(such as theoretical peak performance) are provided by the vendor. Also, these details may not precisely match the real-time metrics and hence we have to recompute it by microbenchmarking the hardware.   

We have come across work on microbenchmarking Fermi architecture in \cite{resios2011}, \cite{Wong2010} and \cite{Kothapalli2009}. When we carried out the microbenchmarking, to the best of our knowledge there was not much work done on Kepler architecture. Also, we did not come across a study which computed  minimum number of ready warps required to achieve peak throughput through microbenchmarking. Similarlly, We did not find a study which quantified kernel launch overhead through microbenchmarking.   

With these perspectives in mind, we present the following in this chapter:
\begin{itemize}
\item Microbenchmarks are developed for GPU architecture which calculate the peak performance values of computing instructions belonging to PTX instruction set. 
\item A model to estimate kernel launch overhead which has significant impact on performance of Kernel.
\item A Global memory latency Model which predicts the latency of instruction based on access stride 
\end{itemize}

We begin with understanding the instructions presented in PTX representation of CUDA kernel. 

\section{PTX ISA}
PTX stands for Parallel Thread Execution \cite{PTX}, a low-level parallel thread execution virtual machine and instruction set architecture (ISA) for GPU applications.  PTX is a stable ISA which spans multiple GPU generations. PTX is also a machine-independent ISA for C/C++ and other target compilers. Because of these features, PTX is extensively used for hand-coding, and architecture tests. 

\subsection{Generating PTX file}
CUDA code is compiled into PTX code by adding the --ptx option in the nvcc statement while building. This statement results in the generation of a .ptx file. PTX is an intermediate code representation, and not an executable. It is very similar to assembly code, but it is not the actual code that is executed in the GPU, because it has to pass through a few more intermediate steps. The .ptx file achieved after compilation consists of the code of each kernel that is defined in the corresponding .cu file.

A basic PTX code consists of the following statements: Comments, Directives beginning with '.', Computing instructions,  Global memory access statements, Shared memory access statements, Synchronization statements, Branch statements, Labels, and Return statements. We scan through the PTX code and categorise each of these statements. 

We analyse computing instructions and memory access instructions seperately since the approach for microbenchmarking the two differs significantly. We first discuss the details of microbenchmarking Computing instructions.  
\section{Computing Instructions} \label{sec:computing}
Computing instructions are arithmetic and logical instructions which can be integer operations ( such as {\tt add, mul}), floating point operations (such as {\tt fma}), or special instructions (such as {\tt sqrt}). We have developed microbenchmarks to compute latency, throughput and peakwarps for computing instruction. 

To compute latency, we launch a single thread with two dependent instructions of the same instruction type to ensure that the GPU cannot issue the next instruction until the previous instruction has finished. Before executing instructions, the start time $T_s$ is recorded. Then we execute a set of 256 instructions to ensure that the total execution time is substantial enough to be recorded. We record the time after the execution $T_e$ and measure the execution time $T=T_e-T_s$. We repeat this process $N$ times and accumulate the execution time of each run ($T_{tot} ~+= T$). Finally, we  compute the average latency of the instruction as $\mathcal{L}=T_{tot}/(2*256*N)$. 
%

\begin{figure}
\begin{minipage}{.5\textwidth}  
\lstset{linewidth = 6.3cm, breaklines=true,basicstyle=\scriptsize} 
\begin{lstlisting}[caption=throughput\_kernel,label=code1,captionpos=b,basicstyle=\tiny]
__global__ void throughput_kernel(int *dummy) {
  // initialization of parameters
    int j= blockIdx.x*blockDim.x + threadIdx.x + blockDim.x*threadIdx.y ;
    for(int i=0;i<innerLoopIter;i++){
	repeat256(b+=a; a+=b;);
    }
    dummy[j]= b+a; //store results to avoid compiler optimization 
}
\end{lstlisting}
\end{minipage}
\quad 
\begin{minipage}{0.48\textwidth} 
\includegraphics[width=\linewidth]{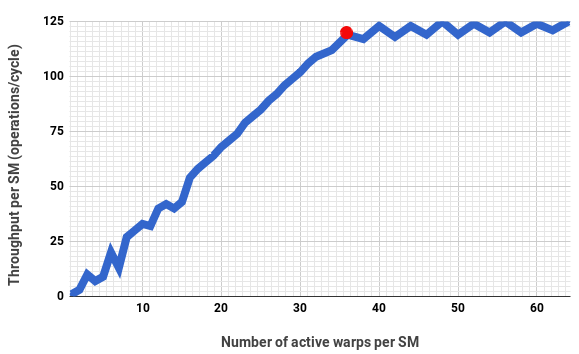}%
\vspace{-0.2cm}      \caption{Throughput for {\tt fma} with ILP=1}
      \label{fig:throughput}
\end{minipage}
\vspace{-0.3cm}
\end{figure}

Throughput and peakwarps value estimation processes are relatively more complex than the latency computation since instruction level parallelism (ILP) plays an important role here. 
In the absence of any independent instructions, it was observed that throughput saturates after a certain Thread Level Parallelism(TLP) value. To get a further increase in throughput, we need independent instructions inside the CUDA kernel available to each scheduler. This can be achieved by adding independent instructions in the kernel; with every new independent instruction, the ILP value is raised by 1. The set of instructions for ILP 1, ILP 2 and ILP 3 shown below, explains this approach. 
\begin{description}
\item[ILP=1] Instructions: b+=a; a+=b;
\item[ILP=2] Instructions: b+=a; a+=b; c+=d; d+=c;
\item[ILP=3] Instructions: b+=a; a+=b; c+=d; d+=c; e+=f; f+=e;
\end{description}
 For illustration, the CUDA kernel snippet for {\tt adds} instruction is shown in listing \ref{code1} for ILP 1. This kernel is invoked from the main program shown in listing~\ref{code2} to compute throughput for this instruction. 
 An important observation is that the peak throughput and the number of active warps required to achieve peak throughput differ for different ILP values. To characterize this phenomenon, we execute this benchmark with different ILP values. For each ILP, we create a different version of the kernel in listing~\ref{code1} by altering line 6  with the code for the corresponding ILP (shown above), keeping the rest of the code in listing~\ref{code1} and listing ~\ref{code2} the same. Throughput and peakwarp computation process for different ILP values for a compute instruction (such as {\tt adds}) is summarized below:
\begin{enumerate}
\item Design multiple versions of a throughput computation kernel, for each ILP. 
\item For each ILP, execute the corresponding version of the kernel. The {\tt throughput\_func} code in Listing \ref{code2} is called by increasing the number of active warps launched, to generate a dataset $D(w,tput)$ by recording throughput per cycle $tput$ for each active warp $w$.
\item For each ILP, we define the peakwarp as
$$\mathcal{PW}=\min\{w~| \mid D(w)-D(w')\mid >\epsilon,\forall w'\in W\}$$ 
This is computed from $D$ that has a set of active warps $W$, and for each $w\in W$, there is a corresponding throughput per cycle $tput=D(w)$. This metric is derived by modifying the standard $\argmax$ function by introducing a small positive threshold parameter $\epsilon$. This, in turn, ignores minor variations in throughput values from the maximum (a small positive threshold $\epsilon$). As seen in Figure \ref{fig:throughput}, $\mathcal{PW}$ (highlighted as a red dot) is the warp value present on the x-axis of the red dot. Note that the throughput value starts oscillating after this highlighted point. 
\end{enumerate}

%
\lstset{breaklines=true}
\begin{lstlisting}[caption=PeakWarps Calculation Code ,frame=tlrb, label=code2,captionpos=b]
int throughput_func(dim3 Db,dim3 Dg){
  cudaEventRecord(start, 0);  throughput_kernel<<<Dg, Db>>>(d_dummy);
  cudaEventRecord(stop, 0);  cudaEventSynchronize(stop);
  cudaEventElapsedTime(&elapsedTime, start, stop);     
  /* return per unit throughput using time, GPU clock speed etc. */
  return throughput_per_cycle;
}
\end{lstlisting}
In order to execute the benchmark programs for measurement, we have taken the following steps to avoid quantification errors.
\begin{itemize}
    \item Compiler optimization has been disabled to zero level to avoid undesirable optimizations done even at zero level, results of  GPU kernel execution were  stored in dummy variables as shown in listing~\ref{code1}. The use of dummy variables to avoid compiler-level optimization is a common practice in designing microbenchmarks \cite{hristeamicro,papadopoulou2009micro}.
    \item  We ensured that only one single thread is executed for measuring the latency values to avoid latency hiding due to parallelism, by  launching the latency kernel with only one thread per block in one grid. 
    \item While computing throughput values, we have taken care that occupancy of SM is always above 90\%. This was ensured by launching 256 threads per block per SM. 
\end{itemize}

\begin{table}[htp]
\centering
\caption{Tesla K20 Compute Instruction Summary}
\small
\label{instruction_summary}
\resizebox{\linewidth}{!}{
\begin{tabular}{|l|l|l|l|l|l|l|l|}
\hline
\multirow{2}{*}{\textbf{Instruction}} & \multirow{2}{*}{$\mathcal{L}$} & \multicolumn{3}{l|}{\textbf{Throughput ($\mathcal{TP}$)}} & \multicolumn{3}{l|}{\textbf{PeakWarps ($\mathcal{PW}$)}} \\ \cline{3-8} 
                                      &                                   & ILP=1         & ILP=2   & ILP=3        & ILP=1        & ILP=2        & ILP=3        \\ \hline
addf,subf,mulf                        & 9                                 & 122          & 128         & 167         & 36          & 20          & 18          \\ \hline
adds,subs,and                         & 9                                 & 120          & 127         & 136         & 36          & 20          & 18          \\ \hline
fma                                  & 10                                & 119          & 95          & 143         & 36          & 20          & 16          \\ \hline
mads                                  & 20                                & 31           & 28          & 25          & 20          & 10          & 8           \\ \hline
muls                                  & 9                                 & 28           & 32          & 32          & 8           & 8           & 8           \\ \hline
divs                                  & 424                               & 2.35         & 2.5         & 2.36        & 32          & 32          & 32          \\ \hline
divf                                  & 894.5                             & 1.066        & 1.0         & 1.02        & 32          & 32          & 32          \\ \hline
sqrt                                  & 359                               & 3.48         & 3.19        & 3.2         & 40          & 40          & 40          \\ \hline
setp                                  & 22                                & 50           & 50          & 50          & 36          & 28          & 28          \\ \hline
cvt                                   & 10                                & 31           & 31          & 31          & 12          & 12          & 12          \\ \hline
mov                                   & 2                                 & 150          &      N/A       &      N/A       & 32          & N/A         & N/A         \\ \hline
\end{tabular}
}
\end{table}
\begin{table}[htp]
\centering
\caption{Measured PTX instruction latencies $\cal L$}
\label{tab:latency-table}
\resizebox{\linewidth}{!}{
\begin{tabular}{|l|l|l|l|l|l|}
\hline
\textbf{NVIDIA GPU ->} & \textbf{Quadro K4200} & \textbf{Tesla K20} & \textbf{Tesla M60} & \textbf{GTX 1050} &  \textbf{Tesla V100} \\ \hline
\textbf{Architecture ->} & Kepler & Kepler & Maxwell & Pascal &  Volta \\ \hline
addf & 10 & 9 & 15 & 15 & 15 \\ \hline
adds & 9 & 9 & 15 & 15 & 15\\ \hline
subf & 10 & 9 & 15 & 15 & 15\\ \hline
subs & 10 & 9 & 15 & 15 & 15\\ \hline
mulf & 9 & 9 & 15 & 15 & 15\\ \hline
muls & 9 & 9 & 86 & 86 & 15\\ \hline
and & 9 & 9 & 15 & 15 & 15\\ \hline
fma & 9 & 10 & 188 & 12 & 232\\ \hline
mads & 18 & 20 & 100 & 15 & 30\\ \hline
divf & 1252 & 894.5 & 1278 & 1398 & 977 \\ \hline
divs & 418 & 424 & 1026 & 503 & 815\\ \hline
cvt & 33 & 10 & 195 & 195 & 218\\ \hline
sqrt & 440 & 359 & 550 & 481 &  487\\ \hline
setp & 22 & 22 & 30 & 30 & 30\\ \hline
mov & 2 & 2 & 51 & 55 & 49 \\ \hline
shared load \& store & 40 & 47 & 38 & 39 & 39\\ \hline
\end{tabular}
}
\end{table}


%
 \subsection{Computing Instruction Latency Results}
The latency values obtained using proposed microbenchmarks are presented in the second column of Table \ref{instruction_summary}. As seen in Table \ref{instruction_summary}, a significant difference is observed  in the results obtained by varying ILP values for each type of instruction. In an ideal scenario, in Tesla K20 SM with four warp schedulers with two instruction dispatchers, 256 instructions can be dispatched every cycle. Observed throughput value without independent instructions (e.g. 120 adds, subs instruction per cycle for ILP=1) suggests that  at least a few of the dispatchers may remain underutilized due to the non-availability of independent instruction from the same warp. We also observed that incrementing ILP after 3 (for most of the instructions) does not significantly affect throughput, which can be correlated to the fact that both the dispatch units of the scheduler are getting fully utilized if there are more than three independent instructions available. We present latency of computing instructions across multiple GPU architectures in Table \ref{tab:latency-table}.

\section{Memory Instructions} \label{sec:memory}
NVIDIA GPU offers multiple memory options, such as global, shared, constant, and texture. 
We focus on only global and shared memory instructions in the current work. We have developed microbenchmarks for quantifying latency and throughput of global and shared memory instructions. We will consider constant and texture memory as a future work. Latency for accessing global memory instructions (global load and store) depends upon the amount of data being accessed at a particular moment to account for the additional waiting time due to resource constraints. Due to its high latency value, global memory access heavily influences the execution time of a CUDA kernel. Therefore, it is crucial to accurately estimate the latency value of global memory access for a useful execution time prediction\cite{Hong2009,resios2011}.  

Among the wide variety of existing approaches, pointer-chasing based microbenchmarking is considered an accurate and popular approach to compute memory instruction latency. The pointer chasing algorithm for GPU is presented in Algorithm  \ref{algo:mlatency_host} for the host (CPU) and Algorithm~\ref{algo:mlatency_device} for the device. In the host algorithm, the array is initialized with stride values then the kernel code is invoked. Inside kernel code in the device algorithm, the start time and end time of the memory access instruction with the pointer chasing approach is recorded. In line 4 of Algorithm~\ref{algo:mlatency_device}, we ensure that the next memory instruction is not executed until the previous instruction is complete. Hence, this approach can record the number of cycles utilized for executing one single memory instruction. An iterator is used to execute the instruction a large number of times to ensure the latency is not too small to be measurable. We store the value of {\tt j} in a dummy variable to avoid any compiler-level optimization. 
\subsection{Pointer Chasing}
A pointer-chasing microbenchmark, first introduced by Saavedra et al. \cite{Saavedra1992} for CPUs, initializes a set of array elements with the index of the next memory access. The distance between two consecutive memory accesses is called stride size. The latency of memory access is the time difference in clock cycles between the memory access issue and the data availability in the processor register. In the pointer-chasing experiment, the complete array is traversed sequentially to record the average memory access latency. This approach was adapted for GPUs as well~\cite{Wong2010,Meltzer2013}. We have modified the approach by~\cite{Meltzer2013} for latency computation of GPU memory instructions.

\subsection{Global Memory Pointer Chasing}
The pointer chasing algorithm for GPU is presented in Algorithm  \ref{algo:mlatency_host} for the host (CPU) and Algorithm~\ref{algo:mlatency_device} for the device. In the host algorithm, the array is initialized with stride values then the kernel code is invoked. Inside kernel code in the device algorithm, the start time and end time of the memory access instruction with the pointer chasing approach is recorded. In line 5 of Algorithm~\ref{algo:mlatency_device}, we ensure that the next memory instruction is not executed until the previous instruction is complete. Hence, this approach can record the number of cycles utilized for executing one single memory instruction. An iterator is used to execute the instruction a large number of times to ensure the latency is not too small to be measurable. We store the value of {\tt j} in a dummy variable to avoid any compiler-level optimization. The code snippet for pointer chasing is available in Appendix \ref{AppendixB}.
%
%
\begin{algorithm}
    \caption{Memory Latency Host (CPU) Algorithm}
    \label{algo:mlatency_host}
    \begin{algorithmic}[1]
   \State initialise stride
    \For{ k= 0 to N }
      \State h\_arr[k]=(k+stride) \% N;
   \EndFor
   \State Copy host array (h\_array) to device ( d\_arr)
    \State memLatKernel<<<Dg, Db>>>(d\_dummy,d\_arr);   \Comment{ Call Latency Kernel }
         \end{algorithmic}
\end{algorithm}

\begin{algorithm}
    \caption{Memory Latency Device (GPU) Algorithm}
    \label{algo:mlatency_device}
    \begin{algorithmic}[1]
     \Procedure{GMmemLatKernel}{d\_dummy, d\_array}
   \State start\_time = clock();
   \State initialise j=0;
\For{ it=0 to iteration }
   \State	j=d\_arr[j] ;
   \EndFor
   \State end\_time = clock();
   \State d\_dummy =j;
   \State latency = (end\_time - start\_time) / iteration \Comment{ average memory latency }
       \EndProcedure
      \end{algorithmic}
      
\end{algorithm}

\subsection{Global Memory Latency Model}
 We observed that in existing studies, the reported latency value for global memory instructions is above 400 cycles for a Kepler architecture \cite{Kothapalli2009,Mei2017}. However, our pointer chasing microbenchmark on Tesla K20 reported the average reported latency 221! 
 If we use the high value reported in the literature (e.g. 580 as mentioned in \cite{Kothapalli2009}), the resultant prediction model can be a gross overestimation. We further noticed that the latency value is dependent on launch parameters, hence it is not fixed. This observation led to the building of a regression model for global memory latency which computes observed latency based on launch parameters. This will ensure that based on the launch configuration, the latency value will vary from the lowest to the maximum observed latency. 
\paragraph*{Data Collection:} We launched the microbenchmark discussed earlier repeatedly to record global load/store instruction latency by varying its launch parameters ($nB$, $nT\_b$). We ensured that we measured the average time taken to execute one instruction per thread by varying the number of threads scheduled. While recording the latency, we disabled the caching of data for memory instructions. We collected a significant number of datapoints to represent the wide range of latency values observed.  Each data point was collected by taking an average of recorded time.  
\begin{enumerate}
    \item Recorded latency values were plotted against the stride of access ($nT\_b \times nB$). As seen in Figure \ref{fig:global_latency}, the plot is non-linear. There are three breakpoints, each delimiting two linear relationships.
    \item A piecewise linear regression model of the form ${\mathcal L}=a\cdot nT\_b\cdot nB+b$ fits the best.  Table \ref{tab:my-globallatency}, presents the four equations for the linear regression model. 
\end{enumerate} 
The proposed latency model's evaluation metrics were MSE: 22.95, RMSE: 4.79, R-Square:  0.99, RSS: 32801.69, and RSS: 4.80. The R-square value (0.9967) observed for this model shows the goodness of fit for this data. 

\begin{table}
\centering
\caption{Latency computation of global memory access instructions for Tesla K20 GPU}
\label{tab:my-globallatency}
\begin{tabular}{|l|l|}
\hline
\textbf{Stride Interval} & \textbf{Global Load/Store Latency } \\ \hline
nT\_b*nB \textless 4096 & $\mathcal{L}= 0.02828\times nT\_b\times nB+220$ \\ \hline
4096\textless nT\_b*nB \textless{}24576 & $\mathcal{L}= 0.004780\times nT\_b\times nB +251.7$ \\ \hline
24576\textless  $nT\_b*nB$ \textless{}991232 &$\mathcal{L}=0.0001679\times nT\_b\times nB+307.8$ \\ \hline
991232\textless nT\_b*nB \textless{}2203648) & $\mathcal{L}=-0.00002529\times nT\_b\times nB+501.8$\\ \hline
\end{tabular}
\end{table}
\begin{table}
    \centering
    \caption{Evaluation of Model for latency of global memory access instructions }
    \label{Table_global_mem}
       \begin{tabular}{|l|l|l|l|l|l|}
            \hline
            MSE & RMSE & \multicolumn{1}{r|}{R-Square} & RSS & RSE & P-value \\ \hline
           22.95 & 4.79 & 0.99 & 32801.69 & 4.80 & 2.20e-16 \\ \hline
       \end{tabular}%
\end{table}

\subsection{Shared Memory Access Latency}
Shared memory is a non-cached memory shared amongst threads within a block. Its latency is much lower than global memory instructions because of its chip location (each SM has a dedicated shared memory space). To avoid long latencies of global memory access, application developers can move the data into and out of shared memory from global memory before and after operation execution.  For shared memory, accesses are issued individually for each thread. We use the pointer chasing approach shown in Algorithm \ref{algo:shared_latency} for shared memory access latency. The approach is similar to the global memory pointer chasing method reported in Algorithm~\ref{algo:mlatency_device}. Here we declare a shared memory array ({\tt shdata[]}) which is first initialized with stride values. In line 8, pointer chasing is utilized to ensure only one instruction per thread is executed. 

Shared memory is divided into equally sized memory modules called banks which are accessed simultaneously whereas global memory is accessed in strides. The latency reported for global memory is at least 100x higher than shared memory. The variations in the reported latency values for shared memory are very low compared to the variations in the global memory. Therefore we did not employ any regression model for shared memory. Results of shared memory latency recorded across architectures are presented in Table \ref{tab:latency-table}. 

\begin{algorithm}
    \caption{Shared Memory Kernel Algorithm}
    \label{algo:shared_latency}
    \begin{algorithmic}[1]
     \Procedure{SMmemLatKernel}{d\_dummy, d\_array}
    \State Declare shdata[] as shared memory array 
    \For{ i=0 to N }
     \State  shdata[i] = d\_array[i]; 
    \EndFor
     \State start\_time = clock();
   \State initialise j=0;
   \For{ it=0 to iteration }
   \State	j=shdata[j] ;
   \EndFor
   \State end\_time = clock();
   \State d\_dummy =j;
   \State latency = (end\_time - start\_time) / iteration \Comment{ average memory latency }
\EndProcedure
\end{algorithmic}
\end{algorithm}

\section{Kernel Launch Overhead}
\begin{figure}[ht]
    \centering
    \begin{minipage}{.5\textwidth}
        \centering
\includegraphics[width=\linewidth]{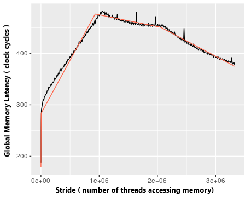}
\caption{Piece-wise model for global instruction latency}
\label{fig:global_latency}
\end{minipage}%
    \begin{minipage}{0.5\textwidth} \centering
\includegraphics[width=\linewidth]{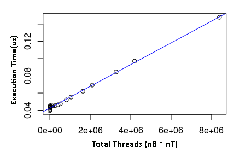}
\caption{Kernel Launch Overhead}
\label{fig:launch_overhead}
\end{minipage}
\end{figure}

A CUDA kernel execution time is impacted by the time to launch a kernel in a GPU, defined here as the \textit{kernel launch overhead}. This is the time consumed just before and after executing the kernel instructions. We have constructed an empirical model to characterize this overhead.  We ran an empty kernel (no instructions) with different configurations(number of threads, threads per block). 
\begin{enumerate}
    \item We recorded the execution time of this empty kernel by changing its launch parameters (number of threads, number of blocks per thread). Resios et al. \cite{resios2011} claimed that kernel launch overhead could be modeled with constants since its value does not change. However, we discerned that the execution time of an empty kernel increases with an increase in the number of threads being launched.
    \item  We built three statistical models to reproduce its behavior and assessed them using the R-square goodness of fit (shown in Table \ref{RsquareModel}). The higher value of the R-square ensures that the model describes the data best. 
    \item The plot of the linear regression model (for Tesla K20 data), which gives maximum R-square ($0.9877$), is depicted in Figure \ref{fig:launch_overhead}. Kernel Launch Overhead ( $l\_overhead$ ) model for Tesla K20 GPU is $0.00002\cdot nB\cdot nT\_b + 1.4489$. Similarly, we built the kernel launch overhead model using linear regression for each GPU architecture under study.
\end{enumerate}
The code snippet of Kernel Launch Overhead is available in Appendix \ref{AppendixB}. 

%


\section{Discussion}
We have designed microbenchmarks for the NVIDIA GPU, which can be used for optimization, simulation, and analysis of GPU software. Microbenchmarking provides a deeper understanding of the execution of computing instruction by quantifying its performance in latency, throughput and peakwarps metrics. Although there are existing studies on microbenchmarking GPU, we have made novel contributions in presenting characteristics of computing and memory instructions. One of the novelty of this work is reporting the change in throughput and peakWarps with different values of ILP. This change in throughput and peakwarps values is utilised for building a performance model in Equation \ref{equationDelay} of Chapter \ref{ch:execTime}  for computing instruction delay.  We have also proposed a novel model for calculating global memory latency which includes microbenchmarking and piece-wise linear regression, which is not carried out in any other study to the best of our knowledge. This model utilises pointer chasing approach for microbenchmarking. A piecewise linear regression model is fit over this results to ensure that the global memory latency is neither overestimated and not underestimated. We also computed kernel launch overhead which influences the performance of a GPU application as well.

Our observed results were validated against other existing studies in microbenchmarking \cite{arafa2019low, cornelis2019pipeline}. Computing instruction and shared memory instruction latency values equal other micro benchmarking studies. Throughput values were found to be fairly close to the theoretical ones for compute capability 3.5, as mentioned in \cite{CUDA2017}. The microbenchmarking model developed in this Chapter is utilised for predicting performance and power prediction in Chapters ~\ref{ch:execTime} and \ref{ch:powerpred}, respectively. Memory instruction execution is impacted by factors such as uncoalesced accesses and data races. We have not considered them in latency calculations. In the future, we would like to explore microbenchmarking all types of memory instructions  in GPU memory hierarchy, focusing on various access factors affecting memory instruction performance.

\let \textcircled=\pgftextcircled
\chapter{Modelling GPU Performance} \label{ch:execTime}

\initial{M}icrobenchmarking approach discussed in Chapter~\ref{ch:microbench} provides the foundation for the execution time prediction model of a GPU kernel. Using the microbenchmark results, we build an analytical model to predict the execution time of a GPU kernel by analyzing the intermediate PTX code of a CUDA kernel, without the need of executing it.  

Performance prediction involves estimating the execution time of executable software or estimating values of other performance factors of a computing machine, such as instruction throughput, cache misses, and so on. In this thesis, we use the term ``performance prediction'' to imply the predicted execution time of a GPU kernel.

In the absence of the internal details of a GPU, we build an abstract GPU model with characteristics gathered from static analysis, one time dynamic analysis and vendor published details which are essential for modelling GPU performance. This abstract GPU model enacts as the actual GPU hardware for referring features in computing certain performance parameters (e.g. instruction latency). It is also useful in building architecture-agnostic model since it helps to study and present common properties between multiple GPU architectures. We now describe this abstract GPU model with which we present the performance prediction model. 

\section{Abstract GPU Model} \label{abstract_model}
For the purpose of estimating the execution time of a kernel, we represent a GPU as $$\mathcal{GPU}=\langle R, \mathcal{A}, T, {\cal PTX}, R_{map}, \mathcal{L}, \mathcal{TP}, \mathcal{BW}, \mathcal{P}\rangle$$
\begin{itemize}
   \item $R$ denotes the following set of GPU resource types for each SM
    i)  $SP$: single precision cores ii) $WS$: warp schedulers iii) $SFU$: special function units iv) $DPU$: double precision units v) $LSU$: load-store units. 
  \item $\mathcal{A}$ denotes the following set of attributes of a GPU:    \\
    i) $access_{sz}$: Represents the number of bytes a GPU accesses for every memory instruction\\
    ii) $nSM$: Number of Streaming Multiprocessors \\
    iii) $nTh\_sm^M$: maximum number of threads per SM \\ 
    v) $L2\_sz$: L2 cache size \\
    vi) $\nu_{gpu}$: GPU clock frequency and \\
    vii) $\nu_{mem}$: memory clock frequency \\
    (viii) $reg\_b^M$: Total number of registers available per block \\
    (ix) $shm\_b^M$: Total number of bytes of shared memory available per block \\ 
  (x) $nB^M$: Maximum thread blocks per SM \\
  (xi) $wSM^M$ : Maximum number of active warps on an SM\\
  (xii) $Sz_{w}$ : Warp size represents number of threads per warp\\
  (xiii) $Sz_{gl}$ : Number of bytes in a single global memory transaction
  
    \item $T$ is the set of all instruction types, namely \{{\it Compute (C), Global Memory (Gm), Shared Memory (Sm), Miscellaneous (M)}\}. Each instruction type is associated with a resource.
    \item ${\cal PTX}$ denotes the set of PTX instructions that can be obtained from the NVIDIA GPU documentation. Each PTX instruction is associated with an instruction type.
    \item $R_{map}:{\cal PTX}\to R$ uniquely maps an instruction to a resource type.
    \item $\mathcal{L}:\mathrm{PTX}\to\mathbb N$ is the latency function that associates a latency of a PTX instruction as a positive integer $l$. Instruction latency stands for the number of clock cycles taken by an instruction to complete its execution.
    \item $\mathcal{TP}$: Represents throughput of computing, global and shared memory instructions.
    \item $\mathcal{BW}$ : Represents bandwidth for global load and store transfers of memory instructions
    \item $\mathcal{P}$: Denotes a set of overheads and penalties, some based on $\mathcal{TP}$. We consider the following: 
i) $l\_overhead$: Kernel Launch Overhead function ii) $gm\_penalty$: Global Memory Penalty function iii) $sm\_penalty$: Shared Memory Penalty function and iv) $cm\_penalty$: Cache Miss Penalty function
\end{itemize}

\begin{table}
\centering
\caption{Notations used for model attributes}
\label{tab:Notations}
\begin{tabular}{|l|l|}
\hline
\multicolumn{2}{|c|}{\textbf{Dynamic Analysis Attributes}}                                  \\ \hline
\textbf{Feature}                      & \textbf{Source of computation} \\ \hline
Latency $\mathcal{L}$                 & microbenchmarking             \\ \hline
Latency for global memory access      & Linear regression Model             \\ \hline
$l\_overhead$  & Linear Regression Model        \\ \hline
$gm\_penalty$       & Exponential Model              \\ \hline
$sm\_penalty$       & Exponential Model              \\ \hline
$cm\_penalty$       & Parameterised Equation         \\ \hline
\multicolumn{2}{|c|}{\textbf{Hardware Attributes}}                       \\ \hline
Device attributes $\mathcal{A}$   & Device Query   \\ \hline
Number of instances for each resource $r\in R$ & Architecture Documentation     \\ \hline
\multicolumn{2}{|c|}{\textbf{Inputs}}                                  \\ \hline
$nB$: Number of thread blocks            & User Input                     \\ \hline
$nT\_b$: Number of threads per block         & User Input                     \\ \hline
$nLoop$: Number of loop iterations              & User Input                     \\ \hline
$reg\_t$: Number of registers per thread &  NVCC Compiler                  \\ \hline
$shm\_b$: Shared memory per block &   NVCC Compiler                     \\ \hline
\end{tabular}
\end{table}

This model is constructed once for a particular GPU architecture. Table \ref{tab:Notations} presents the notations used for model attributes.

\section{Static Analysis} \label{static_analysis}
In order to collect various execution features of a CUDA kernel, we perform a static analysis of a CUDA kernel by compiling the code into a PTX format\cite{PTX} using the {\tt nvcc} compiler. Subsequently, we construct a control-data flow graph for each kernel code in PTX representation. Here, we model each basic block~\cite{Allen1970} as a data flow graph 
$G=\langle V, E\rangle$ 
where each vertex $v\in V$ denotes a PTX instruction and a directed edge $e:u\to v$ denotes that the computation of $v$ is data-dependent on $u$. The edge $e$ is labeled with the latency value $\mathcal{L}(u)$. Furthermore, for each $v\in V$, the $Rmap(v)$ function can be used to get the corresponding resource type. 

The control-data flow graph~\cite{Allen1970}, used to model the entire kernel, is represented as $G_{cfg}=\langle\mathbb{C, E}\rangle$ where each node $B\in \mathbb C$ is a basic block associated with a data-flow graph\cite{Hecht1974,Peterson1973,KOSARAJU1974}. Since the edge of a CDFG represents a control flow, the edge is not labelled with any latency information. A back edge of a CDFG represents a loop. A back-edge is annotated with the loop iteration count $nLoop$. Since the CUDA program follows a structured programming style, it is possible to convert a CDFG as a set of regions~\cite{Wulf1975} where each region is a set of basic blocks, and a back-edge in a CDFG becomes a self-loop of a loop-region~\cite{Wulf1975}. Such a transformation is very common in any compiler during code generation. Hence we do not elaborate on this any further. 

\begin{figure}[htbp]
\includegraphics[width=15cm]{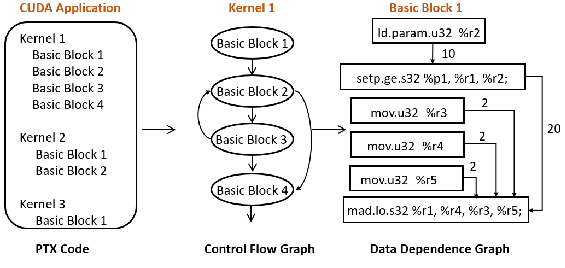}
\caption{Static Analysis of CUDA Code}
\label{fig:static_analysis}
\end{figure}

The static analysis phase is illustrated with an example in Figure \ref{fig:static_analysis}.

We proposed two approaches for modelling GPU performance. The first approach is using parameterised equations to capture Instruction Level Parallelism (ILP) and Thread Level Parallelism (TLP). Second approach for performance modelling is inspired by resource allocation. In the following section, we present the first approach in detail and then elaborate on the approach for building a resource allocation based model for execution time prediction. 

\section{Modelling GPU Performance using ILP \& TLP} \label{pp_approach1}
As discussed in the literature survey (refer to Chapter~\ref{chap:ls}), parallelism is an important factor that influences the performance \cite{resios2011} of a GPU. For all types of instructions, increasing parallelism (either by increasing the number of threads or the number of independent instructions) leads to an increase in performance bounded by hardware constraints. Resios et al. \cite{resios2011} demonstrated  that performance increases linearly with parallelism through experimental study. They demonstrated this phenomenon with transfer overhead, memory bandwidth, and instruction throughput. 

Parallelism in GPU is achievable through two factors: instruction level and thread-level parallelism. At the instruction level, parallelism is achieved by having independent instructions within a thread that hide latency to achieve better performance. At the thread level, with the increasing number of threads, there are more independent instructions hence an increase in performance till the hardware limit is reached. In the past, Resios et al.\cite{resios2011} tried to model GPU behaviour with the parallelism modelling approach. However, they have implemented it on Fermi architecture along with a very naive scheduling algorithm.  

This observation has motivated us to investigate the role of Instruction Level Parallelism (ILP) and Thread Level Parallelism (ILP) in determining a GPU application's performance. Accordingly, we plan to model ILP and ILP using parameterised equations to predict the performance of GPU applications. The details of this approach are presented in Section~\ref{ch:execTime:ispa}.

\subsection{Proposed Model}\label{ch:execTime:ispa}
\begin{figure}
\centering
    \includegraphics[width=15cm]{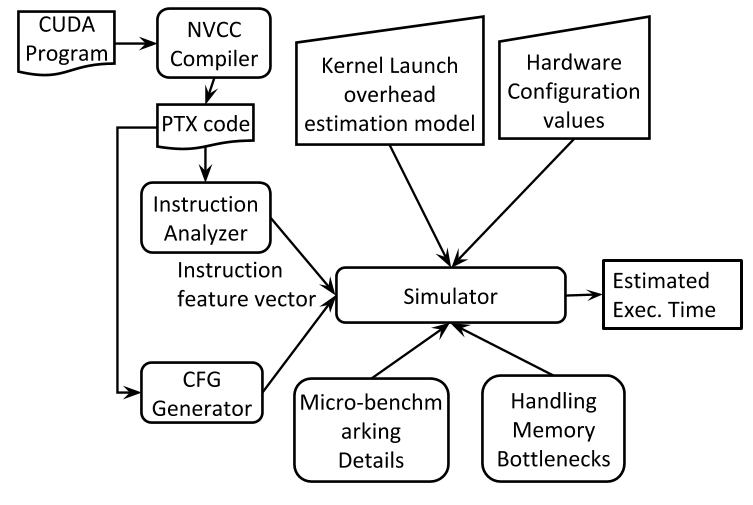}
    \caption{Proposed Execution Time Prediction Model}
    \label{fig:model}
\end{figure}

We present an analytical model for predicting the execution time of CUDA kernel using static analysis of PTX code as shown in Figure \ref{fig:model}, involves two main phases:\\
\indent{\bf 1}. Calculating the execution time of a single thread by adding up individual instruction delays using Delay Computation Algorithm (DCA).\\
\indent{\bf 2}. Calculating the overall execution time of the kernel by modeling the scheduling of threads on Streaming Multiprocessors (SM)using GPU Scheduling Algorithm (GSA). 

Execution Time Model proposed in this work is modelled as a function of these two phases:  
 \begin{equation} \label{eq1}
    t_{kernel} = f(DCA , GSA)
\end{equation} 

In the first phase, we gather the microbenchmarking details for each instruction type along with hardware characteristics. After analyzing the PTX code and building a Control Flow Graph (CFG), these details are used for calculating the delay of each instruction($d_{i}$) using ILP, TLP, instruction features (latency, throughput,peakwarps) and instruction specific features (memory bandwidth, bank conflicts, coalescing). The total delay for the kernel($d_{k}$) is measured by using a delay calculation algorithm that utilizes $d_{i}$ and CFG. 

To execute the threads launched by a kernel, NVIDIA scheduler schedules these threads in batches of waves. The vendors do not disclose the way this scheduling is done. We proposed a simple scheduling algorithm that constitutes the next phase of this work.  

The second phase uses total kernel delay for one thread and simulates the execution of GPU by scheduling the threads to predict kernel execution time($ts_{kernel}$). Kernel launch overhead is added to the execution time computed by this scheduling algorithm. We also calculate the time penalty for each memory bottleneck (b\_penalty) and add it to the total predicted execution time. b\_penalty is summation of $sm\_penalty$ and $gm\_penalty$. 
\begin{equation} \label{eq_2}
t_{kernel}= ts_{kernel}+ l\_overhead +b\_penalty \end{equation}

All the components of these two phases are discussed in detail in section \ref{model_details_section1} and \ref{model_details_section2}.

\subsection{Delay Computation Algorithm} \label{model_details_section1}
In actual GPU execution, PTX code has to pass-through a few more intermediate optimization steps before execution on GPU. However, we do not have access to this optimized code. Hence, we dissected this PTX code to gather attributes and flow of the program which is supplied as the input to the scheduling model. 

\subsubsection{Algorithm for Computing Instruction Delays}
GPU Scheduling model involves calculating the number of GPU clock cycles it would take to execute a single thread entirely. A simplistic model of a kernel program is a sequence of PTX instructions all beginning with the operation name, and a list of register operands if the instruction makes use of operands. The sequence of instructions is read from the PTX code, and one by one, the delay of each instruction is computed using a parametric model proposed by Andreas et al. \cite{resios2011}. Parametric equations from this thesis which are used in this model are discussed in the following sections. 

While the base model remains similar, it requires a significant modification to have an accurate estimation of the execution time. Specifically, we use the base model only to compute instruction delays and then construct a scheduling simulation to calculate the total execution time. In the scheduling algorithm, for computing the number of cycles, we assign the instructions to each dispatching unit serially. However, if there is dependent instruction, it is considered in the next cycle. For brevity and completeness, we have discussed the base model used to compute instruction delays in algorithm \ref{delay_algorithm}. 

\subsubsection{Instruction Level Parallelism}
Within a basic block, the instruction level parallelism factor, which is the effective number of instructions per cycle, is calculated after traversing the block of instructions to count the number of instructions that can be executed independently. The registers used in the instructions are compared to detect dependencies between them.
Here $ILP = \frac{n_{inst}}{n_{cycles}}$, where $n_{inst}$ is the total number of instructions executed in the basic block, and $n_{cycles}$ is the actual count of the number of independent execution cycles for computing number of cycles
we allot the instructions to each dispatching unit serially. However, if there is dependent instruction, it is considered in the next cycle. 

\subsubsection{Thread Level Parallelism}
The instructions in the kernel are taken from the basic block list constructed in the PTX analysis stage. The value of the maximum number of active warps on an SM $wSM^M$ is a hardware upper limit. Since it takes into account the actual kernel launch parameters, this can be taken as the initial thread-level parallelism $TLP$. This is done because the number of blocks launched by a kernel is usually enough to constitute a full wave. For more accuracy, this can be dynamically recalculated at every stage, but in most cases, the hardware limit can be taken as $TLP = wSM^M$. $TLP$  is the number of active warps on an SM. It is a crucial factor to be considered while estimating the execution time of all the instructions.

\subsubsection{Instruction Delay Computation} \label{peakwarps}
Some of the parameterized equations used in this work are influenced by the GPU performance prediction model of Andreas et. al\cite{resios2011}. They have successfully demonstrated these equations for Fermi architecture. We are tuning these equations by improvising them for Kepler architecture.

As stated by their approach, if the number of warps is not enough to reach peak performance, all instructions are executed in parallel, and the latency is hidden because of the thread parallelism.  Once the pipeline of instructions becomes full, the subsequent warps are required to wait for the cores to be freed, and hence stalls occur.  In this case, an extra penalty factor($\frac{Sz_{w}}{\mathcal{TP}}$) is required to measure the effects of the pipeline stalls. Our contribution to the approach by  Andreas et. al\cite{resios2011} is the use of peakWarps and throughput calculated using microbenchmarking. Andreas et. al\cite{resios2011} used the theoretical values for maximum warps per SM and throughput, whereas we computed instruction delay using the observed values computed using microbenchmarking which improves the accuracy of the model.

For a compute instruction $i$,  the delay is computed as 
\begin{equation}d_{i}^c = \left\{
\begin{matrix}
\frac{\mathcal{L}}{ILP \times TLP}    & if ILP \times TLP \leq  peakWarps   \\ \\
\frac{\mathcal{L}}  {ILP \times TLP \times peakWarps} + \frac{Sz_{w}}{\mathcal{TP}} & otherwise \\
\end{matrix}\right.
\label{equationDelay}
\end{equation}
Similarly for a memory access instruction $i$, delay is computed as 

\begin{equation}d_{i}^m = \left\{\begin{matrix}
\frac{\mathcal{L}}{ILP \times TLP} & if ILP \times TLP \leq  peakWarps\\ \\
\frac{\mathcal{L}}{ILP \times TLP} + p^m & otherwise \\
\end{matrix}\right.
\label{equationMemoryDelay}
\end{equation} 
In the case of memory access requests, serialization occurs once the number of memory access requests is too high to be fulfilled at once by the GPU hardware owing to bandwidth and transfer size constraints. Like in the case of computing instructions, a penalty factor ($p^m$) is added to model the serialization effect\cite{resios2011}.

\subsubsection{Memory Access Latency}
The penalty of the memory access instruction differs based on whether the access is a global memory access or a shared memory access because the architecture and speeds of both these types of memory are vastly different. Global memory instructions contribute highly to total instruction delay since the number of clock cycles taken by this instruction are maximum amongst all instructions. In the case of global memory access, the penalty of global memory instruction is given by: \begin{equation}
gm\_penalty = \frac{Sz_{gm} \times c}{BW_{g}}\end{equation}
Coalescing factor, $c$ depends for every benchmark depending upon stride of access which is not possible to analyze from the PTX code. We observed that memory latency of global instruction changes with the change in the stride of access. Based on this observation, we have built a regression model based on global load/store instruction latency against stride of access. This model is explained in Chapter 3 in detail.

In case of shared memory access, penalty is computed as
$$sm\_penalty = \frac{Sz_{w} \times Sz_{sm}}{n_{sb} \times BW_{s}} + \frac{n_{bc} \times size}{BW_{s}}$$
where $Sz_{sm}$ is the number of transferred bytes to/from shared memory.
A shared memory is divided into a limited number of banks $n_{sb}$. The design of shared memory is such that consecutive accesses to the same bank are serialized and cannot be executed in parallel. The number of bank conflicts($n_{bc}$) is also difficult to model by analyzing the PTX code and will be considered in the future work.  We observed latency of 47 cycles for shared memory access instruction using microbenchmarking.


\subsubsection{Delay Calculation Algorithm}
Instructions on GPU cores execute in SIMT fashion. If there is branch divergence within a warp, each of the instructions in both branches is traversed, and the threads which do not take a  branch remain idle during this time, while the remaining threads execute the instruction.

To find the instruction count, an assumption is made that at least one thread takes each branch in every warp, i.e., no branch remains untraversed during the execution of a warp. The total instruction count is obtained by summing up the instructions in sequential basic blocks. If a loop is detected, the instruction count of the looped path is multiplied by the number of iterations, and the computation continues. So, the total delay of the kernel is the sum of delays of all the basic blocks in the kernel. We consider this as the number of cycles that a single thread (or warp) takes to execute on the GPU.

The execution time of a single thread, $t_{thread}$, is given by
\begin{equation}
t_{thread} = \frac{d_{k}}{\nu_{gpu}}\end{equation}

where the overall delay of kernel($d_{k}$) is $d_{k} = \sum_{insts} d_{i}$.

\begin{algorithm}
    \caption{Delay calculation algorithm}
    \label{delay_algorithm}
    \begin{algorithmic}[1]
        \Procedure{Delay calculation}{$\mathcal{GPU , L, TP}$}
        $d_{k} = 0$ 
        \For{each $BasicBlock$ in $Kernel$ }
        \State $d_{B} = 0$
        \For{each $Inst$ in $BasicBlock$}
        \State Calculate $d_{i}$
        \State $d_{B} = d_{B} + d_{i}$
        \EndFor
        \If{$BasicBlock$ has a loop}
        \State $d_{B} = d_{B} * n_{i}$
        \EndIf
        \State $d_{k} = d_{k} + d_{B}$
        \EndFor
        \EndProcedure
        \State \textbf{return} $d_{k}$
    \end{algorithmic}
\end{algorithm}

\subsection{GPU Scheduling Algorithm} \label{model_details_section2}
Since a GPU core can execute one thread at a time, the number of threads that can simultaneously be launched on a GPU is limited by the total number of cores. More threads may be available for execution, but these threads must wait until the cores become available for them to execute. Because these threads wait, they are effectively serialized.  A ``wave'' in our model is the number of threads required to occupy all the cores of the GPU within the limits of resource constraints like shared memory and register requirements. All the threads can be grouped into such waves. Counting the number of such waves leads to the count of the number of serialized sets of threads, which is denoted by the parameter $n_{serial}$.

\subsubsection{Execution time per thread}
The first step to estimate the execution time is to calculate the number of GPU clock cycles required by a thread to execute the kernel. This relies on the premise that each instruction takes different clock cycles, which is again dependent on several microarchitecture hardware factors and program characteristics. The execution time of a single thread say $t_{thread}$, is given by:
\begin{equation}t_{thread} = \frac{\sum_{i\in insts} d_{i}}{\nu_{gpu}}\end{equation}
where the value of $d_i$ depends on the type of instruction.

\subsubsection{Handling Memory Bottlenecks}
Memory bottleneck gives a significant contribution to the error of execution time estimations. It is because memory bottlenecks not only depend on the program but also in the state of the hardware system. Bottlenecks in the performance of a program can be analyzed with the theoretical approach based on an in-depth understanding of the scheduling algorithm. We incorporated one such approach by Luo et al. \cite{luo2011} in this model. It calculates a parameter MPD or memory access parallel degree that represents the maximum warp number that can be executed in parallel at a time. Using MPD, we can calculate the time penalty for each bottleneck and add it to the total predicted execution time.
The global memory bandwidth limits memory transfer speed. So, if too many warps perform
the memory operation, due to limited bandwidth some of the warps will need to wait till the bandwidth become available. Hence this kind of execution will be a serial execution. In order to calculate the upper limit on the number of warps that can execute at the same time, we computed per warp memory bandwidth. Then we obtained the MPD by dividing the GPU bandwidth by the bandwidth of all the warps in execution. Using MPD, we can get the time penalty of the bottleneck which is added to the total execution time at every occurrence of the bottleneck.

\subsubsection{Parameters used in scheduling algorithm}
The kernel launch parameters, i.e. the number of blocks ($nB$) and the number of threads per block ($n_{tb}$), must be given to the scheduling model. The total number of threads launched is calculated from the kernel launch parameters by $nT = n_{tb} * nB$, and a number of warps per thread block is given as $n_{wb} =\frac{n_{tb}}{Sz_{w}}$.
All threads of the same thread block are executed in the same SM. Assuming a balanced load scheduling model, each SM in the GPU gets an approximately equal number of blocks to execute. So the number of blocks assigned per SM is obtained by equally dividing the blocks among all available SMs i.e. $nTh\_sm=\frac{nT}{m_{tSM}*nSM} $. The maximum number of blocks that can execute on an SM in one wave is computed using maximum blocks per SM by $nB^M = nTh\_sm/nSM$. 

\subsubsection{Algorithm for scheduling threads}
CUDA provides an abstraction over how the execution of threads occurs. When it comes to simulating the GPU, terminology differs from that of program launching parameters. In this section, we will discuss the simulation algorithm in terms of the basic unit of execution i.e., warp. Now that the delay information is obtained, the blocks must be scheduled on the GPU such that the overall time can be estimated. The simulation algorithm takes the GPU hardware specifications and the limitations of block count reported by the CUDA Occupancy Calculator as inputs and calculates the value of $n_{serial}$ in order to determine the overall execution time. The warp scheduler in a GPU schedules one instruction from each of the threads in a warp to execute in parallel. This is modeled by assuming that an instruction $Inst$ is executed completely on all warps before the next instruction is issued to all warps. In this manner, all the instructions are executed one by one in all warps. However, this suffices to calculate overall execution time. 

We have assumed that one instruction executes per-cycle for all the benchmarks. This assumption is made after profiling benchmarks and observing its Instruction Per Cycle (IPC) value to be between 0.8 and 1.2. The number of instruction issue cycles ($n_{ic}$) for a particular benchmark is derived from the number of warps issuing instructions per cycle and the number of warps per block. $n_{ic}$ is then used to get the execution time per thread ($t_{thread}$) by multiplying it with the $t_{thread}$. The blocks are allotted to the GPUs by dividing them in a set of warps till the hardware limit of maximum warps per SM($m_{wSM}$) is reached for each SM.  This is one wave of allocation. So, a maximum of $m_{wSM} * nSM$ blocks are allotted in one wave. This continues until all the warps to be scheduled are executed. 

The simulation algorithm is given in Algorithm \ref{simulation}. In each wave, the execution time is calculated by a product of the maximum number of blocks scheduled($num_{blocksRound}$) by delay for one thread. The total execution time is the sum of individual wave's execution times. After all possible full waves are executed the execution time of any remaining blocks is added to the calculated execution time. The last wave is not a full wave as they are not enough blocks to occupy all the SMs. In the end, total execution time$t_{kernel}$ is computed by adding kernel launch overhead and memory bottleneck penalty to execution time computed by the scheduling of threads$ts_{kernel}$ in line 21 of the algorithm. 
\begin{algorithm}
    \caption{GPU Scheduling Algorithm}
    \label{simulation}
    \begin{algorithmic}[1]
        \Procedure{SCHEDULER}{$\mathcal{GPU}$, $d_{k}$}
         \State $ts_{kernel} = 0$
         \State $n_{ic} = \frac{nT_{b}}{n_{cSM}}$
        \State $t_{thread} = \frac{d_{k}}{\nu_{gpu}}$
        \State $t_{thread} = t_{thread} * n_{ic}$
        \For{each $ThreadBlock$}
        \State $num_{blocksRound} = 0$
        \State $currentSM = 0$
        \State $SMcounters[0..num_{SMs}] = 0$
        \While{$num_{blocksRound} <= nB^M$}
        \State $SMcounters[currentSM] = SMcounters[currentSM] + 1$
        \State $num_{blocksRound} = num_{blocksRound} + 1$
        \If{$currentSM == num_{SMs} - 1$}
        \State $currentSM = 0$
        \Else
        \State $currentSM = currentSM + 1$
        \EndIf
        \EndWhile
        \State $ts_{kernel} = ts_{kernel} + maximum(num_{blocksRound}) * t_{thread}$
        \EndFor
       
        \State $t_{kernel}= ts_{kernel} + l\_overhead + b\_penalty$
          \State \textbf{return} $t_{kernel}$ 
        \EndProcedure
    \end{algorithmic}
\end{algorithm}

\subsection{Experimentation Details \& Data Collection} \label{section_experimentation}
The proposed model is implemented using Java. Apart from PTX code as the primary input, the model takes in configuration files for the GPU hardware characteristics and some kernel-specific parameters which cannot be predicted (launching parameters, loop iterations) but are known to an application developer. The ratio of the number of active warps to the maximum allowed number of warps on an SM is known as the occupancy of each SM. We used NVIDIA's CUDA Occupancy Calculator \cite{Nvidia2017} to get the maximum number of blocks and warps which are taken as input for the model, as these hardware limits are required during the simulation phase. 

This work was implemented for the NVIDIA Tesla K20 GPU which consists of 13 streaming multiprocessors. Each SM of this Kepler architecture features 192 single-precision CUDA cores, 32 load/store units, 32 Special Function Units (SFUs), 64 double-precision units, and four warp schedulers (each warp scheduler has two dispatch units). 
The execution time of a kernel greatly depends on the underlying hardware characteristics of the GPU. NVIDIA's CUDA SDK includes a \textit{deviceQuery} program which, on execution, reports a list of the hardware characteristics of the GPU being used. Some parameters are obtained from the specifications of the GPU\cite{NvidiaWhitepaper2014}. The CUDA SDK also includes a \textit{bandwidthTest} program which is repeatedly executed to find the average global memory bandwidth. 

\subsection{Results} \label{section_results}
\begin{figure}
\centering
    \includegraphics[width=10cm]{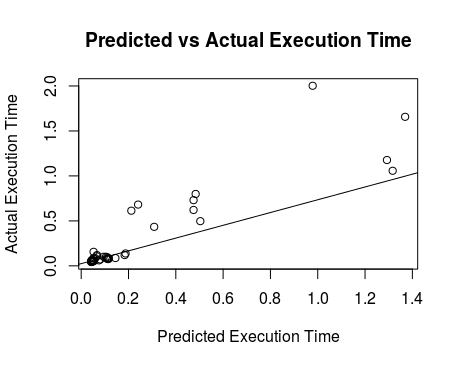}
    \caption{Execution Time Prediction Model}
    \label{fig:exectimeplot}
\end{figure}

Results are shown for this proposed model by analyzing and scheduling 45 benchmark kernels. These benchmarks are from CUDA toolkit\cite{NvidiaCUDAToolkit} and Rodinia \cite{Che2009}. Benchmarks can be classified into popular parallel dwarfs based on their functionality. The predicted execution time is compared against the actual execution time of kernels. The actual execution time for each kernel is obtained by running them on the Tesla K20 GPU using CUDA events. The mean absolute error obtained is 26.86\%. The plot of actual vs. predicted execution time for these benchmarks is shown in Figure \ref{fig:exectimeplot}. R square value for this regression plot is 0.8245, and the RSS value is 0.928. The results for the benchmarks under study are presented in Table \ref{gpu_model1}. 

\begin{table}
\caption{Results of GPU Scheduling Algorithm}
\label{gpu_model1}
\resizebox{\linewidth}{!}{
\begin{tabular}{|l|l|l|l|}
\hline
\textbf{Benchmark}      & \textbf{Kernel}              & \textbf{\begin{tabular}[c]{@{}l@{}}Execution   Time\\      (Actual)\end{tabular}} & \textbf{\begin{tabular}[c]{@{}l@{}}Execution   Time\\  ( Predicted )\end{tabular}} \\ \hline
Binomial   Options      & BinomOpt                     & 5541.7                                                                            & 4967                                                                               \\ \hline
Black   Holes           & BlackScholes                 & 544.75                                                                            & 814.83                                                                             \\ \hline
Breadth   First Search  & BFSv1\_65536                 & 60.1                                                                              & 52.12                                                                              \\ \hline
Breadth   First Search  & BFSv1\_4096                  & 47.6                                                                              & 43.07                                                                              \\ \hline
Breadth   First Search  & BFSv1\_1MW                   & 119.9                                                                             & 226.6                                                                              \\ \hline
CFD                     & CFDFlux\_193K                & 613.2                                                                             & 355.999                                                                            \\ \hline
CFD                     & CFDTimeStep\_Missile         & 119.8                                                                             & 81.44                                                                              \\ \hline
CFD                     & CFDTimeStep\_193K            & 106.72                                                                            & 79.83                                                                              \\ \hline
CFD                     & CFDInit\_Missile             & 59.58                                                                             & 54.637                                                                             \\ \hline
CFD                     & CFDInit\_193K                & 58.66                                                                             & 53.89                                                                              \\ \hline
CFD                     & CFDTimeStep\_97K             & 60.54                                                                             & 60.92                                                                              \\ \hline
CFD                     & CFDStepF\_193K               & 61.06                                                                             & 96.181                                                                             \\ \hline
Clock                   & Clock                        & 49.7                                                                              & 53.692                                                                             \\ \hline
Convolution   Seperable & ConvS\_R                     & 1657.22                                                                           & 2422.2                                                                             \\ \hline
Convolution   Text      & ConvT\_R                     & 730.37                                                                            & 787.6                                                                              \\ \hline
Convolution   Text      & ConvT\_C                     & 621.41                                                                            & 787.675                                                                            \\ \hline
Fast   Walsh Transform  & FWT\_M                       & 799.42                                                                            & 775.623                                                                            \\ \hline
Gaussian                & GaussF2\_2048                & 1355.52                                                                           & 1134.2                                                                             \\ \hline
Gaussian                & GaussF2\_4096                & 5365.5                                                                            & 4651.41                                                                            \\ \hline
Gaussian                & GaussF1\_2048                & 334.144                                                                           & 299.3                                                                              \\ \hline
Gaussian                & GaussF1\_1000                & 45.024                                                                            & 43.074                                                                             \\ \hline
K   means               & KmeansInvert                 & 7715                                                                              & 3035.6                                                                             \\ \hline
K   means               & KmeansPoint                  & 2981.41                                                                           & 1247.89                                                                            \\ \hline
LUD                     & LUDPeri                      & 55.772                                                                            & 50.66                                                                              \\ \hline
LUD                     & LUDDiag                      & 52.5                                                                              & 47.97                                                                              \\ \hline
Matrix   Multiplication & MatrixMul\_200               & 539.8                                                                             & 487.872                                                                            \\ \hline
Merge   Sort            & MergeMS\_generateSample      & 3310.6                                                                            & 2830.641                                                                           \\ \hline
NW                      & NW2                          & 48.54                                                                             & 44.87                                                                              \\ \hline
NW                      & NW1                          & 43.81                                                                             & 44.83                                                                              \\ \hline
Particle   Filter       & ParFilt\_Norm                & 368.67                                                                            & 330.88                                                                             \\ \hline
saxpy                   & Saxpy                        & 95.968                                                                            & 133.14                                                                             \\ \hline
Scan                    & scanExclusiveShared          & 433.89                                                                            & 436.014                                                                            \\ \hline
Shfl\_scan              & shfl\_initimage              & 84.54                                                                             & 88.373                                                                             \\ \hline
Sobol   Qring           & QR\_inv                      & 156.09                                                                            & 52.424                                                                             \\ \hline
Sobol   Qring           & QR\_QRNG                     & 496.89                                                                            & 503.6                                                                              \\ \hline
Srad                    & Srad1                        & 81.952                                                                            & 66.29                                                                              \\ \hline
Stereo                  & Stereo                       & 2002.4                                                                            & 1256.9                                                                             \\ \hline
Transpose               & Tpose1\_shared   memory copy & 42.465                                                                            & 42.651                                                                             \\ \hline
Transpose               & Tpose0\_simple   copy        & 101.12                                                                            & 116                                                                                \\ \hline
Transpose               & Tpose2\_Naive                & 101.38                                                                            & 129.99                                                                             \\ \hline
Transpose               & Tpose7\_diagonal             & 86.75                                                                             & 149.46                                                                             \\ \hline
Transpose               & Tpose6\_fine-grained         & 78.3                                                                              & 141.12                                                                             \\ \hline
Transpose               & Tpose3\_coalesced            & 83.26                                                                             & 154.9                                                                              \\ \hline
Transpose               & Tpose4\_optimized            & 83.26                                                                             & 156.89                                                                             \\ \hline
Vector   Addition       & VectAdd                      & 2.64                                                                              & 2.4298                                                                             \\ \hline
\end{tabular}
}
\end{table} 

\subsection{Classifying results with respect to parallel dwarfs}
Parallel dwarfs represent the pattern of computation and communication \cite{asanovic2006}. 13 Dwarfs were proposed as representatives of parallel applications for investigating parallel programming models as well as architectures.  Analysing the results to know which type of kernel patterns prediction results are precisely close would be insightful. This could pinpoint the issues we need to investigate further. 

To understand how these dwarfs assist in analyzing results, we classified these benchmarks into parallel dwarfs\cite{asanovic2006}\cite{Asanovic2009}. The mean absolute error for each dwarf obtained is shown in Table \ref{dwarf}. It was observed that for Dynamic Programming dwarf which is represented by Needleman-Wunsch (NW) benchmark kernels, the mean absolute error is 4.93\% followed by Dense Linear Algebra dwarfs(e.g., Matrix multiplication, Gaussian elimination).  Dynamic programming dwarfs are memory latency limited whereas Dense Linear Algebra is computationally limited. Kernels represented by Sparse linear algebra are bound by both compute and memory and show a mean absolute error of 29.41. 

Spectral method dwarfs(e.g., Fast Walsh Transform), which are bounded by memory latency, have the highest mean absolute error of 39.46\%. All other dwarfs considered in this work are memory bound except Monte Carlo, which is problem dependent. With this analysis, we claim that our proposed model works better for computationally bound kernels than memory-bound kernels.

\begin{table}
    \centering
    \caption{Dwarf and its Mean Absolute Error}
    \label{dwarf}
    \begin{tabular}{|l|l|l|}
        \hline
        \textbf{Dwarf}        & \textbf{No. of Kernels} & \textbf{Mean Absolute Error} \\ \hline
        Unstructured Grid     & 9                       & 29.07\%                         \\ \hline
        Structured Grid       & 6                       & 26.58\%                         \\ \hline
        Spectral Method       & 1                       & 39.46\%                         \\ \hline
        Sparse Linear Algebra & 2                       & 29.41\%                         \\ \hline
        Monte Carlo           & 3                       & 33.88\%                         \\ \hline
        Graph Traversal       & 3                       & 26.91\%                         \\ \hline
        Dynamic Programming   & 2                       & 4.93\%                          \\ \hline
        Dense Linear Algebra  & 19                      & 26.47\%                         \\ \hline
    \end{tabular}
\end{table}

\subsection{Limitations of proposed algorithm} \label{limitations_approach1}
As discussed earlier, predicting the execution time of a CUDA kernel using static analysis with very high accuracy is a hard problem. From this perspective, the observed mean absolute error of 26.86\% is a reasonable performance of the proposed model. While investigating the prediction results, we observed that the model underestimates the execution time in many cases as seen in Figure \ref{fig:exectimeplot}. When we revisited our approach, we concluded that the usage of parameters such as Instruction Level Parallelism (ILP) as well as Thread Level Parallelism (TLP) in Equation \ref{equationDelay} and \ref{equationMemoryDelay} makes this delay value too low.

Considering a fully pipelined architecture with Instruction level parallelism, the delay due to an instruction execution is a small value; however, it may not be as low as the delay computed using \ref{equationDelay} and \ref{equationMemoryDelay} in some cases. Especially in cases where kernel code is executed with a small number of threads executing an instruction, we still have to wait for this instruction to complete its execution (instruction latency) on the allocated resource if there are not concurrently enough executing threads and instructions. If this latency value is divided by a large value based on the equation, the computed delay is lower than the actual delay. Hence considering the resource allocation approach can help to improve upon this underestimation. In the first approach, the model does not consider GPU hardware resource allocation while estimating the delay except for the maximum number of parallel executing warps. GPU features resources such as Single Precision Units, Load Store Units, Special Function Units, etc. which cater to different types of instructions. Scheduling the threads and instructions based on details of hardware resource availability associated with the instruction type  may be a more intuitive way of designing a more robust and efficient model. 

In the second model we present in the next section, we consider another perspective for modelling  Instruction Level Parallelism (ILP) and Thread Level Parallelism (TLP) models by adapting a resource allocation approach. In this approach, we model ILP by modelling the instruction mapping to resources using the possibility of parallel execution of instructions. TLP is modelled using hardware resource constraints on the parallel execution of threads. We reutilize the static analysis of the CUDA code of building basic blocks and control flow graph from the first approach. We then schedule these topologically sorted instructions by considering hardware constraints. We present this model in detail in the next section. 

\section{Modelling GPU Performance using Resource Allocation} \label{PP_approach2}

\begin{figure}[htbp]
\centering
\includegraphics[width=15cm]{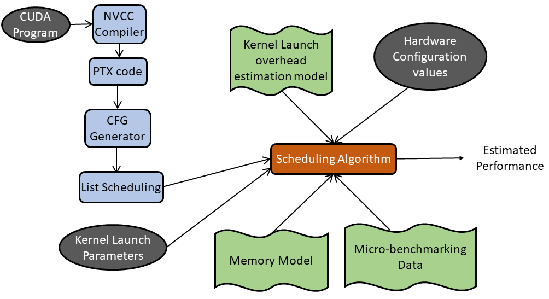}
\caption{Proposed performance prediction model}
\label{proposed_model_fig}
\end{figure}

Performance prediction involves estimating the execution time of a computing machine or estimating the execution of other performance factors (such as cache misses). This work uses the term ``performance prediction'' to imply the predicted execution time of a GPU kernel. The proposed performance prediction model takes a CUDA program as input and estimates its execution time, as explained in Figure~\ref{proposed_model_fig}. 

We have re-utilised some of the pre-processing steps and data used in the first approach. That is why Figure~\ref{proposed_model_fig} and Figure \ref{fig:model} have significant similarities since the steps of pre-processing using PTX code by static analysis are identical in both approaches. Both models use kernel launch parameters as user input and employ microbenchmarking data. However, both the models differ significantly in modelling memory behaviour and GPU scheduling algorithm approach.  

The overall process consists of three main phases:
\begin{enumerate}
    \item We compile the program to obtain the corresponding PTX format\cite{PTX}. We also consider inputs related to kernel launch parameters, namely the grid size (number of blocks) and the block size (number of threads per block). Then a static analysis is performed on the PTX code.
    \item We develop a set of models by collecting the hardware details of the GPU under consideration by running the device query application provided by NVIDIA\cite{NvidiaCUDAToolkit} only once. We also run various microbenchmark programs to collect several other parameters required for the performance prediction. 
    \item We perform scheduling of the PTX code using the above information to estimate the total delay. Computing CUDA kernel execution time (t\_kernel) using our proposed model is summarized in the following equation: 
    \begin{align}
d_{total}=& d_{kernel} + l\_overhead + gm\_penalty+sm\_penalty+cm\_penalty \label{eqn:dtotal}\\
t_{kernel}=&\dfrac{d_{total}}{\nu_{gpu}} \label{eqn:tkernel}   
\end{align} 
\end{enumerate}

Equation~\ref{eqn:dtotal} sums up the delay due to the execution of instructions and the penalties, which significantly contribute to the total execution time of the kernel.  We compute the simulated delay of a kernel ($d_{kernel}$) by analyzing the PTX code using Algorithm \ref{algo:b1}-\ref{algo:b3} discussed in Section \ref{sec:schedulingmodel}. To this delay, we add launch overhead,  global memory penalty, shared memory penalty, and cache penalty, which significantly contribute to the total execution time of a kernel. These penalties are independent in terms of the resources being used; hence can be summed up together. We compute these kernel penalty factors by running microbenchmarks designed as discussed in detail in Section \ref{dynamic_analysis}. 

\subsection{Analysing loops and branch divergence}
 The overall instruction count takes into account a loop estimation (based on the number of iterations) and branch divergence. If a loop is identified, the instruction delay of the looped path needs to be multiplied with the input provided by the user (number of loop iterations), and the calculation proceeds. We have to rely on users for loop iterations since most of the practical applications are non-affine \cite{Feautrier1996}, and the problem of predicting the number of loop iterations of non-affine benchmarks is yet to be solved.

A GPU SM executes each instruction in a Single Instruction Multiple Thread (SIMT) model, where threads in one warp (set of 32 threads) execute the same instruction. When an instruction is a branch type, all the threads in a warp may not take the same branch, thereby causing a branch-divergence~-\cite{COOK2013}. Like the loop, it is hard to predict the branch divergence.   

The proposed approach assumes that each branch is taken by at least one thread in every warp. In other words, we assume that no branch remains untraversed during the warp execution.  This implies that the thread taking the branch will execute the instruction while the remaining
threads will execute a NO-OP. Hence the execution time will be computed based on the longest path in CDFG. However, the final instruction delay can be imprecise in case the longest branch is never chosen during a warp execution. 
\subsection{Scheduling Model}  \label{sec:schedulingmodel}
Using the static analysis data discussed in the previous section, we develop a scheduling model which captures behavioural characteristics of a GPU. Since vendors do not publish crucial insights such as pipeline details, branch prediction, and caching behaviour, we make some assumptions about GPU operation to implement GPU scheduling. First, we discuss resource allocation, which plays a vital role in scheduling. 

\subsubsection{Resource Allocation}
As mentioned in Section \ref{abstract_model}, $\cal GPU$ comprises of a set of resources $R$.  Algorithm~\ref{algo:b1} performs scheduling and resource allocation of instruction $u$ in a SIMT fashion to compute the delay. During this process, the algorithm allocates multiple instances of the resource type for $u$. 

For instance, suppose that an {\tt add} instruction uses the resource type ``single-precision adder''. The algorithm will allocate all single-precision adders (say 32) in the GPU while scheduling the {\tt add} instruction. In doing so, multiple warps may become active simultaneously if the number of available resources is sufficiently large. 
In reality, the scheduling mechanism and warp management are pretty complex since instructions dispatched for executions can be a mixture belonging to multiple warps. Modeling such intricacies is impossible without knowing the exact scheduling strategy employed by an NVIDIA GPU. We have not modeled such behavior in Algorithms~\ref{algo:b1}-\ref{algo:b3}.  

\subsubsection{Notion of a Wave}\label{sec:waves}
The number of threads ready for execution can be more than the maximum number of threads (per SM) allowed in the GPU. In such a case, the proposed scheduler executes the threads in batches. We refer to this batched execution as "waves'' in the WORK. One wave of execution in our model implies the number of threads that can launch at a time on a GPU. The execution time observed for one wave is then multiplied by the number of waves launched to obtain the total execution time. 

\begin{figure}[htbp]
\centering
\includegraphics[width=9cm]{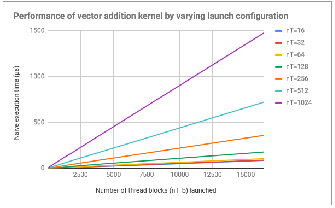}
\caption{Effect of launch parameters on the performance of kernel}
\label{fig_wave}
\end{figure} 

To compute the number of waves, we experimented with running a vector addition kernel with different $nB$ (number of blocks) and $nT\_b$ (threads per block) values as launch parameters for a Tesla K20 GPU, which has 13 SMs. The result has been depicted in Figure \ref{fig_wave}. We observed, for example, with $nB=15$, SM0 gets block 0 and block 13, SM1 gets block 1 and block 14, and the remaining SMs get one block each, irrespective of $nT\_b$ values. Intuitively, this implies that the number of waves can be calculated as $waves=\left\lceil nB/nSM\right\rceil$. Such an approach was also proposed by Kothapalli et al. \cite{Kothapalli2009}. However, a closer observation at Figure \ref{fig_wave} reveals that this intuitive formulation does not hold as explained next.

Suppose the K20 GPU takes t units of time to execute 13 blocks (each SM executes one block) of the vector addition kernel. In that case, the GPU should take close to 2t units of time when we attempt to launch 26 blocks if this intuitive computation was to be reasonably accurate. However, none of the nT\_b lines in Figure 3 exhibits this behaviour. For instance, the actual execution time for nB=5000 is not double the execution time for nB=2500. Thus, we cannot conclude that in general, if we compute the execution time for one wave, it can be then multiplied by 
$\left\lceil nB/nSM\right\rceil$ to get the total execution time.

In the proposed approach, we first compute the total number of threads per SM to be scheduled ($nTh\_Schd$) as $\dfrac{nB*nT\_b}{nSM}$. Next, based on the register and shared memory requirement, we compute the maximum number of threads per SM ($nTh\_sm^M$) for one wave.  Using $nTh\_Schd$ and $nTh\_sm^M$, Algorithm \ref{algo:b3} computes the number of waves.

In each wave, we use a resource allocation technique for modelling the execution of threads on a GPU, which is loosely inspired by \cite{Lam1988}. 

\subsection{Other Assumptions} \label{assumption}
There are few other assumptions, listed as follows:
\begin{itemize}
\item The GPU is superscalar architecture; at least one instruction will complete its execution every cycle after the pipeline is filled\cite{Shuaiwen2013}. To model this, we have divided the latency of an instruction into pipeline latency ($\mathcal{L}_{pipeline}$) and instruction latency ($\mathcal{L}$) during instruction scheduling in line 6 of Algorithm~\ref{algo:b1}. 
\item The number of waves depends on the availability of resources required for the type of instruction. Therefore, we compute waves during the scheduling of threads.  
\item One block is scheduled on one SM irrespective of the number of threads per block. Once all the SMs are assigned one block each, the next round of assignments will start from the first SM. This phenomenon was confirmed with experimentation. To accommodate this behaviour, we compute nTh\_schd in line no. 2 of Algorithm \ref{algo:b3} by dividing the number of blocks ($nB$) by the number of SMs ($nSM$). 
\item It is not possible to simulate the instructions in the actual order that they follow on a GPU since multiple warps are launched one after the other. A warp scheduler can select $n$ warps and dispatches $m$ independent instructions every cycle based on the architectural resource constraints\cite{Kepler2012}. Hence instruction in multiple warps does not follow a particular order if more than one warp is scheduled.  However, implementing this behaviour is extremely challenging. Here, the model assumes that the instructions for one warp are executed completely before the next warp resumes execution as seen in Algorithm \ref{algo:b1} line 15, where delay for each instruction is added based on the resource being used. 
\end{itemize}

\subsection{Scheduling Algorithm}\label{sec:sched_algo}
We wish to find the optimal order of execution of an instruction for which we implement an algorithm, loosely based on list scheduling~\cite{Rau1981, Lam1988}. The scheduling honours two constraints, namely:
\begin{enumerate}
    \item \textit{Precedence Constraints:} The scheduler uses data-flow graph $G$ associated with a basic block while scheduling operation and ensures that the minimum delay between two nodes with an edge $(u,v)$ is the latency($\mathcal{L}(u)$) of the node u. The scheduler also honours the control flow across blocks by considering the control flow graph $G_{cfg}$ of the entire kernel in Algorithm~\ref{algo:b2}.
    \item \textit{Resource Constraints:} Since the GPU computation follows a SIMT model; one instruction is executed by multiple threads, each on a different data element. Even when the scheduler can schedule a large number of threads to execute an instruction, the execution is finally constrained by the total number of resources available for that instruction. For instance, if $N$ threads are ready to execute an addition operation and there are $R[SP]$ number of single-precision cores available (per SM), the addition will be performed in $\left\lceil\frac{N}{R\left[SP\right]}\right\rceil$ batches. The proposed scheduling algorithm considers this resource constraint during the computation of execution cycles.
\end{enumerate}
\begin{algorithm}
    \caption{Scheduling a block}
    \label{algo:b1}
    \begin{algorithmic}[1]

    \Procedure{ScheduleBlock}{$\mathcal{GPU}$, $G=\langle V,E\rangle$, $nTW$}
    \State $V' \leftarrow$ \Call{Toposort}{$G$}
    \State $Sch\leftarrow []$; $Rsv\leftarrow [][]$;\Comment{$Sch$ keeps track of schedule of each node and $Rsv$ is the resource reservation table, initialized with 0}
    \State $delay\leftarrow0 $; 
    \For{$v\in V'$}
        \State $t\leftarrow\max_{(u\to v)}\left\{Sch[u]+\mathcal{L}(u)+\mathcal{L}_{pipeline}\left(\left\lceil\cfrac{nTW}{R[R_{map}(u)]}\right\rceil-1\right) \right\}$
       \State $j \leftarrow R_{map}(v)$
       \State $d\leftarrow \mathcal{L}(v)+\mathcal{L}_{pipeline}\left(\left\lceil\frac{nTW}{R[R_{map}(v)]}\right\rceil-1\right)$
       \State Let $(t1\geq t)$ be the index such that $Rsv[t1,j]\cdots Rsv[t1+(d-1),j]$ are all 0.
        \State $Sch[v] \leftarrow t1$; $i \leftarrow 0$; 
       \For{$i<d$}
          \State $Rsv[t1+i,j] \leftarrow 1$; $i$++;
   
       \EndFor
           \State $delay \leftarrow Sch(v)+d$;
    \EndFor
    \State \textbf{return} $delay$ 
    \EndProcedure
    \end{algorithmic}
\end{algorithm}
\begin{algorithm}
    \caption{Scheduling a CFG}
    \label{algo:b2}
    \begin{algorithmic}[1]

    \Procedure{ScheduleCFG}{$\mathcal{GPU},G_{cfg}\leftarrow\langle\mathbb{C, E}\rangle, nTW$}
    \State $BB' \leftarrow$ \Call{Toposort}{$G_{cfg}$}
    \State $delay \leftarrow\left[~\right]$
    \For{$G\in BB'$}
       \State $d \leftarrow max_{e=G_p\to G}\left\{delay[G_p]\right\}$
       \State $d_{G} \leftarrow $\Call{ScheduleBlock}{$\mathcal{GPU},G, nTW, ngM, nShM$}
       \If{$G$ has back-edge}
         \State $d_G \leftarrow d_G*nLoop$
       \EndIf
       \State $delay[G] \leftarrow d+d_G$
       \State $d_{cfg} \leftarrow  delay[G]$
    \EndFor
    \State \textbf{return} $d_{cfg}$\Comment{Delay of the CFG}
    \EndProcedure
    \end{algorithmic}
\end{algorithm}

\begin{algorithm}
    \caption{Scheduling a Kernel}
    \label{algo:b3}
    \begin{algorithmic}[1]
    \Procedure{ScheduleKernel}{$\mathcal{GPU},G_{cfg}=\langle\mathbb{C, E}\rangle$, $nB, nT\_b, reg\_t, shmb$} 
   
    \State \textit{nTh\_schd} $\leftarrow\left\lceil\frac{nB}{nSM}*nT\_b\right\rceil$; 
    \State $waves\leftarrow 0$;
    \State $nT h\_sm\leftarrow nTh\_schd > nTh\_sm^M  ? $ $  nTh\_sm^M : nTh\_schd $;
 \State  $maxR\leftarrow reg\_t \times nTh\_sm$ > $reg\_b^M$ ?  $ \frac{reg\_b^M}{reg\_t \times nT\_b}$ $\times nT\_b$ : $nTh\_sm$ ;
  \State    $maxSh\leftarrow (shmb \times \frac{nTh\_sm}{nT})$ > $shm\_b^M$ ? $\frac{shm\_b^M}{shmb} \times nT\_b$  : $nTh\_sm$  ;
    \State    $maxB \leftarrow (shmb \times \frac{nTh\_sm}{nT})$ > $nB^M$ ? $nB^M \times nT\_b$  : $nTh\_sm$  ;
 \State  $nTh\_sm \leftarrow min(maxR,maxSh, maxB, nTh\_sm)$ ;
 \While{\textit{nTh\_schd} $\geq 0$}
     \State  $\textit{nTh\_wave} \leftarrow  nTh\_sm$;
     \State $d_{kernel} \leftarrow d_{kernel} + $ \Call{ScheduleCFG}{$\mathcal{GPU}, G_{cfg}$, {\it nTh\_wave}};
   \State $nTh\_schd \leftarrow  nTh\_schd-~nTh\_sm^M$;
      \State $waves \leftarrow waves+1$;
    \EndWhile
     \State populate the values of $nShM$ and $nGM$;
    \State \textbf{return} ($d_{kernel}$, waves, $nGM$, $nShM$); \Comment{Kernel Delay }
    \EndProcedure
    \end{algorithmic}
\end{algorithm}

\subsubsection{Counting Memory Instruction}
In addition to the scheduling of instructions, it is required to count the total number of global and shared memory access instructions, denoted by $nGM$ and $nShM$ respectively, present in the control flow graph of a kernel. These two counts are used to estimate global memory and shared memory access penalties,  described in the next section. This counting is performed while scheduling each instruction in a basic block (during Step 6 of Algorithm~\ref{algo:b1}). We do not show this simple counting mechanism here for brevity. We highlight that Algorithm~\ref{algo:b3} obtains these two counts in step 10.

\subsubsection{Illustration}
\label{ex:latency}
\begin{figure}[htbp]
\centering
\includegraphics[width=18cm]{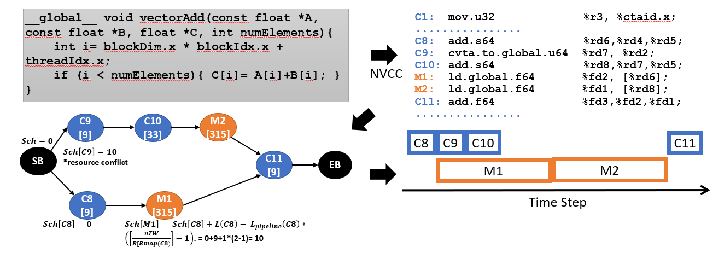}
\caption{Illustration of Algorithm~\ref{algo:b1}}
\label{fig:latencyex}
\end{figure}
We illustrate how scheduling is performed using the vector addition kernel example introduced in Section~\ref{sec:waves}. As mentioned in Section \ref{assumption}, GPU launches a mixture of several instructions during execution, which we can't model. Hence, the proposed algorithm assumes that {\tt vectorAdd} kernel runs with 100\% occupancy and schedules an instruction making all the resources for this instruction available  (as explained in Section~\ref{sec:sched_algo}). 

Figure~\ref{fig:latencyex} shows a snippet of the corresponding PTX code and the data-dependency graph for the PTX snippet where SB and EB are two dummy nodes denoting the start and end of the basic block respectively. %
Let us consider the target GPU to be Tesla K20, for which $\cal L(${\tt ld}) is 315 cycles, $\cal L(${\tt add})= 9 cycles and $\cal L(${\tt cvta})= 33 cycles. We shall explain how these latency values are derived later. For Tesla K20, there are 192 cores and 32 load-store units, hence $R[SP]=192$ and $R[LSU]=32$. Assume that the launch parameters for this kernel are 13 blocks of 256 threads per block, hence $nTW=256$.

Algorithm~\ref{algo:b1} topologically sorts this graph to schedule the nodes one by one. One such topological order is C8, C9, C10, M1, M2, C11. Since SB is a dummy node, $Sch[C8]$, which represents a time unit (relative to the block) for scheduling C8, is 0. Steps 11 and 12 blocks the resource $SP$ for 10 cycles. In the next iteration, for C9, Step 6 and 7 of Algorithm~\ref{algo:b1} will compute $t=0$. However, C8 has reserved the resource for {\tt add} for 10 cycles, i.e. $Rsv[0,SP]\cdots Rsv[9,SP]=1$. Therefore, C9 can start only at the 10th cycle ($Sch[C9]=10$). This is computed in steps 8 and 9 of the algorithm. 
The memory-bound instruction M1 starts after C8, concurrently with C9. While M2 can also start in parallel to M1, Algorithm ~\ref{algo:b1} will schedule it after M1 since the Load Store Unit ($LSU$) is assumed to be fully occupied by M1 (step 8 and 9 of Algorithm~\ref{algo:b1}). Since C11 is data dependent on M1 and M2,  $\max\{\langle M1\to C11\rangle, \langle M2\to C11\rangle\}$ is computed for its possible schedule. This in turn implies $t\leftarrow\max\{332, 655\}=655$ in step 6 of Algorithm~\ref{algo:b1}. Since there are no other instruction using the resource for {\tt add}, $Rsv[655$,{\tt add}$]=0$ and $Sch[C11]=655$.

\paragraph*{Latency Hiding}
Latency hiding involves overlapping execution times of instruction with execution time of another~\cite{volkov2016}. Algorithm~\ref{algo:b1} implements latency hiding while ensuring resource constraints. In this example, C9 and M1 are independent, and their resources $R_{map}(C9)=SP$ and $R_{map}(M1)=LSU$ are different; hence they are scheduled at the same time as shown in Figure~\ref{fig:latencyex}. When both instructions start simultaneously, the C9 instruction's latency is hidden by the execution of M1. In fact, the latency of both C9 and C10 are hidden using M1.

Next, we highlight another type of execution latency due to instruction pipelining. In our model, an instruction $v$ is executed in  $nTW/R[R_{map}[v]]$ number of batches. Except for the first batch, the latency of $v$ in one batch is hidden by the latency in the next batch. We assume the latency of $v$ is 1 when it is perfectly pipelined. Let us consider the schedule of M1, which starts after C8 in Figure~\ref{fig:latencyex}. Though ${\cal L}(C8)=9$, $\left\lceil nTW/R[SP]\right\rceil=\lceil256/192\rceil=2$. With instruction pipelining, C8 will take $9+1*(2-1)=10$ cycles. Since $Sch[C8]=0, Sch[M1]= 0+9+1=10$th cycle. 
\subsection{Dynamic Analysis} \label{dynamic_analysis}
A kernel execution time is impacted by the time to launch a kernel in a GPU, popularly known as the \textit{kernel launch overhead}. This is the time consumed just before and after executing the kernel instructions. We have constructed an empirical model to characterize this overhead. 

NVIDIA GPUs have a complex memory hierarchy comprising global memory, shared memory, texture memory, and constant memory. Memory access instructions need to be handled differently because the delay caused by them is not only influenced by instruction parallelism but also due to resource constraints. Our proposed memory access model has three parts i) the Access latency model and ii) the Access Penalty model iii) the Cache miss penalty model.

\subsubsection{Memory Access Penalty Model}
Memory instructions have unpredictable higher delays due to other factors such as access patterns in the case of global memory and bank conflicts for shared memory. We need to factor in this behaviour of memory instructions when computing the total delay of a CUDA kernel. The computing latency of one global and shared memory instruction is not sufficient to model the phenomenon of access patterns or bank conflicts. To accommodate this behaviour while calculating the delay, we have taken the approach of introducing penalty factors associated with memory access instructions. These penalties are summed up with the delay obtained from the scheduler to get the total execution time as shown in Eq. \ref{eqn:dtotal}.  

GPU offers multiple memory options such as global, shared, constant, and texture. We consider global and shared memory penalties in our model since constant and texture memory is more concerned with graphics applications. We focus on General Purpose GPU Computing (GPGPU). We have illustrated the memory design for a CUDA thread in Figure \ref{fig_memory_design}.

\begin{figure}[htb]
    \centering
\includegraphics[width=0.7\linewidth]{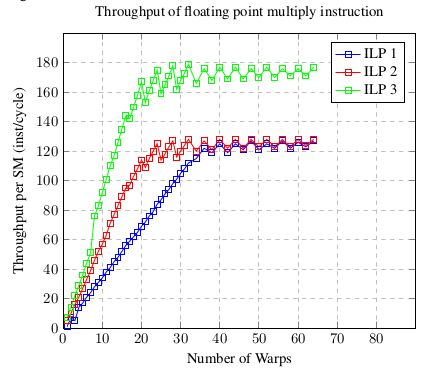}
\caption{Model for Computing Global Memory Throughput}
\label{fig_throughput}
\end{figure}

We built penalty models for global memory instructions (Eq. \ref{eqn:global}), shared memory instructions (Eq. \ref{eqn:shared}) and cache behaviour (Eq. \ref{eqn:cache}) of GPU. Each of the penalty models is described in detail in the following subsections. 
\begin{table}
\caption{Modelling result for dynamic analysis parameter across different statistical functions}
\label{RsquareModel}
\resizebox{\linewidth}{!}{
\begin{tabular}{|l|l|l|l|l|l|l|}
\hline
\multicolumn{1}{|c|}{\multirow{2}{*}{\textbf{Parameter}}} & \multicolumn{2}{c|}{\textbf{Linear Regression}} & \multicolumn{2}{c|}{\textbf{Logarithmic Growth}} & \multicolumn{2}{c|}{\textbf{Exponential Growth}} \\ \cline{2-7} 
\multicolumn{1}{|c|}{}                                    & R\_square                 & RMSE                & R\_square               & RMSE                   & R\_square               & RMSE                   \\ \hline
Kernel Launch Overhead                                    & 0.9877                    &            13.15         & 0.3240                  & 97.5958                & 0.8398                  & 47.95                  \\ \hline
Shared Memory Throughput                                  & 0.2472                    &                 181202.3     & 0.8360                  & 83413.19               & 0.9782                  & 41347.23               \\ \hline
Global Memory Throughput                                  & 0.1896                    &            23075.85         & 0.8798                  & 8885.28                & 0.9528                  & 5568.59                \\ \hline
                                    & \multicolumn{3}{c|}{R\_square}                                            & \multicolumn{3}{c|}{RMSE}                                                 \\ \hline
Cache Penalty                       & \multicolumn{3}{c|}{0.9045}                                              & \multicolumn{3}{c|}{3797.42} \\  \hline
\end{tabular}
}
\end{table}

\begin{table}
\caption{Exponential Models for Global \& Shared Memory Throughput}
\label{tab:exponential_model}
\resizebox{\linewidth}{!}{
\begin{tabular}{|l|l|l|}
\hline
\textbf{GPU Architecture} & \textbf{Global Memory Throughput} $R_{global}$ & \textbf{Shared Memory Throughput} $R_{shr}$\\ \hline
Tesla K20 & 76363.8*(1.04-exp(-0.00021342*ngM) & 823761.8*(1 - exp(-0.000013830*nsM) \\ \hline
Quadro K4200 & 68145.9*(1.03-exp(-0.00027584*ngM) & 70628.9019*(1.03-exp(-0.000301068*nsM) \\ \hline
Tesla M60 & 71453.6*(1.22-exp(-0.000033469*ngM) & 1618990*(1-exp(-0.0000055295*nsM) \\ \hline
GTX 1050 & 47244.33*(1 - exp(-0.001223*ngM) & 613743.6*(1.12-exp(-0.00000694*nsM) \\ \hline
\end{tabular}
}
\end{table}

\subsubsection{Global Memory Penalty}
In the case of global memory, an essential factor that influences its performance is the coalescing effect which we cannot model during scheduling. When a GPU SM executes a memory instruction, the SM creates memory requests and switches to another warp until all the requested memory values of this warp are available\cite{HongTr2009,resios2011}. Our model incorporates this phenomenon by adding an extra delay as a penalty value. This penalty value is a factor of the number of global memory requests $nGM$, derived from Algorithm~\ref{algo:b3}. The global memory penalty is derived as follows:

\indent 1. We collected the memory throughput data (the rate at which data can be read or stored in global memory) by running an application with global load/store instructions. The kernel was launched with different launch parameters (number of blocks and number of threads per block) to ensure that by varying the number of instructions, throughput is recorded. 

\indent 2. We build a model for the throughput by attempting to fit linear, logarithmic, and exponential growth functions. R-square for each of these models is tabulated in  Table \ref{RsquareModel}. From the result, we conclude that an exponential growth function $$\mathcal{TP}(gm)=a\times(b-e^{-c\times  \mathcal{L}(gm)})$$ that represents global memory transactions has the highest R-square value and least RMSE value. This exponential growth model is shown in Figure \ref{fig_throughput} for Tesla K20. We have discussed how we tested the efficacy of the global memory throughput model in Appendix \ref{AppendixB}.

\indent 3. Global memory penalty is determined by first computing the number of memory accesses using load-store units ($\frac{nB\cdot nT\_b}{R[LSU]}$). Then we measure the time taken for each access by dividing $accessGm_{sz}$ with its $\mathcal{TP}(gm)$.

\indent 4. This penalty value for one global memory instruction is multiplied by the number of global memory instructions ($nGM$) obtained from the scheduling algorithm to obtain the penalty value for all the global memory load and store instructions. 

We compute the global memory penalty by the following equation:
\begin{equation} \label{eqn:global}
gm\_penalty=\frac{nB\cdot nT\_b}{R[LSU]}\times \frac{accessGm_{sz}}{\mathcal{TP}(gm)}\times nGM
\end{equation}
\subsubsection{Shared Memory Penalty}
Execution of a CUDA kernel is also affected by bank conflicts in the case of shared memory access. We have modelled this by adding a penalty value for shared memory accesses. The penalty value is computed as the throughput that gets calculated using the observed throughput affected by bank conflicts\cite{resios2011}. The estimated delay for these shared memory requests ($nShM$) can be computed by dividing the access size by this observed throughput. 

The shared memory penalty is computed using the following steps:

\indent 1. Similar to global memory throughput behaviour, shared memory throughput grows by common factors over equal intervals. To model shared memory throughput  ($\mathcal{TP}(sm)$)  against the number of shared memory load/store transactions, we tested with linear, logarithm, and exponential growth models whose R-square value is provided in Table \ref{RsquareModel}. Amongst them, the exponential growth model $$\mathcal{TP}(sm)=a\times(b-e^{-c\times \mathcal{L}(sm)})$$  produces the maximum R-square value and the least RMSE value which ensures goodness of fit. 
 
\indent 2. We then calculate the penalty for one shared memory instruction by computing the number of times shared memory instructions are scheduled per SM-- $\frac{nB\cdot nT\_b}{R[LSU]\cdot nSM}$. Next, we compute the time taken for scheduling each shared load or store instruction ($\frac{access_{sz}}{\mathcal{TP}(sm)}$). Here,  $accessShm_{sz}$ stands for the number of Bytes accessed during each shared memory load and store instruction.

\indent 3. In the final step, we use the number of shared memory instructions counted (nShM) in the scheduling algorithm to obtain a shared memory penalty for all the shared load and store instructions scheduled. 

The sm\_penalty is computed as follows: 
\begin{equation} \label{eqn:shared}
sm\_penalty=\frac{nB\cdot nT\_b}{R[LSU]\cdot nSM}\times \frac{accessShm_{sz}}{\mathcal{TP}(sm)}\times nShM
\end{equation}

\subsubsection{Cache Miss Penalty}
Modelling accurate Cache behaviour is an unattainable task since it depends on multiple runtime factors. NVIDIA GPU L2-Cache stores a part of the global memory, which is the primary source of instruction delay. We approximate this behaviour by computing an L2 cache miss penalty using available architecture and program details to capture the behaviour of a GPU. The cache penalty is calculated using the following steps:

\indent 1. In order to estimate the number of times the Cache is full and a miss can lead to a penalty (which we model as $cm\_penalty$), we use the notion of ``waves'' computed in Algorithm~\ref{algo:b1}, \ref{algo:b2} and \ref{algo:b3}.

\indent 2. The notation $L2\_sz$ represents the L2 Cache size which is derived from the device query. 
 
The total number of global memory instructions $nGM$ ( statically computed during Scheduling ) is divided by available Cache memory, which provides us with the approximate number of Cache misses.

\indent 3. This value is then used along with global memory latency ( calculated using microbenchmarks, shown in Table~\ref{tab:my-globallatency}) in order to compute the Cache Penalty for the misses that can occur during kernel execution.  

The cache miss penalty function is modeled using the following equation:
\begin{equation} \label{eqn:cache}
cm\_penalty=\dfrac{nT\_b\cdot nB\cdot nGM}{\frac{waves~\cdot~ L2\_sz}{access_{sz}}}\times \mathcal{L}(gm)
\end{equation}

We validated our proposed equation by comparing it against the value obtained by profiling many benchmarks and recording the L2 Cache miss rate using the {\tt nvprof} profiling tool\footnote{https://docs.nvidia.com/cuda/profiler-users-guide/index.html}. Based on this miss rate, the actual cache penalty was calculated for validation. The actual penalty value is compared with the cache penalty derived from the above equation. The R-square value and RMSE for this actual versus predicted value are provided in Table \ref{RsquareModel} to justify our claim.

\subsection{Results} \label{experimental_results}
We developed a Java-based predictor tool that analyses a PTX code, takes as input kernel launch parameters (number of blocks, number of threads per block, number of loop iterations), and dynamic analysis model results to predict the execution time of a CUDA kernel. To collect dynamic analysis results, we perform microbenchmarking for collecting the latency value of each instruction type. Other runtime features (gm\_penalty, sm\_penalty) are collected by building models based on data recorded using {\tt nvprof}.  We have also recorded our tool's time to perform the prediction in Table \ref{tab:results}. On average, our prediction tool takes 66.48ms for CUDA kernel execution time prediction. The observed minimum prediction time is 1.031ms, and the maximum prediction time is 206.62ms when we performed the analysis on an Intel i3-5005U CPU with 8GB memory, using JVM version 15.0.1.

\subsection{Experiment Setup}
CUDA benchmarks used in this study belong to different areas of application such as vector operations (e.g., Vector addition, Saxpy), matrix operation (e.g., Transpose, Matrix multiplication), mathematical problems (e.g., BFS, Gaussian), and advanced operations (e.g., BlackScholes). We used the applications from CUDA toolkit \cite{NvidiaCUDAToolkit} and Rodinia Testbench\cite{Che2009}.

While building the model, we designed our experiment and verified results with Kepler architecture using the Tesla K20 GPU. We also verified results with the Tesla K4200 GPU, which is also based on Kepler Architecture. To investigate the extent to which the proposed model is architecture-agnostic, we repeated the experiment on Maxwell (Tesla M60) and Pascal (GTX 1050) architectures. The architecture feature value is mentioned in Table \ref{tab:hardwarevalue-table}.

\begin{table}
\centering
\caption{Hardware Features values for GPUs under consideration}
\label{tab:hardwarevalue-table}
\begin{tabular}{|l|l|l|l|l|}
\hline
GPU Hardware Features   & Quadro K4200 & Tesla K20 & Tesla M60 & GTX 1050 \\ \hline
Architecture            & Kepler       & Kepler    & Maxwell   & Pascal   \\ \hline
Compute Capability      & 3.0          & 3.5       & 5.2       & 6.1      \\ \hline
Number of SMs           & 7            & 13        & 16        & 5        \\ \hline
Number of  cores Per SM & 192          & 192       & 128       & 128      \\ \hline
GPU Clock Rate(MHz)     & 706          & 784       & 1178      & 1493     \\ \hline
Memory Clock Rate(MHz)  & 2700         & 2600      & 2505      & 3504     \\ \hline
L2 Cache (Bytes)        & 524288       & 1310720   & 2097152   & 524288   \\ \hline
\end{tabular}
\end{table}

\begin{table}
\centering
\caption{Benchmarks Under Study}
\label{tab:benchmakrs}
\resizebox{\textwidth}{!}{ 
\begin{tabular}{|l|l|l|l|}
\hline
\textbf{Benchmark} & \textbf{Kernel Name} & \textbf{Source} & \textbf{Dwarf Type} \\ \hline
Breadth-First Search & Kernel1 & Rodinia Benchmark Suite & Graph Traversal \\ \hline
Breadth-First Search & Kernel2 & Rodinia Benchmark Suite & Graph Traversal \\ \hline
Convolution Seperable & convolutionRowsKernel & CUDA Toolkit & Structured Grid \\ \hline
Convolution Seperable & convolutionColumnsKernel & CUDA Toolkit & Structured Grid \\ \hline
Convolution Texture & convolutionRowsKernel & CUDA Toolkit & Structured Grid \\ \hline
Convolution Texture & convolutionColumnsKernel & CUDA Toolkit & Structured Grid \\ \hline
Srad & srad\_cuda\_1 & Rodinia Benchmark Suite & Structured Grid \\ \hline
Srad & srad\_cuda\_2 & Rodinia Benchmark Suite & Structured Grid \\ \hline
Hotspot & calculate\_temp & Rodinia Benchmark Suite & Structured Grid \\ \hline
Merge Sort & mergeSortSharedKernel & CUDA Toolkit & Backtrack and Branch-and-Bound \\ \hline
Merge Sort	&	generateSampleRanks  & CUDA Toolkit & Backtrack and Branch-and-Bound \\ \hline
Merge Sort	&	mergeRanksAndIndices  & CUDA Toolkit & Backtrack and Branch-and-Bound \\ \hline
Merge Sort	&	mergeElementaryIntervals  & CUDA Toolkit & Backtrack and Branch-and-Bound \\ \hline
Needleman-Wunsch & needle\_cuda\_shared\_1 & Rodinia Benchmark Suite & Dynamic Programming \\ \hline
Scan & scanExclusiveShared & CUDA Toolkit & Sparse Linear Algebra \\ \hline
Sobol Qring & sobolGPU\_kernel & CUDA Toolkit & Monte Carlo \\ \hline
CFD Solver & compute\_step\_factor & Rodinia Benchmark Suite & Unstructured Grid \\ \hline
CFD Solver & cuda\_compute\_flux & Rodinia Benchmark Suite & Unstructured Goverheadrid \\ \hline
CFD Solver & initialize\_variables & Rodinia Benchmark Suite & Unstructured Grid \\ \hline
CFD Solver & cuda\_time\_step & Rodinia Benchmark Suite & Unstructured Grid \\ \hline
Clock & timedReduction & CUDA Toolkit & Unstructured Grid \\ \hline
BlackScholes & BlackScholesBodyGPU & CUDA Toolkit & Dense Linear Algebra \\ \hline
LU Decomposition & lud\_diagonal & Rodinia Benchmark Suite & Dense Linear Algebra \\ \hline
LU Decomposition & lud\_perimeter & Rodinia Benchmark Suite & Dense Linear Algebra \\ \hline
LU Decomposition	&	lud\_internal & Rodinia Benchmark Suite & Dense Linear Algebra \\ \hline
K means	&	KmeansInvert  & Rodinia Benchmark Suite & Dense Linear Algebra \\ \hline
K means	&	KmeansPoint  & Rodinia Benchmark Suite & Dense Linear Algebra \\ \hline
Matrix Multiplication & matrixMulCUDA & CUDA Toolkit & Dense Linear Algebra \\ \hline
Gaussian & Fan1 & Rodinia Benchmark Suite & Dense Linear Algebra \\ \hline
Gaussian & Fan2 & Rodinia Benchmark Suite & Dense Linear Algebra \\ \hline
Transpose & transposeNaive & CUDA Toolkit & Dense Linear Algebra \\ \hline
Transpose & transposeCoarseGrained & CUDA Toolkit & Dense Linear Algebra \\ \hline
Transpose	&	Diagonal	& CUDA Toolkit & Dense Linear Algebra \\ \hline
Transpose	&	NoBankConflicts	& CUDA Toolkit & Dense Linear Algebra \\ \hline
Transpose	&	FineGrained	& CUDA Toolkit & Dense Linear Algebra \\ \hline
Transpose	&	CoarseGrained	& CUDA Toolkit & Dense Linear Algebra \\ \hline
Transpose	&	copy	& CUDA Toolkit & Dense Linear Algebra \\ \hline
Transpose	&	copySharedMem	& CUDA Toolkit & Dense Linear Algebra \\ \hline
Vector Addition & vectorAdd & CUDA Toolkit & Dense Linear Algebra \\ \hline
Stereo Disparity & Stereo Disparity & CUDA Toolkit & Dense Linear Algebra \\ \hline
Saxpy & saxpy & CUDA Toolkit & Dense Linear Algebra \\ \hline
Fast Walsh Transform	&	fwtBatch1 & CUDA Toolkit & Dense Linear Algebra \\ \hline
Fast Walsh Transform	&	fwtBatch2 & CUDA Toolkit & Dense Linear Algebra \\ \hline
\end{tabular}
}
\end{table}

\subsection{Data Collection}
We collected microbenchmarking data by running benchmarking applications for each type of instruction. These applications were run only once on each architecture. Similarly, we ran applications to collect global memory throughput and shared memory throughput for the different number of instructions by profiling applications containing this instruction using \verb+nvprof+. Hardware attributes were collected by running the Device Query application provided in CUDA Samples\cite{NvidiaCUDAToolkit}.

To test and verify our proposed model, we ran many popular benchmarks, some having more than one kernel as mentioned in Table \ref{tab:benchmakrs} on each architecture, and recorded its execution time using \verb+nvprof+. Some of the kernels can be launched with different launch parameters, and some of them require fixed launch parameters. We created $108$ configurations of these benchmarks, considering all permissible launch configurations. Some of these benchmarks can operate on different datasets, and for some, the sample size is restricted by the dataset provided along with the benchmark designers. On average, we have four samples for each configuration, and we have more than $400$ data samples to test the model. The results have been presented in Table~\ref{tab:results} and Table~\ref{tab:my-table-dwarf-results}.

\subsection{Results for Kepler Architecture} \label{Sec:keplerresults}
First, we tested our model on the Kepler architecture-based Tesla K20 GPU comprising of 13 streaming multiprocessors. Each SM comprises of 192 Single-precision CUDA cores, 64 Double-precision units, 32 Special Function units, 32 Load-Store units, and 4 Warp schedulers\cite{Kepler2012}. To evaluate the accuracy of our prediction approach, we have used both Mean Percentage Error (MPE) and Mean Absolute Percentage Error (MAPE). 

\begin{figure}[ht]
\centering
\includegraphics[width=12cm]{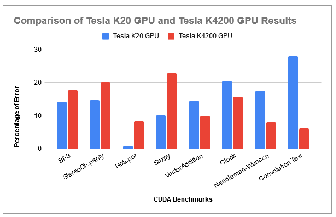}
\caption{Error observed for two GPUS of Kepler Architecture}
\label{fig_kepler_results}
\end{figure}

The recorded MPE and MAPE for the Tesla K20 GPU are $-8.88$\% and $28.3$\%, respectively. Next, we tested this model on Quadro K4200 GPU, which also has the Kepler architecture. To test this model, we tuned it to the hardware characteristics of the Quadro K4200. The MPE and MAPE values for this machine are $-5.66$\% and $29.4$\%, respectively.  We have illustrated the absolute percentage of error in prediction for some of the popular benchmarks for the two Kepler GPUs in Figure \ref{fig_kepler_results}. 

\subsection{Results across GPU Architectures} \label{Sec:architectureresults}
For all the GPUs under consideration for this study, we tuned our performance prediction model for each architecture by first collecting the architectural details using the Device Query application and the vendor documentation, computing the latency values, and finally constructing the memory access models described earlier.
%


\begin{figure}
  \begin{subfigure}{0.5\textwidth}
   \includegraphics[width=15cm]{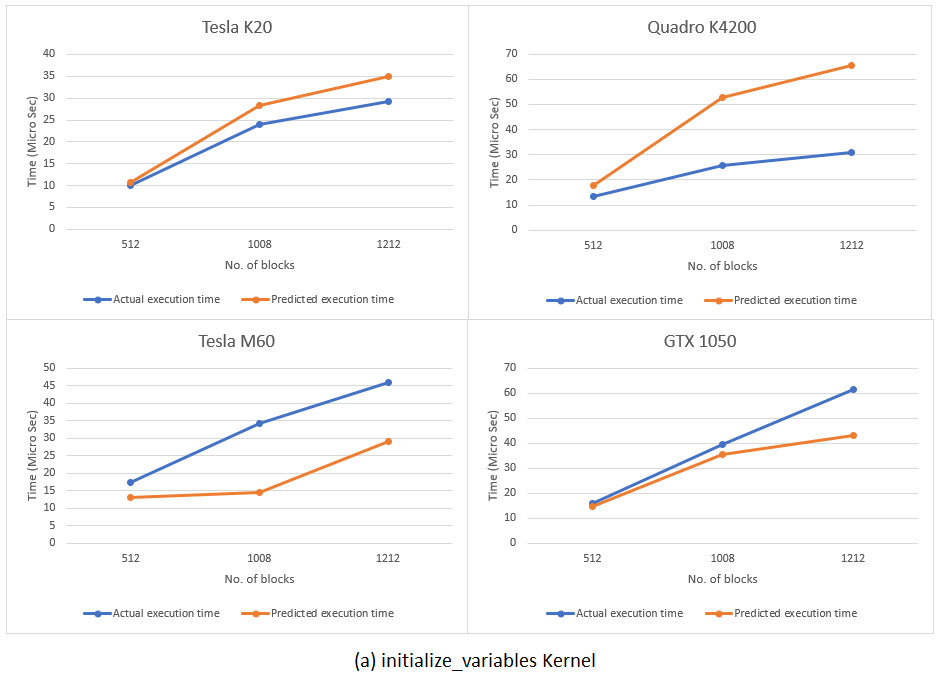}
 \end{subfigure}
 \linebreak
   \begin{subfigure}{0.5\textwidth}
   \includegraphics[width=15cm]{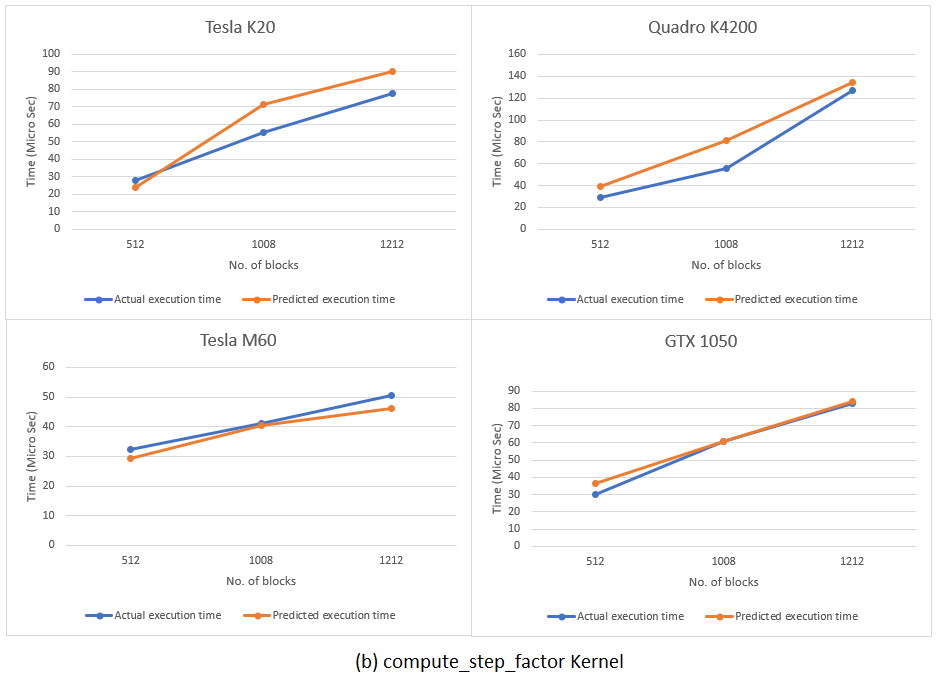}
  \end{subfigure}
\end{figure}
  
\begin{figure}
   \begin{subfigure}{0.5\textwidth}
   \includegraphics[width=15cm]{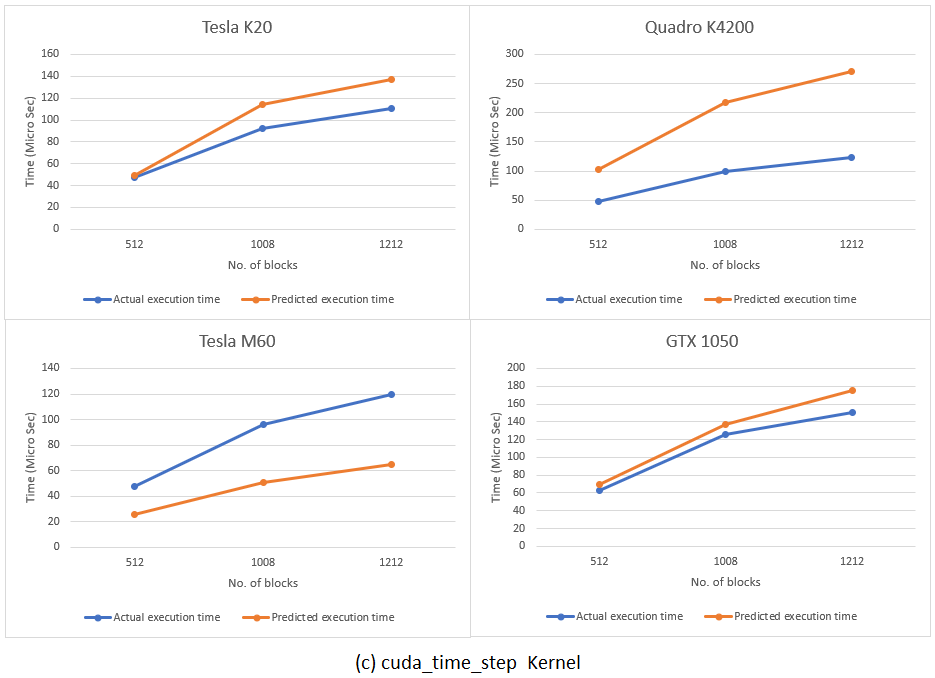}
    \end{subfigure} 
     \linebreak
  \begin{subfigure}{0.5\textwidth}
    \includegraphics[width=15cm]{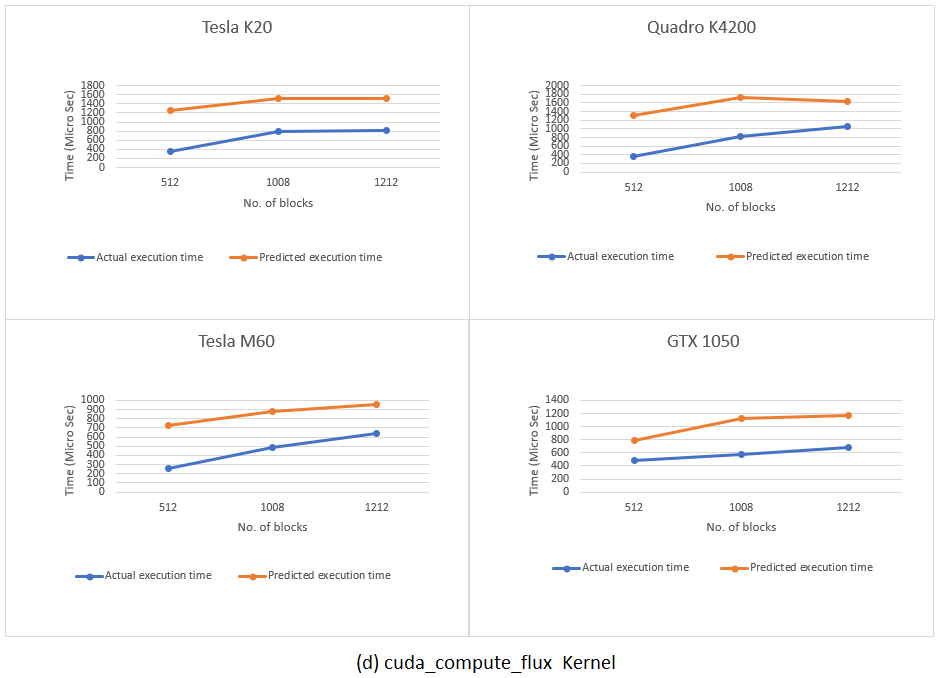}
     \end{subfigure}
     \caption{Actual vs predicted execution time of CFD Solver for multiple GPU architectures}
     \label{fig_cfd_results}
\end{figure}

\begin{table}
\caption{Actual vs Predicted Execution Time ($\mu s$) }
\label{tab:results}
\resizebox{\textwidth}{!}{ 
\begin{tabular}{|l|l|l|l|l|l|l|l|l|l|}
\hline
\multirow{3}{*}{\textbf{Benchmark}} & \multirow{3}{*}{\textbf{Kernel}} & \multicolumn{8}{c|}{\textbf{GPU Architecture}} \\ \cline{3-10} 
 &  & \multicolumn{2}{l|}{\textbf{Tesla K20}} & \multicolumn{2}{l|}{\textbf{Quadro K4200}} & \multicolumn{2}{l|}{\textbf{Tesla M60}} & \multicolumn{2}{l|}{\textbf{GTX 1050}} \\ \cline{3-10} 
 &  & \textbf{Actual} & \textbf{Predicted} & \textbf{Actual} & \textbf{Predicted} & \textbf{Actual} & \textbf{Predicted} & \textbf{Actual} & \textbf{Predicted} \\ \hline
CFD Solver	&	StepFactor	&	55.39	&	71.28	&	55.84	&	80.95	&	41.09	&	40.54	&	61.10	&	60.89	\\ \hline
CFD Solver	&	Flux	&	820.19	&	1,519.40	&	1,050.90	&	1,638.00	&	338.30	&	1,519.40	&	631.60	&	1,130.14	\\ \hline
CFD Solver	&	InIt	&	29.15	&	34.86	&	30.84	&	65.61	&	45.76	&	29.07	&	61.50	&	35.52	\\ \hline
CFD Solver	&	TimeStep	&	47.17	&	49.03	&	47.07	&	103.40	&	47.49	&	25.59	&	62.80	&	69.93	\\ \hline
Clock	&	timedReduction	&	5.10	&	4.05	&	8.02	&	6.75	&	4.15	&	4.05	&	6.17	&	5.48	\\ \hline
Convolution Seperable	&	convolutionRowsKernel	&	1,665.50	&	2,801.80	&	1,751.90	&	1,937.43	&	1,765.77	&	2,801.80	&	2,890.90	&	2,959.94	\\ \hline
Convolution Seperable	&	columns	&	1,165.50	&	985.57	&	1,751.90	&	1,205.50	&	2,483.50	&	2,057.84	&	1,651.08	&	1,284.50	\\ \hline
Convolution Texture	&	convolutionRowsKernel	&	620.85	&	794.89	&	1,271.50	&	1,352.65	&	698.57	&	794.89	&	1,596.70	&	1,253.06	\\ \hline
Convolution Texture	&	convolutionColumnsKernel	&	681.04	&	794.89	&	1,286.80	&	1,352.65	&	691.53	&	794.89	&	1,594.40	&	1,253.06	\\ \hline
Breadth-First Search	&	Kernel1	&	14.98	&	17.14	&	20.87	&	17.14	&	14.98	&	10.37	&	16.98	&	20.73	\\ \hline
Breadth-First Search	&	Kernel2	&	15.14	&	17.46	&	11.87	&	19.23	&	11.01	&	17.14	&	12.31	&	20.73	\\ \hline
BlackScholes	&	BlackScholesBodyGPU	&	547.12	&	819.18	&	631.51	&	819.18	&	660.25	&	819.18	&	813.50	&	970.27	\\ \hline
Fast Walsh Transform	&	fwtBatch1	&	459.66	&	390.25	&	486.91	&	520.62	&	571.66	&	414.85	&	684.03	&	720.84	\\ \hline
Fast Walsh Transform	&	fwtBatch2	&	879.38	&	770.58	&	1,367.70	&	1,280.54	&	569.74	&	410.80	&	1,389.20	&	1,148.25	\\ \hline
Gaussian	&	Fan1	&	2.14	&	1.91	&	4.54	&	2.07	&	5.65	&	2.91	&	2.82	&	1.44	\\ \hline
Gaussian	&	Fan2	&	326.63	&	147.93	&	533.50	&	251.47	&	374.26	&	147.93	&	549.41	&	189.12	\\ \hline
Hotspot	&	calculate\_temp	&	6.98	&	7.05	&	5.99	&	6.50	&	6.24	&	5.75	&	5.02	&	4.80	\\ \hline
K means	&	KmeansInvert	&	7,566.60	&	6,565.74	&	12,934.00	&	11,194.97	&	7,483.20	&	5,065.74	&	9,354.88	&	9,190.17	\\ \hline
K means	&	KmeansPoint	&	2,909.70	&	2,515.04	&	3,024.20	&	2,861.97	&	2,994.30	&	2,065.74	&	3,791.30	&	3,655.93	\\ \hline
LU Decomposition	&	lud\_diagonal	&	16.29	&	14.04	&	20.06	&	11.17	&	16.13	&	15.30	&	20.26	&	16.44	\\ \hline
LU Decomposition	&	lud\_perimeter	&	15.95	&	20.24	&	12.48	&	18.55	&	5.45	&	20.24	&	2.75	&	10.09	\\ \hline
LU Decomposition	&	lud\_internal	&	93.98	&	81.59	&	110.86	&	70.86	&	4.45	&	2.03	&	1.47	&	0.89	\\ \hline
Matrix Multiplication	&	matrixMulCUDA	&	6,997.75	&	5,825.05	&	8,600.28	&	5,583.24	&	4,917.40	&	5,825.05	&	8,514.90	&	9,099.98	\\ \hline
Merge Sort	&	mergeSortSharedKernel	&	2,184.26	&	3,102.00	&	5,683.00	&	5,014.00	&	1,946.60	&	2,645.00	&	4,459.00	&	5,728.00	\\ \hline
Merge Sort	&	generateSampleRanks	&	183.33	&	150.60	&	257.88	&	234.86	&	146.78	&	118.28	&	122.30	&	92.50	\\ \hline
Merge Sort	&	mergeRanksAndIndices	&	8.38	&	7.17	&	24.42	&	21.58	&	5.34	&	3.74	&	7.55	&	7.14	\\ \hline
Merge Sort	&	mergeElementaryIntervals	&	1,114.10	&	910.58	&	1,723.60	&	1,526.89	&	589.74	&	328.60	&	1,412.20	&	1,283.60	\\ \hline
Needleman-Wunsch	&	needle\_cuda\_shared\_1	&	18.75	&	15.45	&	15.49	&	14.24	&	13.63	&	15.45	&	10.59	&	7.82	\\ \hline
Needleman-Wunsch	&	needle\_cuda\_shared\_2	&	22.24	&	16.99	&	15.52	&	15.63	&	13.51	&	16.99	&	7.39	&	8.55	\\ \hline
Saxpy	&	saxpy	&	79.18	&	71.09	&	160.11	&	123.51	&	78.30	&	71.09	&	93.23	&	77.55	\\ \hline
Scan	&	scanExclusiveShared	&	361.83	&	678.01	&	501.37	&	887.07	&	425.17	&	678.01	&	894.11	&	622.25	\\ \hline
Sobol Qring	&	sobolGPU\_kernel	&	110.66	&	115.24	&	104.80	&	120.76	&	119.33	&	115.24	&	150.80	&	151.54	\\ \hline
Srad	&	srad\_cuda\_1	&	11.26	&	14.84	&	18.56	&	26.04	&	8.42	&	14.84	&	14.53	&	19.20	\\ \hline
Srad	&	srad\_cuda\_2	&	13.46	&	11.56	&	20.66	&	19.56	&	8.26	&	6.56	&	7.74	&	6.86	\\ \hline
Stereo Disparity	&	Stereo Disparity	&	379.60	&	322.96	&	286.10	&	343.86	&	363.46	&	322.96	&	295.36	&	302.09	\\ \hline
Transpose	&	transposeNaive	&	100.83	&	182.47	&	429.98	&	261.22	&	146.52	&	182.47	&	374.05	&	150.47	\\ \hline
Transpose	&	transposeCoalesced	&	78.05	&	234.03	&	160.10	&	319.02	&	166.94	&	573.03	&	124.56	&	213.61	\\ \hline
Transpose	&	Diagonal	&	83.17	&	69.75	&	162.21	&	144.69	&	93.23	&	74.75	&	128.64	&	122.18	\\ \hline
Transpose	&	NoBankConflicts	&	77.35	&	64.32	&	158.81	&	131.41	&	72.72	&	54.32	&	107.07	&	113.57	\\ \hline
Transpose	&	FineGrained	&	73.41	&	63.23	&	132.86	&	138.76	&	121.20	&	83.23	&	92.81	&	156.19	\\ \hline
Transpose	&	CoarseGrained	&	74.88	&	58.65	&	147.90	&	119.13	&	301.14	&	70.65	&	476.12	&	450.06	\\ \hline
Transpose	&	copy	&	62.15	&	53.03	&	87.10	&	57.89	&	66.94	&	52.28	&	93.02	&	82.56	\\ \hline
Transpose	&	copySharedMem	&	68.96	&	79.28	&	385.53	&	401.20	&	274.42	&	119.80	&	423.20	&	451.20	\\ \hline
Vector Addition	&	vectorAdd	&	2.12	&	1.81	&	2.20	&	1.98	&	1.82	&	1.49	&	1.65	&	1.39	\\ \hline
\multicolumn{2}{|c|}{\textbf{Mean Percentage Error}} & \multicolumn{2}{l|}{$-8.88\%$} & \multicolumn{2}{l|}{$-5.66\%$} & \multicolumn{2}{l|}{$-10.64\%$} & \multicolumn{2}{l|}{$-3.94\%$} \\ \hline
\multicolumn{2}{|c|}{\textbf{Mean Absolute Percentage Error}} & \multicolumn{2}{l|}{28.3\%} & \multicolumn{2}{l|}{29.4\%} & \multicolumn{2}{l|}{47.8\%} & \multicolumn{2}{l|}{28.5\%} \\ \hline
\multicolumn{2}{|c|}{\textbf{Execution Time of Predictor Tool}} & \multicolumn{8}{|l|}{Total: 2925ms \hspace{35pt}  Average: 66.48ms  \hspace{35pt}  Min: 1.031ms  \hspace{35pt} Max: 206.62ms} \\ \hline
\end{tabular}
}
\end{table}

\begin{table}
\caption{Mean Absolute Percentage Error for each dwarf across different GPU architecture}
\label{tab:my-table-dwarf-results}
\begin{center}
\resizebox{\textwidth}{!}{ 
\begin{tabular}{|l|l|l|l|l|}
\hline
\textbf{Dwarf Name}     & \textbf{Tesla K20} & \textbf{Quadro K4200} & \textbf{Tesla M60} & \textbf{GTX 1050} \\ \hline
Dynamic Programming   	&	20.6	&	4.39	&	19.56	&	20.93	 \\ \hline
Dense Linear Algebra 	&	29.95	&	27.92	&	51.75	&	34.67	 \\ \hline
Sparse Linear Algebra	&	87.38	&	76.93	&	59.47	&	30.41	 \\ \hline
Backtrack and Branch-and-Bound 	&	23.15	&	10.93	&	32.42	&	16.86	 \\ \hline
Graph Traversal   	&	14.87	&	39.64	&	43.23	&	45.24	 \\ \hline
Monte Carlo  	&	4.14	&	15.23	&	3.43	&	0.49	 \\ \hline
Structured Grid 	&	25.05	&	15.34	&	29.88	&	16.49	 \\ \hline
Unstructured Grid 	&	31.61	&	69.82	&	87.09	&	28.81	 \\ \hline
\end{tabular}
}
\end{center}
\end{table}

We have executed the benchmarks described in Table~\ref{tab:benchmakrs} on all the GPUs to understand our approach's generalizability across different architectures. We ran each kernel for multiple launch configurations by changing its grid and block size. For each configuration, we varied the dataset sizes wherever possible. For some benchmarks, the prediction accuracy is high (error<10\%), whereas, for some others, there have been over and underestimating the prediction. For instance, the transpose coalesced kernel in the Transpose benchmark suffers from the maximum overestimation across all architectures. Upon further investigation, we find that bank conflicts, coalescing, and caching are the significant causes of the overestimation.
The MPE and MAPE values for these two GPUs have been provided in Table~\ref{tab:results}.

In order to highlight various aspects of our estimation model, we chose the Computational Fluid Dynamics (CFD) solver benchmark taken from the Rodinia test suite and plotted the results in Figure \ref{fig_cfd_results}. CFD is a reasonably complex and popular algorithm used in many engineering as well as research fields. The benchmark comprises of four kernels. We have executed these kernels on a problem size of 97K data points. 

Figure~\ref{fig_cfd_results} depicts the results of all four GPU architectures. The graph shows that our model predicts execution time for the three kernels (initialize, compute\_step\_factor, and time\_step) with reasonable accuracy. We ran CFD benchmarks for three different grid size configurations (506 blocks, 1008 blocks, and 1212 blocks). Results show that our prediction model is scalable since it works decently across all three configurations. However, the prediction error is higher for the compute\_flux kernel. We profiled compute\_flux kernel to understand the reason for the prediction error. We found that kernel execution was affected due to branch divergence, which we have not considered in our current model. 

\begin{table}
\caption{Throughput Comparison: Our approach vis a vis Volkov's[20] approach}
\label{tab:volkov}
\resizebox{\textwidth}{!}{ 
\begin{tabular}{|l|l|r|r|r|r|r|r|}
\hline
\multicolumn{2}{|l|}{\textbf{Benchmark Details}}  & \multicolumn{3}{c|}{\textbf{Tesla K20 (Arithmetic Throughput Values)}} & \multicolumn{3}{c|}{\textbf{Tesla M60 (Arithmetic Throughput Values)}} \\ \hline
\textbf{Benchmark} &
  \textbf{Kernel} &
  \multicolumn{1}{l|}{\textbf{\begin{tabular}[c]{@{}l@{}}Volkov\cite{volkov2016} Model\end{tabular}}} &
  \multicolumn{1}{l|}{\textbf{\begin{tabular}[c]{@{}l@{}}Proposed Model\end{tabular}}} &
  \multicolumn{1}{l|}{\textbf{\begin{tabular}[c]{@{}l@{}}Actual \end{tabular}}} &
  \multicolumn{1}{l|}{\textbf{\begin{tabular}[c]{@{}l@{}}Volkov\cite{volkov2016} Model\end{tabular}}} &
  \multicolumn{1}{l|}{\textbf{\begin{tabular}[c]{@{}l@{}}Proposed Model\end{tabular}}} &
  \multicolumn{1}{l|}{\textbf{\begin{tabular}[c]{@{}l@{}}Actual \end{tabular}}} \\ \hline
Vector   Addition       & vectorAdd               & 6.481       & 6.4107      & 5.4733      & 5.5907      & 7.7875      & 6.3755      \\ \hline
Matrix   Multiplication & matrixMulCUDA           & 39.0696     & 1.5933      & 1.3263      & 23.7688     & 0.4621      & 0.5474      \\ \hline
Hotspot                 & calculate\_temp         & 5.2443      & 7.0464      & 7.1171      & 4.682       & 7.8197      & 7.2056      \\ \hline
Gaussian                & Fan1                    & 1.2217      & 2.7528      & 2.4569      & 1.0399      & 1.5576      & 0.8022      \\ \hline
Saxpy                   & saxpy                   & 6.6365      & 36.1331     & 32.4413     & 4.0374      & 21.6778     & 19.6817     \\ \hline
MergeSort               & mergeRanksAndIndices    & 21.6338     & 19.4705     & 16.6591     & 15.6288     & 26.7592     & 18.7414     \\ \hline
Transpose               & transposeCoalesced      & 17.8371     & 15.1779     & 45.5105     & 10.8516     & 4.5358      & 15.5694     \\ \hline
Transpose               & Diagonal                & 16.4114     & 47.6405     & 39.9534     & 9.9842      & 32.2877     & 25.8877     \\ \hline
CFDSolver               & Flux                    & 38.7485     & 3.9998      & 7.4097      & 23.5734     & 2.9726      & 13.3509     \\ \hline
Clock                   & timedReduction          & 17.2784     & 20.1448     & 15.9973     & 12.0472     & 14.3252     & 13.98       \\ \hline
BlackScholes            & BlackScholesBodyGPU     & 37.5466     & 33.7858     & 50.5861     & 22.8423     & 25.2992     & 31.389      \\ \hline
Needleman-Wunsch        & needle\_cuda\_shared\_1 & 0.1886      & 0.3139      & 0.2587      & 0.1621      & 0.2552      & 0.2893      \\ \hline
\end{tabular} }
\end{table}

\begin{figure}[htbp]
\centering
\includegraphics[width=12cm]{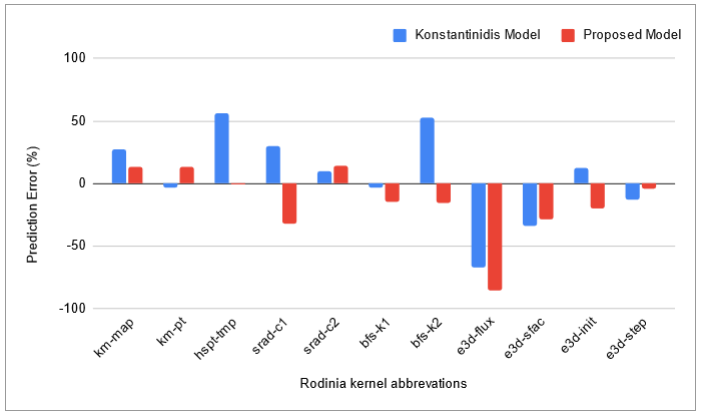}
\caption{Prediction error of the  Rodinia kernels on Tesla K20 GPU for Konstantidis Vs Proposed model}
\label{fig:konstanditis}
\end{figure}

\subsection{Analysis}

We have tested the model across three architectures: Kepler, Maxwell, and Pascal, whose details are provided in Table \ref{tab:hardwarevalue-table}. Each architecture differs in the number of SMs, cores per SM, clock rate, and Cache size. Benchmarks considered in this study are mentioned in Table \ref{tab:benchmakrs}. These benchmarks are classified into Parallel Dwarfs to investigate our results for a particular type of communication and computation pattern in a parallel algorithm. We presented results for two GPUs belonging to Kepler architecture in section \ref{Sec:keplerresults}. Errors observed in Figure\ref{fig_kepler_results} suggest that the benchmarks show acceptable precision for a given architecture. Saxpy, Clock, Vector Addition, and Convolution Text benchmarks have a very minute difference in their observed error. This implies that our model predicts reasonably for two GPU models belonging to the same architecture but with different performance capabilities. The insights we derived from the experiment are as follows:
\begin{description}
\item[Accuracy:]
In Table \ref{tab:results}, we presented results for benchmarks across GPU architecture. Analysis of results across architecture affirms that our model works reasonably across all three architectures under consideration: Kepler, Maxwell, and Pascal. The result shows that the prediction accuracy is the best for Kepler architecture, and then closely followed by Pascal architecture. In the experiment, the prediction accuracy of Maxwell architecture is the worst. 

Based on these results and error values across four GPU architectures, we are hopeful that the model will be able to predict with a comparable accuracy when applied to a future GPU architecture.

\item[Best cases:]
  We observed that the prediction model works best for benchmarks belonging to Dynamic Programming, Structured Grid, Backtrack and Branch-and-Bound, and Dense Linear Algebra, as suggested in Table \ref{tab:my-table-dwarf-results}. Vector Addition, Saxpy, and Matrix Multiplication belong to the Dense Linear Algebra dwarf and are computationally bound\cite{Asanovic2009}. Based on the result shown in Table~\ref{tab:results} and Table~\ref{tab:my-table-dwarf-results}, it appears that our model works well for computationally bound benchmarks. We have also observed that the prediction model works well for benchmarks with short execution time, such as Vector Addition and Needleman-Wunsch, and benchmarks that take a long time to execute, such as Matrix Multiplication and Black-Scholes. Therefore, the accuracy of our model is not strongly dependent on the execution time.
 
Upon profiling Structured Grid and Dynamic Programming benchmarks, we observed that global memory bandwidth mostly bounds these benchmarks. It appears that our empirical models related to global memory access have been effective in predicting latency values of memory-bound instructions of these benchmarks.

 Monte Carlo benchmarks are neither compute-bound nor memory-bound. Their execution characteristics are highly dependent on the nature of the problem, as discussed in~\cite{asanovic2006}. Although our model effectively predicts Monte Carlo benchmarks, we cannot claim our prediction's generality for such benchmarks.

\item[Worst cases:]
As observed in Table \ref{tab:my-table-dwarf-results}, Sparse Linear Algebra dwarf gives the worst results, followed by Unstructured grids. Sparse Linear Algebra dwarf consists of indirect and irregular memory access patterns \cite{Malossi2014}, and hence predicting its execution time is a complex problem. As per Asanovic et al. \cite{asanovic2006} report, which introduced parallel dwarfs, unstructured grids could be loosely interpreted as a sparse linear algebra problem. Hence challenges of sparse linear algebra prediction apply to unstructured grids as well. 

The result shows that the \emph{Fan1} kernel of the Gaussian benchmark is underestimated for all the architectures. Fan1 has several {\tt if} statements that can lead to branch divergence. We also observed that it has $16.34\%$ stalls due to data requests. We have not been able to model such a stalling phenomenon in our prediction model as yet. \emph{TransposeCoarseGrained} kernel of Transpose benchmark is overestimated in most cases. In a study by Greg et al. \cite{transpose}, it is reported that the performance bottleneck for this kernel lies in writing data to the transposed location in global memory. We found that this kernel's memory access pattern is uncoalesced, which we have not factored in our model.

\item[Bandwidth prediction:]
We have also observed that the effective bandwidth (GB/s) of \emph{TransposeCoarseGrained} is very low for this kernel. It appears that our model is not able to incorporate the impact of low effective bandwidth. Vector Addition, Saxpy, and Matrix Multiplication, for which the prediction model worked well, had a very high bandwidth utilization. 

\item[Comparison with other approaches]
Lemeire et al. \cite{Lemeire2016} proposed a set of equations based on microbenchmark applications to calculate the issue and completion latency (denoted by $\Delta$), unlike ours. For the sake of comparison, we have considered the $\Delta$ values of arithmetic instructions (9 cycles) for Kepler architecture with ours. We observed that our results (shown in Table~\ref{tab:latency-table}) indeed match that of Lemeire. Both these approaches\cite {volkov2016}\cite{Lemeire2016} have not been tested well using real-life applications. 

Volkov et al. \cite{volkov2016} proposed an analytical model to predict arithmetic and memory throughput, whereas our model predicts end-to-end execution time. Our proposed model also exhibits the exponential behavior of memory throughput. For comparison, we used our prediction model to generate arithmetic throughput. We applied our approach and Volkov's \cite{volkov2016}] model on a set of kernels presented in our work and computed arithmetic throughput values. Table~\ref{tab:volkov} shows a comparison of arithmetic throughput with the Volkov\cite{volkov2016} model, which indicates that throughput computed using our approach is close to the actual throughput. We have considered Kepler and Maxwell architectures since Volkov's work\cite{volkov2016} also includes these two architectures.  

In contrast to Konstantinos et al.\cite{Konstantinidis2017AQR} improvised roofline model, our approach predicts the execution time of a CUDA kernel without using any runtime data of this kernel; we believe that the use of runtime data of a kernel to predict its execution time largely defeats the purpose. Figure \ref{fig:konstanditis} shows the prediction error (not absolute error) for benchmarks that are common to Konstantinidis \cite{Konstantinidis2017AQR} model and ours. When we compare the performance of Konstantinidis's approach\cite{Konstantinidis2017AQR} vis a vis ours, the average absolute error reported by them is $27.66\%$, whereas MAPE values reported by our model is $28.3\%$ for Tesla K20 GPU, which they also use. Therefore, our prediction based on the static data is not too far off compared to the result by Konstantinidis et al.'s runtime data. We have shown a comparative study of our results with existing approaches in Table \ref{tab:existing approaches}. As seen in the table considering the size of the dataset used for testing and the variation in benchmark characteristics, our model shows decent results.

\begin{table}
\caption{Comparison with existing approaches}
\label{tab:existing approaches}
\resizebox{\textwidth}{!}{%
\begin{tabular}{|l|l|l|l|l|}
\hline
\textbf{Reference}                                         & \textbf{Type of   Approach}        & \textbf{Analysis Type} & \textbf{Dataset Size} & \textbf{Accuracy}                                                                                         \\ \hline
Resios et al. \cite{resios2011}           & Parameterized Model                & Static             & 2                     & \begin{tabular}[c]{@{}l@{}}Precision:   Transpose (1.64)  , \\ 2D Convolution   filter(1.89)\end{tabular} \\ \hline
Kothapalli et   al. \cite{Kothapalli2009} & Analytical model                   & Static             & 3                     & \begin{tabular}[c]{@{}l@{}}Matrix Multiplication   (31.25\%  ),\\  List Ranking (12.50\% )\end{tabular}   \\ \hline
Hong et al. \cite{Hong2010}               & Analytical model                   & Static             & 6                     & MAPE : 13.3\%                                                                                             \\ \hline
Elias et al. \cite{Konstantinidis2017AQR} & Roofline  Model using runtime data & Runtime            & 30                    & MAPE: 27.66\%                                                                                             \\ \hline
Amaris et al.    \cite{amaris2015}        & Intuitive   BSP based Model        & Static             & 2                     & \begin{tabular}[c]{@{}l@{}}Within   0.8 to 1.2 times \\ the actual execution time\end{tabular}            \\ \hline
Luo et al.  \cite{luo2011}                & Analytical   Model                 & Runtime            & 6                     & Average   accuracy of 90\%                                                                                \\ \hline
Amaris et al. \cite{amaris2015}           & Machine   Learning Model           & Runtime            & 2                     & \begin{tabular}[c]{@{}l@{}}Within 0.5 and 1.5 times \\ the actual execution time\end{tabular}             \\ \hline
Our Approach                                               & Analytical Model                   & Static             & 45                    & MAPE: 28.3\%, MPE: -8.88\%                                                                                \\ \hline
\end{tabular}%
}
\end{table}
\end{description}

\section{Discussion}
In this chapter, we described, evaluated, and analyzed two performance prediction models. The first model is based on using parametrised equations for modelling parallelism using ILP \& TLP in order to compute execution time. In order to improve upon the benchmarks which underestimate the execution time, we proposed a second method based on the hardware resource allocation approach. We reutilised initial preprocessing steps, microbenchmarking results, launch the overhead model from the first model, and rebuilt the GPU scheduling algorithm. A more complex one-time dynamic analysis based memory model was also introduced. Both the proposed models were tested and validated on a large number of benchmarks. We have demonstrated that the proposed model is easily tunable to newer architectures. Performance prediction is a hard problem, especially for parallel computers and GPUs hence the accuracy of the proposed model is a novel contribution to performance prediction research .

\let \textcircled=\pgftextcircled
\chapter{Power Prediction}
\label{ch:powerpred}

\initial{A}s described in objective 2, the power consumption of a computing device is intricately related to its hardware features. Consequently, there has been a significant body of work to investigate the relationship between hardware features and power consumption~\cite{Fan2019}\cite{Guerreiro2018}\cite{Lim2015}\cite{Lucas2013}. However, our goal is to develop a power prediction model based on code features and minimal runtime information so that the model can be used for power consumption prediction without running a CUDA program. To the best of our knowledge, not much work has been done that uses such an approach.

\section{Why a machine learning based approach for power prediction? } \label{sec:whymlforPower}
Most analytical power prediction models use operational characteristics of the GPU and underlying electrical properties to predict the power consumed by a GPU. Such a model shows a good performance for the system whose operational properties have been taken into account while building the analytical model.  Also, we did not come across any algorithmic approach to understand the relationship between power consumption and program characteristics for GPU applications. Also, quantifying the power consumption of GPU at the instruction level is not capturable. Because of this, we have employed a machine-learning-based approach to model the complex relationships between application, architecture features, and power consumption.

Data-driven power prediction model is an alternate approach that estimates the power consumption influenced by various operating components of a GPU that is highly non-linear. 
Due to the inherent generality in data-driven approach, a data-driven model with sufficient data points has the potential to capture the complexity arising out of the diversity of different components of a GPU acting together to execute a code. Of course, the model must be adequately trained with a large size and diversity of the dataset used for learning. The generality of a data-driven approach lies in the fact the model does not require any functional properties or physical characteristics of the components inside a GPU, nor does it require any minute architectural details of a GPU. In our approach, we have consciously tried to increase this diversity by choosing benchmarks with different characteristics, which could be a reason for the robustness of the model demonstrated in the paper.

\section{Initial Investigation}
Before finalising the features for our power prediction model, we performed an initial study that included features derived from only static analysis of code and hardware features. We use a data-driven approach wherein we consider all the features which we think might contribute to power consumption and then systematically choose the final subset of features.
\subsection{Features Considered}
We considered features that can be extracted from program analysis that can lead to an increase in power consumption. Most of these features are self-explanatory, details of derived features such as inst\_issue\_cycles are available in section \ref{Feature_identification} for details. Program Feature Extraction Algorithm is presented in section \ref{pfea}. The sixteen such features considered in this initial study are listed and described in Table \ref{table_all_features_initial_model}.
\begin{table}[]
\caption{Features Considered for feature selection}
\centering
\label{table_all_features_initial_model}
\begin{tabular}{|l|l|}
\hline
\textbf{Notation} & \textbf{Description} \\ \hline
block\_size & Number of threads per block \\ \hline
grid\_size & Number of blocks \\ \hline
comp\_inst\_kernel & Number of computation instruction in kernel \\ \hline
comp\_inst\_sim & Number of simulated computation instructions \\ \hline
glob\_inst\_kernel & Number of global memory instruction in kernel \\ \hline
glob\_inst\_sim & Number of simulated global memory instructions \\ \hline
inst\_issue\_cycles & Instruction Issue Cycles \\ \hline
misc\_inst\_kernel & Number of miscellaneous instruction in kernel \\ \hline
misc\_inst\_sim & Number of miscellaneous instructions \\ \hline
occupancy & \begin{tabular}[c]{@{}l@{}}Ratio of active warps on an SM to maximum\\ number of active warps supported by the SM\end{tabular} \\ \hline
shar\_inst\_kernel & Number of shared memory instruction in kernel \\ \hline
shar\_inst\_sim & Number of simulated computation instructions \\ \hline
total\_threads & Total number of threads launched \\ \hline
total\_inst & Total number of instructions simulated \\ \hline
waves & Number of waves of blocks executed on SMs \\ \hline
\end{tabular}
\end{table}

\subsection{Feature Engineering}
\begin{figure}[ht]
\centering
\caption{Pearson Correlation Coefficient}
\label{fig_pearson_IUCC}
\includegraphics[width=\linewidth]{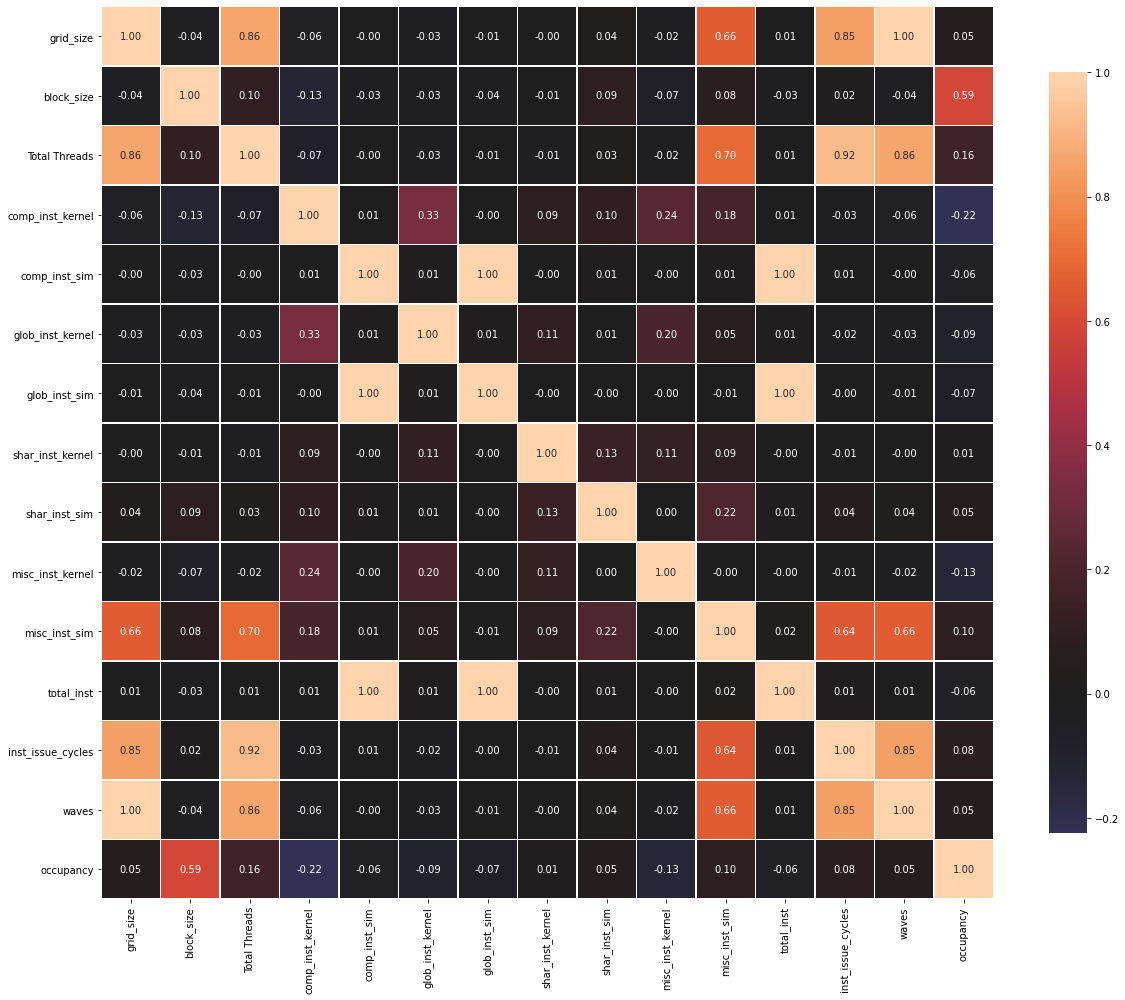}
\end{figure}

We then utilised the Pearson correlation coefficient to observe the relationship of attributes among themselves and discard highly correlated features. We have discussed Pearson correlation analysis in detail in section \ref{featureEngineering}. With the resulting heat-map of the Pearson correlation coefficient in Figure \ref{fig_pearson_IUCC}, we can observe the following:
\begin{itemize}
    \item Features like waves, grid\_size and Total Threads are highly correlated with the feature inst\_issue\_cycles with a correlation coefficient of 0.85,0.92 and 0.85 respectively. Hence we drop these features from our model and only keep inst\_issue\_cycles.
    \item comp\_inst\_sim, glob\_inst\_sim, and total\_inst are highly correlated with each other with a correlation coefficient of 1. So we consider only glob\_inst\_sim for our model and drop the other two features.  
    
\end{itemize}

\begin{table}
\caption{Features Considered for initial case study}
\centering
\label{table_final_features_initialmodel}
\begin{tabular}{|l|l|}
\hline
\textbf{Notation} & \textbf{Source} \\ \hline
block\_size & User Input \\ \hline
comp\_inst\_kernel & PFEA \\ \hline
glob\_inst\_kernel & PFEA  \\ \hline
glob\_inst\_sim & PFEA \\ \hline
inst\_issue\_cycles & PFEA\\ \hline
misc\_inst\_kernel & PFEA \\ \hline
misc\_inst\_sim & PFEA \\ \hline
occupancy & CUDA Occupancy Calculator \\ \hline
shar\_inst\_kernel & PFEA \\ \hline
shar\_inst\_sim & PFEA \\ \hline
\end{tabular}
\end{table}

\subsection{Model Building \& Results}
 Decision Trees \cite{Breiman1983ClassificationAR} builds the regression or classification model based on a tree-based structure, containing decision nodes and leaf nodes. The cost of using a decision tree is logarithmic in the number of data points used to train the model. Decision Trees could provide a basic idea for evaluating the performance of other complex Tree-Based models on the power consumption data. We use the sklearn library for building our Decision Tree model, which uses the modified version of the CART (Classification and Regression Trees) algorithm. We train the model by tuning various hyper-parameters such as minimum samples required to split an internal node, maximum depth of the tree, and maximum features required for the best split and employing pruning which avoids the overfitting of the model.

\subsection{Result \& Analysis} \label{analysingresults_IUCC}
We used three essential metrics to validate our model based on the mean of 5-fold cross-validation scores: $R^{2}$ score, Root Mean Squared Error (RMSE), and Mean Absolute Error (MAE). We have tabulated results for these metrics in Table \ref{result_table} for all three GPU machines under consideration. 

$R^{2}$ score for all three architectures are decent, and hence we claim our model produces accurate predictions. RMSE values for all three architectures also suggest that our model is robust and accurate. As seen in Figure \ref{fig_predict}, for all of the popular benchmarks seen our predicted power is quite precise against measured value.  

\begin{table}[]
\centering
\caption{Results across GPU architectures}
\label{result_table}
\begin{tabular}{|l|r|r|r|}
\hline
\textbf{Architecture} & \multicolumn{1}{l|}{\textbf{$R^{2}$ score}} & \multicolumn{1}{l|}{\textbf{RMSE}} & \multicolumn{1}{l|}{\textbf{MAE}} \\ \hline
Tesla K20             & 0.8199                                      & 10.8832                            & 5.7835                            \\ \hline
Tesla M60             & 0.8533                                      & 7.3676                             & 3.3581                            \\ \hline
Tesla V100            & 0.8973                                      & 12.2025                            & 5.4570                            \\ \hline
\end{tabular}
\end{table}

Through this initial study, we attempt to answer the following Research Questions (RQ) based on the observed results and model analysis: 

\textbf{RQ1: Is it possible to build an architecture agnostic model to predict the power consumed by a CUDA program with tolerable error and accuracy?  } 
As seen in the experimental results, we validated our model across three different GPU architectures: Kepler, Maxwell, Volta.  All three validation metrics, i.e. $R^{2}$ score, RMSE, and MAE results, suggest that our model is architecture agnostic since it performs efficiently across all three architectures. Our model was tested for a heterogeneous mixture of CUDA kernel benchmarks with varying memory access patterns, launch configuration, applicability and execution bound type. This answers RQ1 affirmatively.    

\begin{figure}[ht]
 \centering
 \includegraphics[width=12cm]{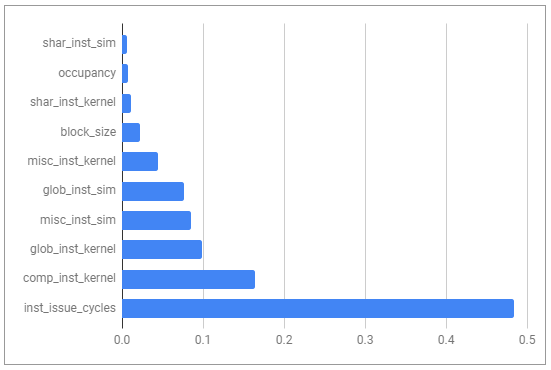}
 \caption{Relative feature importance for Tesla K20}
 \label{fig_fi_k20}
 \end{figure}

\textbf{ RQ2: Which features of the CUDA program considered in this work impact most significantly to the power consumption of GPU?} \\
The observed power prediction model result suggests that we can relate power consumed by CUDA kernel and program features. We analyze which of the program features considered in this work contribute the most to power prediction. Based on the feature importance analysis using Decision Tree for Tesla K20 as shown in Figure\ref{fig_fi_k20}, we found that instruction inst\_issue\_cycle contribute the most to power consumption across three architectures.


Global memory access instructions have a significant effect on power consumption. Both instruction count (glob\_inst\_kernel) and a simulated number of instructions ( glob\_inst\_sim) have ranked higher in all three architectures. This supports the observation by Song et al. \cite{Shuaiwen2013} on the influence of global memory on power consumption. Computing instructions count (comp\_inst\_kernel) also affects the power prediction similar to miscellaneous instruction count (misc\_inst\_kernel) and the number of simulated miscellaneous instructions (misc\_inst\_sim). Some of the computing instructions ({\tt fmadd, divf}) have much higher latency compared to others ({\tt add, mul}) \cite{Alavani2018}. As an extension to this work, we can explore the effect of the latency of each type of instruction on GPU power consumption. 

The effectiveness of the number of blocks launched (block\_size) follows features, as mentioned earlier. The number of blocks launched can represent the number of SMs activated during CUDA kernel execution lifetime. This is because each new block is mapped to one SM. However, in our observation, it has affected power consumption moderately. 



We presented this study in our earlier publication \cite{Alavani2020}. Although the observed result study is decent, we strongly believed that the inclusion of dynamic analysis features collected using architecture specific details can make a more reliable and accurate model. Also, exploring
other machine learning techniques which can improve
the observed prediction error is important to build a better accurate model. We also performed an initial study of using the runtime feature for execution time, power, and energy prediction which is presented in Appendix \ref{AppA}.   

Hence to further strengthen the model we propose an advanced study, in which more features from the architectural perspective which can be collected using one-time dynamic analysis to build a more robust and accurate prediction model with the help of newer machine learning algorithms.

\section{Proposed Model} \label{powermodel}
In order to build a power consumption prediction model, it is required to understand the hardware intricacies of the GPU. Furthermore, it is also necessary to understand the run-time behavior of a CUDA application. However, publicly available information on NVIDIA GPUs do not disclose crucial hardware execution details such as instruction latencies, cache behavior, memory access details, scheduling strategy, resource allocation, and so forth, that influence GPU power consumption. 

\begin{figure}[]
\centering
\includegraphics[width=15cm]{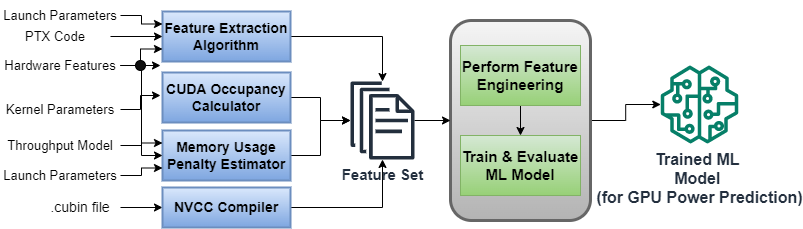}
\caption{Overview of Power Prediction Modeling Approach}
\label{fig:overview}
\end{figure}

We have developed a machine-learning-based power prediction model that combines i) CUDA program characteristics obtained from program analysis ii) GPU architectural features obtained from GPU specification provided by the Vendor and those derived from operational data (by running a set of microbenchmarks). An overview of the approach is shown in Figure \ref{fig:overview}.

We will be using the abstract GPU model described in section \ref{abstract_model} of Chapter \ref{ch:execTime} for presenting the proposed power prediction model. The notations in this model are used to explain some of the features collected for prediction. 

We first discuss in detail how features were chosen and collected for building this power prediction model. The next section presents a comprehensive approach of features considered from different dimensions.

\section{Feature Identification \& Acquisition } \label{Feature_identification}
Power consumed by a GPGPU is influenced by a complex interplay among the i) GPGPU architecture-specific nuances, ii) CUDA program characteristics, its runtime behavior, and iii) launch strategy (determined by launch parameters). However, publicly available information on NVIDIA GPUs does not disclose crucial hardware execution details such as instruction latencies, cache behavior, memory access details, scheduling strategy, resource allocation, and so forth that influence GPU power consumption. To circumvent this issue, we propose a machine learning-based power prediction model that uses features derived from i) GPU architectural features and ii) CUDA program characteristics obtained from program analysis. Machine learning algorithms are being effectively used to predict architecture nuances for modern architectures \cite{Nagasaka2010}\cite{Chen2011}. 

We now discuss how various features are derived in the following subsection. 
\subsection{Registers per thread, Shared memory per block} 
Registers per thread and shared memory per block impact the occupancy of SM, which in turn influences hardware utilization. Past studies~\cite{Hong2010,jatala2017greener} have shown that these attributes play a crucial role in power consumption. Hence we consider both features in our work. 

We generated these values at compile time using \verb+nvcc --cubin+ to generate a .cubin file~\footnote{https://docs.nvidia.com/cuda/cuda-binary-utilities/index.html}. A .cubin file is a CUDA binary file that is a ELF-formatted consisting of CUDA executable code sections as well as other details such as symbols, relocators, debug info, etc. It contains information on how many registers and shared memory per block are expected to be used for the code during runtime.

\subsection{Occupancy}
Occupancy measures the ratio of the actual number of warps executed on an SM to the maximum number of warps allowed for an SM. We consider occupancy as a feature in our model to understand the effect of thread-level parallelism (TLP) on GPU power consumption. Studies have also proved that GPU power consumption is affected by Occupancy \cite{Alnori2018,Shuaiwen2013}. The occupancy value considered in this work is theoretical occupancy which captures the limit for occupancy restricted by the kernel launch configuration and the capabilities of the CUDA device. Achieved occupancy is measured during the runtime of code, and the two values may slightly differ. However, to achieve our goal of prediction using static analysis, we consider theoretical occupancy in our work. 

Since we do not wish to execute our code, we utilize NVIDIA's CUDA Occupancy Calculator to compute theoretical occupancy for a CUDA program. CUDA Occupancy Calculator utilizes the number of threads per block (block\_size), number of registers used per thread (reg\_thread), number of shared memory used by a block of thread (shmem\_block), and CUDA Compute Capability of the device under consideration.

We utilize CUDA Occupancy Calculator \footnote{https://docs.nvidia.com/cuda/cuda-occupancy-calculator/index.html}, which reports the theoretical maximum occupancy that can be achieved on the execution of a particular kernel by taking in four inputs: 
\begin{itemize}
    \item number of threads per block
    \item number of registers used per thread
    \item number of shared memory used by a block of thread
    \item CUDA Compute Capability of the device under consideration
    \end{itemize}

Some architecture features are possible to derive without any detailed internal knowledge of a GPU . We describe these features in detail.

\subsection{Instruction issue cycles}
An NVIDIA GPU uses a Fetch Decode Schedule (FDS) unit to issue instructions to threads for execution. Naturally, an FDS unit contributes to the energy consumed by a GPU. In order to model the role of an FDS in power consumption, we have proposed a feature called Instruction issue cycles (\emph{inst\_issue\_cycle}), which characterizes the number of batches of instructions issued. This is computed by dividing the total instructions issued for a kernel with a hardware limit of the number of instructions allowed per cycle. We use publicly available hardware details such as the number of warp schedulers (nWS) and number of dispatch units (nDU) per SM to compute \emph{inst\_issue\_cycle}. The dispatch unit is responsible for taking a given instruction, its warp ID, and the thread mask and calculating the thread IDs that will actually execute the instruction. nDU then sends each thread instruction applicable functional unit by checking its availability.
%
Issuance of instructions is dependent on the total threads launched, total instructions, and warp scheduling policy by the GPU scheduler. We can extract total threads launched (\emph{total\_threads}) and total instructions (\emph{total\_inst}) using user-defined parameters and the feature extraction algorithm (Algorithm \ref{algo:b1} and \ref{algo:b2}) respectively. Since the warp scheduling and idle cycles of the warp scheduler are difficult to predict, we assume that the warp is always full and doing useful work. The warp size (\emph{warp\_size}), which is typically 32 threads per warp, can be obtained by querying the GPU only once. We calculate \emph{inst\_issue\_cycle} using the following formula:
\begin{equation} \label{eqn:inst_issue_cycle}
inst\_issue\_cycle= \dfrac{total\_threads}{nWS\cdot warp\_size } \cdot \dfrac{total\_inst}{nDU}
\end{equation}
$\dfrac{total\_threads}{nWS\cdot warp\_size }$ computes the number of batches of warps scheduled on GPU. We multiply this value by the number of batches of instruction scheduled  $\dfrac{total\_inst}{nDU}$. Since GPU is a pipelined architecture, we assume each batch of instruction is issued in one cycle.

\subsection{Memory Access Penalties}
The work by Nagasaka et al.~\cite{Nagasaka2010} and Shuaiwen~\cite{Shuaiwen2013} have demonstrated that GPU memory utilization is a significant contributor to power consumption. We model memory access overheads (that eventually contribute to power consumption) as penalties. In order to compute these penalty values, we require $\mathcal{TP}(gm)$ and $\mathcal{TP}(sm)$ throughput values. We have created empirical models for these two throughputs using microbenchmarking data. Details of these model creations and validations are in chapter \ref{ch:execTime}.
\subsubsection{Cache Penalty} 
 Global memory accesses are always routed through the L2 caches, and it is shared by all SMs. There is a significant delay in instruction execution if there is a cache miss. Cache penalty describes the delay after a cache miss occurs. Modeling accurate cache behaviour using program analysis is an impossible task since it depends on multiple runtime factors. We approximate this behaviour by computing an L2 cache miss penalty using available architecture and program details to approximate the behaviour of a GPU Cache for global memory data quantitatively. L2\_sz describes the L2 Cache size collected using device query. We use the waves and \emph{glob\_inst\_sm} computed in feature extraction Algorithm~\ref{algorithm:b1}, and \ref{algorithm:b2}. Using the global memory latency ({L}(gm)), we calculate cache penalty as:
\begin{equation} \label{equation:cache}
cache\_penalty=\dfrac{total\_threads\cdot glob\_inst\_sm}{\frac{waves~\cdot~ L2\_sz}{access_{sz}}}\times \mathcal{L}(gm)
\end{equation}
\subsubsection{Global Memory Penalty}
Since global memory usage plays a crucial role in power consumption, we create a global memory penalty as a feature that quantitatively determines global memory usage. To do so, we first count the number of memory accesses using the number of load-store units per SM ( R[LSU] ), i.e., $\frac{total\_threads}{R[LSU]}$. Then, we calculate the time required by each such access ($\frac{access\_{sz}}{\mathcal{TP}(gm)}$). We multiply the number of accesses by the time required to get a penalty for one global memory instruction. This penalty is then multiplied to global memory instruction per SM (\emph{glob\_inst\_sm}) to compute the global memory penalty by the following equation:
\begin{equation} \label{equation:global}
glb\_penalty=\frac{total\_threads}{R[LSU]}\times \frac{access\_{sz}}{\mathcal{TP}(gm)}\times glob\_inst\_sm
\end{equation}

\subsubsection{Shared Memory Penalty}
Shared memory instructions also significantly affect the power consumption of GPU, especially when kernel execution experiences a significant number of bank conflicts \cite{Shuaiwen2013}. Hence we also consider the shared memory penalty feature to model this effect on power. Then we compute the penalty for one shared memory instruction by counting the number of times shared memory instructions are occurring on SM ($\frac{total\_threads}{R[LSU]\cdot nSM}$). Here, nSM stands for the number of SMs in a GPU architecture and can be derived from a device query. We then multiply this penalty value for one instruction by the time taken for each of these shared memory instructions ($\frac{access_{sz}}{\mathcal{TP}(sm)}$) where \emph{access\_{sz}} stands for the number of bytes per access of shared memory. In the final computing shared memory penalty step, we multiply the number of shared memory instructions per SM generated using a feature extraction algorithm.
The \emph{sm\_penalty} is computed as follows: 
\begin{equation} \label{equation:shared}
sm\_penalty=\frac{total\_threads}{R[LSU]\cdot nSM}\times \frac{access\_{sz}}{\mathcal{TP}(sm)}\times  shar\_inst\_sm
\end{equation}
\subsection{Program Feature Extraction Algorithm} \label{pfea}
The Program Feature Extraction Algorithm (PFEA) described in Algorithm \ref{algorithm:b1} and \ref{algorithm:b2}, enacts GPU execution behavior based on static analysis of CUDA code and generates additional features for the dataset. Depending upon the number of times the kernel code is run based on launch configuration, the number of executed instructions, their total latency, and their average latency are computed. We have modified the GPU scheduling Algorithms \ref{algo:b1},\ref{algo:b2}, and \ref{algo:b3} presented in Chapter 4 in order to extract these features. GPU Scheduling approach in both the algorithms is same. We utilise variables to store the instruction details which are used as features for power modelling.   

The algorithm works as follows. The user provides CUDA kernel PTX along with launch parameters \emph{block\_size} (nT), \emph{grid\_size} (nB) and the number of loop iterations (nLoop) as inputs to FeatureExtraction Algorithm \ref{algorithm:b2}. Based on the maximum number of threads that can reside per SM ($nTh_m$) and the total number of threads that are scheduled per SM (\emph{nTh\_schd}), the number of threads per wave (\emph{nTh\_wave}) is calculated in line 6. The iteration continues till \emph{nTh\_schd} is zero, which ensures that all the threads are scheduled and complete their execution. A CUDA kernel is further divided into Control Flow Graph (CFG), which is popularly used to model the kernel application \cite{Hecht1974}. In line 7, using GenerateCFG, we apply topological sort on CFG to get sorted basic blocks (BB'). For each graph present in basic block BB' we call the CollectBBFeatures function, which extracts features for each basic block.   

In Algorithm \ref{algorithm:b1} CollectBBFeatures function, for each basic block (BB) of CFG ($G_{cfg}$) code, each instruction represents a vertex in a topologically sorted order. Each vertex v in the basic block represents an instruction. Algorithm \ref{algorithm:b1} calculates the count of each instruction type and its latency using $\mathcal{I}(\mathcal{T}(v))$ (line 6) and  $\mathcal{L}(\mathcal{T}(v))$ (line 7) respectively, and the number of global load, global store instructions, and branch instructions using  $\mathcal{N}(\mathcal{T}(v))$ (line 15).  We include the global load and store instructions separately since many researchers suggest that global memory instruction impacts power consumption significantly \cite{Nagasaka2010}\cite{Shuaiwen2013}.  Using number of instructions executed ($\mathcal{I}(\mathcal{T}(v))$.count) and total instruction latency ($\mathcal{L}(\mathcal{T}(v))$.total) populated using algorithm, we compute  \emph{avg\_comp\_lat}, \emph{avg\_misc\_lat}, \emph{avg\_shar\_lat}, and \emph{avg\_glob\_lat}. In line 18, we generate the array with feature count $c_{bb}$ and return to Algorithm \ref{algorithm:b2}. In Algorithm \ref{algorithm:b2}, if there is a loop, the feature count for a basic block is multiplied by the number of loop iterations. We then return the features calculated in CFG by adding them to the total feature (f\_{kernel}), which consists of all n number of features collected during the lifespan of a CUDA kernel. 

\begin{algorithm}[ht]
    \caption{Collect features of basic Block}
    \label{algorithm:b1}
    \begin{algorithmic}[1]
    \Procedure{CollectBBFeatures}{$\mathcal{GPU}$, $G=\langle V,E\rangle$, $nTW$}
    \State $c_{bb}=[n-1]$ , f=0;
    \State $V'$=\Call{Toposort}{$G$}
         \For{$v\in V'$}
       \State Read( $\mathcal{T}(v)$ )
             \State $\mathcal{I}(\mathcal{T}(v))$.count++;
             \State $\mathcal{L}(\mathcal{T}(v))$.total=$\mathcal{L}(\mathcal{T}(v))$.total+$\mathcal{L}(\mathcal{T}(v))$;
      \State Read( $\mathcal{N}(v)$ )
              \State $\mathcal{N}(\mathcal{T}(v))$.count++; 
     \EndFor
        \For{each $\mathcal{T}$}
        \State $c_{bb}[f]=\mathcal{I}(\mathcal{T}).$count; $f++;$
        \State $c_{bb}[f]= \mathcal{L}(\mathcal{T})$.total; $f++;$
         \If{$\mathcal{N} \epsilon (branch, global load, global store)$ }
             \State $c_{bb}[f]=\mathcal{N}(\mathcal{T})$.count; $f++;$
         \EndIf
     \EndFor
    \State \textbf{return} $c_{bb}$; \Comment{Feature count for Basic Block}
    \EndProcedure
    \end{algorithmic}
\end{algorithm}

\begin{algorithm}[ht]
    \caption{Program Feature Extraction Algorithm (PFEA) }
    \label{algorithm:b2}
    \begin{algorithmic}[1]
    \Procedure{featureExtraction}{$\mathcal{GPU},G_{cfg}=\langle\mathbb{C, E}\rangle$, $nB, nT, nLoop$} 
     \State $f_{kernel}=[n]$; \Comment{ Feature array for Kernel of size n  }
    \State \textit{nTh\_schd} $=\left\lceil\frac{nB}{nSM}*nT\right\rceil$;
    \State waves=0;
    \While{\textit{nTh\_schd} $\geq 0$}
      \State \textit{nTh\_wave} = \textit{ nTh\_schd}~$>nTh_m$~?~$nTh_m$ : ~\textit{nTh\_schd};
      \State  $BB'$= \Call{GenerateCFG}{$\mathcal{GPU}, G_{cfg}$, {\it nTh\_wave}};
      \For{$G\in BB'$}
       \State $c_{BB}=$\Call{CollectBBFeatures}{$\mathcal{GPU},G, nTW$};
       \If{$G$ has back-edge}
          \For{$f=0$ to $n-1$}
          \State $c_{BB}[f]=c_{BB}[f]*nLoop$
          \EndFor
           \EndIf
      \EndFor
         \For{$f=0$ to $n-1$ }
           \State $f_{kernel}[f]=f_{kernel}[f]+ c_{BB}[f]$; $f++$; 
        \EndFor
           \State {\it nTh\_schd} = {\it nTh\_schd}~$-~nTh_m$;
          \State waves=waves+1;
    \EndWhile
          \State $f_{kernel}[n-1]$=waves;
   \State \textbf{return} \textbf($f_{kernel}$); \Comment{Feature array collected for Kernel  }
    \EndProcedure
    \end{algorithmic}
\end{algorithm}
Table~\ref{tab:all_features} shows all the features collected using the algorithm. 
\begin{table}[htb]
\caption{Features considered for feature selection. Highlighted rows indicate features that are finally selected by data-driven methods (section \ref{featureEngineering})}
\label{tab:all_features}
\small
\centering
\resizebox{\linewidth}{!}{
\begin{tabular}{|l|l|l|}
\hline
\multicolumn{1}{|c|}{\textbf{Name}} & \multicolumn{1}{c|}{\textbf{Description}}                                                                                          & \multicolumn{1}{c|}{\textbf{Source}} \\ \hline
\rowcolor[HTML]{C0C0C0} 
\textbf{avg\_comp\_lat}                 & Average computation instruction latency                                                                                            & PFEA          \\ \hline
\rowcolor[HTML]{C0C0C0} 
\textbf{avg\_glob\_lat}                 & Average global memory instruction latency                                                                                          & PFEA          \\ \hline
avg\_misc\_lat                          & Average miscellaneous instruction latency                                                                                          & PFEA          \\ \hline
\rowcolor[HTML]{C0C0C0} 
\textbf{avg\_shar\_lat}                 & Average shared memory instruction latency                                                                                          & PFEA          \\ \hline

\rowcolor[HTML]{C0C0C0} 
\textbf{branch}                         & Number of branch instructions                                                                                                      & PFEA          \\ \hline
\rowcolor[HTML]{C0C0C0} 
\textbf{comp\_inst\_kernel}             & Number of computation instruction in kernel                                                                                        & PFEA          \\ \hline
comp\_inst\_sm                          & Number of computation instructions per SM                                                                                          & PFEA          \\ \hline
comp\_lat\_sm                           & Total Computation instruction latency per SM                                                                                       & PFEA          \\ \hline
\rowcolor[HTML]{C0C0C0} 
\textbf{glob\_inst\_kernel}             & Number of global memory instruction in kernel                                                                                      & PFEA          \\ \hline
glob\_inst\_sm                          & Number of global memory instructions per SM                                                                                     & PFEA          \\ \hline
glob\_lat\_sm                           & Total global memory instruction latency per SM                                                                                     & PFEA          \\ \hline
\rowcolor[HTML]{C0C0C0} 
\textbf{glob\_load\_sm}                 & Number of global load instructions per SM                                                                                           & PFEA          \\ \hline
\rowcolor[HTML]{C0C0C0} 
\textbf{glob\_store\_sm}                & Number of global store instructions per SM                                                                                         & PFEA          \\ \hline
\rowcolor[HTML]{C0C0C0} 
\textbf{misc\_inst\_kernel}             & Number of miscellaneous instruction in kernel                                                                                      & PFEA          \\ \hline
misc\_inst\_sm                          & Number of miscellaneous instructions per SM                                                                                        & PFEA          \\ \hline
misc\_lat\_sm                           & Total miscellaneous instruction latency per SM                                                                                     & PFEA          \\ \hline
shar\_inst\_kernel                      & Number of shared memory instruction in kernel                                                                                      & PFEA          \\ \hline
shar\_inst\_sm                          & Number of shared memory instructions per SM                                                                                        & PFEA          \\ \hline
shar\_lat\_sm                           & Total shared memory instruction latency per SM                                                                                     & PFEA          \\ \hline
sm\_active                              & Number of SMs activated                                                                                                            & PFEA          \\ \hline
n\_warps                                & Number of warps per SM                                                                                                             & PFEA          \\ \hline
waves                                   & Number of waves of blocks executed on SMs                                                                                          & PFEA          \\ \hline
total\_threads                          & Total number of threads launched                                                                                                   & $grid\_size \times block\_size$               \\ \hline
\rowcolor[HTML]{C0C0C0} 
\textbf{inst\_issue\_cycles}            & Instruction Issue Cycles                                                                                                           &  Equation \ref{eqn:inst_issue_cycle}   \\ \hline
\rowcolor[HTML]{C0C0C0} 
\textbf{cache\_penalty}                 & Time delay due to cache behaviour                                                                                                  & Equation \ref{equation:cache}              \\ \hline
glb\_penalty                            & Time delay due to global memory instructions                                                                                       & Equation \ref{equation:global}               \\ \hline
sh\_penalty                             & Time delay due to shared memory instructions                                                                                       & Equation \ref{equation:shared}              \\ \hline

\rowcolor[HTML]{C0C0C0} 
\textbf{occupancy}                      & \begin{tabular}[c]{@{}l@{}}The ratio of active warps on an SM to maximum\\ number of active warps supported by the SM\end{tabular} & CUDA Occupancy Calculator            \\ \hline
\rowcolor[HTML]{C0C0C0} 
\textbf{reg\_thread}                    & Registers per thread                                                                                                               & Cubin File                           \\ \hline

\rowcolor[HTML]{C0C0C0} 
\textbf{shmem\_block}                   & Shared memory per block                                                                                                            & Cubin File                           \\ \hline
\rowcolor[HTML]{C0C0C0} 
\textbf{block\_size}                    & Number of threads per block                                                                                                        & Provided by user                     \\ \hline
grid\_size                              & Number of blocks                                                                                                                   & Provided by user                     \\ \hline
\end{tabular}
}
\end{table}

\begin{figure}[ht]
\centering
\includegraphics[width=\linewidth]{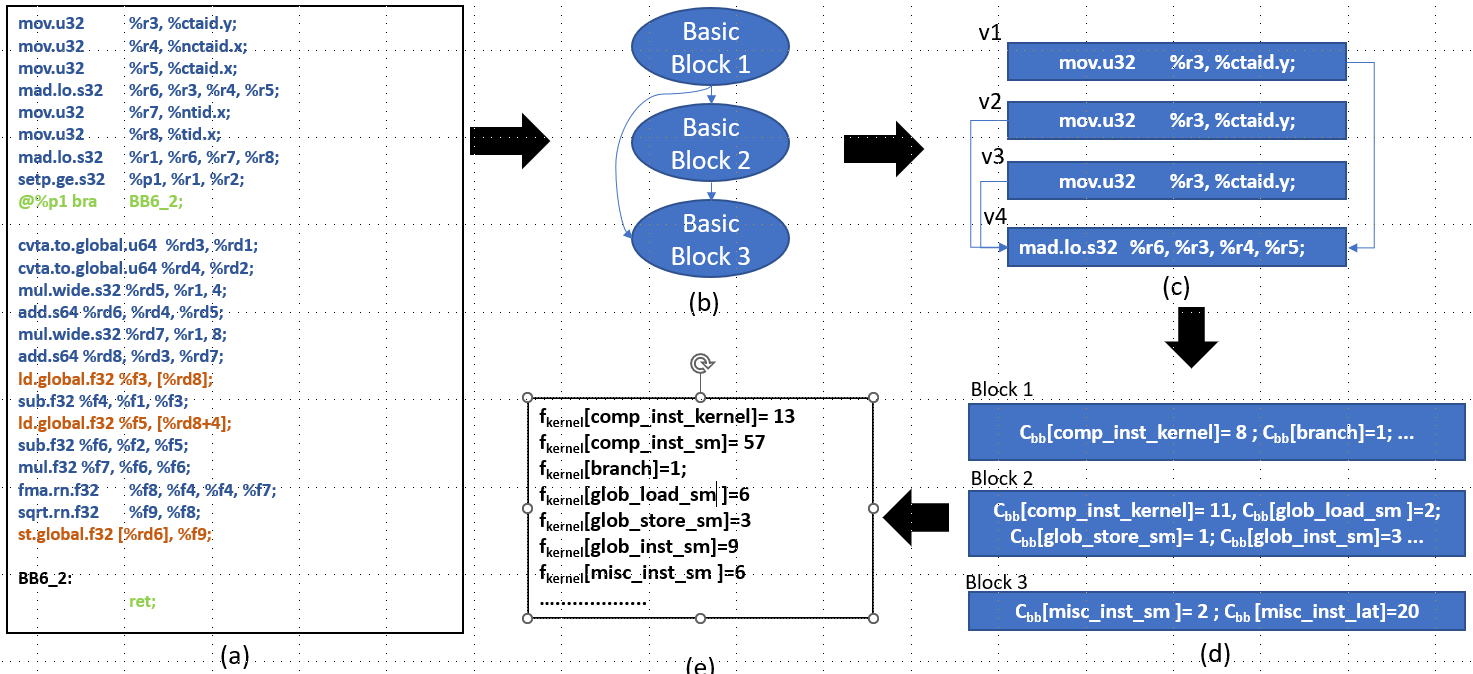}
\caption{Example of Program Feature Extraction Algorithm}
\label{pfea_example}
\end{figure}

\subsubsection{Illustration of PFEA}
We illustrate how the Program Feature Extraction algorithm extracts features with an example of neural network ({\tt nn} kernel) shown in Fig. \ref{pfea_example}. We consider a Tesla K20 GPU, with 13 SMs for this illustration. In order to analyze a CUDA application such as {\tt nn} kernel, we take its PTX code as the input (Fig. \ref{pfea_example}(a)). This code is converted into a control flow graph of CFG ($G_{cfg}$) of basic blocks in Fig.\ref{pfea_example} (b) as presented in line 7 of Algorithm \ref{algo:b2}. A Control Flow Graph (CFG) is popularly used to model the kernel application \cite{Hecht1974}. Each vertex of a basic block is a PTX instruction and an edge between two vertices denotes data dependency between them, shown in Fig.\ref{pfea_example}(c).

 Algorithm \ref{algo:b1} CollectBBFeatures function first performs a topological sort of the vertices of a basic block so as to honor the data dependencies. Algorithm \ref{algo:b1} calculates the count of each instruction type and its latency using $\mathcal{I}(\mathcal{T}(v))$ (line 6) and  $\mathcal{L}(\mathcal{T}(v))$ (line 7) respectively, and the number of global loads, global store instructions, and branch instructions using  $\mathcal{N}(\mathcal{T}(v))$ (line 15). We include the global load and store instructions separately since many researchers suggest that global memory instruction impacts power consumption significantly \cite{Nagasaka2010}\cite{Shuaiwen2013}.

Algorithm  \ref{algo:b2} uses a feature array (f\_{kernel}) to store the features for the CUDA kernel being analyzed. 
The user provides CUDA kernel PTX along with launch parameters \emph{block\_size} (nT), \emph{grid\_size} (nB), and the number of loop iterations (nLoop) as inputs to FeatureExtraction Algorithm \ref{algo:b2}. Let us consider the number of blocks (nB) is 78, and the number of threads per block (nT) is 1024. In line 3 of algorithm \ref{algo:b2}, we calculate the number of threads to be scheduled per SM, i.e., $nTh\_schd= \frac{nt\cdot nB}{SM}=6144$ . We then calculate the number of threads per wave (nTh\_wave) by based on the maximum threads per SM limit (2048 threads for Tesla K20).  In this case, $nTh\_wave=2048$, (line 6 of Algorithm \ref{algo:b2}). 

The algorithm iterates till $nTh\_schd=0$, which ensures that all the threads are scheduled and complete their execution. In this example, the algorithm iterates three times ($\lceil\frac{nTh\_schd}{nTH\_wave} \rceil$) to cover all the threads. In each iteration, Algorithm \ref{algo:b1} is called repeatedly for each basic block, to generate basic block features. 
If the basic block has a back-edge which means it is a loop, the feature count of this basic block is multiplied by the number of loop iterations $nLoop$, (supplied by the user), as shown in line 12 of Algorithm \ref{algo:b2}.

In the illustrated example, there is no loop. In the final step shown in Fig.\ref{pfea_example}(e), we add the features collected during each wave and calculate the total features for a CUDA kernel. Since there were three waves as per the launch parameters, $f_{kernel}[comp\_inst\_sm]$ is the addition of computing instructions in the kernel($f_{kernel}[comp\_inst\_kernel]$) thrice (once during each wave) hence equals to 57. Similarly, $\mathcal{L}(\mathcal{T}(v))$ in Algorithm \ref{algo:b1} represents latency of each instruction type (which have been calculated for a given architecture using microbenchmarking). The algorithm keeps accumulating the latency for each instruction and stores it in  ($f_{kernel}[total\_comp\_latency])$. Average computing instruction latency (($f_{kernel}[avg\_comp\_lat]$)) is then computed as \\$f_{kernel}[total\_comp\_latency]/f_{kernel}[comp\_inst\_kernel]$. 

\subsection{Dataset Creation} \label{dataset_collection}
We have created the dataset from three popular benchmark suites: the Rodinia \cite{Che2009}, CUDA SDK \footnote{https://docs.nvidia.com/cuda/cuda-samples/index.html}, and the Tango GPU \cite{TangoGPU}. We have created this dataset for three GPU architectures, namely Tesla K20, Tesla M60, and Tesla V100. Hardware specifications of all three GPUs are mentioned in section IV of the Supplementary File.

To create the dataset we ran these benchmarks with different launch configurations (grid size, block size) and measured the power consumption of each benchmark using NVIDIA Management Library (NVML) \footnote{https://docs.nvidia.com/pdf/NVML\_API\_Reference\_Guide.pdf} and Unified Power Profiling Application Programming Interface (UPPAPI) \cite{Chadha2016}. NVML is a C-based programmatic interface for monitoring and managing various hardware states within NVIDIA Quadro and Tesla Line GPUs. Both NVML and UPPAPI libraries use on-chip sensors for power measurement. NVML is an underlying library for the NVIDIA-supported nvidia-smi tool which is thread-safe to make simultaneous NVML calls from multiple threads. You call your CUDA kernel within a function call to NVML and UPPAPI. When the kernel code executes, the GPU power consumed is stored in a text file.  These libraries use on-chip sensors for power measurement. Collected power values are stored in a file, and an average value is recorded as measured power for the kernel. The standard deviation of data collected for some popular benchmarks is presented in Table \ref{standard deviation}. 

The UPPAPI library\cite{Chadha2016} is used to compute the actual power consumed by an application at runtime. NVIDIA provides a command-line tool, nvidia-smi, for monitoring GPU power. However, for measuring the power consumption of an application, we need to insert a probe in the application itself; hence we utilised UPPAPI. The prediction model is evaluated against the actual power consumed measured using UPPAPI. UPPAI library can compute the power consumed by a GPU as well as the host CPU. Briefly, the measurement approach is guided by the following equation presented in \cite{Chadha2016}:

\begin{equation}
 P_{act}(t_i)= Pm(t_i)+ C \times \frac{Pm(t_{i+1}) \times Pm(t_{i-1})} {t_{i+1} - t_{i-1}}     
\end{equation}

where $t_i$ denotes the current time. $t_{i-1} $ and $t_{i+1}$ represent the preceding and succeeding time instances. $P$ stands for the power readings, wherein $P_{m}$ is the power collected from the probe of the sensors, and $P_{act}$ is calculated as corrected power. 

This equation increases the accuracy of the sensor data by adding a capacitive component. This component is calculated by multiplying a constant, C (empirically derived from being 0.84), with the slope of the point under consideration. An empirical evaluation of the approach has been described in~\cite{Chadha2016}.

\subsection{Likelihood of noise in the dataset}
Dataset creation can be susceptible to noise. Noise can have a significant impact  on the results of a machine learning algorithm in terms of accuracy,
size of the model, and the time taken to build the model \cite{gupta2019dealing}. Since we are not dealing with classification, the question of label noise does not arise. We believe that the likelihood of attribute noise in the dataset is minimal and does not impact the prediction model because of the following reasons.
\begin{itemize}
     \item The architecture-related properties are obtained using reliable/certified tools. Therefore there is very little chance that there will be any noise here.
    \item Program characteristics are extracted using a tool built in-house. Some of the program characteristics are structural properties of the code. It is not possible to introduce errors in obtaining these characteristics.
    \item We have used a set of regression-based models to derive properties related to the runtime characteristics of the program. Examples are cache miss penalty, global memory access penalty, etc. This dataset can have some noise related to wrong information. If these regression models introduce noise, the noise will be present in all tuples of the dataset. It is not possible that some data tuples will have noise of this type and another data tuple will not have noise. 
    \item We have measured the power consumed by a CUDA kernel from onboard sensors via APIs developed by NVIDIA. This is not a likely source of the noise.
\end{itemize}
This dataset and feature extraction code is made available \footnote{ https://github.com/santonus/energymodel}.

\subsection{Other Assumption} \label{power_assumptions}
There are a few other assumptions with respect to the PFEA algorithm and data collection, listed as follows:
\begin{itemize}
\item The NVML library used for power data collection is designed by NVIDIA. Hence it is expected its results are considered accurate since it has access to monitor and manage various states within NVIDIA GPU. 
\item  The number of waves depends on the availability of resources required for the type of instruction. Therefore, we compute waves during the scheduling of threads. 
\item One block is scheduled on one SM irrespective of the number of threads per block. Once all the SMs are assigned one block each, the next round of assignments will start from the first SM. 
\item Since predicting the idle cycles of the warp scheduler is difficult, we assume that the warp is always full and doing useful work. 
\item L1 cache coexists along with the shared memory. In the thesis by Sunpyo Hong \cite{hong2012modeling}, the contribution of the L1 cache to GPU power consumption is close to that of shared memory. And since we have considered five features based on shared memory usage, we did not use L1 cache as a feature separately since modeling accurate cache behavior statically is a hard problem. 
\item L2 is LLC in all three GPU architectures under consideration. Similar to L1, modeling accurate L2 cache behavior is not possible. Hence in order to represent the L2 cache's impact on power consumption, we utilized a feature cache Penalty for L2 which predicts the cache miss penalty. 
\item  Although considering the impact usage of explicit
synchronization primitives in a program is a very crucial problem, we have not considered the impact of synchronization primitives since predicting it statically is a hard problem. Since this problem cannot be addressed statically, we plan to study this as a future work using runtime data. 
     
\end{itemize}

\begin{table}
\caption{Standard deviation of power values collected}
\label{standard deviation}
\begin{tabular}{|c|c|c|c|}
\hline
\textbf{Benchmark} & \textbf{Data Points} & \textbf{Average Power} & \textbf{Standard Deviation} \\ \hline
vecADD             & 194                  & 60.57                  & 1.1830                      \\ \hline
clock              & 18                   & 100.29                 & 1.7796                      \\ \hline
NN                 & 53                   & 48.12                  & 0.2668                      \\ \hline
Particle filter    & 630                  & 62.22                  & 1.5285                      \\ \hline
Cifarnet           & 206176               & 64.31                  & 0.1359                      \\ \hline
\end{tabular}
\end{table} 

\section{Feature Engineering} \label{featureEngineering}
Out of all the features considered in this study (refer Table \ref{tab:all_features}), we systematically select the features that are likely to predict power consumption accurately. 
While choosing the final features for the model, we followed the following approach.\\
\subsection{Derived Attributes:} We preferred to use derived attributes as this is one of the ways to inject expert knowledge into the model. For instance, instead of using total latency and total instructions per SM, we used average latency, computed from these two attributes. Consequently, we dropped \emph{comp\_inst\_sm}, \emph{glob\_inst\_sm}, \emph{misc\_inst\_sm}, \emph{shat\_inst\_sm} (total instructions per SM) and  \emph{comp\_lat\_sm}, \emph{glob\_lat\_sm}, \emph{misc\_lat\_sm}, \emph{shat\_lat\_sm} (total latency per SM) features and included only the average latencies.\\
%
\subsection{Non-correlated Attributes}
Having highly correlated features in our model does not improve the model's performance, and on the other hand, they may mask the interactions among various features. This suggests we include only one of the highly correlated features in our model. We use the correlation analysis on the remaining twenty-four features to observe the relationship of attributes among themselves. Correlation analysis computes the strength of the association between two variables and the direction of the relationship. Measurement of the strength of the relationship is done with the value of the correlation coefficient which varies between +1 and -1. A value of ± 1 indicates that two variables are completely associated. The strength can be observed with the correlation coefficient value going towards 0, the relationship between the two variables will be weaker. The direction of the relationship is indicated by the sign of the coefficient; a $+$ sign indicates a positive relationship and a $-$ sign indicates a negative relationship.  

We consider two types of correlation coefficients: the Pearson coefficient and the Kendall rank coefficient which we discuss one after the other. The Pearson correlation coefficient describes the linear relationship between two features. If two features have a stronger linear correlation, their coefficient will be as high as 1. Such two features with complete linear correlation can be expressed by each other, which means that the information represented by both has no difference. Removing the highly correlated features can help to improve the accuracy of the model due to less data information loss \cite{zheng2019xgboost}. 

\begin{figure}
\centering
\includegraphics[width=\linewidth]{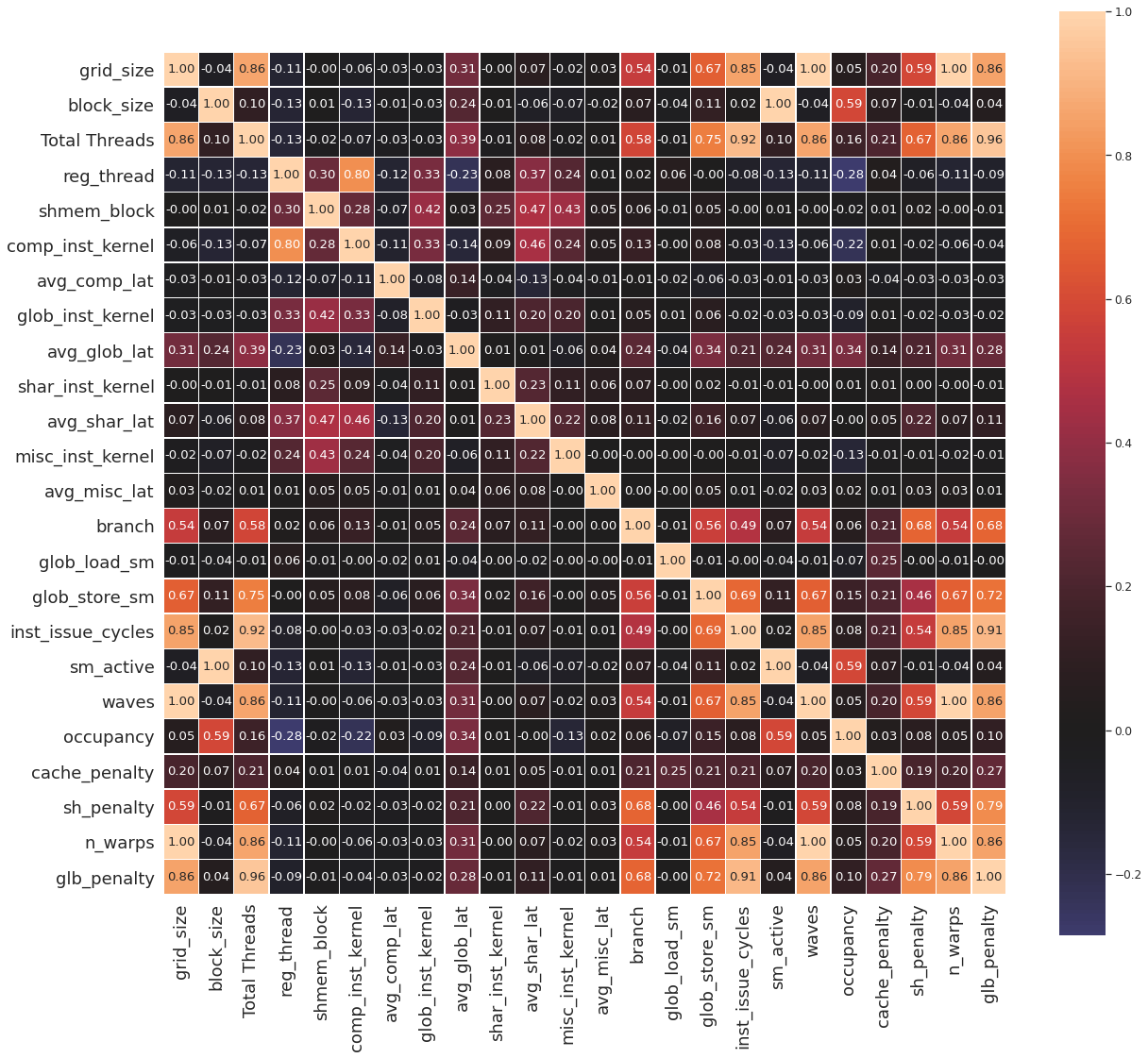}
\caption{Pearson Correlation Coefficient Heatmap}
\label{fig_pearson}
\end{figure}

As seen in the heat-map representation of Pearson coefficient in Figure \ref{fig_pearson} we can finalize feature selection by reading the heat-map as:
\begin{itemize}
    \item The correlation between two variables(x and y) can be obtained from the cell whose row corresponds to x and column corresponds to y or vice versa.
    \item The colour map towards the right end of the heat map represents the range of values for the coefficient represented by them, although their exact values are mentioned in their corresponding cells.
    \item Generally absolute values of correlation greater than 0.85 are considered highly correlated while values close to zero are considered independent variables.
\end{itemize}

Hence we considered a correlation coefficient of 0.85 to remove the most highly correlated features. 
Using the resulting heatmap of Pearson correlation coefficient, we drop 
\begin{itemize}
\item \emph{sm\_active} as it is strongly correlated to \emph{block\_size} with correlation coefficients $1.0$. We chose \emph{block\_size} since we observed its impact on feature importance through dynamic analysis available in Appendix \ref{AppA}

\item \emph{inst\_issue\_cycles} is highly correlated with \emph{glb\_penalty} (0.91), \emph{n\_warps} (0.85), \emph{waves} (0.85), \emph{grid\_size} (0.85), and \emph{total\_threads} (0.92). 
\emph{inst\_issue\_cycles} loosely represents active cycles which we found to have higher feature importance through dynamic analysis available in Appendix \ref{AppA}.
\end{itemize}


Kendall coefficient is more robust to outliers than Pearson \cite{abdi2007kendall}. Also, the Pearson coefficient has an assumption that the variables are normally distributed. However, this may not always hold. So we also utilize the Kendall coefficient for our analysis which helps us find further monotonic relationships among the features apart from linear relationships highlighted by Pearson analysis. Kendall’s correlation coefficient utilises pairs of observations and determines the strength of association depending on concordance and discordance between the pairs. We utilise the Kendall coefficient over Spearman due to the small size of the dataset. P values for the Kendall coefficient are more accurate with smaller sample sizes \cite{abdi2007kendall}. 

\begin{figure}
\centering
\includegraphics[width=\linewidth]{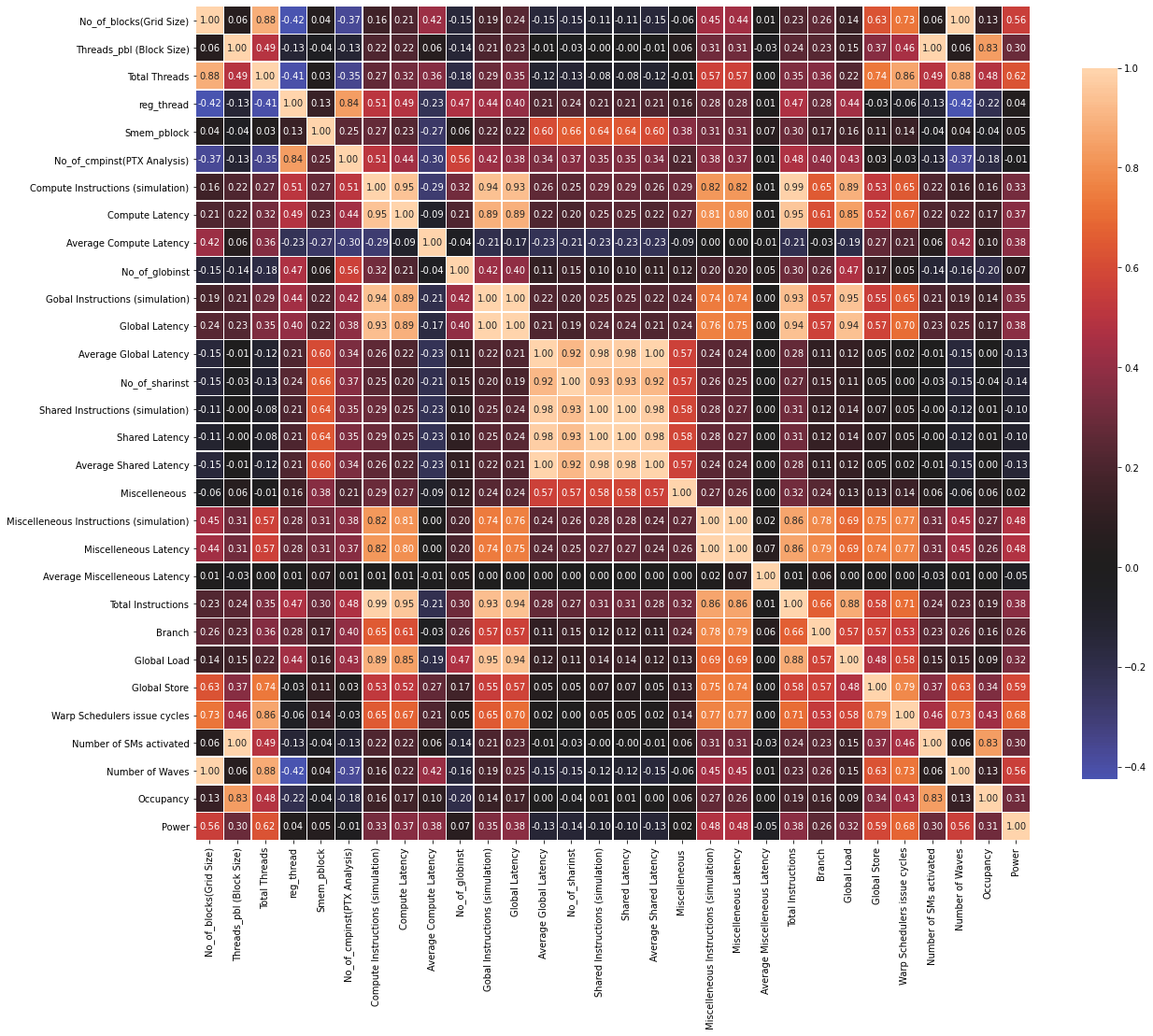}
\caption{Kendall Correlation Coefficient Heatmap}
\label{fig_kendall}
\end{figure}

We perform the Kendall correlation analysis on the new feature set (eighteen features) after discarding the features based on the Pearson coefficient. Kendall coefficient is presented as a heat map in Figure \ref{fig_kendall}. \\ Based on the heat map generated using the Kendall correlation coefficient, we observe that 

\begin{itemize}
\item \emph{avg\_shar\_lat} is highly correlated with \emph{shar\_inst\_kernel} (0.89) and \emph{sh\_penalty} (0.94). Since   \emph{avg\_shar\_lat}  is a derived feature we chose it over \emph{shar\_inst\_kernel}. 
\end{itemize}

Therefore we drop the two features and keep \emph{avg\_shar\_lat} in the final set of features. Through correlation analysis, we removed six highly correlated features from the set of twenty-four features. The final sixteen features after correlation analysis are expected to have no or low relationship and interaction amongst features.   

\subsection{Impactful Attributes}
We have used Random Forest Regressor feature importance with sixteen features and then systematically removed the features which have no impact on the prediction accuracy. In this process, we have removed the \emph{avg\_misc\_lat} feature. 

 \begin{figure}[ht]
\centering
\includegraphics[width=\linewidth]{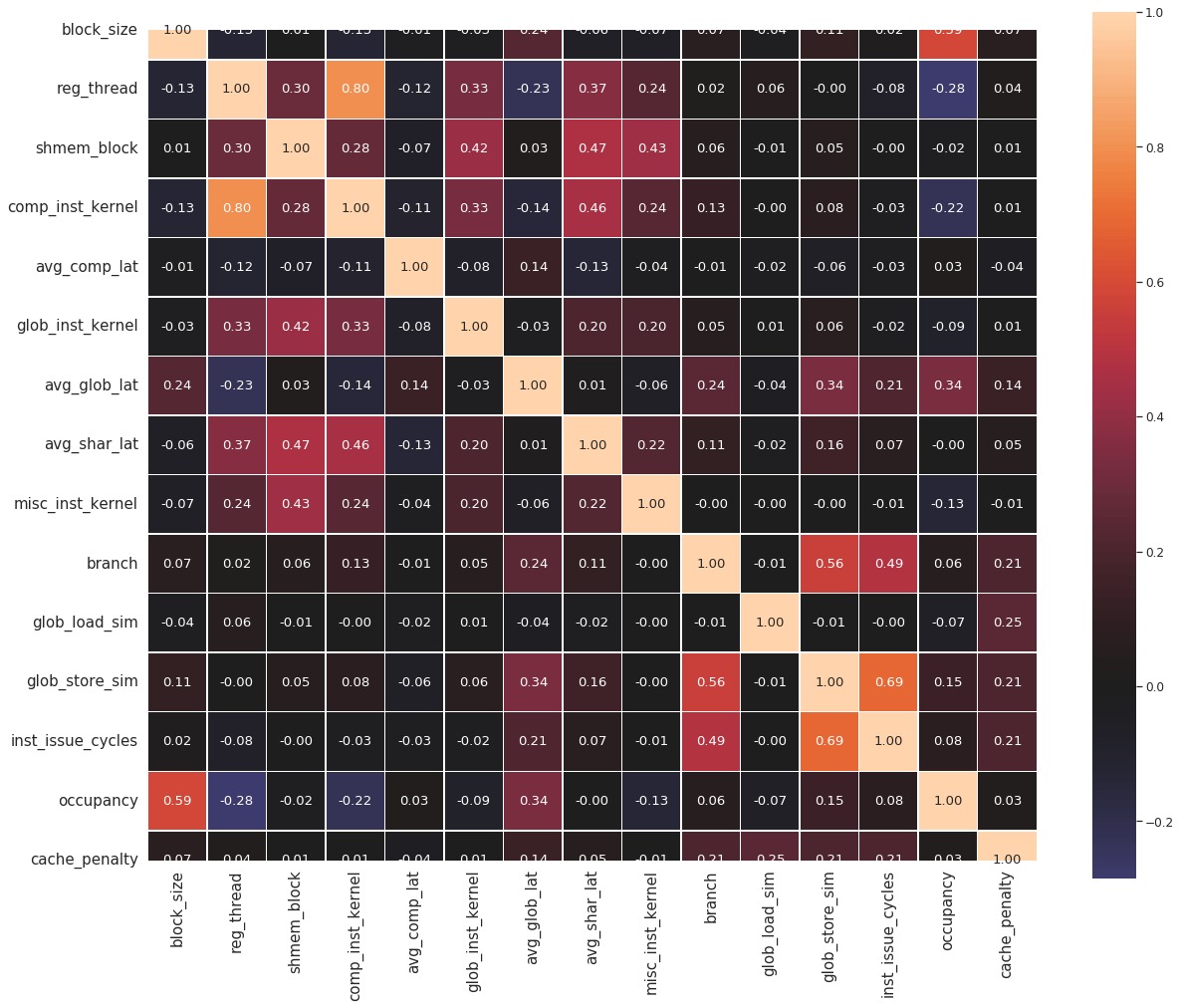}
\caption{Pearson Correlation Coefficient heatmap of final set of features}
\label{fig_peason_final}
\end{figure}

\begin{figure}[ht]
\centering
\includegraphics[width=\linewidth]{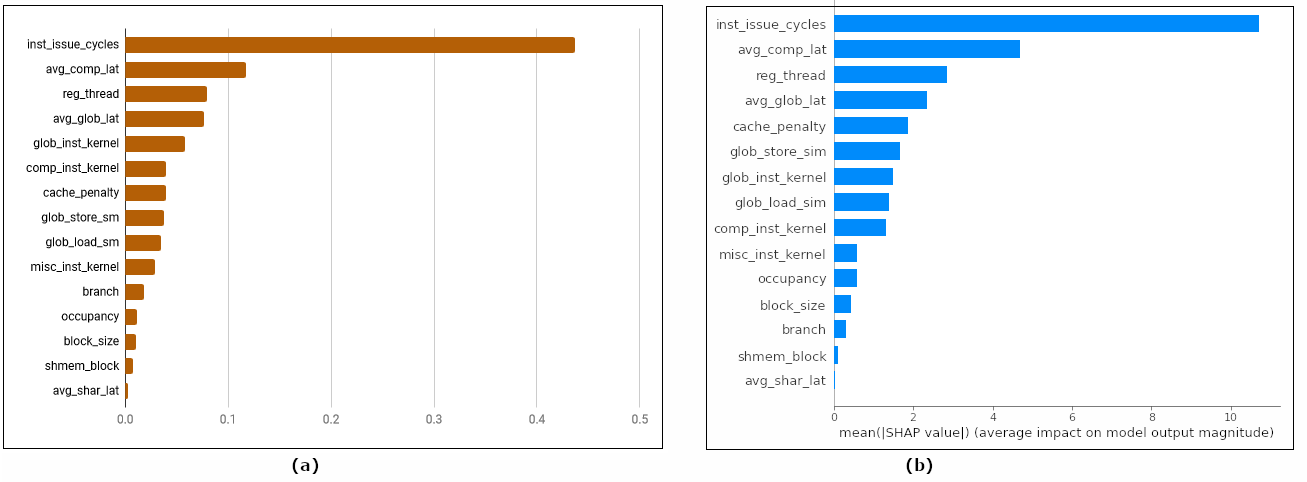}
\caption{Feature Importance Plot using (a) Random Forest  (b) SHAP}
\label{shap_fi}
\end{figure}

We also utilized  SHAP (Shapley Additive explanations) \cite{Lundberg2017} to verify the features which contribute the most to power prediction. Comparison of Random Forest and SHAP feature importance results for Tesla K20 GPU are presented in Figure \ref{shap_fi}. As seen in the figure, these results are closely consistent since the top four features and two least important features are the same, with a minor difference in order for other features. Pearson coefficient results for these features as seen in Figure \ref{fig_peason_final}, which confirms a very low correlation between these final fifteen features. The final set of features is highlighted in Table \ref{tab:all_features}. 
%

\section{Model Evaluation Metrics}
We will be using three essential metrics to validate our model based on the mean of 5-fold cross-validation scores: $R^{2}$ score, Root Mean Squared Error (RMSE), and Mean Absolute Error (MAE). We discuss each of these metrics in detail.




\section{Model Selection \& Evaluation} \label{model_construction}
We undertook a systematic study of classical ML and Neural Network models to arrive at the final model based on the model mechanics and performance. We first investigated whether a linear relationship between predictor variables (pairwise) and between the response and predictor variables can be established. As evident in the plots of the predictor features versus response, i.e., power ( Figure \ref{featureVsPower} ), there is no linear relationship between input features and GPU power consumption. If a relationship between two variables is not linear, the rate of increase or decrease can change as one variable changes, causing a complex pattern in the data. A nonlinear function or approach might better model this complex trend between feature and power. However, to rule out the possibility of any linear regression model, we performed multi-linear regression on the dataset. The $R^{2}$ score turned out to be $-0.1919$ for the multi-linear regression model, an indicator of poor model fitting on the experimental data.  

This observation led us to explore more complex, non-linear regression techniques, as presented in the following section. As shown later in the chapter, {\em Catboost and XGBoost} are the two State-of-the-Art models that are eventually chosen based on the performance and relevance of w.r.t to selected features. 
\begin{figure}
 \begin{subfigure}{0.5\textwidth}
    \centering\includegraphics[width=8cm]{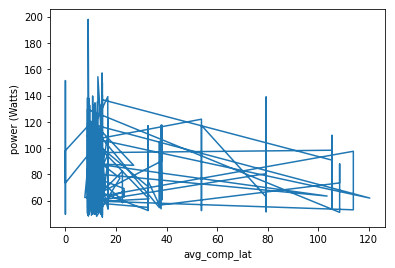}
\end{subfigure}
\begin{subfigure}{0.5\textwidth}
    \centering\includegraphics[width=8cm]{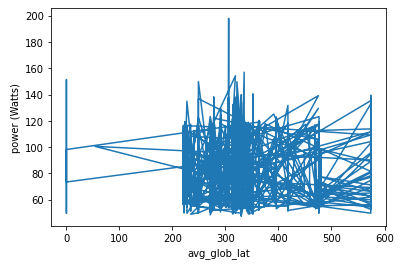}
  \end{subfigure}
  \newline
   \begin{subfigure}{0.5\textwidth}
    \centering\includegraphics[width=8cm]{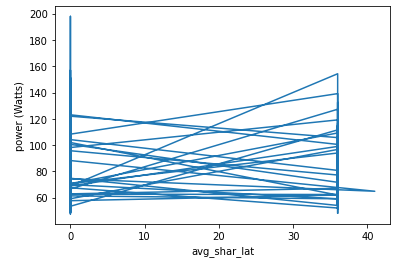}
    \end{subfigure} 
  \begin{subfigure}{0.5\textwidth}
    \centering\includegraphics[width=8cm]{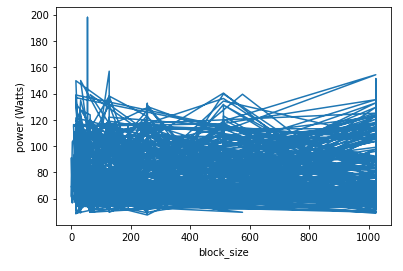}
     \end{subfigure}
\newline
   \begin{subfigure}{0.5\textwidth}
    \centering\includegraphics[width=8cm]{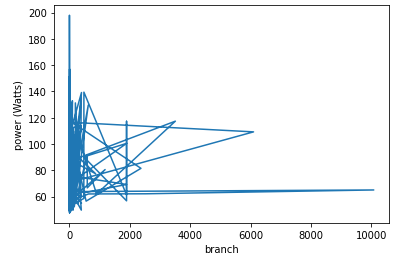}
    \end{subfigure}
      \begin{subfigure}{0.5\textwidth}
    \centering\includegraphics[width=8cm]{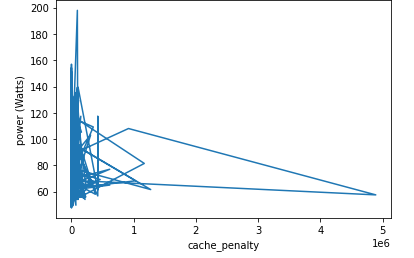}
     \end{subfigure}
\newline
    \begin{subfigure}{0.5\textwidth}
    \centering\includegraphics[width=8cm]{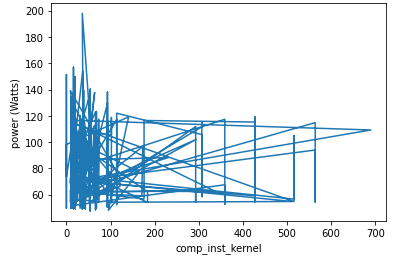}
    \end{subfigure}
      \begin{subfigure}{0.5\textwidth}
    \centering\includegraphics[width=8cm]{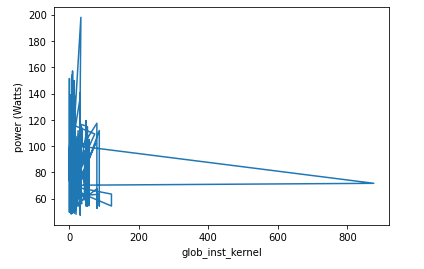}
     \end{subfigure}
\end{figure}

\begin{figure}
    \begin{subfigure}{0.5\textwidth}
    \centering\includegraphics[width=8cm]{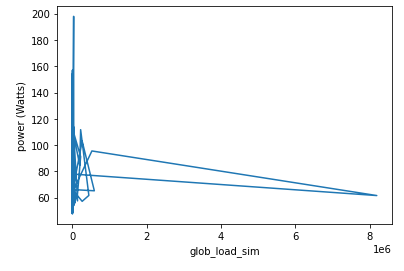}
    \end{subfigure}
\begin{subfigure}{0.5\textwidth}
    \centering\includegraphics[width=8cm]{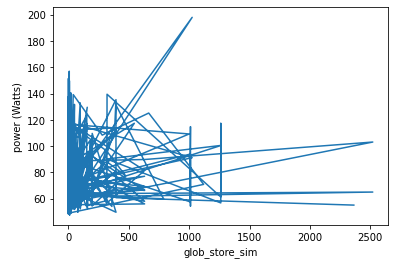}
     \end{subfigure}
\newline
     \begin{subfigure}{0.5\textwidth}
    \centering\includegraphics[width=8cm]{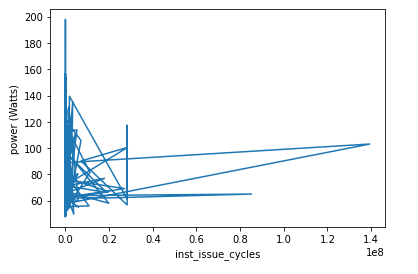}
    \end{subfigure}
      \begin{subfigure}{0.5\textwidth}
    \centering\includegraphics[width=8cm]{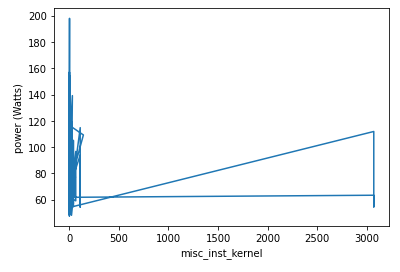}
     \end{subfigure}
\newline
\begin{subfigure}{0.5\textwidth}
    \centering\includegraphics[width=8cm]{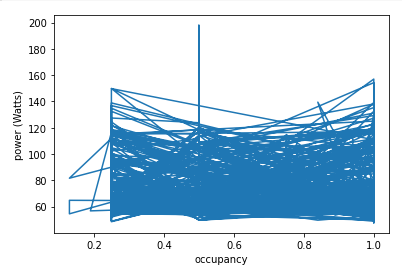}
    \end{subfigure}
      \begin{subfigure}{0.5\textwidth}
    \centering\includegraphics[width=8cm]{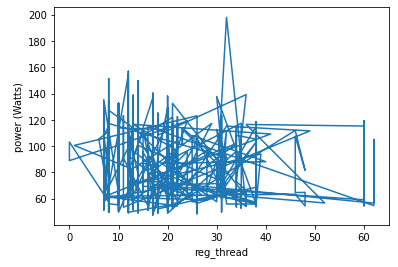}
     \end{subfigure}
\newline
      \begin{subfigure}{0.5\textwidth}
    \centering\includegraphics[width=8cm]{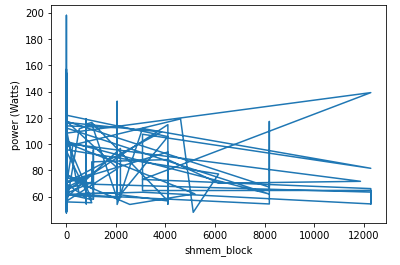}
    \end{subfigure}
\newline
   \caption{Relationship between features and power consumption}
   \label{featureVsPower}
\end{figure}
\subsection{ANN Experiments \& Analysis}
Among the wide variety of available methods and algorithms, Artificial Neural Networks(ANN) \cite{hill1994artificial} is one of the most popular methods while modeling non-linear relationship models on complex patterns in data, especially in prediction problems. 
We extensively experimented with ANN by changing the ANN architecture, dataset size, and learning rates. We have used multiple feature selection methods:\\ i) Feature engineering, described in Section~\ref{featureEngineering} and Table~\ref{tab:all_features} \\
ii) Variational autoencoder (VAE) with 15 latent features and \\
iii) Auto-feature selection (without VAE). 

\begin{figure}[ht]
\centering
\includegraphics[width=\linewidth]{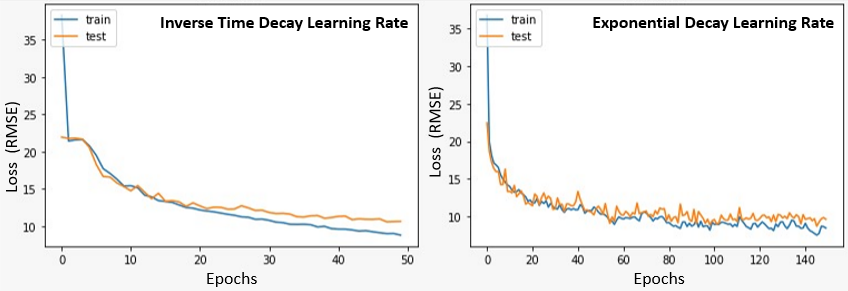}
\caption{Loss function plot for Adaptive Learning rate ANN}
\label{loss_function_ann}
\end{figure}

We trained ANN with three hidden layers for two different layer configurations. To investigate whether ANN performs better with a larger dataset, we ran the experiments for a smaller dataset (990 tuples) and a larger dataset (1810 tuples).
Details of ANN experiments are presented in Table \ref{tab:ann}.  Our findings are as follows:\\
\indent 1. We observed that ANN performs most efficiently for 64,64,128 neurons in three layers.\\
\indent 2. When training an ANN model, it is often useful to lower the learning rate as the training progresses and observe its effect on model performance. In addition to the uniform learning rate (first two result rows of Table~\ref{tab:ann}, we used the adaptive learning rate method with i) exponential decay and ii) inverse time decay learning rate schedules. For all three learning rate methods, we have used the features selected from the feature engineering method. 
The inverse time decay model was the most precise amongst all the ANN models trained, and it was the only method that performed better on the larger dataset. \\
\indent 3. From the loss function plot for both the methods shown in Figure \ref{loss_function_ann}, validation loss is seen to be greater than the training loss as the number of epochs increases. This is an evidence of over-fitting. Moreover, the training plots don't exhibit uniform smoothness and monotonicity. 
So, the loss is likely to experience non-smooth behavior.\\
\indent 4. 
VAE with 15 latent features (number of features are kept the same as the feature engineering approach) with {\em inverse time decay}, shows an unacceptably low $R^2$ score as shown in Table~\ref{tab:ann}. Increasing the number of latent features did not yield any significant improvement. The auto feature selection approach performed better than VAE based approach, as shown in Table~\ref{tab:ann}, and the $R^2$ values are comparable to the feature engineering based approach.

\subsection{SVR Based Model}
Support Vector Machine Regressor (SVR) \cite{parbat2020python} considers the presence of non-linearity in the data and provides a proficient prediction mode. After experimenting with popular kernels for SVR, we observed that the Radial Basis function (RBF) kernel gives the most precise model. However, an $R^{2}$ score of 0.7041 for the RBF kernel for the Tesla K20 GPU is still not acceptable for a power prediction model.

\begin{table}
\caption{ANN Model Experiment Results}
\label{tab:ann}
\resizebox{\linewidth}{!}{
\begin{tabular}{|llllll|}
\hline
\multicolumn{2}{|c|}{\textbf{Model Parameters}}                   & \multicolumn{2}{c|}{\textbf{Smaller dataset}}          & \multicolumn{2}{c|}{\textbf{Larger dataset}} \\ \hline

\multicolumn{1}{|l|}{\begin{tabular}[c]{@{}l@{}}ANN Architecture\\ (3 hidden layers)\end{tabular}} &
  \multicolumn{1}{l|}{\begin{tabular}[c]{@{}l@{}}Learning\\ rate\end{tabular}} &
  \multicolumn{1}{l|}{Epochs} &
  \multicolumn{1}{l|}{$R^{2}$ score} &
  \multicolumn{1}{l|}{Epochs} &
 $R^{2}$ score\\ \hline
 \multicolumn{6}{|c|}{\textbf{15 features selected using feature engineering, uniform learning rate}}      \\ \hline
\multicolumn{1}{|l|}{5,5,5}     & \multicolumn{1}{l|}{0.01}       & \multicolumn{1}{l|}{130} & \multicolumn{1}{l|}{0.6226} & \multicolumn{1}{l|}{135}      & 0.5003       \\ \hline
\multicolumn{1}{|l|}{64,64,128} & \multicolumn{1}{l|}{0.01}       & \multicolumn{1}{l|}{150} & \multicolumn{1}{l|}{0.7011} & \multicolumn{1}{l|}{120}      & 0.6791       \\ \hline
\multicolumn{6}{|c|}{\textbf{15 features selected using feature engineering, Exponential decay learning rate}}                                                                                                        \\ \hline
\multicolumn{1}{|l|}{64,64,128} & \multicolumn{1}{l|}{0.9}        & \multicolumn{1}{l|}{175} & \multicolumn{1}{l|}{0.7246} & \multicolumn{1}{l|}{150}      & 0.7057       \\ \hline
\multicolumn{6}{|c|}{\textbf{15 features selected using feature engineering, Inverse time decay learning rate}}                                                                                                       \\ \hline
\multicolumn{1}{|l|}{64,64,128} & \multicolumn{1}{l|}{0.5}        & \multicolumn{1}{l|}{100} & \multicolumn{1}{l|}{0.7418} & \multicolumn{1}{l|}{50}       & 0.7737       \\ \hline
\multicolumn{6}{|c|}{\textbf{VAE for feature extraction (latent features:15), Inverse time decay learning rate}}                                                                                    \\ \hline
\multicolumn{1}{|l|}{64,64,128} & \multicolumn{1}{l|}{step decay} & \multicolumn{1}{l|}{}    & \multicolumn{1}{l|}{0.3270} & \multicolumn{1}{l|}{}         & 0.00802      \\ \hline
\multicolumn{6}{|c|}{\textbf{Auto feature selection (without VAE), uniform learning rate}}                                                                                                               \\ \hline
\multicolumn{1}{|l|}{5,5,5}     & \multicolumn{1}{l|}{0.01}       & \multicolumn{1}{l|}{150} & \multicolumn{1}{l|}{0.6081} & \multicolumn{1}{l|}{150
}         &     0.53970         \\ \hline
\multicolumn{1}{|l|}{64,64,128} & \multicolumn{1}{l|}{0.01}       & \multicolumn{1}{l|}{150} & \multicolumn{1}{l|}{0.7084} & \multicolumn{1}{l|}{120}      & 0.6484       \\ \hline
\end{tabular}
}
\end{table}

Since the multi-linear regression model such as ANN and SVR cannot accurately map the power consumption to input features and are shown to over-fit, we conclude that these models are not sufficient for the power prediction problem. Additionally, Neural network regressors are data-hungry and, therefore, usually fail to model reasonably well in the absence of abundant data.

\subsection{Random Forest Based Regressors}
A Random Forest (RF) \cite{biau2012analysis} is a meta-estimator that aggregates many decision trees with some helpful modifications.
\begin{itemize}
    \item The number of features that can be split at each node is limited to some percentage of the total. This ensures that the ensemble model does not rely too heavily on any individual feature and makes fair use of all potentially predictive features.
   \item Each tree draws a random sample from the original data set when generating its splits, adding a further element of randomness that prevents overfitting.
\end{itemize}

When the data is limited and has a significant number of feature attributes, tree-based regression models have proved to be highly efficient  \cite{Chen2011}\cite{basak2019predicting}\cite{khaidem2016predicting}. As the next step, we consider tree-based methods like Decision Trees, which model well on non-linear relationships. The resulted in  $R^{2}$ score of 0.8447 for decision trees for the Tesla K20 GPU shows promising results for tree-based approaches. We further explore advanced tree-based algorithms, which are known to perform better than decision trees.

Random forests (RF) offer higher resolution in the feature space since the trees are not pruned and split the feature space into smaller regions. RF utilizes a random subsample of rows at each node, and a distinct sample of features is selected for splitting. Since RFs are learned on diverse samples, they are endowed with greater diversity. Each tree in RF produces an individual prediction, and the final prediction is an average of the prediction of all trees. The averaging ensures smoothness in training and validation loss and, along with RF's randomness, improves its accuracy and reduces overfitting. 

RF has some advantages compared to ANN in specific cases, including their robustness and benefits in cost and time. RF is also particularly advantageous in terms of interpretability \cite{AHMAD201777} in Energy consumption data. Random forest is found to be  a better choice than a decision Tree or Line-fit regressors, as empirically observed from the results seen in Table \ref{power_result_table}. For Tesla K20, $R^2$ score of RF is 0.8899. we considered other popular tree-based ensemble machine-learning algorithms to ascertain the most accurate and efficient algorithm in GPU power prediction. 

Ensemble learning is the process by which multiple models, such as classifiers or experts, are strategically generated and combined to solve a particular computational intelligence problem. Ensemble learning is primarily used to improve the (classification, prediction, function approximation, etc.) performance of a model or reduce the likelihood of an unfortunate selection of a poor one. \cite{dong2020survey, zhang2012ensemble}. These algorithms include:  ExtraTrees Regressor\cite{Geurts2006}, Gradient Boosting Regressor \cite{Friedman2000}, XGBoost Regressor \cite{chen2016xgboost,zhang2020predicting}, and Catboost \cite{Dorogush2018CatBoostGB} to predict GPU power consumption. These algorithms cover a breadth of ensemble approaches in predicting continuous variables. XGBoost turned out to be the most efficient and accurate predictor model for power consumption. We first discuss each of these techniques and then present experiments, results, and analyses of these approaches. 

\subsection{ExtraTrees Regressor}
Extremely Randomized Trees Classifier(Extra Trees Classifier) is an ensemble learning technique that aggregates the results of multiple de-correlated decision trees collected in a “forest” to output its classification result. Although there are similarities between Random Forest Classifier and Extra Trees, both differ in constructing the decision trees in the forest. Random Forest uses bootstrap replicas to subsample the input data with replacement, whereas Extra Trees uses the complete data sample. Random Forest chooses the optimum split, while Extra Trees chooses it randomly. However, once the split points are selected, the two algorithms determine the best one between all the subsets of features. Therefore, Extra Trees adds randomization but still has optimization. 

The Extra Trees algorithm is faster than Random Forest in terms of computational cost and execution time. This is because although the whole procedure is the same, but it randomly chooses the split point and does not calculate the optimal one. As seen in Table \ref{power_result_table}, Extra Trees performs close (Kepler) and slightly better (Maxwell and Volta) than Random Forest Regressor. 

\subsection{Gradient Boosting Regressor}
Gradient Boosting Machines (GBM) is one of the popular ensemble methods based on creating multiple weak models that are combined to get better performance using the whole group. Decision trees are used as the weak learner, especially regression trees that output real values for splits and whose output can be added together, allowing subsequent model outputs to be added. This output corrects the residuals in the predictions\cite{Natekin2013}. 

The two significant differences between the gradient-boosting trees and the random forests are training and output decision. In gradient boosting, we train the trees sequentially, one tree at a time, each to correct the errors of the previous ones. Whereas in Random Forest, we construct the trees independently. Hence we can train an RF in parallel but not the former. The other difference, i.e., how they output decisions, is different since the trees in a random forest are independent. Hence RF can determine their outputs in any order, whereas gradient-boosting trees run in a fixed order. Therefore sequence cannot change, which restricts Gradient Boosting to sequential evaluation. Employing GBM improves the $R^{2}$ score compared to RF, which motivates us to try a more powerful technique based on GBM, i.e., XGBoost. 

\subsection{XGBoost Regressor}
XGBoost (Extreme Gradient Boosting) is an optimized type of Gradient Boosting algorithm. However, there are some key differences between the two, which make XGBoost faster and more accurate technique than Gradient Boosting. GBM does not use regularization\cite{tian2022comprehensive} whereas XGBoost employs L1 and L2 regularization, which improves model generalization. XGBoost has an additional advantage since its training is very fast and can be parallelized/distributed across clusters. XGBoost proved to be the most accurate model for the power prediction problem, as evident from results presented in Table \ref{power_result_table}. Another efficient and  powerful technique based on  gradient boosting is Catboost.

\subsection{CatBoost}
CatBoost is an advancement to Gradient Boosting in how it deals with
high cardinality categorical variables.  For low cardinality categorical variables, CatBoost uses one-hot encoding \cite{rodriguez2018beyond}. CatBoost makes use of binary decision trees as base predictors \cite{Dorogush2018CatBoostGB, Prokhorenkova2018}.  CatBoost has proved to be most efficient compared to other machine learning algorithms in the past studies \cite{Abdullahi2020,samat2021gpu} especially for classification problems \cite{,al2019comparison,hancock2020catboost}.  However, it was observed that XGBoost performs slightly better than CatBoost for power prediction using the features under study.  We validate that the superior performance of CatBoost and XGBoost are discussed further in \ref{analysingresults}.

\begin{table}
\caption{Machine Learning model and its hyperparameters}
\label{tab:my-hyperparameters}
\resizebox{\linewidth}{!}{
\begin{tabular}{lllllllll}
\cline{1-2} \cline{4-5}
\multicolumn{2}{|c|}{\cellcolor[HTML]{C0C0C0}\textbf{Random Forest}}     & \multicolumn{1}{l|}{} & \multicolumn{2}{c|}{\cellcolor[HTML]{C0C0C0}\textbf{XGBoost}}       &                       & \multicolumn{2}{c}{\textbf{}}                                                 &                               \\ \cline{1-2} \cline{4-5}
\multicolumn{1}{|l|}{n\_estimators}        & \multicolumn{1}{l|}{500}    & \multicolumn{1}{l|}{} & \multicolumn{1}{l|}{n\_estimators}      & \multicolumn{1}{l|}{500}  &                       &                                          &                                    &                               \\ \cline{1-2} \cline{4-5}
\multicolumn{1}{|l|}{max\_depth}           & \multicolumn{1}{l|}{14}     & \multicolumn{1}{l|}{} & \multicolumn{1}{l|}{max\_depth}         & \multicolumn{1}{l|}{6}    &                       &                                          &                                    &                               \\ \cline{1-2} \cline{4-5} \cline{7-8}
\multicolumn{1}{|l|}{min\_samples\_split}  & \multicolumn{1}{l|}{2}      & \multicolumn{1}{l|}{} & \multicolumn{1}{l|}{learning\_rate}     & \multicolumn{1}{l|}{0.05} & \multicolumn{1}{l|}{} & \multicolumn{2}{c|}{\cellcolor[HTML]{C0C0C0}\textbf{Extra Trees}}             &                               \\ \cline{1-2} \cline{4-5} \cline{7-8}
\multicolumn{1}{|l|}{max\_features}        & \multicolumn{1}{l|}{auto}   & \multicolumn{1}{l|}{} & \multicolumn{1}{l|}{max\_depth}         & \multicolumn{1}{l|}{6}    & \multicolumn{1}{l|}{} & \multicolumn{1}{l|}{n\_estimators}       & \multicolumn{1}{l|}{220}           &                               \\ \cline{1-2} \cline{4-5} \cline{7-8}
\multicolumn{1}{|l|}{min\_samples\_leaves} & \multicolumn{1}{l|}{1}      & \multicolumn{1}{l|}{} & \multicolumn{1}{l|}{min\_child\_weight} & \multicolumn{1}{l|}{1}    & \multicolumn{1}{l|}{} & \multicolumn{1}{l|}{max\_depth}          & \multicolumn{1}{l|}{16}            &                               \\ \cline{1-2} \cline{4-5} \cline{7-8}
                                           &                             & \multicolumn{1}{l|}{} & \multicolumn{1}{l|}{colsample\_bytree}  & \multicolumn{1}{l|}{0.7}  & \multicolumn{1}{l|}{} & \multicolumn{1}{l|}{min\_samples\_split} & \multicolumn{1}{l|}{2}             &                               \\ \cline{1-2} \cline{4-5} \cline{7-8}
\multicolumn{2}{|c|}{\cellcolor[HTML]{C0C0C0}\textbf{CatBoost}}          & \multicolumn{1}{l|}{} & \multicolumn{1}{l|}{subsample}          & \multicolumn{1}{l|}{0.7}  & \multicolumn{1}{l|}{} & \multicolumn{1}{l|}{max\_features}       & \multicolumn{1}{l|}{auto}          &                               \\ \cline{1-2} \cline{4-5} \cline{7-8}
\multicolumn{1}{|l|}{depth}                & \multicolumn{1}{l|}{8}      & \multicolumn{1}{l|}{} & \multicolumn{1}{l|}{silent}             & \multicolumn{1}{l|}{1}    & \multicolumn{1}{l|}{} & \multicolumn{1}{l|}{min\_sample\_leaves} & \multicolumn{1}{l|}{1}             &                               \\ \cline{1-2} \cline{4-5} \cline{7-8}
\multicolumn{1}{|l|}{iterations}           & \multicolumn{1}{l|}{270}    &                       &                                         &                           &                       &                                          &                                    &                               \\ \cline{1-2}
\multicolumn{1}{|l|}{loss\_function}       & \multicolumn{1}{l|}{RMSE}   &                       & \multicolumn{2}{c}{\textbf{}}                                       &                       & \multicolumn{2}{c}{\textbf{}}                                                 &                               \\ \cline{1-2} \cline{7-8}
\multicolumn{1}{|l|}{logging\_level}       & \multicolumn{1}{l|}{Silent} &                       &                                         &                           & \multicolumn{1}{l|}{} & \multicolumn{2}{c|}{\cellcolor[HTML]{C0C0C0}\textbf{Gradient Boosting}}       & \multicolumn{1}{c}{\textbf{}} \\ \cline{1-2} \cline{4-5} \cline{7-8}
                                           &                             & \multicolumn{1}{l|}{} & \multicolumn{2}{c|}{\cellcolor[HTML]{C0C0C0}\textbf{SVR}}           & \multicolumn{1}{l|}{} & \multicolumn{1}{l|}{n\_estimators}       & \multicolumn{1}{l|}{250}           &                               \\ \cline{1-2} \cline{4-5} \cline{7-8}
\multicolumn{2}{|c|}{\cellcolor[HTML]{C0C0C0}\textbf{Decision Tree}}     & \multicolumn{1}{l|}{} & \multicolumn{1}{l|}{C}                  & \multicolumn{1}{l|}{200}  & \multicolumn{1}{l|}{} & \multicolumn{1}{l|}{max\_depth}          & \multicolumn{1}{l|}{5}             &                               \\ \cline{1-2} \cline{4-5} \cline{7-8}
\multicolumn{1}{|l|}{max\_depth}           & \multicolumn{1}{l|}{14}     & \multicolumn{1}{l|}{} & \multicolumn{1}{l|}{epsilon}            & \multicolumn{1}{l|}{1}    & \multicolumn{1}{l|}{} & \multicolumn{1}{l|}{learning\_rate}      & \multicolumn{1}{l|}{0.17}          &                               \\ \cline{1-2} \cline{4-5} \cline{7-8}
\multicolumn{1}{|l|}{samples\_split}       & \multicolumn{1}{l|}{5}      & \multicolumn{1}{l|}{} & \multicolumn{1}{l|}{gamma}              & \multicolumn{1}{l|}{0.5}  & \multicolumn{1}{l|}{} & \multicolumn{1}{l|}{max\_features}       & \multicolumn{1}{l|}{auto}          &                               \\ \cline{1-2} \cline{4-5} \cline{7-8}
\multicolumn{1}{|l|}{criterion}            & \multicolumn{1}{l|}{mse}    & \multicolumn{1}{l|}{} & \multicolumn{1}{l|}{kernel}             & \multicolumn{1}{l|}{rbf}  & \multicolumn{1}{l|}{} & \multicolumn{1}{l|}{criterion}           & \multicolumn{1}{l|}{friedman\_mse} &                               \\ \cline{1-2} \cline{4-5} \cline{7-8}
\end{tabular}
}
\end{table}

\subsection{Model Building \& Training: Implementation Details}
Machine Learning models work much better if the values of the features in the dataset are relatively on a similar scale. The tree-based models are scale-invariant. However, we scaled the data since it is a useful technique to accelerate the calculations. We use the MinMaxScalar technique here since it preserves the shape of the distribution of the features, which, in turn, preserves the information embedded in the original data. The obtained dataset is shuffled to remove skewness from the dataset.

The tree-based bagging and boosting models are constructed using the Tree Regression method provided by the scikit-learn (Version: 0.24.1). 
Simple cross-validation uses the same data for hyperparameter tuning and evaluates the model's performance. This may lead to a biased evaluation of the model. So we consider using nested cross-validation where the scores of each hyperparameter combination are computed on all the splits and the best parameter combination is thus obtained. This best parameter combination is used to calculate scores on all the splits again. 

The parameters tuned for these regression models are mentioned in Table \ref{tab:my-hyperparameters}. 
\emph{n\_estimators} denotes the number of trees used in the model, \emph{max\_depth} represents the maximum depth a tree can attain, \emph{min\_samples\_split} indicates the minimum number of samples to split an internal node, max\_features denotes the number of features to consider for best split, and the split criterion is used to measure the quality of split (can be MSE or MAE). The Support Vector Regression model with the Radial Basis Function kernel is built using the scikit-learn library. We use the nested cross-validation technique mentioned earlier to tune parameters like kernel co-efficient, regularization parameter, and the epsilon distance from the actual value for which no penalty is associated. 

The Keras library (Version: 2.4.3) is utilized to build the ANN model. We employ the incremental approach for constructing the ANN model \cite{jahn2020artificial}, and finally, come up with three fully connected hidden layers of 64, 64, and 128 neurons each. It uses the Adam optimizer for performing the gradient descent. We utilize the dropout technique and regularization to prevent overfitting. We use nested cross-validation for optimal parameters like the number of epochs, batch size, and learning rate by observing the training and validation loss. 5-fold cross-validation is used for splitting training and testing data.

\paragraph{Model Validation}
We use nested cross-validation for optimal parameters like the number of epochs, batch size, and learning rate by observing the training and validation loss. 
We use sklearn's k-fold cross-validation (CV) method\footnote{\url{https://scikit-learn.org/stable/modules/cross_validation.html}
} for splitting training and testing data. A k-fold CV ensures that prior knowledge about the test set may not ``leak'' into the model and evaluation metrics report generalization performance. A test set is held out for final evaluation, but the validation set is no longer needed when doing a CV. The training set is split into $k$ smaller sets, and for each of the $k$ folds, a model is trained using $k-1$ of the folds as training data, and subsequently, the resulting model is validated on the remaining part of the data implying the remaining part is used as a test set to compute a performance measure such as accuracy. We set k=5 for 5-fold cross-validation.

 In 5-fold cross-validation, after shuffling the dataset randomly, we split the dataset into 5 folds. For each unique fold, we considered one fold as a test data set and took the remaining folds as a training data set. Then, our model is fit
into the training set and evaluated on the test set. Moreover, we ran our model 5 times and reported the mean and standard deviation of the performance. As we observe from Section 7, the performance is robust as the standard deviation from the mean performance is low, on all 5 runs. 





\begin{table}
\centering
\small
\caption{Validation score for machine learning algorithms across GPU architectures}
\label{power_result_table}

\begin{tabular}{|l|l|r|r|r|}
\hline
\textbf{Architecture} & \textbf{Regression Model} & \multicolumn{1}{l|}{\textbf{$R^{2}$ score}} & \multicolumn{1}{l|}{\textbf{RMSE}} & \multicolumn{1}{l|}{\textbf{MAE}} \\ \hline
\multirow{7}{*}{\textbf{Tesla K20}} & Random Forest & 0.8899 & 7.9265 & 4.4177 \\ \cline{2-5} 
 & Extra Trees & 0.8893 & 7.9438 & 3.9346  \\ \cline{2-5} 
 & Gradient Boosting & 0.9025	&	7.4657	&	3.7849\\ \cline{2-5} 
 & \textbf{XG Boost} & \textbf{ 0.9101}	&	\textbf{7.1592}	&	\textbf{3.4816} \\ \cline{2-5} 
  & \textbf{CatBoost} & \textbf{0.9067}	&	\textbf{7.3001}	&	\textbf{3.8308} \\ \cline{2-5} 
 & Decision Tree & 0.8447	& 9.3899		&  5.007 \\ \cline{2-5} 
 & SVR & 0.7041	&	13.0078	&	8.4063 \\ \cline{2-5} 
 & ANN & 0.7011	&	12.1048	&	7.4597 \\ \hline
\multirow{7}{*}{\textbf{Tesla M60}} & Random Forest & 0.9398	&	4.5547	&	2.2427\\ \cline{2-5} 
 & Extra Trees & 0.9557	&	3.9610	&	1.9362 \\ \cline{2-5} 
 & Gradient Boosting & 0.9415	&	4.3645	&	2.0169 \\ \cline{2-5} 
 & \textbf{XG Boost} & \textbf{0.9544}	&	\textbf{3.9172}	&	\textbf{1.7588} \\ \cline{2-5} 
   & \textbf{CatBoost} & \textbf{0.9548}	&	\textbf{3.89382}	&	\textbf{1.8252} \\ \cline{2-5} 
    & Decision Tree &  0.8762	& 6.6781		&  3.7151 \\ \cline{2-5} 
 & SVR & 0.6021	&	12.0735	&	8.7283 \\ \cline{2-5} 
 & ANN & 0.7314	&	8.8217	&	5.9993 \\ \hline
\multirow{7}{*}{\textbf{Tesla V100}} & Random Forest & 0.9415	&	8.8127	&	3.7447\\ \cline{2-5} 
 & Extra Trees & 0.9526	&	7.9428	&	3.1989 \\ \cline{2-5} 
 & Gradient Boosting & 0.9595	&	7.4883	&	3.2475 \\ \cline{2-5} 
 & \textbf{XG Boost} & \textbf{0.9646}	&	\textbf{6.9968}	&	\textbf{2.9028} \\ \cline{2-5} 
   & \textbf{CatBoost} & \textbf{0.9543}	&	\textbf{7.8954}	&	\textbf{3.1766} \\ \cline{2-5} 
    & Decision Tree &  0.8978	& 10.9053		& 4.4655 \\ \cline{2-5} 
 & SVR & 0.6435	&	22.6876	&	13.5012 \\ \cline{2-5} 
 & ANN & 0.7134	&	18.3762	&	11.0164 \\ \hline
\end{tabular}

\end{table}

\begin{figure}[ht]
\centering
\includegraphics[width=\linewidth]{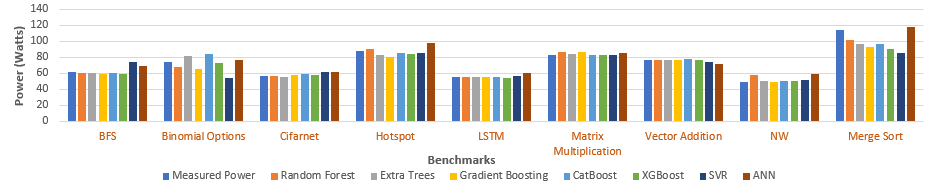}
\caption{Measured Vs Predicted power consumption for popular benchmarks}
\label{fig_predict}
\end{figure}

\begin{figure}[ht]
\centering
\includegraphics[width=\linewidth]{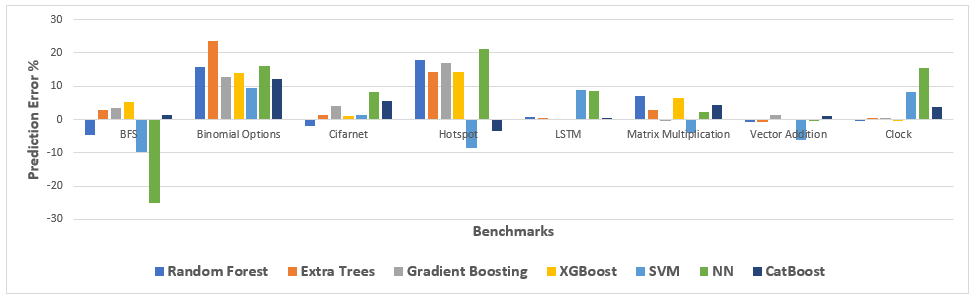}
\caption{Error in predicted power for popular benchmarks}
\label{fig_errorinpredict}
\end{figure}

%

\section{Final Model Analysis}  \label{analysingresults}
\noindent We present here measured versus predicted power consumption of some of the popular benchmarks belonging to all three benchmark suites in Figure \ref{fig_predict} for all the machine learning models considered in this study. For a better presentation, we chose a small set to present in Figure \ref{fig_predict} and not all the benchmarks considered in this study. We have presented the error in prediction for the same benchmarks in Figure \ref{fig_errorinpredict}. 

\begin{figure}[ht]
\centering
\includegraphics[width=\linewidth]{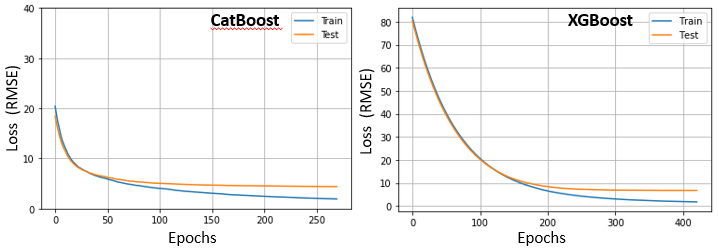}
\caption{Loss function plot for most precise tree-based methods}
\label{loss_function_catboostXGBoost}
\end{figure}

\noindent \textbf{Analysis of Results:} Experimental results suggest that the XGBoost model outperforms all the other algorithms for all three GPU architectures under study. CatBoost Loss plot follows its XGBoost counterpart in terms of a monotonically decreasing trend. Loss function plots for the models assure that the two models are training well since both the models exhibit smooth, monotonic descent, as seen in Figure\ref{loss_function_catboostXGBoost}. Test Loss plot follows the training loss plot closely; hence, the model generalizes well on new data. This is in sharp contrast with ANN and other non-tree regressors (see Figure\ref{loss_function_ann}). The numerical difference between training and test loss is small enough to justify the better fitting capability of XGB, compared to any other regressor, including Catboost. Hence we evaluate and compare the performance of these two tree-based models 
in Table \ref{tab:catboostVsXgBoost} (shown for Tesla K20 GPU for brevity). 

We further analyze the CatBoost and XGBoost model vs. ANN-based models on the basis of loss plots (Figure \ref{loss_function_ann} and Figure \ref{loss_function_catboostXGBoost}). ANN overfits very early, around 20 epochs. At $20$ epochs, the validation  $R^{2}$ score  for inverse decay ANN is $0.6943$ (train loss: $12.51$, validation loss: $13.13$), and for exponential decay ANN, it is $0.5839$  (train loss: $11.77$, validation loss: $12.97$). It gets worse beyond 20 epochs with fluctuating loss curve (\textit{Final train loss:8.82, validation loss:11.656 }). The CatBoost model starts overfitting around $60$ epochs, with $R^{2}=0.839$ (train loss: 7.397, validation loss: 7.55), much higher than ANN. XGBoost clearly outperforms all the models. It achieves the final  $R^{2}$ score of ANN ($0.773$) at $165$ epochs 
(train loss: 8.802, validation loss:8.86) without overfitting. When XGBoost starts to slightly overfit around $200$ epochs, its validation  $R^{2}$ score reaches $0.848$ (train loss: $6.04$, validation loss: $7.54$), which is significantly higher than the ANN model's final $R^2$ score. Among tree-based techniques, the Decision tree also overfits early i.e. at depth=4, with validation $R^{2}=0.641$ (train loss:$11.79$, validation loss:$14.21$). 
Furthermore, please notice that the loss is oscillating and doesn't decay smoothly. It also saturates after a certain number of epochs (50 for inverse decay and 140 for exponential decay), unlike in Fig. \ref{loss_function_catboostXGBoost}, where tree-based methods demonstrate smooth, monotonically  decaying loss. We ran the XGBoost model for Tesla V100 five times to test its robustness, and the observed results are presented in Table \ref{tab:xgboost_3arch}. This proves that our model is  performing well, not by accident and the prediction accuracy is consistent irrespective of changes in  testing and training data.

\begin{table}[]
\centering
\caption{ XGBoost results for all three architectures after 5 runs }
\label{tab:xgboost_3arch}
\begin{tabular}{|l|l|l|l|l|}
\hline
\textbf{Architecture} & \textbf{$R^{2}$ score} & \textbf{RMSE}         & \textbf{MAE}        & \textbf{MAPE}      \\ \hline
\textbf{Kepler}       & $0.901 \pm 0.0083$     & $7.4919 \pm 0.0.3194$ & $3.6740 \pm 0.1126$ & $4.5784\pm 0.1809$ \\ \hline
\textbf{Maxwell}      & $0.9418 \pm 0.0.0056$  & $4.4000 \pm 0.1702$   & $1.8780 \pm 0.0634$ & $2.9278\pm 0.0857$ \\ \hline
\textbf{Volta}        & $0.9553 \pm 0.0056$    & $7.7712 \pm 0.4081$   & $3.0521 \pm 0.1365$ & $3.9900\pm 0.0732$ \\ \hline
\end{tabular}
\end{table}

We relate the response, i.e. the power consumed by the CUDA kernel and program features and understand its features contribution using feature importance. As seen in Table \ref{tab:catboostVsXgBoost}, we can conclude that \emph{inst\_issue\_cycles}, \emph{avg\_comp\_lat}, \emph{reg\_thread}, \emph{glob\_inst\_kernel}, and \emph{glob\_store\_sm} are the \textit{top five} features, having high feature importance in both the models. Since we have used Shapley during the feature selection process, described in Section~\ref{featureEngineering}, these features are marked important by Shapley as well.

\begin{table}[ht]
\caption{XGBoost Vs CatBoost Evaluation }
\label{tab:catboostVsXgBoost}
\centering
\resizebox{\linewidth}{!}{
\begin{tabular}{|lc|c|c|llll}
\cline{1-4}
\multicolumn{2}{|l|}{\textbf{Evaluation Metric}}                                                                                         & \multicolumn{1}{l|}{\textbf{XGBoost}}                                                                                                                                                                                                                                                                                    & \multicolumn{1}{l|}{\textbf{Catboost}}                                                                                                                                                                                                                                                                                   &  &  &  &  \\ \cline{1-4}
\multicolumn{2}{|l|}{\textbf{$R^2$ score}}                                                                                                   & 0.9101                                                                                                                                                                                                                                                                                                                   & 0.9067                                                                                                                                                                                                                                                                                                                   &  &  &  &  \\ \cline{1-4}
\multicolumn{2}{|l|}{\textbf{RMSE}}                                                                                                      & 7.1592                                                                                                                                                                                                                                                                                                                   & 7.3001                                                                                                                                                                                                                                                                                                                   &  &  &  &  \\ \cline{1-4}
\multicolumn{2}{|l|}{\textbf{MAE}}                                                                                                       & 3.4816                                                                                                                                                                                                                                                                                                                   & 3.8308                                                                                                                                                                                                                                                                                                                   &  &  &  &  \\ \cline{1-4}
\multicolumn{1}{|l|}{\textbf{Smaller dataset}} & \multirow{2}{*}{\textbf{\begin{tabular}[c]{@{}c@{}}Memory\\  consumption\end{tabular}}} & 323.23 MiB                                                                                                                                                                                                                                                                                                               & 241.52 MiB                                                                                                                                                                                                                                                                                                               &  &  &  &  \\ \cline{1-1} \cline{3-4}
\multicolumn{1}{|l|}{\textbf{Larger dataset}}  &                                                                                         & 162.68 MiB                                                                                                                                                                                                                                                                                                               & 214.86 MiB                                                                                                                                                                                                                                                                                                               &  &  &  &  \\ \cline{1-4}
\multicolumn{1}{|l|}{\textbf{Smaller dataset}} & \multirow{2}{*}{\textbf{\begin{tabular}[c]{@{}c@{}}Execution\\  time\end{tabular}}}     & 3.29 s                                                                                                                                                                                                                                                                                                                   & 10.3s                                                                                                                                                                                                                                                                                                                    &  &  &  &  \\ \cline{1-1} \cline{3-4}
\multicolumn{1}{|l|}{\textbf{Larger dataset}}  &                                                                                         & 6.4s                                                                                                                                                                                                                                                                                                                     & 11.8s                                                                                                                                                                                                                                                                                                                    &  &  &  &  \\ \cline{1-4}
\multicolumn{2}{|p{1.5in}|}{\textbf{Feature Importance Ranking} (Consistent across all the three GPU architectures)}                                                                                & \multicolumn{1}{l|}{\begin{tabular}[c]{@{}l@{}}1. glob\_store\_sm \\ 2. inst\_issue\_cycles\\ 3. avg\_comp\_lat\\ 4. misc\_inst\_kernel\\ 5. glob\_inst\_kernel\\ 6. reg\_thread\\7. comp\_inst\_kernel\\8. avg\_glob\_lat\\9. cache\_penalty\\10. avg\_shar\_lat\\11. occupancy\\ 12. shmem\_block\\13. branch\\14. glob\_load\_sm\\15. block\_size\end{tabular}} & \multicolumn{1}{l|}{\begin{tabular}[c]{@{}l@{}} 1. avg\_comp\_lat\\ 2.  inst\_issue\_cycles\\3. reg\_thread\\4. glob\_inst\_kernel\\ 5. comp\_inst\_kernel\\6. avg\_glob\_lat\\7. misc\_inst\_kernel\\ 8. glob\_store\_sm\\9. branch\\10. glob\_load\_sm\\11. cache\_penalty\\ 12. occupancy\\13. avg\_shar\_lat\\14. block\_size\\15. shmem\_block\end{tabular}} &  &  &  &  \\ \cline{1-4}
\end{tabular}
}
\end{table}

\begin{figure}
\centering
\includegraphics [width=\linewidth]{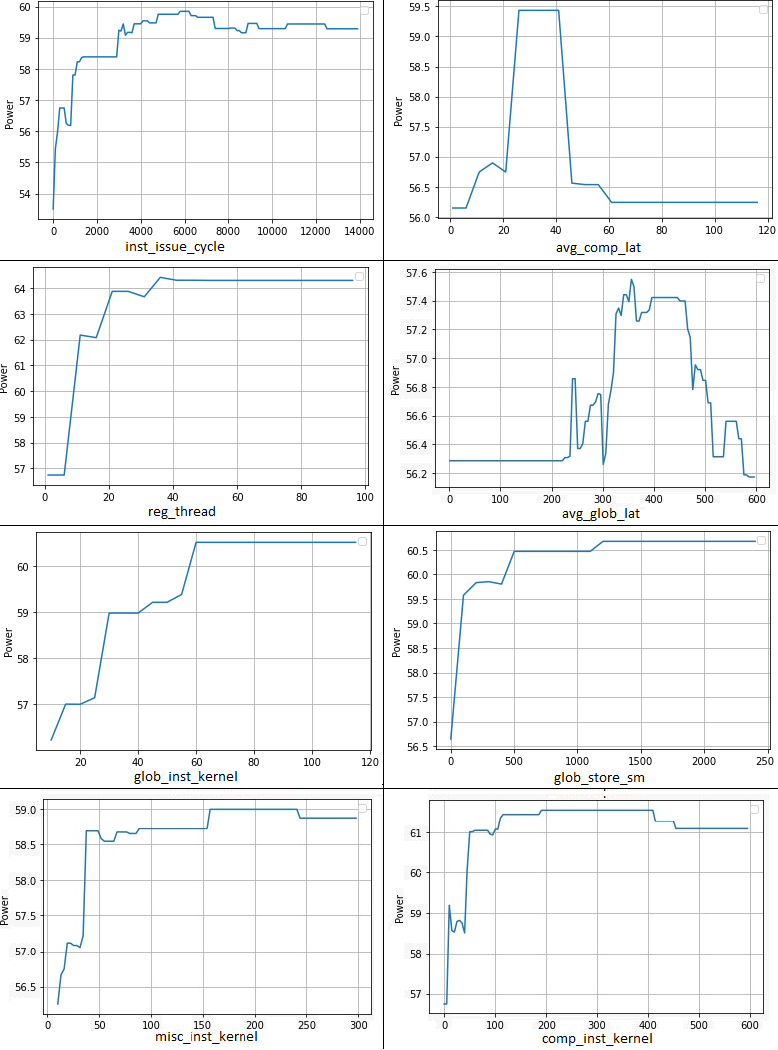} 
\caption{Plots for crucial program features' dependency on the power consumption}
\label{fig_impFeat}
\vspace{-0.2in}
\end{figure}

We now consider these important features that influence the response  
and aim to provide an explanation for the variation observed in the response variable corresponding to changes in the predictor variables, one at a time. It is also assumed that the important features have little pairwise interaction between them. In the nonlinear model fitted by XGBoost, we froze values of all (important) features except for one feature under consideration. As the values of that feature are varied and other features are kept constant, the change in the response variable is plotted. This will measure the impact of each crucial feature against predicted power in Figure \ref{fig_impFeat}. As expected, these top-ranked features are indeed influential in explaining the variations in the response variable. Note that, since there is no closed-form regression model available, this is intuitively a reasonable check of the behavior of the response variable against the top-ranked predictor variables.

\subsection{Analyzing Influential Features from Architecture Viewpoint}
The feature importance analysis found that \emph{inst\_issue\_cycle} contributes the most to power consumption across all three architectures.
Since the FDS unit is employed in issuing instructions, inst\_issue\_cycle reasonably represents the number of times FDS is activated. Hong et al. \cite{Hong2010} noted that the FDS unit consumes higher power than other components because all instructions utilize this unit. Thus, our finding that this feature is the most significant feature certainly conforms to the findings~\cite{Hong2010}. 
Since it is difficult for developers to control the number of instructions issued for a particular task, a plausible option would be to focus on writing efficient code with a lesser number of instructions to do the task. 

The feature \emph{avg\_comp\_lat} is the average latency of all computing instructions executed, including floating-point instructions such as {\tt maddf}, {\tt divf}. It is observed from Figure~\ref{fig_impFeat}, \emph{avg\_comp\_lat} impacts power prediction when its value exceeds ten cycles. This is the case when an application constitutes floating-point instruction since the latency of other integer computing instructions ({\tt add, sub}) is significantly lower (section III in supplementary file). Our results {\em differ with} that of Hong et al. \cite{Hong2010}'s finding, which concluded that there is no considerable difference between instruction and floating-point unit's power consumption.

The next important feature denotes the number of registers per thread (\emph{reg\_thread}). This is in concurrence with the findings reported by Greener ~\cite{jatala2017greener}. As register file size increases, the leakage power increases; hence limited use of registers is advisable for building power-efficient applications. 

The number of global memory instructions per kernel (\emph{glob\_inst\_kernel}) and the average global memory latency (\emph{avg\_glob\_lat}) also contribute to power prediction—particularly global store instruction (\emph{glob\_store\_sm}), which follows \emph{glob\_inst\_kernel}. Nagasaka et al. \cite{Nagasaka2010} concluded that the acquisition of global memory access counts is crucial for accurate GPU power modeling. According to runtime hardware unit analysis by Hong et al. \cite{Hong2010}, global memory instructions consume a significant amount of power, much more than floating-point benchmarks. This supports our finding that global memory contributes significantly to power consumption. To reduce the increase in power consumption due to global memory, one can replace global memory instructions by using shared memory as frequently as possible. Observing the difference between the feature importance of \emph{glob\_store\_sm} over \emph{glob\_load\_sm} suggests that writing to the global memory is more power consuming than reading from it.

The \emph{block\_size} and \emph{shmem\_block} are the least important features consistently across all the feature importance techniques employed. Observing feature importance of features related to shared memory ( \emph{shmem\_block,  avg\_shar\_lat}) suggests that its utilization has the most negligible impact on power prediction compared to all other instruction types. With respect to \emph{block\_size}, we can conclude that the \emph{block\_size} is not as crucial as long as total threads ($grid\_size \times block\_size$) launched is higher which is used to compute \emph{inst\_issue\_cycle}.

\begin{figure}
  \begin{subfigure}{0.5\textwidth}
    \centering\includegraphics[width=8cm]{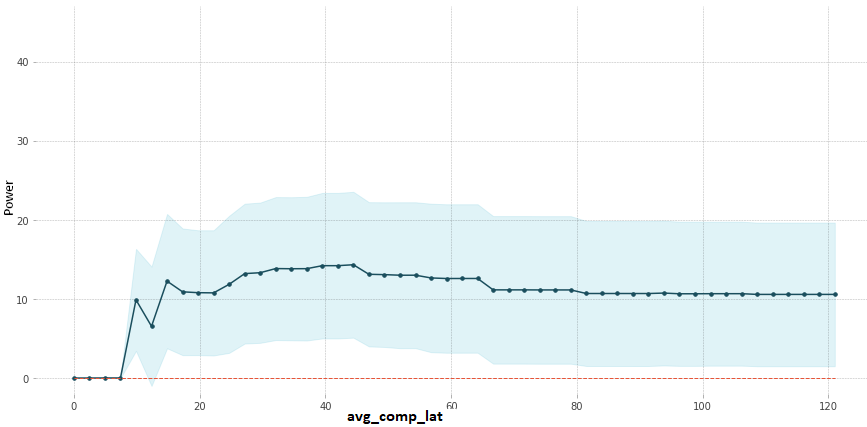}
       \end{subfigure}
  \begin{subfigure}{0.5\textwidth}
    \centering\includegraphics[width=8cm]{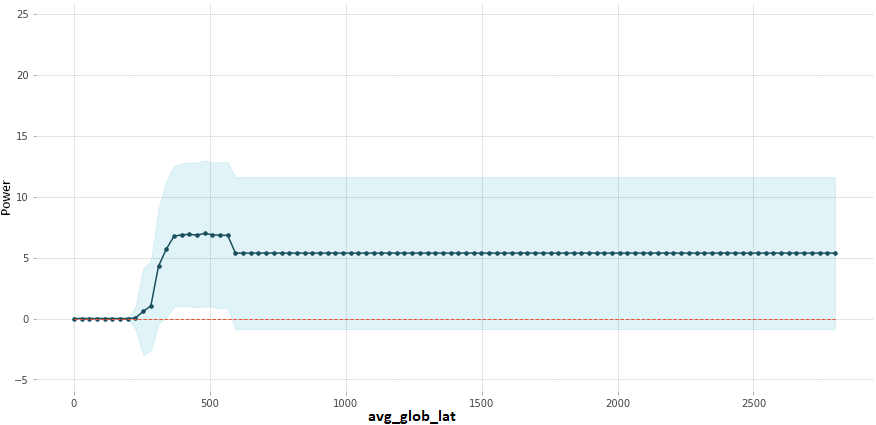}
  \end{subfigure}
   \begin{subfigure}{0.5\textwidth}
    \centering\includegraphics[width=8cm]{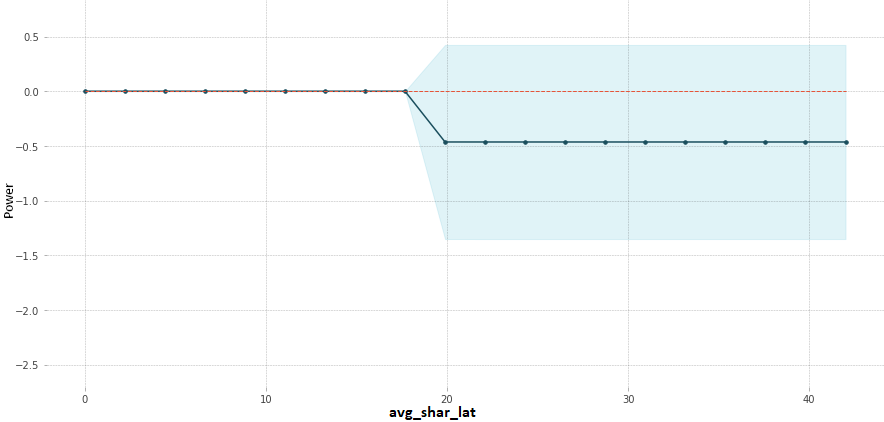}
    \end{subfigure} 
  \begin{subfigure}{0.5\textwidth}
    \centering\includegraphics[width=8cm]{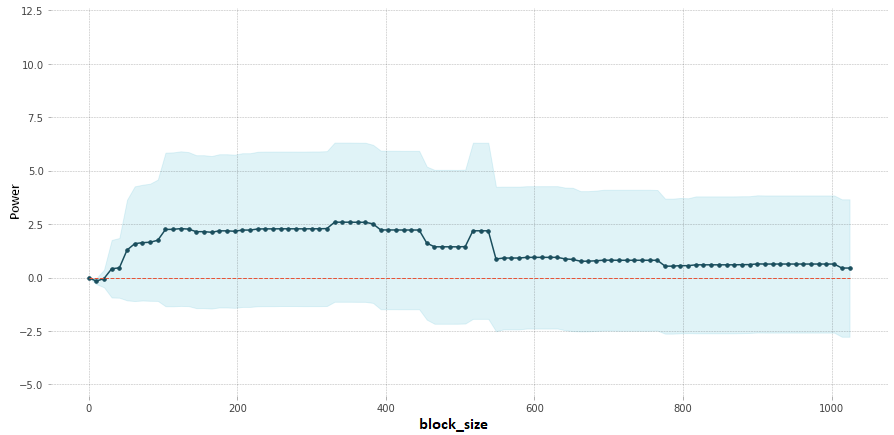}
     \end{subfigure}
   \begin{subfigure}{0.5\textwidth}
    \centering\includegraphics[width=8cm]{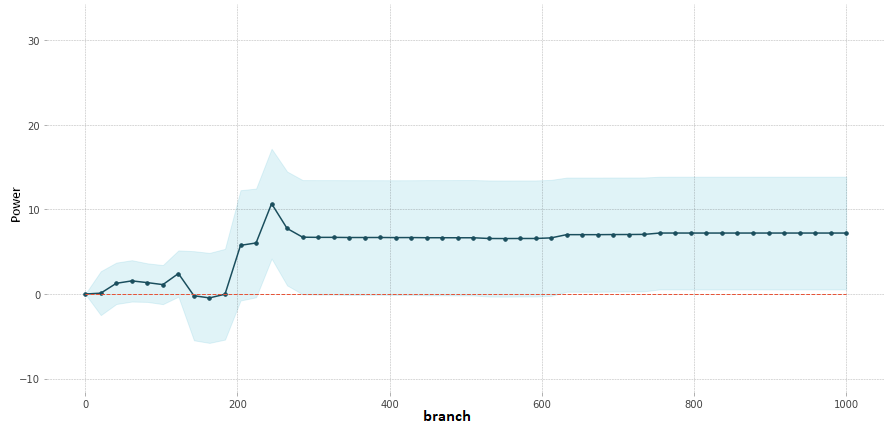}
    \end{subfigure}
      \begin{subfigure}{0.5\textwidth}
    \centering\includegraphics[width=8cm]{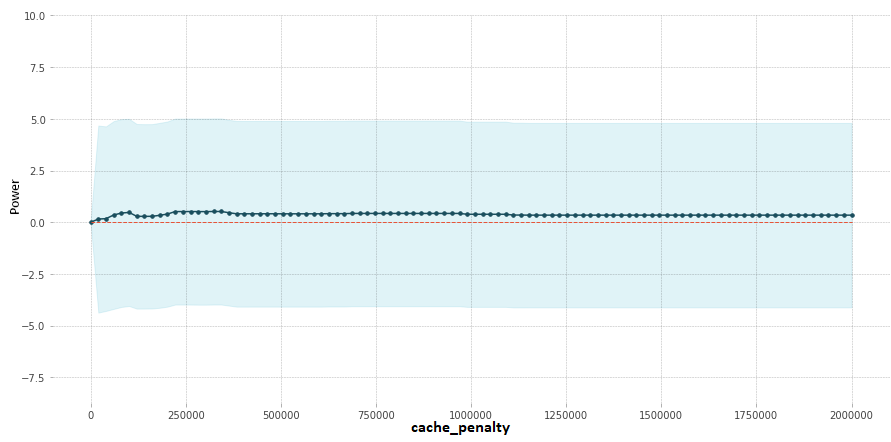}
     \end{subfigure}
    \begin{subfigure}{0.5\textwidth}
    \centering\includegraphics[width=8cm]{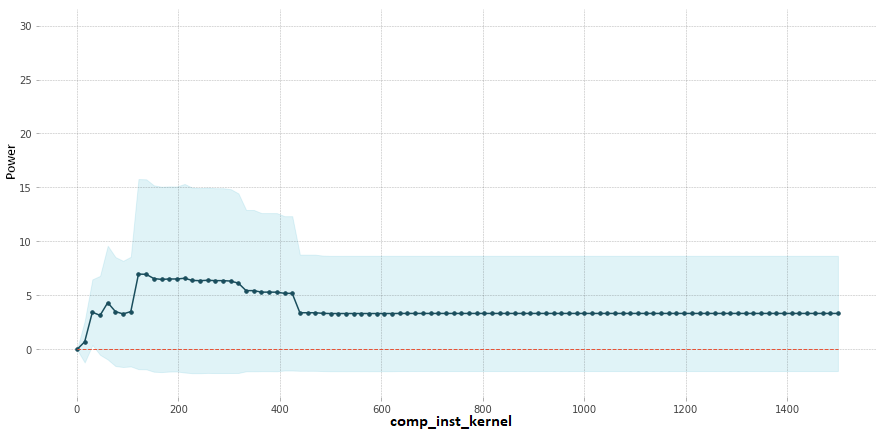}
    \end{subfigure}
      \begin{subfigure}{0.5\textwidth}
    \centering\includegraphics[width=8cm]{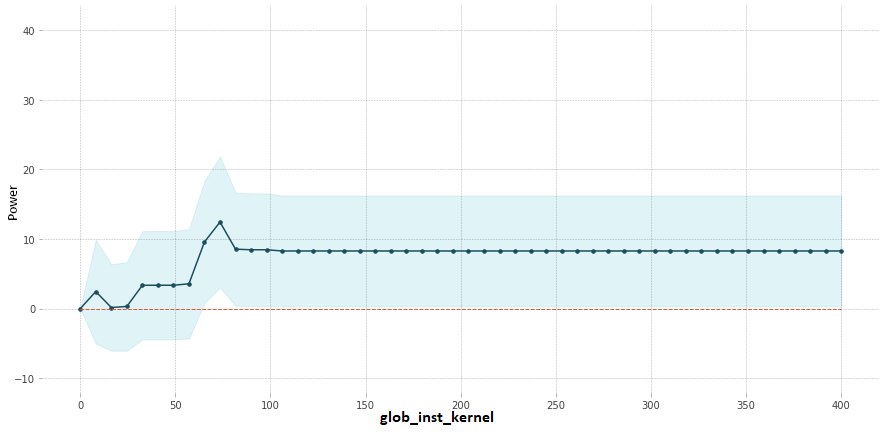}
     \end{subfigure}
    \begin{subfigure}{0.5\textwidth}
    \centering\includegraphics[width=8cm]{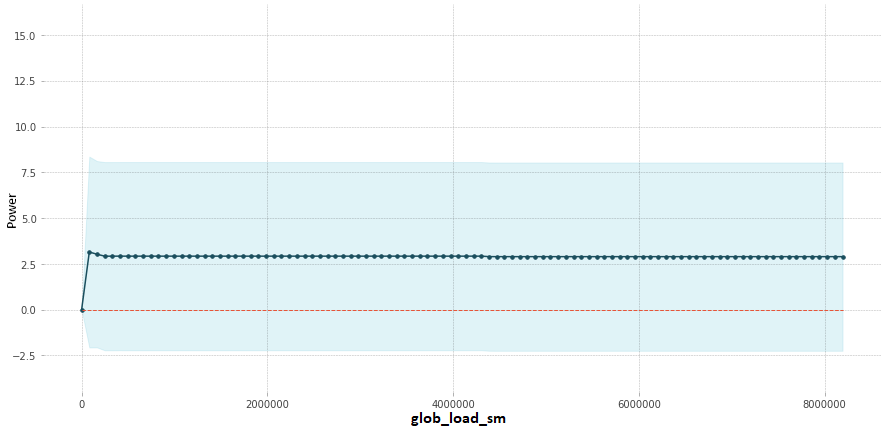}
    \end{subfigure}
\begin{subfigure}{0.5\textwidth}
    \centering\includegraphics[width=8cm]{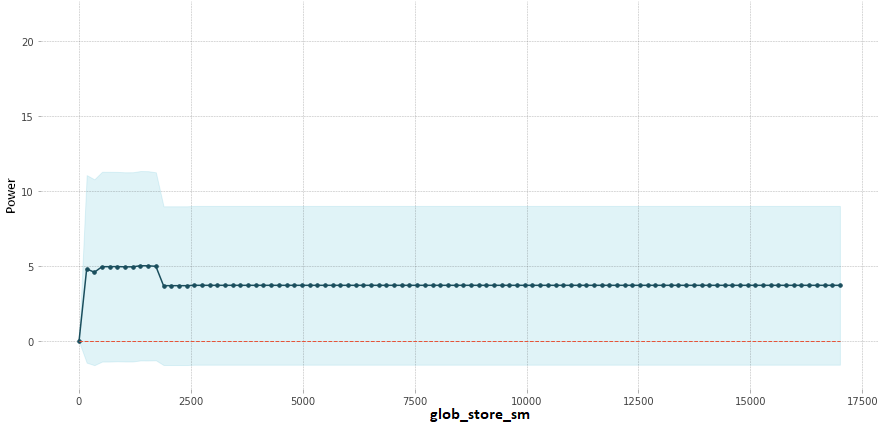}
     \end{subfigure}
  \end{figure}
    \begin{figure} 
     \begin{subfigure}{0.5\textwidth}
    \centering\includegraphics[width=8cm]{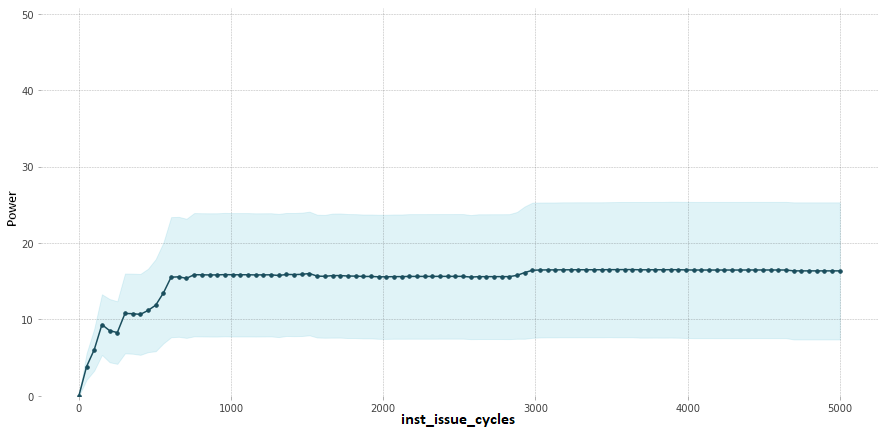}
 
    \end{subfigure}
      \begin{subfigure}{0.5\textwidth}
    \centering\includegraphics[width=8cm]{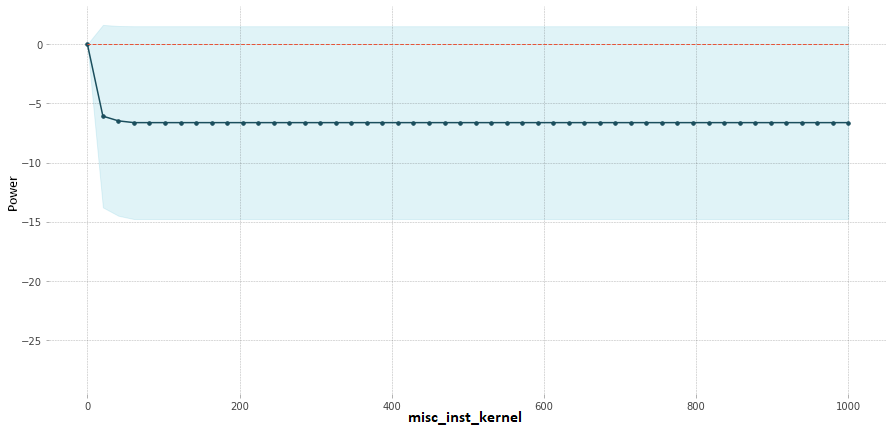}
     \end{subfigure}

\begin{subfigure}{0.5\textwidth}
    \centering\includegraphics[width=8cm]{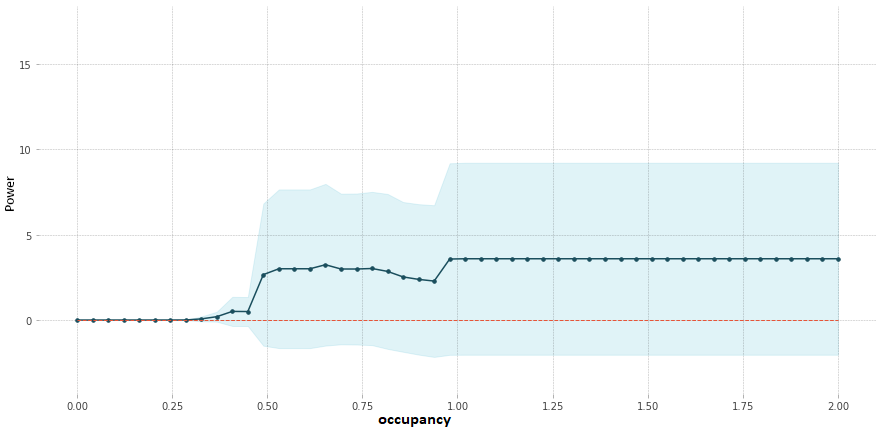}
    \end{subfigure}
      \begin{subfigure}{0.5\textwidth}
    \centering\includegraphics[width=8cm]{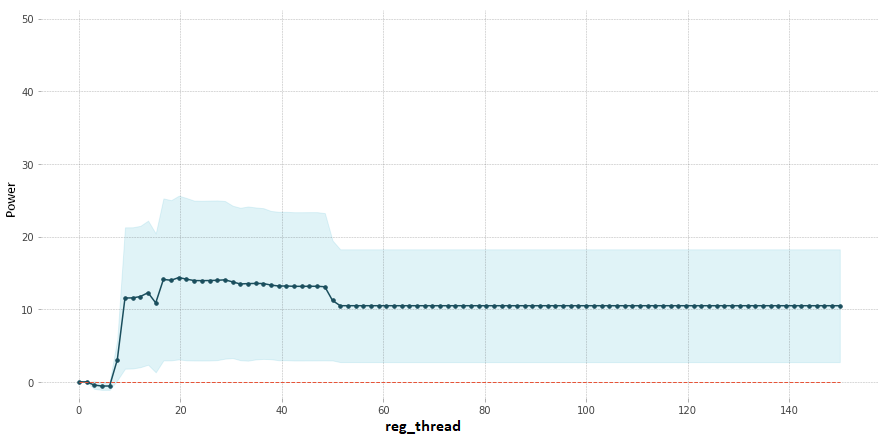}
     \end{subfigure}
      \begin{subfigure}{0.5\textwidth}
    \centering\includegraphics[width=8cm]{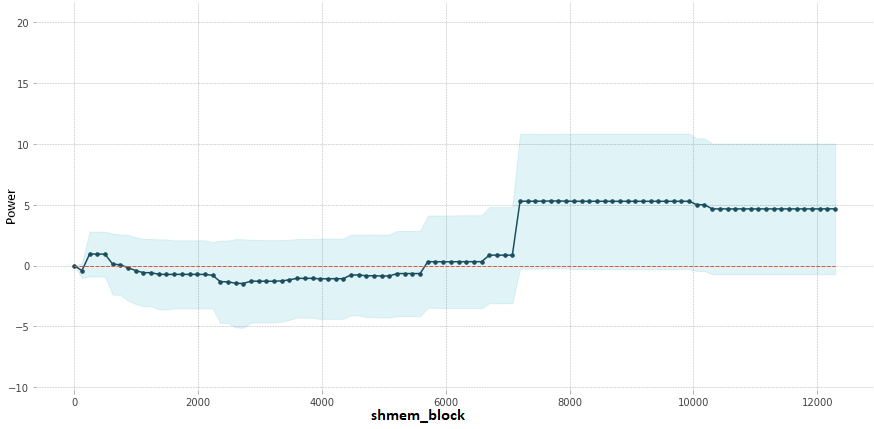}
    \end{subfigure}
   \caption{Partial Dependency Plot (PDP) of all features of XGBoost}
   \label{pdp}
\end{figure}

\subsection{Outlier Detection}
Outliers affect statistical measures and data distributions due to the misleading representation of the underlying data and relationships. Hence removing outliers from training data is necessary for modeling results in a better fit of the data. This in turn results in more skillful predictions and better accuracy of results. In the case of the power prediction model constructed from program features, outliers can occur if a CUDA benchmark's power consumed is lowered or higher compared to the rest of the data points with similar characteristics. For e.g. a vector addition benchmark consists of three launch configurations i.e grid size, block size ( c1: (256,1024), c2: (512,512), c3: (1024,	256)). Total threads are the same for all three configurations and hence most of the features for all three configurations are equal. However, it was observed that power consumption for c1 is 109.91 W, c2 is 112.08 W and c3 is 112.49304 W. In this scenario, our outlier detection method should highlight c1 as an outlier. 

If such outliers exist, they can affect the accuracy of the model.  Hence we would like to perform outlier detection in order to avoid any outlier which may influence the prediction accuracy of our model. As seen earlier, there is no linear relationship between features and prediction variables. Also, due to the large number of features, visual inspection is not possible for outlier detection. Hence we employ HDBSCAN (Hierarchical Density-Based Spatial Clustering of Applications with Noise ) \cite{Campello2013} which is a Density-Based Clustering algorithm \cite{Ester1996,Kriegel2011}. The HDBSCAN algorithm performs DBSCAN over varying epsilon values and integrates the result to find a clustering that gives the best stability over epsilon. Therefore HDBSCAN can discover clusters of varying densities (unlike DBSCAN). This makes HDBSCAN more robust to parameter selection. 

\begin{figure}
\centering
\includegraphics{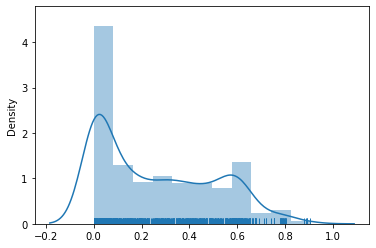} 
\caption{Outlier score Density using HDBSCAN}
\label{fig_outliers}
\end{figure}

The HDBSCAN library supports the GLOSH outlier detection algorithm and does so within the HDBSCAN clustering class \cite{campello2015hierarchical}. Figure \ref{fig_outliers} shows the outlier density score using HDBSCAN. The algorithm detected ninety-seven outliers. However, after removing the outliers, the $R^2$ score for XGBoost using the Tesla K20 dataset is lower ($R^2$ score: 0.8942) than the original dataset ($R^2$ score:0.9101). CatBoost results after outlier removal are $R^2$ score: 0.8937 which was higher ($R^2$ score: 0.9067)  before outlier removal. Hence we conclude that outlier removal is not contributing to building a more robust model. Results after outlier detection for all the machine learning techniques are tabulated in Table \ref{tab:_outlier}.

\begin{table}
\caption{Validate score for machine learning techniques after outlier removal }
\label{tab:_outlier}
\centering
\begin{tabular}{|l|l|l|l|}
\hline
\textbf{Model}    & \textbf{$R^{2}$ score} & \textbf{RMSE} & \textbf{MAE} \\ \hline
Random Forest     & 0.8777                 & 10.2092       & 1.8252       \\ \hline
Extra Trees       & 0.8768                 & 8.2609        & 4.4506       \\ \hline
Gradient Boosting & 0.8875                 & 9.7673        & 3.9698       \\ \hline
XGBoost           & 0.8942                 & 5.9352        & 3.6002       \\ \hline
CatBoost          & 0.8937                 & 6.8369        & 3.9394       \\ \hline
Decision Tree     & 0.8451                 & 10.8890       & 4.9394       \\ \hline
SVR               & 0.6747                 & 18.39         & 11.015       \\ \hline
ANN               & 0.5737                 & 14.084        & 9.46         \\ \hline
\end{tabular}
\end{table}

\subsection{Analysing Prediction Model using Parallel Dwarfs}
Parallel dwarfs represent the communication and computation pattern of a parallel application \cite{Asanovic2009}. To further analyse the predicted power results of the model built, we choose a set of benchmarks for each type of dwarf under study. We ran these benchmarks for multiple launch configurations and predicted their power consumption using the machine learning models trained. 
The observed Mean Absolute  Error (MAE) for each dwarf for the Tesla K20 machine is presented in Table \ref{tab:dwarf}. 

    \indent\textbf{Best cases:} It is observed that the proposed model works best for predicting the power of benchmarks belonging to dense linear algebra, dynamic programming, and unstructured grid dwarfs. Dense linear algebra benchmarks are compute-bound applications, whereas dynamic programming and unstructured grid are memory latency bound. Hence we claim that our model works for both compute and memory-bound kernels. 
    
    \indent\textbf{Worst cases:} We observe that benchmarks belonging to Graph Traversal, Branch and Bound, and Structured Grid dwarfs have higher MAE. The branch and bound benchmark's performance limit is problem-specific. Structured grid benchmarks have high spatial locality. Benchmarks belonging to graph traversal dwarf include indirect table lookup and less computation. We conclude that our model may not produce precise results for benchmarks having these characteristics.

\begin{table}
\small
\caption{Mean Absolute Error across Parallel Dwarfs}
\label{tab:dwarf}
\resizebox{\linewidth}{!}{
\begin{tabular}{|l|r|r|r|r|r|r|}
\hline
\textbf{Model}               & \multicolumn{1}{l|}{\textbf{\begin{tabular}[c]{@{}l@{}}Branch \\ and Bound\end{tabular}}} & \multicolumn{1}{l|}{\textbf{\begin{tabular}[c]{@{}l@{}}Graph   \\ Traversal\end{tabular}}} & \multicolumn{1}{l|}{\textbf{\begin{tabular}[c]{@{}l@{}}Dense Linear \\ Algebra\end{tabular}}} & \multicolumn{1}{l|}{\textbf{\begin{tabular}[c]{@{}l@{}}Structured \\ Grid\end{tabular}}} & \multicolumn{1}{l|}{\textbf{\begin{tabular}[c]{@{}l@{}}Unstructured\\ Grid\end{tabular}}} & \multicolumn{1}{l|}{\textbf{\begin{tabular}[c]{@{}l@{}}Dynamic\\ Programming\end{tabular}}} \\ \hline
\textbf{Random   Forest}     & 11.31                                                                                     & 13.17                                                                                      & 4.45                                                                                          & 14.31                                                                                    & 0.33                                                                                      & 8.33                                                                                        \\ \hline
\textbf{Extra   Trees}       & 16.17                                                                                     & 19.06                                                                                      & 3.66                                                                                          & 13.52                                                                                    & 0.18                                                                                      & 0.3                                                                                         \\ \hline
\textbf{Gradient   Boosting} & 19.51                                                                                     & 14.52                                                                                      & 4.6                                                                                           & 25.74                                                                                    & 0.43                                                                                      & 0.05                                                                                        \\ \hline
\textbf{XGBoost}             & 21.14                                                                                     & 14.85                                                                                      & 3.91                                                                                          & 24.03                                                                                    & 0.3                                                                                       & 0.96                                                                                        \\ \hline
\textbf{CatBoost}            & 14.42                                                                                     & 23.4                                                                                       & 7.22                                                                                          & 6.29                                                                                     & 3.28                                                                                      & 2.65                                                                                        \\ \hline
\textbf{SVR}                 & 29.34                                                                                     & 25.86                                                                                      & 5.89                                                                                          & 3.69                                                                                     & 1.37                                                                                      & 1.96                                                                                        \\ \hline
\textbf{ANN}                 & 15.91                                                                                     & 32.91                                                                                      & 4.74                                                                                          & 4.8                                                                                      & 5.96                                                                                      & 10.04                                                                                       \\ \hline
\end{tabular}
}
\end{table}

\section{Discussion}
In this chapter, we present a GPU power prediction model based on program analysis employing machine learning algorithms. We utilize hardware details to generate the features considered in this study without running the application. Machine learning models considered in this study include Random Forest Regressor, ExtraTrees Regressor, Gradient Boosting Regressor, XGBoost Regressor, CatBoost Regressor, SVR, and ANN. We tested these models for our hypothesis across three GPU machines: Tesla K20, Tesla M60, and Tesla V100. We observed that tree-based regression methods produce the most accurate power prediction model than SVR and ANN results. We observed that \emph{instruction issue cycle} is the most crucial feature across models since it represents the FDS unit that is utilized by all instructions; hence it is a constant contributor irrespective of other benchmark parameters. Benchmarks with floating-point instructions and global memory instructions also consume a significant amount; therefore, their usage has to be limited to build more energy-efficient applications. The use of registers in an application is also crucial and needs to be minimally utilized for energy efficiency. Thus, the final set of 'influential features' obtained through a 'data-driven' approach explains the variations in the response (power) well and is consistent with the domain-expertise.
This study has given rise to further GPU power modeling field investigations. We can utilize this study to predict the energy consumption of GPU. One can also dig deeper to understand how a program can be refactored using the presented observations in this work. This refactoring should be effective in such a way that it consumes lower power without compromising performance. Future work is also to analyze and improve prediction for benchmarks belonging to Graph Traversal, Branch and Bound, and Structured Grid dwarfs. 
\let \textcircled=\pgftextcircled
\chapter{Tool Design}
\label{ch:tool_design}

\initial{D}esign and implementation-level tools for analyzing and predicting the performance of GPU-accelerated applications are scarce. Although performance modeling approaches for GPUs exist, their complexity makes them virtually impossible to use at the time of coding to quickly analyze the performance of an application and obtain easy-to-use, readable feedback. This is why although GPUs are significant performance boosters in many HPC domains, performance prediction is still based on extensive benchmarking, and performance bottleneck analysis remains a non-systematic, experience-driven process. 

The main goal of developers of GPU is to tune their code extensively to obtain optimal performance, making efficient use of different resources available on the GPU. However, an application running on an HPC infrastructure should perform optimally, and at the same time, it needs to consume the least energy required to run the application. A developer must be equipped with tools to understand the energy consumption of GPUs before launching the application. Researchers are working on building tools to assist developers in sustainable computing \cite{pawanr2021, Couto2020, SCI}. On similar grounds, i.e., understanding and reducing energy usage, we propose an energy prediction tool that will aid developers in understanding the Performance and Power of GPU applications, which can be used for a-priory energy consumption analysis. 

We developed the Energy Estimation Tool, a prediction tool built as an Eclipse plugin that predicts the performance and power consumption of an application without running it. The plugin tool also provides a detailed report on which features contribute to the power consumption of an application. The Energy Estimation Tool plugin is validated for three GPU architectures: Kepler, Maxwell, and Volta. We believe the tool presented in this thesis will provide useful guidance on developing a GPU application and understanding the application's power and performance profile. The architecture and usage details of the tool are discussed in the next section.

\section{Tool Architecture and Implementation} \label{Design}
 
 \begin{figure}
  \centering
  \includegraphics[width=10cm]{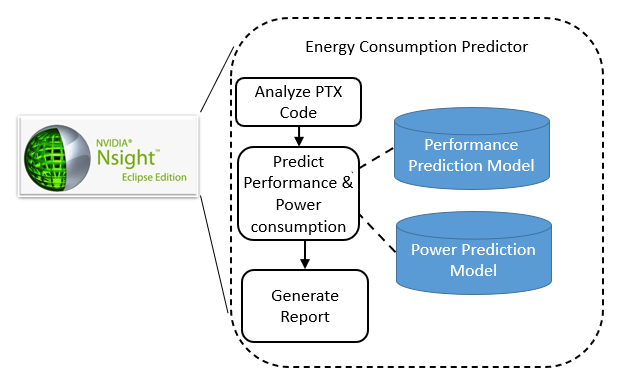}
  \caption{Overview of Energy Estimation Tool}
 \label{Fig:tool_overview}
\end{figure}

\begin{figure}
  \centering
  \includegraphics[width=15cm]{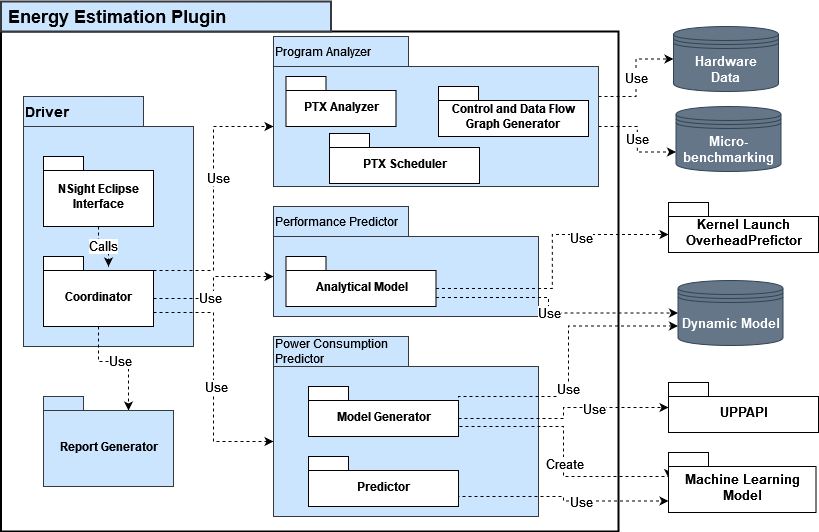}
  \caption{Functional Architecture of Energy Estimation Tool}
 \label{Fig:sysDiscription}
\end{figure}

The tool reported in this work is built as a plugin that leverages the NVIDIA NSight IDE for CUDA program compilation. The plugin analyzes the PTX code generated by the {\tt nvcc} compiler and generates the features supplied to a prediction model. Figure \ref{Fig:tool_overview} shows the execution flow of the Energy Estimation Tool. The power prediction model generates a power consumption report which includes details of features contributing to power prediction. Figure \ref{Fig:sysDiscription} presents the functional architecture of the Energy Estimation Tool in detail. The plugin comprises of various modules described in Figure \ref{Fig:sysDiscription}. We discuss each of these modules in detail as follows.

 \subsection{Program Analyzer}
The Program Analyzer consists of three submodules. Firstly, the PTX analyzer dissects the PTX code and extracts instruction information. Then Control and Data Flow Generator utilize this information to build a control-data flow graph (CDFG) \cite{Allen1970}. Finally, the CDFG graph is generated using the PTX scheduler module to produce the features representing GPU's runtime behavior. 
 
The PTX scheduler uses device characteristics that represent the device's hardware configuration and execution limits. The scheduler additionally needs application characteristics that cannot be derived from static analysis. These include the grid size, block size, and the number of loop iterations. The PTX scheduler generates a set of features  using  the algorithm presented in Chapter 5. User needs to provide {\em Compute Capability} of GPU, the {\em number of threads per block}, and the {\em occupancy} value, calculated using CUDA Occupancy Calculator. The PTX Analyzer module generates all the other features from the PTX code  and microbenchmarking data.

\subsection{Execution Time Predictor}
This model utilises the PTX Analyzer details and schedules instructions to compute the performance of the CUDA kernel. This module utilizes kernel launch overhead model along with a dynamic analysis model to predict the execution time. Details of our approach for performance prediction are provided in Chapter \ref{ch:execTime} section \ref{PP_approach2}.

\subsection{Power Consumption Predictor}
This module is responsible for predicting the power consumed by a CUDA kernel. We used a machine learning approach to train the prediction model, built from the features extracted by the Program Analyzer module. Details of the power prediction model are discussed in detail in Chapter \ref{ch:powerpred}.

\subsection{Report Generator}
 Based on the prediction using the chosen machine learning technique, the plugin reports the performance and power consumption of the application under study in $\mu$s and Watts respectively. Energy is computed by multiplying the results obtained from the two modules. 
 
Power prediction is further analysed using the LIME tool \cite{Ribeiro2016}. LIME is a model-agnostic tool and provides the user details on why this prediction was derived or which variables contributed to the prediction. We firmly believe this gives a deeper insight into the application's power consumption for the application developer to refactor the code to make it more power-efficient. 
 
Based on the regression technique chosen and other application details supplied by the user, Energy Estimation Tool predicts the performance and power consumption of the GPU application. Figure \ref{Fig:energy_result} shows the result window.  We utilize LIME to further dissect the prediction made by the power prediction tool and present a detailed report. We discuss the working of the Energy Estimation Tool in further detail using a case study in section \ref{case_study}.

\section{Plugin Usage Details}
\subsection{Prerequisite}
\begin{figure}[ht]
  \centering
  \includegraphics[width=15cm]{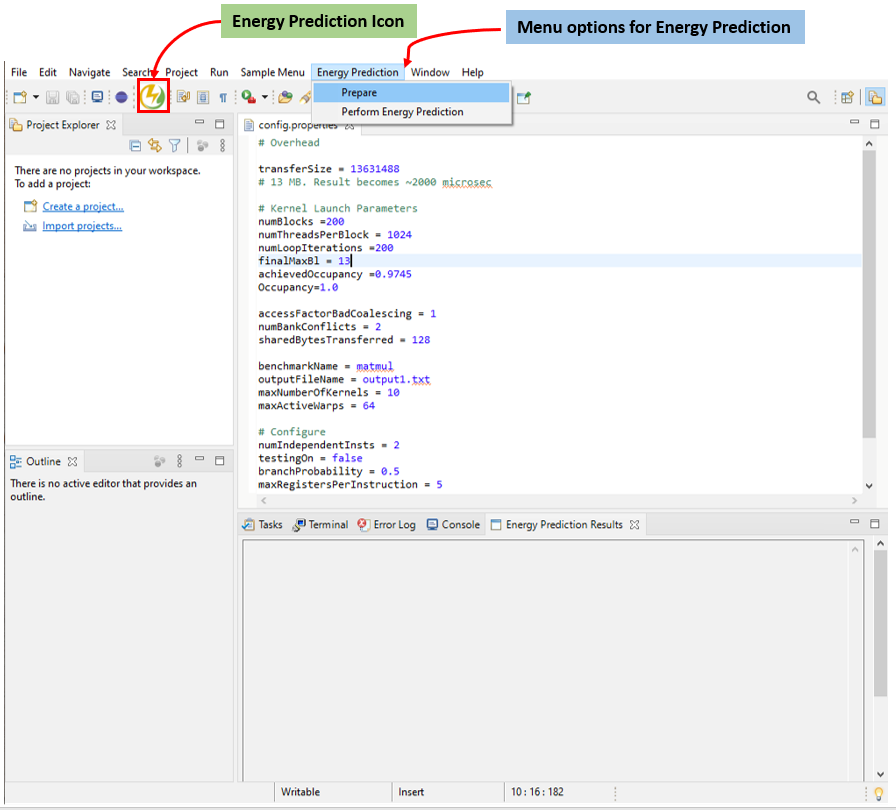}
  \caption{Energy Estimation Tool}
 \label{Fig:energy_plugin}
\end{figure}

Once the NVIDIA Nsight Eclipse IDE \footnote{https://developer.nvidia.com/nsight-eclipse-edition} is started, the plugin calls a
method to check if {\tt NVCC} is installed on the user’s machine or not. If not,
it freezes the plugin and prompts the user to install {\tt NVCC}.  If {\tt NVCC} is installed, then some executable files are extracted to a temporary location by the plugin. These executables are needed for querying and running microbenchmarks for the user’s hardware. The method for extracting these files is called on Eclipse start-up. Figure \ref{Fig:energy_plugin} shows the plugin window.  

\subsection{Plugin Setup}
The user can now enter the ”plugin setup” phase. It is in this phase that the executables are run for querying and running microbenchmarks for the user’s GPU. This phase might take some time when executed for the first time. The output of these executables are stored as text files on the user’s machine, and after the
first time, the preparation phase consists of just reading from those text files, which is faster. Note that the plugin generates some models during this phase, that are not stored as text files. So, every time the user wants to use the plugin, he has to ”Setup” at least once.

\subsection{User Input}
After the ”Preparation” phase is done, the user may proceed to provide the required details of the benchmark in a config file. These details include launch parameters and the number of loop iterations as seen in Figure \ref{Fig:energy_plugin}. We have also taken other details of benchmarks that we think are relevant in future developments of the plugin.  

\subsection{Result Generation}
\begin{figure}[ht]
  \centering
  \includegraphics[width=15cm]{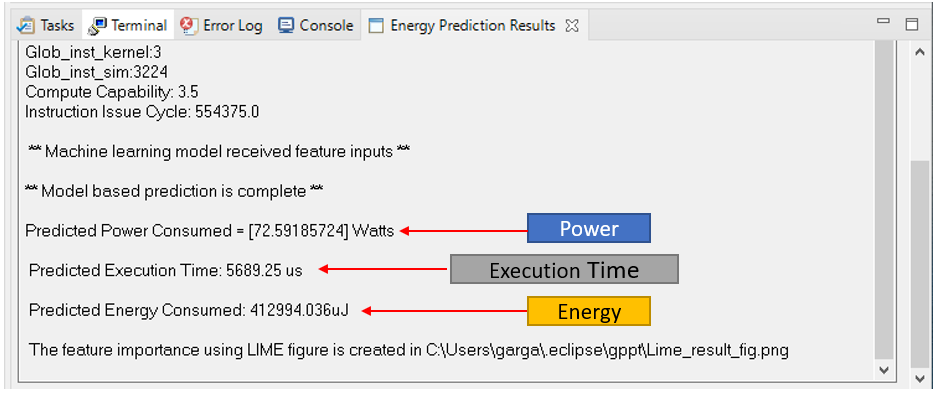}
  \caption{Energy Estimation Result}
 \label{Fig:energy_result}
\end{figure}

After the required parameters have been filled, the user can perform energy estimation by selecting ”Energy Estimation>Perform Energy Estimation” on the menu bar. One can also click the plugin icon on the menu bar to perform this task. The results will be displayed on the custom console. Figure \ref{Fig:energy_result} shows the result generated in a  window. For every power prediction result, the LIME Analysis figure is generated with a message in a dialogue box that shows the location of the generated figure as seen in Figure \ref{Fig:lime}. 

\begin{figure}[htbp]
  \centering
  \includegraphics[width=15cm]{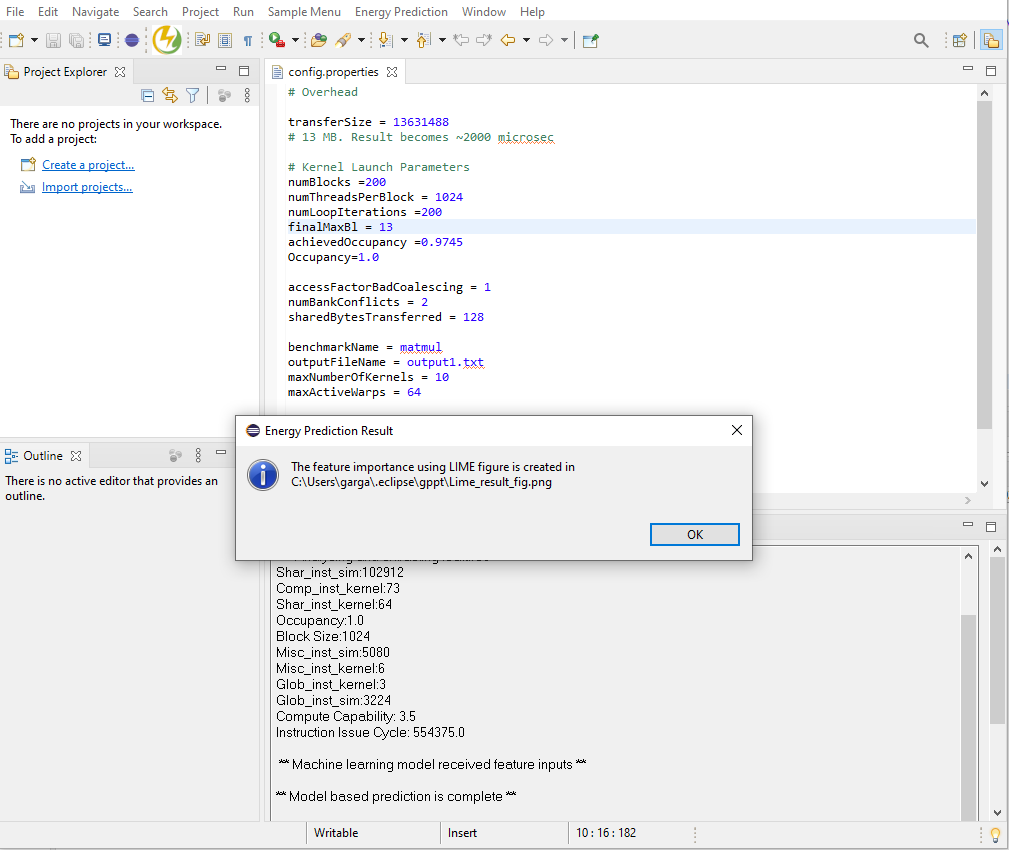}
  \vspace{-0.3cm}
  \caption{LIME output generation in Plugin}
  \label{Fig:lime_output}
\end{figure}

\subsection{Case Study: Matrix Multiplication} \label{case_study}
Matrix multiplication is a prevalent problem
in GPU computing with its application in multiple domains. It contains both global and shared memory instructions and is a memory-bound application. We have presented execution time prediction results for Matrix multiplication in Chapter 4. In Table \ref{tab:energyresults}, we present energy prediction results of matrix multiplication when it is run with 1024 blocks of 1024 threads per block each on all three architectures under study. 

\begin{table}[]
\caption{Results for Matrix Multiplication Across architectures}
\label{tab:energyresults}
\resizebox{\linewidth}{!}{
\begin{tabular}{|l|rrr|rrr|}
\hline
\multirow{2}{*}{} & \multicolumn{3}{c|}{Actual Execution}                                                                        & \multicolumn{3}{c|}{Predicted Results}                                                                       \\ \cline{2-7} 
                  & \multicolumn{1}{l|}{Execution Time (us)} & \multicolumn{1}{l|}{Power(W)}  & \multicolumn{1}{l|}{Energy (uJ)} & \multicolumn{1}{l|}{Execution Time (us)} & \multicolumn{1}{l|}{Power(W)}  & \multicolumn{1}{l|}{Energy (uJ)} \\ \hline
Tesla K20         & \multicolumn{1}{r|}{6997.75}             & \multicolumn{1}{r|}{83.17}    & 582002.86                      & \multicolumn{1}{r|}{5689.25}             & \multicolumn{1}{r|}{83.28} & 473800.74                      \\ \hline
Tesla M60         & \multicolumn{1}{r|}{4917.4}              & \multicolumn{1}{r|}{84.99}  & 417929.82                      & \multicolumn{1}{r|}{5617.52}             & \multicolumn{1}{r|}{86.21}     & 484286.39                      \\ \hline
Tesla V100        & \multicolumn{1}{r|}{8514.9}              & \multicolumn{1}{r|}{137.26} & 1168755.17                    & \multicolumn{1}{r|}{8945.25}             & \multicolumn{1}{r|}{138.16}    & 1235875.74                       \\ \hline
\end{tabular}
}
\end{table}

\begin{figure}[htbp]
  \centering
  \includegraphics[width=15cm]{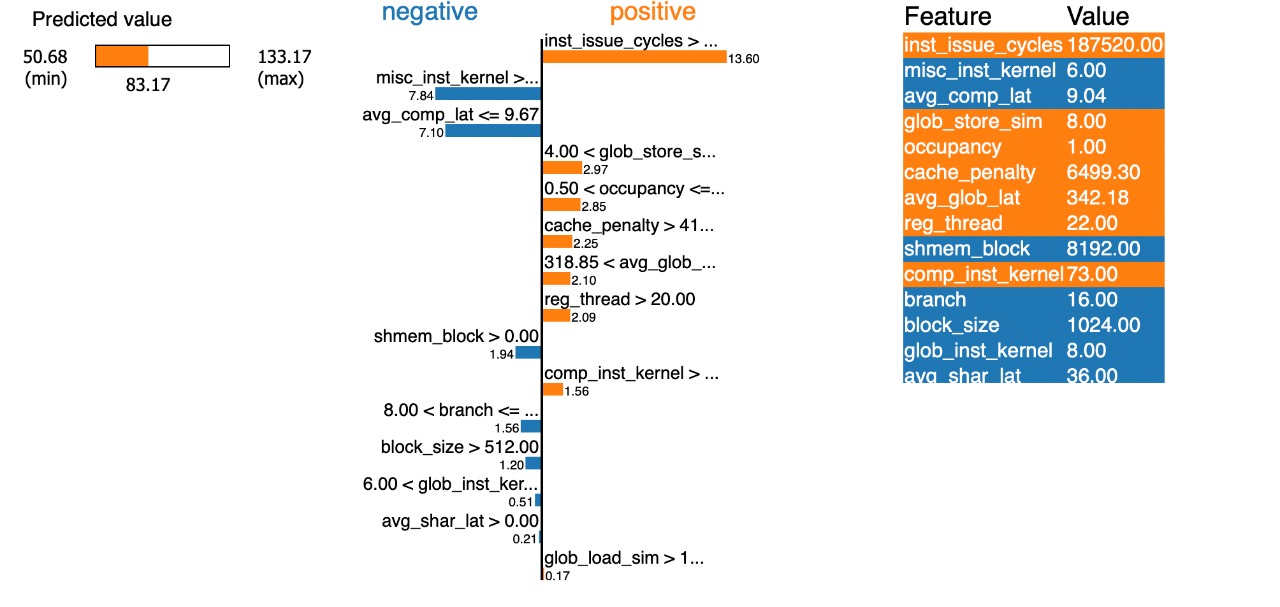}
  \vspace{-0.3cm}
  \caption{Power prediction analysis using LIME}
  \label{Fig:lime}
\end{figure}

We also present the LIME report for matrix multiplication prediction by XGBoost on Kepler architecture in Figure \ref{Fig:lime}. LIME  provides details of which features contributed to the prediction. As seen in Figure  \ref{Fig:lime}, the y-axis shows the features while the x-axis shows the relative strength of each feature. The positive value  (orange color) shows that the feature supports or increases the prediction value. In contrast, the negative value (blue color) negatively affects or decreases the prediction value. In this case, shared memory instruction does not contribute significantly to power consumption as seen from the contribution of shmem\_block and avg\_shar\_lat features. To develop a more power-efficient application, a developer needs to employ shared memory instructions over global memory instructions as much as feasible. The impact of global memory instructions on power consumption is depicted by how avg\_glob\_lat, glob\_store\_sim contribute to power prediction. Register per thread also significantly contributes to power prediction; hence one can conclude that reducing register usage can help to reduce power consumption in this benchmark. Reducing the number of instructions executed is critical since the inst\_issue\_cycles feature contributes most to power prediction as observed in Chapter \ref{ch:powerpred}.   

\section{Comparison of results of Power \& Execution Time Prediction}
\begin{figure}[htbp]
  \centering
  \includegraphics[width=15cm]{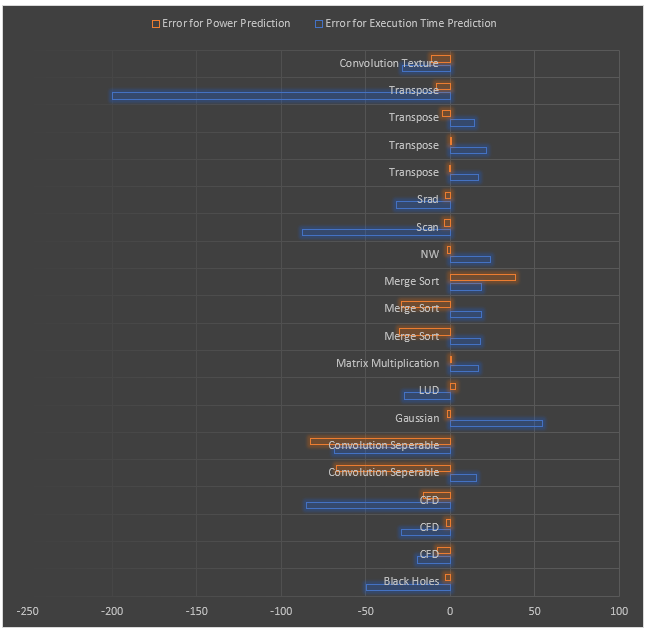}
  \vspace{-0.3cm}
  \caption{Comparison of results of Power \& Execution Time Prediction}
  \label{Fig:exectimepower}
\end{figure}

In this section, we have compared the high error results for execution time prediction with their power prediction results. To do this, we have chosen 20 benchmarks whose execution time prediction error is greater than 15\%. We have presented the comparison for these results in Figure \ref{Fig:exectimepower}. The details of results for this comparison are given in Table \ref{tab:my-powerperformance}. As seen in the plot, apart from the Convolution Seperable benchmark, errors for all other benchmarks are not consistent for execution time and power. The power prediction model seems to perform better in comparison with execution time prediction. 
When we analyzed the Convolution Separable benchmark in detail to under why both the models fail to predict, we observed that the difference between instructions issued (18420093) and instructions executed (11943936)
 for this benchmark is quite high for convolutionRowsKernel. This could be due to hardware-level optimizations which our model failed to predict. The power prediction model seems to perform better in comparison with execution time prediction.

\begin{table}[]
\caption{Comparison of Performance \&  Power results for selected benchmarks}
\label{tab:my-powerperformance}
\resizebox{\textwidth}{!}{%
\begin{tabular}{|r|r|r|r|r|r|r|}
\hline
\multicolumn{1}{|l|}{\textbf{Sr No.}} & \multicolumn{1}{l|}{\textbf{Benchmark}}  & \multicolumn{1}{l|}{\textbf{Kernel Name}}  & \multicolumn{1}{l|}{\textbf{Actual Performance}} & \multicolumn{1}{l|}{\textbf{Predicted Performance}} & \multicolumn{1}{l|}{\textbf{Actual Power}} & \multicolumn{1}{l|}{\textbf{Predicted Power}} \\ \hline
1                                     & Black Holes                              & BlackScholes                               & 547.12                                           & 819.18                                              & 65.40066                                   & 67.28                                         \\ \hline
2                                     & CFD                                      & CFDInit                                    & 29.15                                            & 34.86                                               & 68.0969                                    & 73.34959                                      \\ \hline
3                                     & CFD                                      & CFDComputeStepF                            & 55.39                                            & 71.28                                               & 89.8506                                    & 91.64792                                      \\ \hline
4                                     & CFD                                      & CFDFlux                                    & 820.19                                           & 1,519.40                                            & 48.1210                                    & 55.588272                                     \\ \hline
5                                     & Convolution Seperable                    & ConvS\_C                                   & 1,165.50                                         & 985.57                                              & 61.4910                                    & 102.682365                                    \\ \hline
6                                     & Convolution Seperable                    & ConvS\_R                                   & 1,665.50                                         & 2,801.80                                            & 59.7500                                    & 109.073555                                    \\ \hline
7                                     & Gaussian                                 & GaussF2                                    & 326.63                                           & 147.93                                              & 99.92654                                   & 101.66327                                     \\ \hline
8                                     & LUD                                      & LUDPeri                                    & 15.95                                            & 20.24                                               & 63.539                                     & 61.495842                                     \\ \hline
9                                     & Matrix Multiplication                    & MatrixMul                                  & 6,997.75                                         & 5,825.05                                            & 95.5720                                    & 95.3723                                       \\ \hline
10                                    & Merge Sort                               & MergeMS\_generateSample                    & 183.33                                           & 150.60                                              & 89.7283                                    & 116.52137                                     \\ \hline
11                                    & Merge Sort                               & mergeRanksAndIndices                       & 1,114.10                                         & 910.58                                              & 59.2251                                    & 76.29441                                      \\ \hline
12                                    & Merge Sort                               & mergeSortSharedKernel                      & 1,114.10                                         & 910.58                                              & 113.4631                                   & 69.39528                                      \\ \hline
13                                    & NW                                       & NW                                         & 22.24                                            & 16.99                                               & 49.6300                                    & 50.31136                                      \\ \hline
14                                    & Scan                                     & scanExclusiveShared                        & 361.83                                           & 678.01                                              & 122.45502                                  & 126.607155                                    \\ \hline
15                                    & Srad                                     & Srad                                       & 11.26                                            & 14.84                                               & 56.3509                                    & 57.872116                                     \\ \hline
16                                    & Transpose                                & Tpose\_optimized                           & 77.35                                            & 64.32                                               & 75.8094                                    & 76.264854                                     \\ \hline
17                                    & Transpose                                & Tpose\_coarse-grained                      & 74.88                                            & 58.65                                               & 76.7434                                    & 76.43441                                      \\ \hline
18                                    & Transpose                                & Tpose\_Naive                               & 62.15                                            & 53.03                                               & 84.9380                                    & 88.74651                                      \\ \hline
19                                    & Transpose                                & Tpose\_coalesced                           & 78.05                                            & 234.03                                              & 77.0491                                    & 83.31207                                      \\ \hline
20                                    & \multicolumn{1}{l|}{Convolution Texture} & \multicolumn{1}{l|}{convolutionRowsKernel} & 620.85                                           & 794.89                                              & 65.556                                     & 72.84                                         \\ \hline
\end{tabular}%
}
\end{table}

\section{Discussion} \label{conclusion}
We presented an eclipse plugin that combines execution time and a power prediction model developed in this thesis to predict the energy consumption of  a CUDA kernel. This energy prediction plugin is made available here\footnote{ https://github.com/santonus/energymodel}.  It provides easy to use design assistance tool for GPGPU developers which can be used for understanding the energy consumption of an application without executing it. The plugin is developed so that it can be extended to multiple architectures and analyse different types of CUDA benchmarks. We strongly believe that this tool is a novel step in making GPUs a sustainable computing alternative and can be further extended to find the energy hotspots in the code. These energy hotspots should then be refactored to build energy-efficient applications. 

\let \textcircled=\pgftextcircled
\chapter{Conclusion and Future Directions}
\label{ch:conclusion}

\initial{T}hrough this dissertation, we attempted to answer the research questions on the prediction of performance and power of GPU applications using static and one-time dynamic analysis. We now conclude by summarizing the significance of our
work and laying directions for future research.

\section{Summary}
The energy estimation of an application can be analyzed by studying its performance and power consumption. Hence we divided our research goal into two subproblems. In the first subproblem, we presented an analytical performance prediction model for NVIDIA GPU architecture based on static and dynamic analysis of a CUDA kernel. The static analysis includes PTX code analysis to construct control and data flow information. Additionally, we performed the dynamic analysis to obtain runtime behavioral features like instruction latency, memory throughput, and kernel launch overhead as a one-time activity. Using the information obtained from the static and dynamic analysis, we perform the scheduling of the kernel to estimate its execution time. 

The second subproblem is to present a GPU power prediction model based on program analysis. We utilize hardware details to generate the features considered in this study without running the application. We employed machine learning algorithms to learn from these features and perform the prediction. Machine learning models considered in this study include {\em Random Forest Regressor, ExtraTrees Regressor, Gradient Boosting Regressor, XGBoost Regressor, CatBoost Regressor, SVR, and ANN}. We tested these models for our hypothesis across three GPU machines: Tesla K20, Tesla M60, and Tesla V100. We observed that tree-based regression methods produce the most accurate power prediction model than SVR and ANN results. It was found that \emph{inst\_issue\_cycle} is the most crucial feature across models since it represents the FDS unit which is utilized by all instructions; hence it is a constant contributor irrespective of other benchmark parameters. Benchmarks with floating-point instructions and global memory instructions also consume a significant amount of power; therefore, their usage has to be limited to build more energy-efficient applications. The use of registers in an application is also crucial and needs to be minimally utilized for energy efficiency. Thus, the final set of 'influential features' obtained through a 'data-driven' approach explains the variations in the response (power) well and are consistent with the domain expertise.

We strongly believe that these observations from performance and power prediction can be further extended to build a model that finds energy hotspots. Application developers can then refactor their code to build an energy-efficient application for GPUs.

\section{Directions for Future Work}

Following are some of the possible improvements in our prediction approach:

\begin{itemize}
\item Our approach does not perform loop iteration prediction. Currently, we rely on the user to provide loop iteration count since most of the loops inside a kernel tend to be data-dependent, and non-affine \cite{Feautrier1996}. The problem of predicting the number of loop iterations of non-affine benchmarks is yet to be solved.

\item  Branch divergence in warps is a source of error because the delays of instructions need to be summed up for all executed instructions. An assumption is made that in every warp, every branch is taken by at least one thread. This means that the thread taking the branch will execute the instruction while the remaining threads execute a no-operation. The final instruction execution time may be inaccurate if some branch is never taken during a warp execution.

\item 
We have attempted to model cache penalty with a simple parameterized equation. Modeling cache misses is quite challenging, and we have not come across any well-established model for a-priory cache misses so far. In our future work, we aim to study the cache miss phenomenon in a GPU more deeply.

\item Since register usage can significantly impact the power as well as the performance of a kernel \cite{jatala2017greener,Vishwesh2016}; one needs to understand its effect by analyzing kernels for different numbers of register utilized. We plan to consider the effect of register utilization on both power and performance prediction as part of our future work.

\item We have modelled the impact of coalesced memory access through a global memory penalty factor. While we have considered coalescing, we have not taken uncoalesced memory access in our model. Furthermore, we have not explored whether it is possible to determine the coalesced or uncoalesced access patterns only through the static analysis of a kernel.

\item Adding more architecture-specific attributes in power modelling such as features generated by dynamic analysis of CUDA kernel in power prediction. E.g. including details of memory access patterns, shared memory bank conflicts, and a better model for capturing cache behavior. 

\item We need to study and analyze how a program can be refactored using the presented observations in this work. This refactoring should be effective in such a way that it consumes lower energy without compromising on performance. At the same time, this refactoring should not affect performance since it will defeat the purpose of utilizing performance-accelerating machines. 

\item We need to analyse the relationship between communication and computation patterns of parallel dwarfs with the performance and power prediction. E.g., in power prediction, study and improve prediction for benchmarks belonging to Graph Traversal, Branch and Bound, and Structured Grid dwarfs. 
The branch and bound benchmark's performance limit is problem-specific. Structured grid benchmarks have high spatial locality. Benchmarks belonging to graph traversal dwarf include indirect table lookup and less computation. We need to delve deeper into benchmarks having these characteristics and develop solutions for improving prediction and incorporating them into the model.

\item We can investigate a hybrid modeling approach for execution time prediction, which includes executing a benchmark once on a GPU and collecting its runtime parameters. One can use this data to predict the execution of any other launch configuration for this benchmark.

\item GPUs are very popular for building performance-hungry machine learning libraries ( NVIDIA cuDNN \cite{chetlur2014cudnn}, Tensorflow \cite{pang2020deep} )which sometimes execute for multiple days for big data applications. We can build energy prediction models for GPUs specifically for popular Machine Learning specific libraries. This will be an important contribution to sustainable machine learning computing. 

\item  We have not considered the rare occasions when the application program uses explicit
synchronization primitives. These forced synchronization operations that act across a warp or a block may have an impact on the ability of analytical methods to predict the run time. We would like to model this behavior in our model.

\item Newer generations of CUDA hardware have introduced additional architectural and programming features such as tensor cores, multi-kernel execution, special instructions etc. We need to study and incorporate the impact of these features. 

\item  We would like to utilize an adaptive learning approach for power prediction using machine learning techniques such as reinforcement learning and transfer learning. Through these techniques, systems will learn continuously during application's runtime. 

\item Apart from NVIDIA GPUs, there are many other popular multicore manycore architectures such as Xeon Phi, and AMD GPUs. We can adapt the proposed approach to other multi-core and many-core architectures and observe its usability and effectiveness.

\item Lastly, a common concern in static analysis-based power and performance models is that we utilize PTX code to perform a-priory prediction. However, in using PTX code, we believe in the sometimes untenable hypothesis, which assumes that all the instructions listed will execute. Static analysis predictive models have the disadvantage of the inability to know a priori how many instructions will be executed. Theoretical results have proven this to be invincible— e.g., Turing's work on the halting problem, where it is proven that it is undecidable to know whether a given program and input will run forever. Hence there is a need to understand how a number of instructions executed can be more accurate in static analysis-based models based on PTX code. 

\end{itemize}
\pagestyle{empty}
%
\addcontentsline{toc}{chapter}{Biographies}
%
%
%

%
\backmatter
\SingleSpacing

\refstepcounter{chapter}
\bibliography{parallelpgm}

\begin{thebibliography}{100}

\bibitem{asanovic2006}
K.~Asanovic, R.~Bodik, B.~C. Catanzaro, J.~J. Gebis, P.~Husbands, K.~Keutzer,
  D.~A. Patterson, W.~L. Plishker, J.~Shalf, S.~W. Williams, {\em et~al.},
  ``The landscape of parallel computing research: A view from berkeley,'' tech.
  rep., Technical Report UCB/EECS-2006-183, EECS Department, University of
  California, Berkeley, 2006.

\bibitem{Gianniti2018}
E.~Gianniti, L.~Zhang, and D.~Ardagna, ``Performance prediction of gpu-based
  deep learning applications,'' in {\em International Symposium on Computer
  Architecture and High Performance Computing (SBAC-PAD)}, 2018.

\bibitem{Lattuada2022}
M.~Lattuada, E.~Gianniti, D.~Ardagna, and L.~Zhang, ``Performance prediction of
  deep learning applications training in gpu as a service systems,'' {\em
  Cluster Computing}, vol.~25, p.~1279–1302, 2022.

\bibitem{Huang2009}
S.~Huang, S.~Xiao, and W.~Feng, ``On the energy efficiency of graphics
  processing units for scientific computing,'' in {\em IEEE International
  Symposium on Parallel \& Distributed Processing (IPDPS)}, 2009.

\bibitem{Tesla}
A.~Romero, ``Tesla ai day 2021 review — part 3: Project dojo. tesla’s new
  supercomputer.''
  https://towardsdatascience.com/tesla-ai-day-2021-review-part-3-project-dojo-teslas-new-supercomputer-715d102dbb29,
  2021.

\bibitem{Bridges2016}
R.~A. Bridges, N.~Imam, and T.~M. Mintz, ``Understanding gpu power: A survey of
  profiling, modeling, and simulation methods,'' {\em ACM Computing Surveys},
  vol.~49, pp.~41:1--41:27, 2016.

\bibitem{Mittal2014}
S.~Mittal and J.~S. Vetter, ``A survey of methods for analyzing and improving
  gpu energy efficiency,'' {\em ACM Comput. Surv.}, vol.~47, pp.~19:1--19:23,
  2014.

\bibitem{Puig2018}
M.~Pi~Puig, L.~De~Giusti, and M.~Naiouf, ``Are gpus non-green computing
  devices?,'' {\em Journal of Computer Science and Technology}, vol.~18,
  p.~e17, 2018.

\bibitem{Collange2009}
C.~Collange, D.~Defour, and A.~Tisserand, ``Power consumption of gpus from a
  software perspective,'' in {\em Lecture Notes in Computer Science},
  vol.~5544, pp.~914--923, 2009.

\bibitem{couto2014}
M.~Couto, T.~Car{\c{c}}{\~a}o, J.~Cunha, J.~P. Fernandes, and J.~Saraiva,
  ``Detecting anomalous energy consumption in android applications,'' in {\em
  Brazilian Symposium on Programming Languages}, pp.~77--91, 2014.

\bibitem{Couto2020}
M.~Couto, D.~Maia, J.~a. Saraiva, and R.~Pereira, ``On energy debt: Managing
  consumption on evolving software,'' in {\em Proceedings of the 3rd
  International Conference on Technical Debt}, 2020.

\bibitem{SCI}
G.~S. Foundation, ``Software carbon intensity (sci) specification.''
  \url{https://greensoftware.foundation/projects/}, 2021.

\bibitem{Madougou2016}
S.~Madougou, A.~Varbanescu, C.~de~Laat, and R.~van Nieuwpoort, ``The landscape
  of gpgpu performance modeling tools,'' {\em Parallel Computing}, vol.~56,
  p.~18–33, 2016.

\bibitem{Asanovic2009}
K.~Asanovic, R.~Bodik, {\em et~al.}, ``A view of the parallel computing
  landscape,'' {\em Communications of ACM}, vol.~52, pp.~56--67, 2009.

\bibitem{nielson2004principles}
F.~Nielson, H.~R. Nielson, and C.~Hankin, {\em Principles of program analysis}.
\newblock Springer Science \& Business Media, 2004.

\bibitem{kleimenov2021method}
A.~A. Kleimenov and N.~N. Popova, ``A method for prediction execution time of
  gpu programs,'' {\em Computational nanotechnology}, vol.~8, pp.~38--45, 2021.

\bibitem{sahni1996performance}
S.~Sahni and V.~Thanvantri, ``Performance metrics: Keeping the focus on
  runtime,'' {\em IEEE Parallel \& Distributed Technology: Systems \&
  Applications}, vol.~4, pp.~43--56, 1996.

\bibitem{zhou2021gpa}
K.~Zhou, X.~Meng, R.~Sai, and J.~Mellor-Crummey, ``Gpa: A gpu performance
  advisor based on instruction sampling,'' in {\em 2021 IEEE/ACM International
  Symposium on Code Generation and Optimization (CGO)}, 2021.

\bibitem{Chen2012}
W.~S. Hui~Chen, ``Power measuring and profiling: State of the art,'' {\em
  Handbook of Energy-Aware and Green Computing}, 2012.

\bibitem{Shuaiwen2013}
S.~Song, C.~Su, B.~Rountree, and K.~Cameron, ``A simplified and accurate model
  of power-performance efficiency on emergent gpu architectures,'' in {\em
  International Parallel and Distributed Processing Symposium (IPDPS)}, 2013.

\bibitem{Benedict2016}
S.~Benedict, R.~R.S., and S.~A. Alex, ``Energy and performance prediction of
  cuda applications using dynamic regression models,'' in {\em India Software
  Engineering Conference (ISEC)}, 2016.

\bibitem{Bakhoda2009}
A.~Bakhoda, G.~L. Yuan, W.~W.~L. Fung, H.~Wong, and T.~M. Aamodt, ``Analyzing
  cuda workloads using a detailed gpu simulator,'' in {\em IEEE International
  Symposium on Performance Analysis of Systems and Software (ISPASS)}, 2009.

\bibitem{Leng2013}
J.~Leng, T.~Hetherington, A.~ElTantawy, S.~Gilani, N.~Kim, T.~Aamodt, and
  V.~Janapa~Reddi, ``Gpuwattch: enabling energy optimizations in gpgpus,'' {\em
  ACM SIGARCH Computer Architecture News}, vol.~41, p.~487–498, 2013.

\bibitem{Wang2011}
Y.~Wang and N.~Ranganathan, ``An instruction-level energy estimation and
  optimization methodology for gpu,'' in {\em International Conference on
  Computer and Information Technology (CIT)}, 2011.

\bibitem{NvidiaProfile}
N.~Corporation, ``Profiler user's guide.''
  \url{http://docs.nvidia.com/cuda/profiler-users-guide/}, 2022.

\bibitem{Purnomo2010}
B.~Purnomo, N.~Rubin, and M.~Houston, ``Ati stream profiler: A tool to optimize
  an opencl kernel on ati radeon gpus,'' in {\em ACM SIGGRAPH Posters}, 2010.

\bibitem{Malhotra2014}
G.~Malhotra, S.~Goel, and S.~R. Sarangi, ``Gputejas: A parallel simulator for
  gpu architectures,'' in {\em International Conference on High Performance
  Computing (HiPC), IEEE}, 2014.

\bibitem{Malik2011}
S.~{Malik} and F.~{Huet}, ``Virtual cloud: Rent out the rented resources,'' in
  {\em International Conference for Internet Technology and Secured
  Transactions(ICITST)}, 2011.

\bibitem{Hong2010}
S.~Hong and H.~Kim, ``An integrated gpu power and performance model,'' in {\em
  Proceedings of the 37th Annual International Symposium on Computer
  Architecture (ISCA), ACM}, 2010.

\bibitem{ikram2022investigating}
M.~J. Ikram, M.~E. Saleh, M.~A. Al-Hashimi, and O.~A. Abulnaja, ``Investigating
  the effect of varying block size on power and energy consumption of gpu
  kernels,'' {\em The Journal of Supercomputing}, pp.~1--21, 2022.

\bibitem{boughzala2020predicting}
D.~Boughzala, L.~Lef{\`e}vre, and A.-C. Orgerie, ``Predicting the energy
  consumption of cuda kernels using simgrid,'' in {\em International Symposium
  on Computer Architecture and High Performance Computing (SBAC-PAD)}, 2020.

\bibitem{CUDA2017}
N.~Corporation, ``Cuda c++ programming guide (design guide) v11.7.0.''
  \url{https://docs.nvidia.com/cuda/pdf/CUDA_C_Programming_Guide.pdf}, 2022.

\bibitem{Wong2010}
H.~Wong, M.-M. Papadopoulou, M.~Sadooghi-Alvandi, and A.~Moshovos,
  ``Demystifying gpu microarchitecture through microbenchmarking,'' in {\em
  IEEE International Symposium on Performance Analysis of Systems \& Software
  (ISPASS)}, 2010.

\bibitem{k20builtinsensor}
M.~Burtscher, I.~Zecena, and Z.~Zong, ``Measuring gpu power with the k20
  built-in sensor,'' in {\em Proceedings of Workshop on General Purpose
  Processing Using GPUs, ACM}, GPGPU-7, (New York, NY, USA), pp.~28:28--28:36,
  2014.

\bibitem{Amaris2016}
M.~Amaris, R.~Camargo, M.~Dyab, A.~Goldman, and D.~Trystram, ``A comparison of
  gpu execution time prediction using machine learning and analytical
  modeling,'' {\em IEEE International Symposium on Network Computing and
  Applications (NCA)}, 2016.

\bibitem{Fan2019}
K.~Fan, B.~Cosenza, and B.~Juurlink, ``Predictable gpus frequency scaling for
  energy and performance,'' in {\em International Conference on Parallel
  Processing (ICPP)}, 2019.

\bibitem{Guerreiro2018}
J.~{Guerreiro}, A.~{Ilic}, N.~{Roma}, and P.~{Tomas}, ``Gpgpu power modeling
  for multi-domain voltage-frequency scaling,'' in {\em International Symposium
  on High Performance Computer Architecture (HPCA)}, 2018.

\bibitem{Lim2015}
J.~Lim, N.~B. Lakshminarayana, H.~Kim, W.~Song, S.~Yalamanchili, and W.~Sung,
  ``Power modeling for gpu architecture using mcpat,'' {\em ACM Transactions on
  Design Automation of Electronic Systems (TODAES)}, vol.~19, pp.~1--24, 2014.

\bibitem{Lucas2013}
J.~Lucas, S.~Lal, M.~Andersch, M.~Alvarez-Mesa, and B.~Juurlink, ``How a single
  chip causes massive power bills,'' in {\em International Symposium on
  Performance Analysis of Systems and Software (ISPASS)}, 2013.

\bibitem{Kerr2012}
A.~Kerr, E.~Anger, G.~Hendry, and S.~Yalamanchili, ``Eiger: A framework for the
  automated synthesis of statistical performance models,'' in {\em
  International Conference on High Performance Computing (HIPC)}, 2012.

\bibitem{kumar2008}
R.~Kumar, D.~Marinov, D.~Padua, M.~Parthasarathy, S.~Patel, D.~Roth, M.~Snir,
  and J.~Torrellas, ``Parallel computing research at illinois the upcrc
  agenda,'' {\em University of Illinois, Champaign}, 2008.

\bibitem{wienke2015}
S.~Wienke, T.~Cramer, M.~S. M{\"u}ller, and M.~Schulz, ``Quantifying
  productivity--towards development effort estimation in hpc,'' {\em
  International Conference for High Performance Computing, Networking, Storage
  and Analysis (SC)}, 2015.

\bibitem{NvidiaCUDAToolkit}
N.~Corporation, ``Cuda samples v11.7.0.''
  \url{https://docs.nvidia.com/cuda/cuda-samples/index.html}, 2022.

\bibitem{Hwu2011}
N.~Bell and J.~Hoberock, ``Thrust: A productivity-oriented library for cuda,''
  in {\em GPU computing gems Jade edition}, pp.~359--371, Elsevier, 2012.

\bibitem{Rubin2014}
E.~Rubin, E.~Levy, A.~Barak, and T.~Ben-Nun, ``Maps: Optimizing massively
  parallel applications using device-level memory abstraction,'' {\em ACM
  Transaction Architecture Code Optimization}, vol.~11, pp.~44:1--44:22, 2014.

\bibitem{You2015}
Y.-P. You, H.-J. Wu, Y.-N. Tsai, and Y.-T. Chao, ``Virtcl: A framework for
  opencl device abstraction and management,'' in {\em ACM SIGPLAN Symposium on
  Principles and Practice of Parallel Programming (PPoPP)}, 2015.

\bibitem{Steuwer2013}
M.~Steuwer, M.~Friese, S.~Albers, and S.~Gorlatch, ``Introducing and
  implementing the allpairs skeleton for programming multi-gpu systems,'' {\em
  International Journal of Parallel Programming}, vol.~42, pp.~601--618, 2013.

\bibitem{Marques2013}
R.~Marques, H.~Paulino, F.~Alexandre, and P.~D. Medeiros, ``Algorithmic
  skeleton framework for the orchestration of gpu computations,'' in {\em
  European Conference on Parallel Processing}, 2013.

\bibitem{Kurzak2013}
J.~Kurzak, P.~Luszczek, M.~Faverge, and J.~Dongarra, ``Lu factorization with
  partial pivoting for a multicore system with accelerators,'' {\em IEEE
  Transactions on Parallel and Distributed Systems (IEEE TPDS)}, vol.~24,
  pp.~1613--1621, 2013.

\bibitem{anderson2012}
M.~J. Anderson, D.~Sheffield, and K.~Keutzer, ``A predictive model for solving
  small linear algebra problems in gpu registers,'' in {\em International
  Parallel \& Distributed Processing Symposium (IPDPS)}, 2012.

\bibitem{Li2013}
D.~Li and M.~Becchi, ``Deploying graph algorithms on gpus: An adaptive
  solution,'' in {\em IEEE International Symposium on Parallel Distributed
  Processing (IPDPS)}, 2013.

\bibitem{Sarkar2012}
S.~Sarkar, S.~Mitra, and A.~Srinivasan, ``Reuse and refactoring of gpu kernels
  to design complex applications,'' in {\em International Symposium on Parallel
  and Distributed Processing and Applications (ISPA)}, 2012.

\bibitem{collobert2002}
R.~Collobert, S.~Bengio, and J.~Mariethoz, ``Torch: a modular machine learning
  software library,'' {\em Technical Report IDIAP-RR 02-46, IDIAP}, 2002.

\bibitem{Betts2015}
A.~Betts, N.~Chong, A.~F. Donaldson, J.~Ketema, S.~Qadeer, P.~Thomson, and
  J.~Wickerson, ``The design and implementation of a verification technique for
  gpu kernels,'' {\em ACM Transactions on Programming Languages and Systems},
  vol.~37, pp.~10:1--10:49, 2015.

\bibitem{Li2010}
G.~Li and G.~Gopalakrishnan, ``Scalable smt-based verification of gpu kernel
  functions,'' in {\em ACM SIGSOFT International Symposium on Foundations of
  Software Engineering (FSE)}, 2010.

\bibitem{Sergey2015}
I.~Sergey, A.~Nanevski, and A.~Banerjee, ``Mechanized verification of
  fine-grained concurrent programs,'' in {\em ACM SIGPLAN Conference on
  Programming Language Design and Implementation (PLDI)}, 2015.

\bibitem{Kothapalli2009}
K.~Kothapalli, R.~Mukherjee, M.~S. Rehman, S.~Patidar, P.~Narayanan, and
  K.~Srinathan, ``A performance prediction model for the cuda gpgpu platform,''
  in {\em International Conference on High Performance Computing (HiPC)}, 2009.

\bibitem{Kirtzic2012}
J.~S. Kirtzic, {\em A Parallel Algorithm Design Model for the Gpu
  Architecture}.
\newblock PhD Thesis, University of Texas at Dallas, 2012.

\bibitem{resios2011}
A.~Resios, {\em GPU Performance Prediction using Parameterized Models}.
\newblock Master’s thesis, Utrecht University, 2011.

\bibitem{luo2011}
L.~Cheng and S.~Reiji, ``An execution time prediction analytical model for gpu
  with instruction-level and thread-level parallelism awareness,'' in {\em
  Summer United Workshops on Parallel, Distributed and Cooperative Processing},
  2011.

\bibitem{Zhang2011}
Y.~{Zhang} and J.~D. {Owens}, ``A quantitative performance analysis model for
  gpu architectures,'' in {\em IEEE International Symposium on High Performance
  Computer Architecture (HPCA)}, 2011.

\bibitem{Baldini2014}
I.~{Baldini}, S.~J. {Fink}, and E.~{Altman}, ``Predicting gpu performance from
  cpu runs using machine learning,'' in {\em IEEE International Symposium on
  Computer Architecture and High Performance Computing ((SBAC-PAD)}, 2014.

\bibitem{ZhangY2011}
Y.~{Zhang}, Y.~{Hu}, B.~{Li}, and L.~{Peng}, ``Performance and power analysis
  of ati gpu: A statistical approach,'' in {\em IEEE International Conference
  on Networking, Architecture, and Storage (NAS)}, 2011.

\bibitem{Ardalani2015}
N.~Ardalani, C.~Lestourgeon, K.~Sankaralingam, and X.~Zhu, ``Cross-architecture
  performance prediction (xapp) using cpu code to predict gpu performance,'' in
  {\em International Symposium on Microarchitecture (MICRO)}, 2015.

\bibitem{chapuis2016gpu}
G.~Chapuis, S.~Eidenbenz, and N.~Santhi, ``Gpu performance prediction through
  parallel discrete event simulation and common sense,'' in {\em International
  Conference on Performance Evaluation Methodologies and Tools}, 2016.

\bibitem{lehnert2016}
C.~Lehnert, R.~Berrendorf, J.~P. Ecker, and F.~Mannuss, ``Performance
  prediction and ranking of spmv kernels on gpu architectures,'' in {\em
  European Conference on Parallel Processing}, 2016.

\bibitem{konstantinidis2015}
E.~Konstantinidis and Y.~Cotronis, ``A practical performance model for compute
  and memory bound gpu kernels,'' in {\em International Conference on Parallel,
  Distributed, and Network-Based Processing (PDP)}, 2015.

\bibitem{remmelg2020}
T.~Remmelg, B.~Hagedorn, L.~Li, M.~Steuwer, S.~Gorlatch, and C.~Dubach,
  ``High-level hardware feature extraction for gpu performance prediction of
  stencils,'' in {\em Annual Workshop on General Purpose Processing Using
  Graphics Processing Unit}, 2020.

\bibitem{Fortune1978}
S.~Fortune and J.~Wyllie, ``Parallelism in random access machines,'' in {\em
  Annual ACM Symposium on Theory of Computing (STOC)}, 1978.

\bibitem{Ryu1990}
K.~W. Ryu, ``Efficient parallel algorithms on the network model,'' tech. rep.,
  Thesis, University of Maryland at College Park, United States, 1990.

\bibitem{Valiant1990}
L.~G. Valiant, ``A bridging model for parallel computation,'' {\em
  Communications of the ACM}, vol.~33, pp.~103--111, 1990.

\bibitem{Culler1993}
D.~Culler, R.~Karp, D.~Patterson, A.~Sahay, K.~E. Schauser, E.~Santos,
  R.~Subramonian, and T.~von Eicken, ``Logp: Towards a realistic model of
  parallel computation,'' in {\em ACM SIGPLAN Symposium on Principles and
  Practice of Parallel Programming (PPOPP)}, 1993.

\bibitem{GIBBONS1998}
P.~Gibbons, Y.~Matias, and V.~Ramachandran, ``The queue-read queue-write
  asynchronous pram model,'' {\em Theoretical Computer Science}, vol.~196,
  pp.~3 -- 29, 1998.

\bibitem{Gibbons1999}
P.~B. Gibbons, Y.~Matias, and V.~Ramachandran, ``The queue-read queue-write
  pram model: Accounting for contention in parallel algorithms,'' {\em Society
  for Industrial and Applied Mathematics (SIAM) Journal on Computing},
  p.~733–769, 1999.

\bibitem{amaris2015}
M.~Amaris, D.~Cordeiro, A.~Goldman, and R.~Y. de~Camargo, ``A simple bsp-based
  model to predict execution time in gpu applications,'' in {\em IEEE
  International Conference on High Performance Computing (HiPC)}, 2015.

\bibitem{Wenhao2012}
W.~Jia, K.~A. Shaw, and M.~Martonosi, ``Stargazer: Automated regression-based
  gpu design space exploration,'' in {\em IEEE International Symposium on
  Performance Analysis of Systems Software}, 2012.

\bibitem{Wu2015}
G.~Wu, J.~L. Greathouse, A.~Lyashevsky, N.~Jayasena, and D.~Chiou, ``Gpgpu
  performance and power estimation using machine learning,'' in {\em IEEE
  International Symposium on High Performance Computer Architecture (HPCA)},
  2015.

\bibitem{Lee2007}
B.~C. Lee, D.~M. Brooks, B.~R. de~Supinski, M.~Schulz, K.~Singh, and S.~A.
  McKee, ``Methods of inference and learning for performance modeling of
  parallel applications,'' in {\em Symposium on Principles and Practice of
  Parallel Programming (PPoPP)}, 2007.

\bibitem{Arafa2019}
Y.~Arafa, A.~Badawy, G.~Chennupati, N.~Santhi, and S.~Eidenbenz, ``Ppt-gpu:
  Scalable gpu performance modeling,'' {\em IEEE Computer Architecture
  Letters}, pp.~55--58, 2019.

\bibitem{Hong2009}
S.~Hong and H.~Kim, ``An analytical model for a gpu architecture with
  memory-level and thread-level parallelism awareness,'' in {\em International
  Symposium on Computer Architecture (ISCA)}, 2009.

\bibitem{Sim2012}
J.~Sim, A.~Dasgupta, H.~Kim, and R.~Vuduc, ``A performance analysis framework
  for identifying potential benefits in gpgpu applications,'' {\em ACM SIGPLAN
  symposium on Principles and Practice of Parallel Programming}, 2012.

\bibitem{Huang2014}
J.~{Huang}, J.~H. {Lee}, H.~{Kim}, and H.~S. {Lee}, ``Gpumech: Gpu performance
  modeling technique based on interval analysis,'' in {\em IEEE/ACM
  International Symposium on Microarchitecture}, 2014.

\bibitem{Diamos2010}
G.~F. Diamos, A.~R. Kerr, S.~Yalamanchili, and N.~Clark, ``Ocelot: A dynamic
  optimization framework for bulk-synchronous applications in heterogeneous
  systems,'' in {\em International Conference on Parallel Architectures and
  Compilation Techniques (PACT)}, 2010.

\bibitem{Baghsorkhi2010}
S.~S. Baghsorkhi, M.~Delahaye, S.~J. Patel, W.~D. Gropp, and W.-m.~W. Hwu, ``An
  adaptive performance modeling tool for gpu architectures,'' in {\em ACM
  SIGPLAN Symposium on Principles and Practice of Parallel Programming
  (PPoPP)}, 2010.

\bibitem{volkov2016}
V.~Volkov, {\em Understanding Latency Hiding on GPUs}.
\newblock PhD thesis, EECS Department, University of California, Berkeley.
  UCB/EECS-2016-143, EECS Department, University of California, Berkeley.
  UCB/EECS-2016-143 2016.

\bibitem{Williams2009}
S.~Williams, A.~Waterman, and D.~Patterson, ``Roofline: An insightful visual
  performance model for multicore architectures,'' {\em Communications of the
  ACM}, vol.~52, p.~65–76, 2009.

\bibitem{Konstantinidis2017AQR}
E.~Konstantinidis and Y.~Cotronis, ``A quantitative roofline model for gpu
  kernel performance estimation using micro-benchmarks and hardware metric
  profiling,'' {\em J. Parallel Distrib. Comput.}, vol.~107, pp.~37--56, 2017.

\bibitem{Li2015}
A.~Li, Y.~Tay, A.~Kumar, and H.~Corporaal, ``Transit: A visual analytical model
  for multithreaded machines,'' in {\em International Symposium on
  High-Performance Parallel and Distributed Computing (HPDC)}, 2015.

\bibitem{Lemeire2016}
J.~{Lemeire}, J.~G. {Cornelis}, and L.~{Segers}, ``Microbenchmarks for gpu
  characteristics: The occupancy roofline and the pipeline model,'' in {\em
  Euromicro International Conference on Parallel, Distributed, and
  Network-Based Processing (PDP)}, 2016.

\bibitem{sarkar2015profile}
S.~Sarkar and S.~Mitra, ``A profile guided approach to optimize branch
  divergence while transforming applications for gpus,'' in {\em Proceedings of
  the 8th India Software Engineering Conference}, pp.~176--185, 2015.

\bibitem{Volkov2008}
V.~Volkov and J.~W. Demmel, ``Benchmarking gpus to tune dense linear algebra,''
  in {\em International Conference for High Performance Computing, Networking,
  Storage and Analysis (SC)}, 2008.

\bibitem{Li2012}
X.~Li and R.~Taylor, ``A micro-benchmark suite for amd gpus,'' in {\em
  International Conference on Parallel Processing Workshops (ICPP)}, 2010.

\bibitem{Mei2017}
X.~Mei and X.~Chu, ``Dissecting gpu memory hierarchy through
  microbenchmarking,'' {\em IEEE Transactions on Parallel and Distributed
  Systems}, vol.~28, p.~72–86, 2017.

\bibitem{Jia2018}
Z.~Jia, M.~Maggioni, B.~Staiger, and D.~P. Scarpazza, ``Dissecting the nvidia
  volta gpu architecture via microbenchmarking,'' in {\em arXiv}, 2018.

\bibitem{Lucas2019}
J.~Lucas and B.~Juurlink, ``Mempower: Data-aware gpu memory power model,'' in
  {\em Architecture of Computing Systems -- ARCS 2019} (M.~Schoeberl,
  C.~Hochberger, S.~Uhrig, J.~Brehm, and T.~Pionteck, eds.), (Cham),
  pp.~195--207, Springer International Publishing, 2019.

\bibitem{Andersch2015}
M.~Andersch, J.~Lucas, M.~A. LvLvarez-Mesa, and B.~Juurlink, ``On latency in
  gpu throughput microarchitectures,'' in {\em IEEE International Symposium on
  Performance Analysis of Systems and Software (ISPASS)}, 2015.

\bibitem{ma2009}
X.~Ma, M.~Dong, L.~Zhong, and Z.~Deng, ``Statistical power consumption analysis
  and modeling for gpu-based computing,'' in {\em ACM SOSP Workshop on Power
  Aware Computing and Systems (HotPower)}, 2009.

\bibitem{Nagasaka2010}
H.~Nagasaka, N.~Maruyama, A.~Nukada, T.~Endo, and S.~Matsuoka, ``Statistical
  power modeling of gpu kernels using performance counters,'' in {\em
  International Green Computing Conference (GREENCOMP)}, 2010.

\bibitem{Jia2015}
W.~Jia, E.~Garza, K.~A. Shaw, and M.~Martonosi, ``Gpu performance and power
  tuning using regression trees,'' {\em ACM Transactions on Architecture and
  Code Optimization (TACO)}, vol.~12, pp.~1--26, 2015.

\bibitem{Weaver2012}
V.~M. Weaver, M.~Johnson, K.~Kasichayanula, J.~Ralph, P.~Luszczek, D.~Terpstra,
  and S.~Moore, ``Measuring energy and power with papi,'' in {\em International
  Conference on Parallel Processing Workshops (ICPPW )}, 2012.

\bibitem{Singh2009}
K.~Singh, M.~Bhadauria, and S.~A. McKee, ``Real time power estimation and
  thread scheduling via performance counters,'' {\em ACM SIGARCH Computer
  Architecture News}, vol.~37, p.~46–55, 2009.

\bibitem{Zhao2013}
Q.~Zhao, H.~Yang, Z.~Luan, and D.~Qian, ``Poigem: A programming-oriented
  instruction level gpu energy model for cuda program,'' in {\em International
  Conference on Algorithms and Architectures for Parallel Processing (ICA3PP)},
  2013.

\bibitem{Abe2012}
Y.~Abe {\em et~al.}, ``Power and performance analysis of gpu-accelerated
  systems,'' in {\em USENIX Conference on Power-Aware Computing and Systems},
  2012.

\bibitem{Sheaffer2004}
J.~W. Sheaffer, D.~Luebke, and K.~Skadron, ``A flexible simulation framework
  for graphics architectures,'' in {\em ACM SIGGRAPH/EUROGRAPHICS conference on
  Graphics hardware}, 2004.

\bibitem{Chen2011}
J.~{Chen}, B.~{Li}, Y.~{Zhang}, L.~{Peng}, and J.~{Peir}, ``Tree structured
  analysis on gpu power study,'' in {\em IEEE International Conference on
  Computer Design (ICCD)}, 2011.

\bibitem{Song2014}
W.~Song, S.~Mukhopadhyay, and S.~Yalamanchili, ``Energy introspector: A
  parallel, composable framework for integrated power-reliability-thermal
  modeling for multicore architectures,'' in {\em IEEE International Symposium
  on Performance Analysis of Systems and Software (ISPASS)}, 2014.

\bibitem{Wang2010}
G.~Wang, Y.~Lin, and W.~Yi, ``{Kernel Fusion : an Effective Method for Better
  Power Efficiency on Multithreaded GPU},'' in {\em IEEE/ACM International
  Conference on Green Computing and Communications}, 2010.

\bibitem{Pool2010}
J.~Pool, A.~Lastra, and M.~Singh, ``An energy model for graphics processing
  units,'' in {\em IEEE International Conference on Computer Design (ICCD)},
  2010.

\bibitem{PTX}
N.~Corporation, ``Parallel thread execution isa version 7.7.''
  \url{https://docs.nvidia.com/cuda/parallel-thread-execution/index.html},
  2022.

\bibitem{hristeamicro}
C.-A.-M. Hristea {\em et~al.}, {\em Micro benchmarks for multiprocessor memory
  hierachy performance}.
\newblock PhD thesis, Massachusetts Institute of Technology, 1997.

\bibitem{papadopoulou2009micro}
M.-M. Papadopoulou, M.~Sadooghi-Alvandi, and H.~Wong, ``Micro-benchmarking the
  gt200 gpu,'' {\em Computer Group, ECE, University of Toronto, Tech. Rep},
  2009.

\bibitem{Saavedra1992}
R.~H. Saavedra-Barrera, {\em CPU Performance Evaluation and Execution Time
  Prediction Using Narrow Spectrum Benchmarking}.
\newblock PhD thesis, EECS Department, University of California, Berkeley,
  1992.

\bibitem{Meltzer2013}
C.~Z. R.~Meltzer and C.~Cecka, ``Micro-benchmarking the c2070,,'' {\em poster
  presented at GPU Technology Conference}, 2013.

\bibitem{arafa2019low}
Y.~Arafa, A.-H.~A. Badawy, G.~Chennupati, N.~Santhi, and S.~Eidenbenz, ``Low
  overhead instruction latency characterization for nvidia gpgpus,'' in {\em
  2019 IEEE High Performance Extreme Computing Conference (HPEC)}, pp.~1--8,
  IEEE, 2019.

\bibitem{cornelis2019pipeline}
J.~G. Cornelis and J.~Lemeire, ``The pipeline performance model: a generic
  executable performance model for gpus,'' in {\em Euromicro International
  Conference on Parallel, Distributed and Network-Based Processing (PDP)},
  pp.~260--265, IEEE, 2019.

\bibitem{Allen1970}
F.~E. Allen, ``Control flow analysis,'' in {\em ACM Symposium on Compiler
  Optimization}, 1970.

\bibitem{Hecht1974}
M.~S. Hecht and J.~D. Ullman, ``Characterizations of reducible flow graphs,''
  {\em Journal of the ACM (JACM)}, vol.~21, p.~367–375, 1974.

\bibitem{Peterson1973}
W.~W. Peterson, T.~Kasami, and N.~Tokura, ``On the capabilities of while,
  repeat, and exit statements,'' {\em Communications of the ACM}, vol.~16,
  pp.~503--512, 1973.

\bibitem{KOSARAJU1974}
S.~R. Kosaraju, ``Analysis of structured programs,'' {\em Journal of Computer
  and System Sciences}, vol.~9, pp.~232 -- 255, 1974.

\bibitem{Wulf1975}
W.~A. Wulf, R.~K. Johnsson, C.~B. Weinstock, S.~O. Hobbs, and C.~M. Geschke,
  ``The design of an optimizing compiler,'' in {\em Carnegie-Mellon University,
  Pittsburgh}, 1975.

\bibitem{Nvidia2017}
N.~Corporation, ``Nvidia cuda occupancy calculator v11.7.0.''
  \url{https://docs.nvidia.com/cuda/cuda-occupancy-calculator/index.html},
  2022.

\bibitem{NvidiaWhitepaper2014}
N.~Corporation, ``Nvidia kepler gk110 gk210 architecture whitepaper version
  1.1.''
  \url{https://www.nvidia.com/content/dam/en-zz/Solutions/Data-Center/documents/NVIDIA-Kepler-GK110-GK210-Architecture-Whitepaper.pdf},
  2014.

\bibitem{Che2009}
S.~Che, M.~Boyer, J.~Meng, D.~Tarjan, J.~W. Sheaffer, S.-H. Lee, and
  K.~Skadron, ``Rodinia: A benchmark suite for heterogeneous computing,'' in
  {\em IEEE International Symposium on Workload Characterization (IISWC)},
  2009.

\bibitem{Feautrier1996}
P.~Feautrier, ``Some efficient solutions to the affine scheduling problem. i.
  one-dimensional time,'' {\em International journal of parallel programming},
  vol.~21, pp.~313--347, 1992.

\bibitem{COOK2013}
S.~Cook, ``Chapter 9 - optimizing your application,'' in {\em CUDA Programming}
  (S.~Cook, ed.), pp.~305 -- 440, Morgan Kaufmann, 2013.

\bibitem{Lam1988}
M.~Lam, ``Software pipelining: An effective scheduling technique for vliw
  machines,'' in {\em ACM SIGPLAN Conference on Programming Language Design and
  Implementation (PLDI)}, 1988.

\bibitem{Kepler2012}
N.~Corporation, ``Nvidia's next generation cuda\textsuperscript{TM} compute
  architecture: Kepler\textsuperscript{TM} gk110 the fastest, most efficient
  hpc architecture ever built,'' {\em NVIDIA}, 2012.

\bibitem{Rau1981}
B.~R. Rau and C.~D. Glaeser, ``Some scheduling techniques and an easily
  schedulable horizontal architecture for high performance scientific
  computing,'' in {\em Annual Workshop on Microprogramming (MICRO 14)}, 1981.

\bibitem{HongTr2009}
S.~Hong and H.~Kim, ``An analytical model for a gpu architecture with
  memory-level and thread-level parallelism awareness,'' in {\em International
  symposium on Computer architecture}, 2009.

\bibitem{Malossi2014}
A.~C.~I. {Malossi}, Y.~{Ineichen}, C.~{Bekas}, A.~{Curioni}, and E.~S.
  {Quintana-Ortí}, ``Performance and energy-aware characterization of the
  sparse matrix-vector multiplication on multithreaded architectures,'' in {\em
  International Conference on Parallel Processing Workshops}, 2014.

\bibitem{transpose}
G.~Ruetsch and P.~Micikevicius, ``Optimizing matrix transpose in cuda.''
  \url{https://www.cs.colostate.edu/cs675/MatrixTranspose.pdf}, 2009.

\bibitem{Breiman1983ClassificationAR}
L.~Breiman, J.~H. Friedman, R.~A. Olshen, and C.~J. Stone, {\em Classification
  and regression trees}.
\newblock Routledge, 2017.

\bibitem{Alavani2018}
G.~{Alavani}, K.~{Varma}, and S.~{Sarkar}, ``Predicting execution time of cuda
  kernel using static analysis,'' in {\em 2018 IEEE Intl Conf on Parallel
  Distributed Processing with Applications (ISPA), IEEE}, pp.~948--955, Dec
  2018.

\bibitem{Alavani2020}
G.~Alavani, J.~Desai, and S.~Sarkar, ``An approach to estimate power
  consumption of a cuda kernel,'' in {\em 2020 IEEE Intl Conf on Parallel
  Distributed Processing with Applications, Big Data Cloud Computing,
  Sustainable Computing Communications, Social Computing Networking
  (ISPA/BDCloud/SocialCom/SustainCom)}, pp.~984--991, 2020.

\bibitem{jatala2017greener}
V.~Jatala, J.~Anantpur, and A.~Karkare, ``Greener: A tool for improving energy
  efficiency of register files,'' {\em arXiv}, 2017.

\bibitem{Alnori2018}
A.~Alnori and K.~Djemame, ``A holistic resource management for graphics
  processing units in cloud computing,'' {\em Electronic Notes in Theoretical
  Computer Science}, vol.~340, pp.~3--22, 2018.

\bibitem{TangoGPU}
A.~{Karki} {\em et~al.}, ``Tango: A deep neural network benchmark suite for
  various accelerators,'' in {\em IEEE International Symposium on Performance
  Analysis of Systems and Software (ISPASS)}, 2019.

\bibitem{Chadha2016}
M.~{Chadha}, A.~{Srivastava}, and S.~{Sarkar}, ``Unified power and energy
  measurement api for hpc co-processors,'' in {\em International Performance
  Computing and Communications Conference (IPCCC)}, 2016.

\bibitem{gupta2019dealing}
S.~Gupta and A.~Gupta, ``Dealing with noise problem in machine learning
  data-sets: A systematic review,'' {\em Procedia Computer Science}, vol.~161,
  pp.~466--474, 2019.

\bibitem{hong2012modeling}
S.~Hong, {\em Modeling performance and power for energy-efficient GPGPU
  computing}.
\newblock PhD thesis, Georgia Institute of Technology, 2012.

\bibitem{zheng2019xgboost}
H.~Zheng and Y.~Wu, ``A xgboost model with weather similarity analysis and
  feature engineering for short-term wind power forecasting,'' {\em Applied
  Sciences}, vol.~9, no.~15, p.~3019, 2019.

\bibitem{abdi2007kendall}
H.~Abdi, ``The kendall rank correlation coefficient,'' {\em Encyclopedia of
  measurement and statistics}, vol.~2, pp.~508--510, 2007.

\bibitem{Lundberg2017}
S.~M. Lundberg and S.-I. Lee, ``A unified approach to interpreting model
  predictions,'' in {\em International Conference on Neural Information
  Processing Systems (NIPS)}, 2017.

\bibitem{hill1994artificial}
T.~Hill, L.~Marquez, M.~O'Connor, and W.~Remus, ``Artificial neural network
  models for forecasting and decision making,'' {\em International journal of
  forecasting}, vol.~10, pp.~5--15, 1994.

\bibitem{parbat2020python}
D.~Parbat and M.~Chakraborty, ``A python based support vector regression model
  for prediction of covid19 cases in india,'' {\em Chaos, Solitons \&
  Fractals}, vol.~138, p.~109942, 2020.

\bibitem{biau2012analysis}
G.~Biau, ``Analysis of a random forests model,'' {\em The Journal of Machine
  Learning Research}, vol.~13, pp.~1063--1095, 2012.

\bibitem{basak2019predicting}
S.~Basak, S.~Kar, S.~Saha, L.~Khaidem, and S.~R. Dey, ``Predicting the
  direction of stock market prices using tree-based classifiers,'' {\em The
  North American Journal of Economics and Finance}, vol.~47, pp.~552--567,
  2019.

\bibitem{khaidem2016predicting}
L.~Khaidem, S.~Saha, and S.~R. Dey, ``Predicting the direction of stock market
  prices using random forest,'' {\em arXiv preprint arXiv:1605.00003}, 2016.

\bibitem{AHMAD201777}
M.~W. Ahmad, M.~Mourshed, and Y.~Rezgui, ``Trees vs neurons: Comparison between
  random forest and ann for high-resolution prediction of building energy
  consumption,'' {\em Energy and Buildings}, vol.~147, pp.~77--89, 2017.

\bibitem{dong2020survey}
X.~Dong, Z.~Yu, W.~Cao, Y.~Shi, and Q.~Ma, ``A survey on ensemble learning,''
  {\em Frontiers of Computer Science}, vol.~14, pp.~241--258, 2020.

\bibitem{zhang2012ensemble}
C.~Zhang and Y.~Ma, {\em Ensemble machine learning: methods and applications}.
\newblock Springer, 2012.

\bibitem{Geurts2006}
P.~Geurts, D.~Ernst, and L.~Wehenkel, ``Extremely randomized trees,'' {\em
  Machine Learning}, vol.~63, p.~3–42, 2006.

\bibitem{Friedman2000}
J.~H. Friedman, ``Greedy function approximation: A gradient boosting machine,''
  {\em Annals of Statistics}, vol.~29, pp.~1189--1232, 2000.

\bibitem{chen2016xgboost}
T.~Chen and C.~Guestrin, ``Xgboost: A scalable tree boosting system,'' in {\em
  International conference on knowledge discovery and data mining (ACM SIGKDD
  )}, 2016.

\bibitem{zhang2020predicting}
X.~Zhang, C.~Yan, C.~Gao, B.~A. Malin, and Y.~Chen, ``Predicting missing values
  in medical data via xgboost regression,'' {\em Journal of Healthcare
  Informatics Research}, vol.~4, pp.~383--394, 2020.

\bibitem{Dorogush2018CatBoostGB}
A.~V. Dorogush, V.~Ershov, and A.~Gulin, ``Catboost: gradient boosting with
  categorical features support,'' {\em ArXiv}, vol.~abs/1810.11363, 2018.

\bibitem{Natekin2013}
A.~Natekin and A.~Knoll, ``Gradient boosting machines, a tutorial,'' {\em
  Frontiers in neurorobotics}, vol.~7, p.~21, 12 2013.

\bibitem{tian2022comprehensive}
Y.~Tian and Y.~Zhang, ``A comprehensive survey on regularization strategies in
  machine learning,'' {\em Information Fusion}, vol.~80, pp.~146--166, 2022.

\bibitem{rodriguez2018beyond}
P.~Rodr{\'\i}guez, M.~A. Bautista, J.~Gonz{\`a}lez, and S.~Escalera, ``Beyond
  one-hot encoding: Lower dimensional target embedding,'' {\em Image and Vision
  Computing}, vol.~75, pp.~21--31, 2018.

\bibitem{Prokhorenkova2018}
L.~Prokhorenkova, G.~Gusev, A.~Vorobev, A.~V. Dorogush, and A.~Gulin,
  ``Catboost: Unbiased boosting with categorical features,'' in {\em
  International Conference on Neural Information Processing Systems (NIPS)},
  p.~6639–6649, 2018.

\bibitem{Abdullahi2020}
A.~Ibrahim, R.~L., M.~Muhammed, R.~Abdulaziz, and G.~A., ``Comparison of the
  catboost classifier with other machine learning methods,'' {\em International
  Journal of Advanced Computer Science and Applications}, vol.~11,
  pp.~738--748, 2020.

\bibitem{samat2021gpu}
A.~Samat, E.~Li, P.~Du, S.~Liu, and J.~Xia, ``Gpu-accelerated catboost-forest
  for hyperspectral image classification via parallelized mrmr ensemble
  subspace feature selection,'' {\em IEEE Journal of Selected Topics in Applied
  Earth Observations and Remote Sensing}, vol.~14, pp.~3200--3214, 2021.

\bibitem{al2019comparison}
E.~Al~Daoud, ``Comparison between xgboost, lightgbm and catboost using a home
  credit dataset,'' {\em International Journal of Computer and Information
  Engineering}, vol.~13, pp.~6--10, 2019.

\bibitem{hancock2020catboost}
J.~T. Hancock and T.~M. Khoshgoftaar, ``Catboost for big data: an
  interdisciplinary review,'' {\em Journal of big data}, vol.~7, pp.~1--45,
  2020.

\bibitem{jahn2020artificial}
M.~Jahn, ``Artificial neural network regression models in a panel setting:
  Predicting economic growth,'' {\em Economic Modelling}, vol.~91,
  pp.~148--154, 2020.

\bibitem{Campello2013}
R.~J. G.~B. Campello, D.~Moulavi, and J.~Sander, ``Density-based clustering
  based on hierarchical density estimates,'' in {\em Advances in Knowledge
  Discovery and Data Mining} (J.~Pei, V.~S. Tseng, L.~Cao, H.~Motoda, and
  G.~Xu, eds.), pp.~160--172, 2013.

\bibitem{Ester1996}
M.~Ester, H.-P. Kriegel, J.~Sander, and X.~Xu, ``A density-based algorithm for
  discovering clusters in large spatial databases with noise,'' in {\em
  International Conference on Knowledge Discovery and Data Mining (KDD)}, 1996.

\bibitem{Kriegel2011}
H.-P. Kriegel, P.~Kröger, J.~Sander, and A.~Zimek, ``Density-based
  clustering,'' {\em Wiley Interdisc. Rew.: Data Mining and Knowledge
  Discovery}, vol.~1, pp.~231--240, 2011.

\bibitem{campello2015hierarchical}
R.~J. Campello, D.~Moulavi, A.~Zimek, and J.~Sander, ``Hierarchical density
  estimates for data clustering, visualization, and outlier detection,'' {\em
  ACM Transactions on Knowledge Discovery from Data (TKDD)}, vol.~10, no.~1,
  pp.~1--51, 2015.

\bibitem{pawanr2021}
S.~Pawanr, G.~K. Garg, and S.~Routroy, ``Development of a transient energy
  prediction model for machine tools,'' {\em Procedia CIRP}, vol.~98,
  pp.~678--683, 2021.

\bibitem{Ribeiro2016}
M.~T. Ribeiro, S.~Singh, and C.~Guestrin, ``“why should i trust you?”:
  Explaining the predictions of any classifier,'' in {\em International
  Conference on Knowledge Discovery and Data Mining (ACM SIGKDD)},
  p.~1135–1144, 2016.

\bibitem{Vishwesh2016}
V.~Jatala, J.~Anantpur, and A.~Karkare, ``Improving gpu performance through
  resource sharing,'' in {\em International Symposium on High-Performance
  Parallel and Distributed Computing (HPDC)}, 2016.

\bibitem{chetlur2014cudnn}
S.~Chetlur, C.~Woolley, P.~Vandermersch, J.~Cohen, J.~Tran, B.~Catanzaro, and
  E.~Shelhamer, ``cudnn: Efficient primitives for deep learning,'' {\em arXiv
  preprint arXiv:1410.0759}, 2014.

\bibitem{pang2020deep}
B.~Pang, E.~Nijkamp, and Y.~N. Wu, ``Deep learning with tensorflow: A review,''
  {\em Journal of Educational and Behavioral Statistics}, vol.~45, no.~2,
  pp.~227--248, 2020.

\bibitem{Volkov2018}
V.~Volkov, ``A microbenchmark to study gpu performance models,'' {\em 23rd ACM
  SIGPLAN Symposium on Principles and Practice of Parallel Programming},
  vol.~53, no.~1, p.~421–422, 2018.

\bibitem{ding2019instructiouctin}
N.~Ding and S.~Williams, ``An instruction roofline model for gpus,'' in {\em
  IEEE/ACM Performance Modeling, Benchmarking and Simulation of High
  Performance Computer Systems (PMBS)}, 2019.

\end{thebibliography}
\bibliographystyle{ieeetr}
\clearemptydoublepage
%
%

\appendix
\chapter{Appendix A} \label{AppA}


\section{Dynamic Analysis-based Model} \label{ch:dynamic_model}
A dynamic model utilises the data collected during the application's runtime to predict the application's execution time and power consumption. Using a dynamic model which utilises runtime data every time you wish to predict defeats the purpose of prediction for energy efficiency. In this study, we develop a dynamic model as a initial study to understand the importance of application features with respect to power consumption and energy. 

In this study, GPU runtime events have been considered as dependent variables along with threads per block (block\_size), register per thread (reg\_t) ,  GPU clock frequency ($\nu_{gpu}$), and  memory clock frequency ($\nu_{mem}$) to create the model. 

\section{GPU Events}
NVIDIA GPU Events are essentially various metrics reported by the NVIDIA hardware to NVIDIA CUPTI library. They capture various run-time details such as the number of memory transactions, number of instructions executed etc. GPU events encompass a variety of metrics that capture details about algorithm and its implementation. Hence, by using events it becomes possible to create application independent models for Power, and Execution Time. These events are listed and described as follows.
\begin{enumerate}
    \item  fb\_subp0\_read\_sectors : Number of DRAM read requests to sub partition 0, increments by 1 for 32 byte access.
    \item  fb\_subp1\_read\_sectors : Number of DRAM read requests to sub partition 1, increments by 1 for 32 byte access
    \item  fb\_subp0\_write\_sectors: Number of DRAM write requests to sub partition 0, increments by 1 for 32 byte access.
    \item  fb\_subp1\_write\_sectors: Number of DRAM write requests to sub partition 1, increments by 1 for 32 byte access.
    \item l2\_subp0\_write\_sector\_misses: Number of write misses in slice 0 of L2 cache. This increments by 1 for each 32-byte access
    \item l2\_subp1\_write\_sector\_misses: Number of write misses in slice 1 of L2 cache. This increments by 1 for each 32-byte access.
    \item l2\_subp2\_write\_sector\_misses:Number of write misses in slice 2 of L2 cache. This increments by 1 for each 32-byte access.
    \item l2\_subp3\_write\_sector\_misses:Number of write misses in slice 3 of L2 cache. This increments by 1 for each 32-byte access.
    \item l2\_subp0\_read\_sector\_misses:Number of read misses in slice 0 of L2 cache. This increments by 1 for each 32-byte access.
    \item l2\_subp1\_read\_sector\_misses:Number of read misses in slice 1 of L2 cache. This increments by 1 for each 32-byte access.
    \item	l2\_subp2\_read\_sector\_misses:Number of read misses in slice 2 of L2 cache. This increments by 1 for each 32-byte access.
    \item	l2\_subp3\_read\_sector\_misses:Number of read misses in slice 3 of L2 cache. This increments by 1 for each 32-byte access.
    \item	l2\_subp0\_write\_l1\_sector\_queries: Number of write requests from L1 to slice 0 of L2 cache. This increments by 1 for each 32-byte access.
    \item	l2\_subp1\_write\_l1\_sector\_queries: Number of write requests from L1 to slice 1 of L2 cache. This increments by 1 for each 32-byte access.
    \item	l2\_subp2\_write\_l1\_sector\_queries: Number of write requests from L1 to slice 2 of L2 cache. This increments by 1 for each 32-byte access.
    \item	l2\_subp3\_write\_l1\_sector\_queries: Number of write requests from L1 to slice 3 of L2 cache. This increments by 1 for each 32-byte access.
    \item	l2\_subp0\_read\_l1\_sector\_queries: Number of read requests from L1 to slice 0 of L2 cache. This increments by 1 for each 32-byte access.
    \item	l2\_subp1\_read\_l1\_sector\_queries: Number of read requests from L1 to slice 1 of L2 cache. This increments by 1 for each 32-byte access.
    \item	l2\_subp2\_read\_l1\_sector\_queries: Number of read requests from L1 to slice 2 of L2 cache. This increments by 1 for each 32-byte access.
    \item	l2\_subp3\_read\_l1\_sector\_queries: Number of read requests from L1 to slice 3 of L2 cache. This increments by 1 for each 32-byte access.
    \item	l2\_subp0\_read\_l1\_hit\_sectors: Number of read requests from L1 that hit in slice 0 of L2 cache. This increments by 1 for each 32-byte access.
    \item l2\_subp1\_read\_l1\_hit\_sectors: Number of read requests from L1 that hit in slice 1 of L2 cache. This increments by 1 for each 32-byte access.
    \item	l2\_subp2\_read\_l1\_hit\_sectors: Number of read requests from L1 that hit in slice 2 of L2 cache. This increments by 1 for each 32-byte access.
    \item	l2\_subp3\_read\_l1\_hit\_sectors: Number of read requests from L1 that hit in slice 3 of L2 cache. This increments by 1 for each 32-byte access.
    \item	l2\_subp0\_total\_read\_sector\_queries: Total read requests to slice 0 of L2 cache. This includes requests from L1, Texture cache, system memory. This increments by 1 for each 32-byte access.
    \item	l2\_subp1\_total\_read\_sector\_queries: Total read requests to slice 1 of L2 cache. This includes requests from L1, Texture cache, system memory. This increments by 1 for each 32-byte access.
    \item	l2\_subp2\_total\_read\_sector\_queries: Total read requests to slice 2 of L2 cache. This includes requests from L1, Texture cache, system memory. This increments by 1 for each 32-byte access.
    \item	l2\_subp3\_total\_read\_sector\_queries: Total read requests to slice 3 of L2 cache. This includes requests from L1, Texture cache, system memory. This increments by 1 for each 32-byte access.
    \item	l2\_subp0\_total\_write\_sector\_queries: Total write requests to slice 0 of L2 cache. This includes requests from L1, Texture cache, system memory. This increments by 1 for each 32-byte access.
    \item	l2\_subp1\_total\_write\_sector\_queries: Total write requests to slice 1 of L2 cache. This includes requests from L1, Texture cache, system memory. This increments by 1 for each 32-byte access.
    \item	l2\_subp2\_total\_write\_sector\_queries: Total write requests to slice 2 of L2 cache. This includes requests from L1, Texture cache, system memory. This increments by 1 for each 32-byte access.
    \item	l2\_subp3\_total\_write\_sector\_queries: Total write requests to slice 3 of L2 cache. This includes requests from L1, Texture cache, system memory. This increments by 1 for each 32-byte access.
    \item	gld\_inst\_32bit: Total number of 32-bit global load instructions that are executed by all the threads across all thread blocks
    \item	gst\_inst\_32bit: Total number of 32-bit global store instructions that are executed by all the threads across all thread blocks.
    \item	thread\_inst\_executed: Number of instructions executed by all threads, does not include replays. For each instruction it increments by the number of threads in the warp that execute the instruction.
    \item	gld\_request: : Number of executed load instructions where the state space is not specified and hence generic addressing is used, increments per warp on a multi- processor. It can include the load operations from global,local and shared state space.
    \item	gst\_request: Number of executed store instructions where the state space is not specified and hence generic addressing is used, increments per warp on a mul- tiprocessor. It can include the store operations to global,local and shared state space.
    \item active\_ cycles: Number of cycles a multiprocessor has at least one active warp.
    \item	l1\_global\_load\_transactions: Number of global load transactions from L1 cache. Increments by 1 per transaction. 
   \item	l1\_global\_store\_transactions: Number of global store transactions from L1 cache. Increments by 1 per transaction. 
   \item	uncached\_global\_load\_transaction: Number of uncached global load transactions. Increments by 1 per transaction. 
  \item	global\_store\_transaction: Number of global store transactions. Increments by 1 per transaction. 
\end{enumerate}

\subsection{Results}
\begin{figure}[ht]
\includegraphics[width=\linewidth]{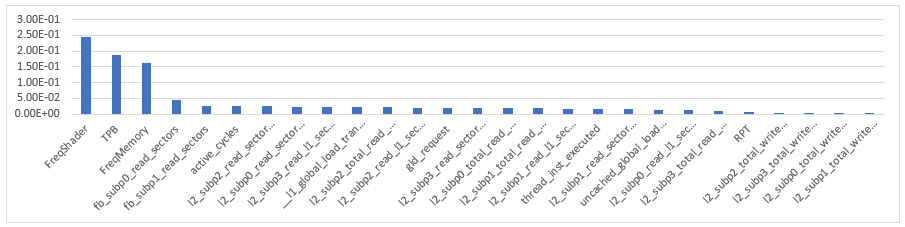}
\caption{Feature Importance for energy prediction}
\label{fig:featImpEnergy}
\end{figure}

\begin{figure}[ht]
\includegraphics[width=\linewidth]{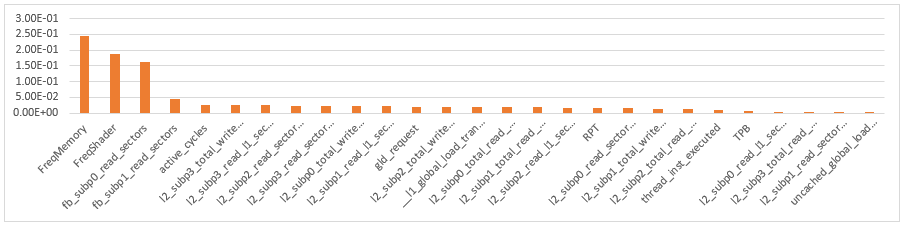}
\caption{Feature Importance for power prediction}
\label{fig:featImpPower}
\end{figure}

The  feature importance using Random Forest for energy prediction is shown in Figure \ref{fig:featImpEnergy} and power prediction in  Figure \ref{fig:featImpPower}.  Runtime memory and shader frequency are the most important features for power prediction. We couldn't utilize these two features in a static analysis-based model since these features are unavailable without executing the code. The feature "active\_cycles" also contributes significantly to power prediction. We have tried to model this feature in the static analysis approach as inst\_issue\_cycles. inst\_issue\_cycles is also one of the most important features in static analysis based model. 

Global memory instructions also show significant impact on power consumption in dynamic analysis which is also seen in static analysis based model. Threads per block (block\_size) is not contributing significantly to power consumption which is also a common observation with static analysis based model. 

We utilised three popular machine learning methods for building the prediction model for execution time, power and energy consumption. Algorithms include Linear Regression , Random Forest and XGBoost whose validation score is presented in Table \ref{tab:dynamicresult}. Random Forest and XGBoost both prove to be efficient in terms of accuracy based on their  $R^{2}$ score for all three metrics. When compared to static analysis based model presented in Chapter \ref{powermodel}, the dynamic analysis model described here is more accurate. This result was expected since the feature utilised are runtime behaviour captured by CUPTI library while executing the application. HOwever, the main goal of this thesis is to predict performance and power consumption of CUDA application without executing it. This dynamic analysis study  gave some insights on relationship of runtime features to power consumption. It was observed that the observations of this dynamic analysis study holds true for the proposed static analysis based model. 

\begin{table}[htbp]
\caption{Validation score for Machine Learning Algorithms}
\label{tab:dynamicresult}
\begin{tabular}{|l|ll|ll|ll|}
\hline
\multirow{2}{*}{} & \multicolumn{2}{l|}{Energy}                  & \multicolumn{2}{l|}{Power}                    & \multicolumn{2}{l|}{Execution Time}         \\ \cline{2-7} 
                  & \multicolumn{1}{l|}{$R^{2}$ score} & RMSE    & \multicolumn{1}{l|}{$R^{2}$ score} & RMSE     & \multicolumn{1}{l|}{$R^{2}$ score} & RMSE   \\ \hline
Linear Regression & \multicolumn{1}{l|}{0.79719}       & 3366204 & \multicolumn{1}{l|}{0.2978}        & 8203.235 & \multicolumn{1}{l|}{0.4327}        & 79.38  \\ \hline
Random Forest     & \multicolumn{1}{l|}{0.9979}        & 568.011 & \multicolumn{1}{l|}{0.9732}        & 1873.3   & \multicolumn{1}{l|}{0.9999}        & 0.0003 \\ \hline
XGBoost           & \multicolumn{1}{l|}{0.9975}        & 1250.51 & \multicolumn{1}{l|}{0.9730}        & 2654.86  & \multicolumn{1}{l|}{0.9993}        & 0.0012 \\ \hline
\end{tabular}
\end{table}

\chapter{Appendix B} 

\label{AppendixB} 

\section{Microbenchmarking} \label{ch:microbenchmark}

\subsection{Kernel Launch Overhead Code Snippet}

\begin{lstlisting}[caption=Kernel Launch Overhead ]{overhead}
__global__ void emptyKernel()
{
   // No instructions are executed
}
// Kernel is launched with different launch configurations 
emptyKernel<<<blocksPerGrid, threadsPerBlock>>>();
\end{lstlisting}

\subsection{Pointer Chasing Benchmark Code Snippet}

\begin{lstlisting}[caption=Pointer Chasing Benchmark ,frame=tlrb]{global_memory}
    int i,x=0;
	for (i = 0; i < innerLoopIterations; i++)
	{
		outerStart = clock();
		startClock = clock();
	//repeat256 instruction makes 256 copies of the instruction.  
		repeat256(x=d_arr[x];);
		stopClock = clock();
		outerStop = clock();
		clockCycleSum += (stopClock - startClock);
		outerClocks += (outerStop - outerStart);
	}
	// saving the results	
	int j= blockIdx.x*blockDim.x*blockDim.y + threadIdx.x + blockDim.x*threadIdx.y ;
	cycleCountDuration[j] = clockCycleSum;
	outerCycles[j] = outerClocks;
	/* *dummyVal variable is introduced so that compiler do not optimise 
	since x is not reused 	*/
	*dummyVal=x;
\end{lstlisting}

\section{Efficacy of Global Memory Throughput Model} \label{sec:efficacyofgmtm}
The global memory throughput model presented in this work is useful in understanding the performance of GPU. 
Global memory instructions are one of the most critical features in contributing to the performance of a GPU. In order to prove the efficacy of the proposed throughput model, we undertake two studies to analyze its efficacy, as discussed further.  

\subsection{Comparison with the throughput model by Volkov et al.}
\begin{figure}
        
         \centering
         \includegraphics[width=\textwidth]{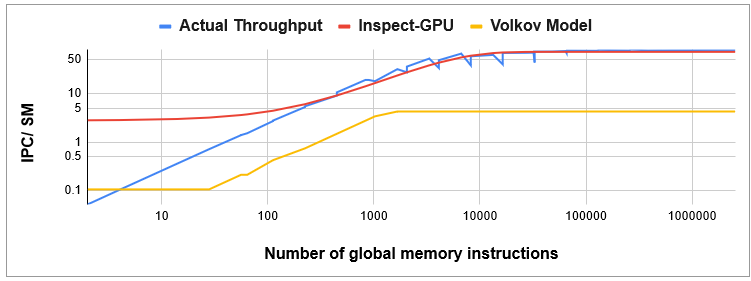}
         \caption{Throughput model vs Volkov model}
         \label{volkovComp}
\end{figure}

We compared our global memory model by comparing it against a popular model developed by Volkov \cite{volkov2016,Volkov2018} which has been used to predict performances against some other existing approaches \cite{Volkov2018}. Volkov \cite{volkov2016}'s unit of throughput representation is IPC/SM, where IPC is instruction per cycle. The following steps were taken to compare our model with the Volkov model:
\begin{enumerate}
    \item We converted the actual throughput and throughput computed using our model to IPC/SM. To do so, we first convert the throughput in GB/s to B/s, then divide this value by the number of bytes per instruction (128 B) \cite{volkov2016} to get the number of instructions per second. To get the number of instructions per cycle, we divide this value by GPU clock frequency. Finally, to get IPC/SM, we divide the number of instructions per Cycle by the number of SMs on the GPU.
    \item Volkov provides us per warp memory throughput using an equation presented in their work \cite{volkov2016}. We convert this throughput value to per instruction by multiplying the number of threads in a warp (32) to get IPC/SM.  We plot the actual throughput, throughput proposed by our model, and the Volkov model in Fig. \ref{volkovComp}.  
\end{enumerate}

\begin{figure}    
      
         \centering
         \includegraphics[width=\textwidth]{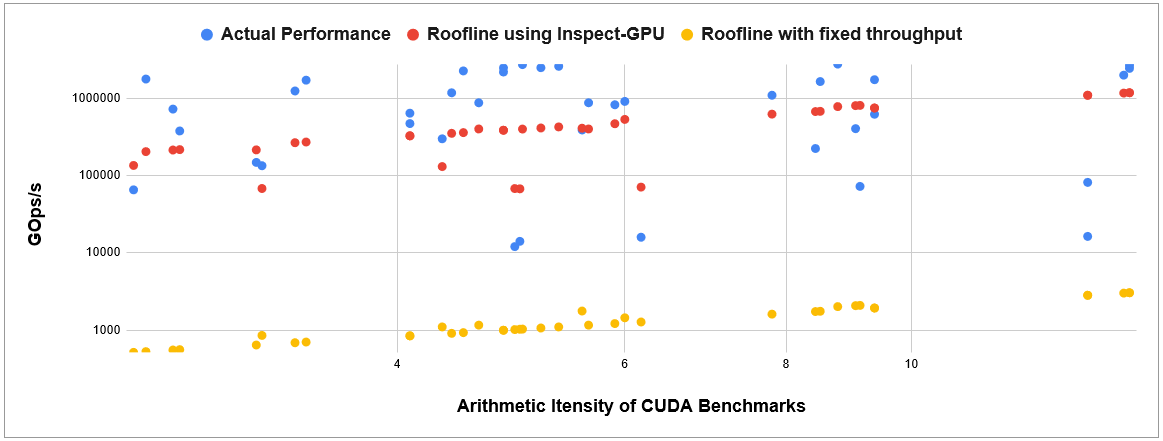}
          \caption{Comparison using Roofline Model}
           \label{roofline}
   
\end{figure}

The Roofline model is used to relate applications performance ($P$) and arithmetic intensity ($I$) to the platform’s peak performance and memory bandwidth \cite{Konstantinidis2017AQR}. The Roofline is a throughput-oriented performance model. Hence we use the Roofline model to observe whether the performance prediction using our memory throughput model is close enough to actual performance. In the Roofline model plot, the x-axis is the arithmetic intensity, and the y-axis is performance. Both the parameters are in log scale. The Roofline formula $P = min(\pi, I\cdot\beta)$ is used to bound $P$ (GOps/s) as a function of machine peak performance ($\pi$), machine peak bandwidth ($\beta$), and arithmetic intensity ($I$) of the application.

We have used 39 CUDA kernels belonging to NVIDIA CUDA Toolkit \footnote{https://docs.nvidia.com/cuda/cuda-samples/index.html} and Rodinia Testbench\cite{Che2009}. For each benchmark, we measure it's actual performance, predicted performance using the Roofline model, and predicted performance using our approach. For each benchmark, we first compute its arithmetic intensity $I$ as the ratio of the number of computing instructions and the number of memory instructions. Next, we perform the following for each benchmark:
\begin{itemize}
    \item We use the throughput (bandwidth) ($\beta$) provided in the vendor specification and compute performance ($P$) using the Roofline formula for this fixed throughput.
    \item we compute $\beta$ using our proposed global memory throughput model, and calculate the predicted performance using the Roofline formula.
    \item  We calculate the benchmark's actual performance by dividing the total number of instructions by its execution time (us) in GOps/s.
\end{itemize}
Each CUDA benchmark is represented by its arithmetic intensity in Figure \ref{roofline}. Actual performance (blue dot) and predicted performance for each benchmark using our throughput model (red dot) and a theoretical value (yellow dot) plotted against each benchmark's arithmetic intensity.

\subsection{Key Takeaways}
Let us consider the popular warp throughput model proposed by Volkov et al. side by side with our approach. Since our model is an exponential one, we notice that the predicted throughput (IPC/SM) of our model is not close to the actual one when the number of memory instructions accessed is small, as shown in Figure~\ref{volkovComp}. However, it is performing well as the number of instructions increases. Irrespective of the number of instructions accessed, the predicted throughput of our model is closer to the actual one than the predicted throughput by Volkov's model. The main reason is that the proposed model is based on actual program execution results. As the number of instructions increases, the number of instructions ready to execute (whose data is fetched) increases, and  hence the throughput increases, which the Volkov model couldn't capture.

Next, let us consider another popular performance prediction approach based on the Roofline model. As shown in Figure~\ref{roofline}, it is quite evident that the prediction of performance using our memory throughput model is closer to actual performance compared to using theoretical throughput provided by Vendor. The Roofline model is a simple performance prediction model and may not be able to capture all the complexities of a multicore architecture like GPU. Many researchers have modified this Roofline model to develop a robust model for performance prediction \cite{Konstantinidis2017AQR,ding2019instructiouctin}. We use this Roofline model to demonstrate that using the proposed throughput, the performance prediction is improved over using a fixed throughput provided in vendor specification as seen in Figure \ref{roofline}. To develop a robust and accurate performance prediction model, one requires more than one metric. Additional metrics such as latency, throughput, concurrency, launch overhead, latency hiding, resource allocation, and so on, at various levels of granularity, starting at the level of individual instruction up to the entire application. To measure all these metrics, the proposed approaches and results in this work will effectively assist the researchers. 
\clearemptydoublepage

\end{document}